%% file: RFtest_final.tex
\documentclass[review]{elsarticle}
\usepackage{lineno}
\usepackage{hyperref}
\modulolinenumbers[5]

\journal{Journal of Computational Statistics and Data Analysis}

\usepackage{rotating}

\usepackage{amsmath}
\usepackage{amssymb}
\usepackage{amssymb,color}
\usepackage{graphicx}
\usepackage{soul}
\usepackage{hyperref}
\usepackage{natbib} 
\usepackage{subcaption}
\usepackage{setspace}
\usepackage{algorithm,algcompatible}
\usepackage{algpseudocode}
\usepackage{tikz}
\usepackage{booktabs}
\usepackage{comment}

\usepackage{epstopdf} 

\usepackage[standard]{ntheorem}

\usepackage{epsfig}

\usepackage{natbib}

\usepackage{tikz}

\algnewcommand\algorithmicto{\textbf{to}}

\DeclareMathOperator*{\argmin}{arg\,min}

\def\func#1{\mathop{\rm #1}}%

\newcommand{\Ind}{\ensuremath{{\mathbb I}}} 

\newcommand{\Var}{\ensuremath{{\mathbb V}}} 
\newcommand{\E}{\ensuremath{{\mathbb E}}} 
\newcommand{\Prob}{\ensuremath{{\mathbb P}}} 

\newcommand{\N}{\ensuremath{{\mathbb N}}}
\newcommand{\R}{\ensuremath{{\mathbb R}}}

\newcommand{\X}{\ensuremath{{\mathcal X}}}

\newcommand{\ceil}[1]{\left\lceil #1 \right\rceil}

\def\func#1{\mathop{\rm #1}}%

\begin{document}

\begin{frontmatter}

\title{On the Use of Random Forest for Two-Sample Testing}

\author{{\large Simon Hediger$^{b}$ \hspace*{2mm}}
{\large Loris Michel$^{a}$ \hspace*{2mm}}
{\large Jeffrey N{\"a}f}$^{a}$ \footnote{\textit{\small Corresponding Author. E-mail address: jeffrey.naef@stat.math.ethz.ch, Address: ETH Zürich, HG G 10.1, Rämistrasse 101, 8092 Zürich.}} \\[3mm]
$^{a}$\textit{\small Seminar for Statistics, ETH Z\"{u}rich, Switzerland} \\
$^{b}$\textit{\small Department of Banking and Finance, University of Zurich, Switzerland}\\
}





\begin{abstract}
Following the line of classification-based two-sample testing, tests based on the Random Forest classifier are proposed. The developed tests are easy to use, require almost no tuning, and are applicable for \emph{any} distribution on $\R^d$. Furthermore, the built-in variable importance measure of the Random Forest gives potential insights into which variables make out the difference in distribution. An asymptotic power analysis for the proposed tests is developed. Finally, two real-world applications illustrate the usefulness of the introduced methodology. To simplify the use of the method, the R-package ``hypoRF'' is provided.
\end{abstract}


\end{frontmatter}
\section{Introduction}

Two-sample testing via classification methods is an old idea tracing back to the work of \cite{Friedman2004OnMG}. Generally speaking, one adapts the output of a classifier to construct a two-sample test. Let $\mathbf{X}_1, \ldots, \mathbf{X}_{n_0}$ and $\mathbf{Y}_1, \ldots, \mathbf{Y}_{n_1}$ be a collection of $\R^d$-valued random vectors, such that $\mathbf{X}_i \stackrel{iid}{\sim} P$ and  $\mathbf{Y}_i \stackrel{iid}{\sim} Q$, where $P$ and $Q$ are some Borel probability measure on $\R^d$. The goal is to test
\begin{align} \label{H0}
H_0: P=Q, \ \ \  H_A:P \neq Q.
\end{align}
Given these iid samples of vectors, we define labels $\ell_i=1$ for each $\mathbf{X}_i$ and $\ell_i=0$ for each $\mathbf{Y}_i$ to obtain the data $\left(\mathbf{Z}_j, \ell_j \right)$, $j=1,\ldots, N$, for $N=n_0+n_1$, and $\mathbf{Z}_j=\mathbf{X}_i$ or $\mathbf{Z}_j=\mathbf{Y}_i$. On this data, we train a classifier $\hat{g}: \R^d \to \{0,1 \}$. If $\hat{g}$ is able to ``accurately'' predict $\ell$ on some test sample, it is taken as evidence against $H_0$. In this work, we assume the data is generated from a mixture distribution
\[
\mathbf{Z}_j \stackrel{iid}{\sim} (1-\pi) P + \pi Q,
\]
such that $n_1 \sim \mbox{Bin}(\pi, N)$, where $\mbox{Bin}$ denotes the Binomial distribution. While our exposition will be valid for general classifiers, we specifically target the use of the Random Forest (RF) classifier in this work. Random Forest is a powerful and flexible method developed by \cite{Breiman2001}, known to have a remarkably stable performance in applications (see e.g. the extensive work of \cite{RFsuperiority}).

This approach to testing was used in scientific applications, especially in the field of neuroscience. We refer to \cite{DPLBpublished} for an excellent literature overview. More recently, a lot of additional work has been produced in this direction in the statistical literature, see e.g., \cite{DPLBpublished}; \cite{Rosen2016};  \cite{facebookguys}; \cite{BORJI2019}; \cite{gagnon-bartsch2019}; \cite{kim2019}; \cite{Cal2020}. The closest relation to our work appears to be the extensive recent work of \cite{DPLBpublished}. Our first out-of-sample test in Section \ref{outofsampletest}, though derived independently, is very closely related to their test in Section 9.1. Moreover, \cite[Proposition 9.1]{DPLBpublished} provide a consistency result for general classifiers under mild assumptions. We add to this discussion, by showing that under imbalance these assumptions nonetheless break down for the Bayes classifier, such that a test based on this classifier is not consistent. \cite{DPLBpublished} also provide a rule of thumb on when to use classification-based tests, as opposed to more fine-tuned statistical tests designed for a specific problem. We extend this discussion by adding a recommendation when to use the RF-based test, as opposed to kernel-based tests, as for instance proposed in \cite{Twosampletest}, \cite{NIPS2012_4727}, \cite{NIPS2015_5685} and \cite{NIPS2016_6148}. These tests are natural competitors to classification-based tests and our work indicates that:
\begin{itemize}
    \item[1.] If the differences between $P$, $Q$ can be found in the marginal distributions, even sparsely so, the RF-based test tends to perform very well. We demonstrate in Section \ref{contamin} that the RF-based test succeeds in an example with marginal differences, that is difficult for kernel-based tests.
    \item[2. ] If the change is mostly found in the dependency structure, or copula, kernel tests like MMD may be preferable. As is demonstrated in \ref{simusec} the RF-based test still has power, but less so than the kernel-based tests.
\end{itemize}
 
In addition, the Random Forest classifier brings two features to the two-sample testing problem: The out-of-bag (OOB) statistics and the variable importance measures. The former is used to increase sample efficiency, compared to a test based on a holdout sample, while the latter provides insights into the source of distributional differences. 

Our work also shares similarities with \cite{Rosen2016}, \cite{gagnon-bartsch2019} and \cite{kim2019}. The work of \cite{gagnon-bartsch2019} focuses on the use of the in-sample classification error as a test statistic in the balanced case. \cite{Rosen2016} focuses attention on the power of different classifier-based test statistics for specific alternatives. They also seem to be the first to propose the use of bootstrap-based classification tests. The work of \cite{kim2019} presents a different approach based on regression and focuses on local testing, i.e. determining where the distributional difference appears.

The next two subsections list our contributions and demonstrate the advantages of our method with a small toy example. Section \ref{framework} introduces the two tests used, the first based on out-of-sample observations and the second on the OOB statistics. It closes with a theoretical insight into the consistency of classifier-based tests. Section \ref{U-stats-Test} extends this theoretical insight into an asymptotic power analysis for a version of the OOB error-based test, using U-statistics theory. Finally, Section \ref{Application} discusses the role of the variable importance measure of the Random Forest and demonstrates the power of our tests with simulated as well as two real-world data sets.

\begin{figure}
\centering
\scalebox{0.4}{
\input{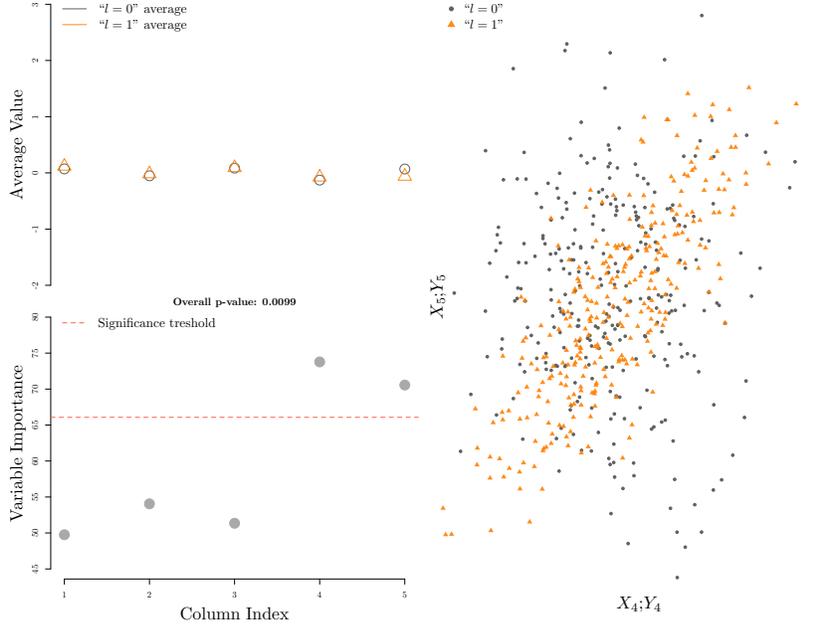}
}
\caption{\textbf{(Intro)} We sampled $300$ observations from a $d=5$ dimensional multivariate normal, with no correlation between the marginals. Likewise $300$ observations were sampled from a multivariate normal, with the last two marginals having a correlation of 0.8. The Random Forest used $500$ trees.}
\label{fig:intro}
\end{figure}

\subsection{Contributions}

Our work differentiates itself from the existing literature in several aspects:

\begin{itemize}
    \item[-] The out-of-sample test based on the class-wise errors in Proposition \ref{Prob1}, though similar to the one in \cite[Proposition 1]{DPLBpublished}, requires less assumptions to conserve the level asymptotically (though \cite{DPLBpublished} focus on a setting, where both the number of observations $N \to \infty$ as well as the dimension $d \to \infty$. In our work, $d$ is assumed to be fixed).
    \item[-]  We show that no test based on the Bayes classifier is consistent for $\pi \neq 1/2$ in Lemma \ref{consistencylemma0}, but that a simple change in the classifier's ``cutoff'' restores consistency.
    \item[-] We utilize the OOB error and variable importance measure in this context to both increase the power of the test and extract more meaning in practice. As shown in simulations, the increase in power with the OOB test is substantial. 
    \item[-] We analyze the asymptotic normality of an OOB error-based test statistic using U-statistics theory and use it to derive an expression for the approximate power of the test in Section \ref{U-stats-Test}.
    \item[-] We provide empirical evidence in Section \ref{contamin}, and in \ref{simusec}, that our test constitutes an important complementary method to powerful kernel-based tests, leading to improved performance in some traditionally difficult examples.
    \item[-] Finally, we provide the \texttt{R}-package hypoRF available on CRAN, with an implementation of the method.
\end{itemize}

\subsection{Motivational example} \label{motexamplesec}

We consider a toy example to demonstrate the proposed methodology underlying the Random Forest classifier two-sample test. We choose $P$ and $Q$ to be five-dimensional multivariate Gaussian probability distributions. The covariance matrix of $P$ is the identity and the distribution $Q$ only differs from $P$ in the last two components between which a positive correlation of $0.8$ is imposed. The OOB statistics-based two-sample test correctly rejects with a $p$-value of $0.0099$ (details are given in Section \ref{finaltest}). Figure \ref{fig:intro} presents a visual summary of the test. The right plot displays the last two components of the sampled points. On the top left, the estimated means, by component and class, indicate that no distributional difference is visible in the margins. The bottom left plot shows the variable importance measure for each component (as presented in Section \ref{varimportance}). We can see that the last two components are picked-up as relevant variables, according to the threshold prescribed by the dotted red line. 

Thus our method correctly rejects in this example and moreover delivers a hint which components might be responsible for the perceived difference in distribution.


\section{Framework} \label{framework}

Let $\mathbf{Z}_1,\ldots, \mathbf{Z}_{N}$ be random vectors with values in $\X \subset \R^d$ and $l_1, \ldots, l_N$ corresponding labels in $\{0,1 \}$, collected in a dataset $D_{N}=\{(\mathbf{Z}_i,l_i)\}^{N}_{i=1}$ with
\[
\mathbf{Z}_i \stackrel{iid}{\sim} (1-\pi) P +  \pi Q.
\]
A sample $\mathbf{Z}_i$ coming from the mixture component $P$ (respectively $Q$) is labeled $l_i=0$ (respectively $l_i=1$). Let $\hat{g}(\mathbf{Z}):=g(\mathbf{Z}, D_{N_{train}})$ be a classifier trained on a subset $D_{N_{train}}$ of size $N_{train} < N $ of the observed data.

Given the setting above, we now present two tests based on the discriminative ability of $\hat{g}$. The first such test uses an independent test set and is very similar to the test proposed by \cite{DPLBpublished}. The second test in Section \ref{finaltest} is entirely new and uses the OOB error to obtain its decision rule.

\subsection{Out-of-sample test}\label{outofsampletest}

Let $N_{test}=N-N_{train}$ be the number of test points. Moreover, $n_{0,j}$ is the number of observations coming from class 0, and $n_{1,j}$ the number of observations from class 1, for $j \in \{train, test \}$. We assume throughout the paper that $n_{0,j} \geq 1, n_{1,j} \geq 1$. If there is no difference in the distribution of the two groups, it clearly holds that
  \[
  \Prob(\ell_i =1 | \mathbf{Z}_i)=\Prob(\ell_i=1)=\pi,
  \]
in other words, $\ell_i$ is independent of $\mathbf{Z}_i$. If $\pi=1/2$, a test can be constructed by considering the overall out-of-sample classification error,
\begin{align*}
   \hat{L}^{(\hat{g})}= \frac{1}{N_{test}} \sum_{i=1}^{N_{test}} \Ind\{ \hat{g}(\mathbf{Z}_i) \neq \ell_i \},
\end{align*}
which under the null hypothesis of equal distributions has $N_{test} \hat{L}^{(\hat{g})} \sim \mbox{Bin}(N_{test}, 1/2)$. Here, $\Ind \{ \hat{g}(\mathbf{Z}_i)  \neq \ell_i \} $ takes the value 1 if $\hat{g}(\mathbf{Z}_i)  \neq \ell_i$ and 0 otherwise. In an effort to extend this principle for general $\pi$, we instead use an approach based on the class-wise errors
\[
\hat{L}_0^{(\hat{g})}=  \frac{1}{n_{0, test}} \sum_{\{i: \ell_i=0 \}}\Ind\{ \hat{g}(\mathbf{Z}_i) \neq 0 \}, \ \  \hat{L}_1^{(\hat{g})}=  \frac{1}{n_{1, test}} \sum_{\{i: \ell_i=1 \}} \Ind\{ \hat{g}(\mathbf{Z}_i) \neq 1 \},
\]
similar to \cite{DPLBpublished}. Define, for $j \in \{0,1\}$, the true class-wise loss for a given classifier $\hat{g}$ as $L_j^{(\hat{g})} = \Prob(\hat{g}(\mathbf{Z}) \neq j| D_{N_{train}}, \ell = j)$. As shown in the proof of Proposition \ref{Prob1}, conditioned on the training data and the number of observations from class $j \in \{0,1 \}$, $n_{j, test} \hat{L}_j^{(\hat{g})}| D_{N_{train}},n_{j, test}  \sim \func{Bin}(n_{j, test}, L_j^{(\hat{g})})$. The loss $L_j^{(\hat{g})}$ depends on the classifier and is generally not known, even under $H_0$. However if $P=Q$, it holds that 
\begin{align*}
    L_0^{(\hat{g})} + L_1^{(\hat{g})} &=  \Prob(\hat{g}(\mathbf{Z})=0| D_{N_{train}}, \ell=1  )  + \Prob(\hat{g}(\mathbf{Z})=1| D_{N_{train}}, \ell=0  )\\
    &=\Prob(\hat{g}(\mathbf{Z})=0| D_{N_{train}}  )  + \Prob(\hat{g}(\mathbf{Z})=1| D_{N_{train}}  )\\
    &=1,
\end{align*}
where we used independence of $\ell$ and $\mathbf{Z}$ when $P=Q$. As a side-note, this shows that $L_0^{(\hat{g})} + L_1^{(\hat{g})}=1$ will be true, as soon as $\ell$ and $\hat{g}(\mathbf{Z})$ are independent. This follows if $P=Q$, but also if $\hat{g}$ negates the dependence between $\ell$ and $\mathbf{Z}$, which essentially means it has no discriminating abilities.

Thus under $H_0$, $L_0^{(\hat{g})}=1-L_1^{(\hat{g})}$. Define for $p \in [0,1]$ the linear combination, $\hat{L}_{p}^{(\hat{g})}:= (1-p) \hat{L}_0^{(\hat{g})} +  p\hat{L}_1^{(\hat{g})}$ and 
\begin{align*}
    \hat{\sigma}_c := 1/2 \sqrt{ \frac{ \hat{L}_{0}^{(\hat{g})} (1-\hat{L}_{0}^{(\hat{g})})}{n_{0,test}} + \frac{\hat{L}_{1}^{(\hat{g})}(1-\hat{L}_{1}^{(\hat{g})})}{n_{1,test}} }.
\end{align*}
Let moreover,
\[
\hat{g}(D_N):=\left( \hat{g}(\mathbf{Z}_1), \ldots, \hat{g}(\mathbf{Z}_{N}) \right).
\]
We are then able to formulate the following decision rule:
\begin{align}\label{Binomialtest}
   \delta_{B}(\hat{g}(D_{N_{test}})) := \Ind \left\{   \hat{L}_{1/2}^{(\hat{g})} - 1/2  < \hat{\sigma}_c \Phi^{-1}(\alpha) + \epsilon_{N_{test}} \right\},
\end{align}
where $\Phi^{-1}(\alpha)$ is the $\alpha$ quantile of the standard normal distribution and $\epsilon_{N_{test}}$ is a decreasing sequence of small non-random numbers. Then 

\begin{proposition} \label{Prob1}
There exists a sequence $\epsilon_{N_{test}}$, such that the decision rule in \eqref{Binomialtest} conserves the level asymptotically, i.e.
\[
\limsup_{N_{test} \to \infty} \Prob \left(\delta_{B}(\hat{g}(D_{N_{test}}))=1 \right) \leq \alpha,
\]
under $H_0:P=Q$.
\end{proposition}

Proposition \ref{Prob1} is related to the first part of Proposition 9.1. in \cite{DPLBpublished}. Note that we did not put any restrictions on how $L_0^{\hat{g}}$, $L_1^{\hat{g}}$ change individually and in particular, we made no assumption on how $N_{train}$ behaves, as $N_{test}$ goes to infinity. The reason for including the sequence $\epsilon_N$ is that, when $N_{train}$ increases with $N_{test}$, boundary cases are possible, in which the variance $L_0^{\hat{g}}(1-L_0^{\hat{g}}) + L_1^{\hat{g}}(1-L_1^{\hat{g}})$ decreases as $1/N_{test}$ or faster, while still being nonzero for finite $N$. In this case the asymptotic normality of $(\hat{L}_{1/2}^{(\hat{g})} - 1/2)/\hat{\sigma}_c$ breaks down and it becomes increasingly difficult to control the behavior of the acceptance probability under the Null. Adding $\epsilon_N$ makes it possible to circumvent this difficulty, albeit at the price of a potential loss in asymptotic power in these boundary cases. If $N_{train}$ grows at the same rate as $N_{test}$, such boundary cases appear unlikely in practice. In fact, for a Random Forest classifier, it rather seems the classifier just outputs the majority class for all $N_{train}$ large enough, such that $\hat{\sigma}_c=0$ and $\hat{L}_{1/2}^{(\hat{g})}=0$, $\hat{L}_{1/2}^{(\hat{g})}=1$ or vice versa. In this case the level is guaranteed, even if $\epsilon_N=0$ for all $N$. We will in the following simply take $\epsilon_{N_{test}}=0$ for the remainder of this paper. The test is summarized in Algorithm \ref{RFTestalg1}.

We briefly highlight the connection between the above decision rule and the one based on the overall classification error $\hat{L}^{(\hat{g})}$, in the case of $\pi=1/2$ and $\epsilon_{N_{test}}=0$. Since, for $\hat{\pi}=n_{1,test}/N_{test}$.
\begin{equation}
    \hat{L}^{(\hat{g})} =  (1-\hat{\pi}) \hat{L}_0^{(\hat{g})} + \hat{\pi}  \hat{L}_1^{(\hat{g})} = \hat{L}^{(\hat{g})}_{\hat{\pi}}, 
\end{equation}
and $\hat{\pi} \to \pi = 1/2$ a.s., it holds that $|\hat{L}^{(\hat{g})} - \hat{L}_{1/2}^{(\hat{g})}| \to 0$, a.s. Consequently, the (unconditional) limiting distribution of $\hat{L}_{1/2}^{(\hat{g})}$ is the same as that of $\hat{L}^{(\hat{g})}$ or,
\[
\frac{ \sqrt{N_{test}}\left( \hat{L}_{1/2}^{(\hat{g})} - 1/2 \right)}{\sqrt{ 1/4 }} \to N(0,1),
\]
under $H_0$. In particular, the asymptotic variance of $\hat{L}_{1/2}^{(\hat{g})}$ under the null is the variance of $\hat{L}^{(\hat{g})}$ and thus one would expect the two tests to behave roughly the same for a large sample size, in the case of $\pi=1/2$. However, as we demonstrate in Section \ref{metrics}, focusing on an equally weighted in-class loss, instead of the overall loss $\hat{L}^{(\hat{g})}$, can be beneficial when $\pi \neq 1/2$.

\begin{algorithm}
\caption{$\text{BinomialTest} \gets \text{function}(Z,\ell,...)$}\label{RFTestalg1}
\begin{algorithmic}[1]
\Require $\mathbf{Z} \in \mathbb{R}^{N \times d}$, $\ell \in \{0,1\}^N$
\State $D_{N_{train}} \gets (\ell_i, \boldsymbol{Z_i})_{i=1}^{N_{train}}$\Comment{random separation of training data}
\State Training of a classifier, $\hat{g}(.)$ on $D_{N_{train}}$
\State $err_0 \gets \frac{1}{n_{0,test}} \sum_{i=N_{train} + 1}^{N}  \Ind{ \{ \ell_i = 0 \} } \Ind{ \{ \hat{g}(\mathbf{Z}_i) \neq 0 \} } $
\State $err_1 \gets \frac{1}{n_{1,test}} \sum_{i=N_{train} + 1}^{N}  \Ind{ \{ \ell_i = 1 \} } \Ind{ \{ \hat{g}(\mathbf{Z}_i) \neq 1 \} } $
\State $err_{1/2} \gets \frac{1}{2} err_0 +  \frac{1}{2} err_1 $\Comment{calculating the out-of-sample classification error}
\State $sig \gets 1/2 \sqrt{err_0 (1-err_0)/n_{0,test} + err_1(1-err_1)/n_{1,test} }$
\If{$sig > 0$}
\State $pvalue \gets \Phi\left( \frac{err_{1/2} - 1/2 }{sig} \right)$
\ElsIf{$sig == 0$}
\State $pvalue \gets \Ind\{ err_{1/2} - 1/2 > 0 \}$
\EndIf
\State \textbf{return} $pvalue$
\end{algorithmic}

\end{algorithm}

Naturally, the split in training and test set is not ideal. For finite sample sizes, one would like to have as many (test) samples as possible to detect differences. At the same time, it would be preferable to have the classifier trained on many data points. This in fact resembles a bias-variance trade-off, similar to what was described in \cite{facebookguys}: Let $g_{1/2}^*$ be the Bayes classifier defined in Section \ref{metrics}. For $\pi=1/2$, there is a trade-off between the closeness of $L^{(\hat{g})}$ to $L^{(g_{1/2}^{*})}$, which may be achieved through a large training set and the closeness of $\hat{L}^{(\hat{g})}$ to $L^{(\hat{g})}$, which is generally only true in large test sets.

\subsection{Out-of-bag test} \label{finaltest}

For the purpose of overcoming the arbitrary split in training and testing, Random Forest delivers an interesting tool: the OOB error introduced in \cite{Breiman2001}. Since each tree is build on a bootstrapped sample taken from $D_{N}$, there will be approximately 1/3 of the trees that are not using the $i$th observation $(\ell_i,\mathbf{Z}_i)$. Thus we may use this ensemble of trees not containing observation $i$ to obtain an estimate of the out-of-sample error for $i$. We slightly generalize this here, in assuming we have an ensemble learner $g$: That is, we assume to have iid copies of a random element $\nu$, $\nu_1, \ldots, \nu_B$, such that each $\hat{g}_{\nu_{b}}(\mathbf{Z}):=g(\mathbf{Z}, D_{N_{train}}, \nu_b)$ is a different classifier. We then consider the average
\begin{align}
    \hat{g}(\mathbf{Z}):=\frac{1}{B} \sum_{b=1}^B  \hat{g}_{\nu_{b}}(\mathbf{Z}).
\end{align}
For $B \to \infty$, this is (a.s.)  $\hat{g}(\mathbf{Z})=\E_{\nu}[\hat{g}_{\nu}(\mathbf{Z})]$. For Random Forest, $\nu$ usually represents the bootstrap sampling of observations and the sampling of variables to consider at each splitpoint for a given tree.

Let as before, $n_0:=\sum_{i=1}^{N} \Ind{\{ \ell_i=0 \} }$ and $n_1:= \sum_{i=1}^{N} \Ind{\{ \ell_i=1 \} }$, with $n_0 \geq 1$, $n_1 \geq 1$. We assume in the following that each $\hat{g}_{\nu_{b}}(\mathbf{Z})$ uses a bootstrapped sample from the original data, as Random Forest does. The class-wise OOB error of such an ensemble of learners trained on $N$ observations is defined as 
\begin{align*}
    \mathcal{E}^{oob}_0 &= \frac{1}{n_0} \sum_{i=1}^{N} \Ind{\{ \ell_i=0 \} } \Ind{ \{\hat{g}_{-i}
(\mathbf{Z}_i) \neq 0 \}},\\
\mathcal{E}^{oob}_1 &= \frac{1}{n_1} \sum_{i=1}^{N} \Ind{ \{ \ell_i=1 \}} \Ind{ \{\hat{g}_{-i}
(\mathbf{Z}_i) \neq 1 \}},\\
\mathcal{E}^{oob}_{p} &= (1-p)   \mathcal{E}^{oob}_0 +   p \mathcal{E}^{oob}_1,
\end{align*}
where $\hat{g}_{-i}$, represents the ensemble of learners not containing the $i^{\text{th}}$ observation for training.

Unfortunately, the test statistic $\mathcal{E}^{oob}_{1/2}$ is difficult to handle; due to the complex dependency structure between the elements of the sum, it is not clear what the (asymptotic) distribution under the null is. For theoretical purposes, we consider in Section \ref{U-stats-Test} a solution based on the concept of U-statistics. Here, we recommend using the OOB error together with a permutation test. See e.g., \cite{permutationtests} or \cite{DPLBpublished}, who use it in conjunction with the out-of-sample error evaluated on a test set: We first calculate the class-wise OOB errors $\mathcal{E}^{oob}_0$, $\mathcal{E}^{oob}_1$ and then reshuffle the labels $K$ times to obtain $K$ permutations, $\sigma_{1}, \ldots, \sigma_{K}$ say. For each of these new datasets $\left(\mathbf{Z}_i , \ell_{\sigma_{k}(i)} \right)_{i=1}^{N}$, $k \in \{1 , \ldots, K \}$, we calculate the OOB errors
\[
\mathcal{E}^{oob, k}_j:=\frac{1}{n_j} \sum_{i=1}^{N} \Ind{\{ \ell_{\sigma_{k}(i)}=j \} } \Ind{ \{\hat{g}_{-i}
(\mathbf{Z}_i) \neq \ell_{\sigma_{k}(i)} \}},
\]
for $j \in \{0,1 \}$. Under $H_0$, $(\ell_{1}, \ldots, \ell_{N})$ and $(\mathbf{Z}_{1}, \ldots, \mathbf{Z}_{N})$ are independent and each $\mathcal{E}^{oob}_{1/2}$ is simply an iid draw from the distribution $F$ of the random variable $\mathcal{E}^{oob}_{1/2}|(\mathbf{Z}_{1}, \ldots, \mathbf{Z}_{N})$. As such we can accurately approximate the $\alpha$ quantile $F^{-1}(\alpha)$ of said distribution by performing a large number of permutations and use the decision rule
\begin{equation}\label{oobdecision}
\delta_{oob}(D_{N})=\left \{ \mathcal{E}^{oob}_{1/2} \leq  F^{-1}(\alpha)\right \}.
\end{equation}
Thus, as in the decision in Equation \eqref{Binomialtest}, the rejection region depends on the data at hand. Nonetheless, the level will be conserved, as proven e.g. in \cite[Theorem 1]{Hemerik2018}.

Heuristically, this procedure will have power under the alternative, as in this case there is some dependence between $(\ell_{1}, \ldots, \ell_{N})$ and $(\mathbf{Z}_{1}, \ldots, \mathbf{Z}_{N})$, formed by the difference in the distribution of the $\mathbf{Z}_i$. The OOB error $\mathcal{E}^{oob}_{1/2}$ will thus be different than the ones observed under permutations.

The whole procedure is described in Algorithm \ref{permutationalg}. We name this test ``hypoRF''.

\begin{algorithm}
\setstretch{1.15}
\caption{$\text{hypoRF} \gets \text{function}(\mathbf{Z},K,...)$}\label{permutationalg}
\begin{algorithmic}[1]
\Require $\mathbf{Z} \in \mathbb{R}^{N \times d}$, $\ell \in \{0,1\}^N, K$
\State $D_{N} \gets (\ell_i, \mathbf{Z}_i)_{i=1}^{N}$
\State $n_{j} \gets  \sum_{i=1}^{N}  \Ind \{ \ell_{\sigma_{k}(i) } =j \}$
\State Training of an ensemble learner $\hat{g}(.)$ on $D_{N}$
\State $OOB_j \gets \frac{1}{n_j} \sum_{i=1}^{N} \Ind{ \{\hat{g}_{-i}(\mathbf{Z}_i)\neq j \}} \Ind \{ \ell_i=j \} $\Comment{calculating the OOB-error for $j \in \{0,1\}$}
\State $OOB_{1/2} \gets 1/2( OOB_0 +  OOB_1 )$
\vspace{5mm}
\For{\texttt{k in 1:K}}
\State $D_{N}^k \gets \left(\ell_{\sigma_{k}(i)}, \mathbf{Z}_i \right)_{i=1}^{N}$\Comment{reshuffle the label}
\State $OOB^k_j \gets \frac{1}{n_{j}} \sum_{i=1}^{N} \Ind{ \{\hat{g}_{-i}(\mathbf{Z}_i) \neq j \}} \Ind \{ \ell_{\sigma_{k}(i) } =j \}$
\State $OOB^k_{1/2} \gets  1/2( OOB_0^k +  OOB_1^k )$\Comment{calculating the OOB-error} 
\EndFor
\vspace{5mm}
\State $mean \gets \frac{1}{K}\sum_{k=1}^K OOB^k_{1/2}$
\State $sig \gets \sqrt{\frac{1}{K-1}\sum_{k=1}^K (OOB^k_{1/2} - mean)^2}$
\If{$sig > 0$} 
\State $pvalue \gets \frac{1}{K+1}\sum_{k=1}^K \left(\Ind\{OOB^k_{1/2}<OOB_{1/2}\}+1 \right)$
\ElsIf{$sig == 0$}
\State $pvalue \gets \Ind\{ OOB_{1/2} - mean > 0 \}$
\EndIf
\State \textbf{return} $pvalue$
\end{algorithmic}
\end{algorithm}

\subsection{What classifier to use} \label{metrics}


The foregoing tests are valid for any classifier $g: \X \to \{0,1 \}$. In practice, most classifiers try to approximate the \emph{Bayes} classifier: Let for $p$, $q$ the densities of $P$, $Q$ 
\begin{equation}
    \eta(\mathbf{z}) := \E[\ell | \mathbf{z} ] =  \frac{\pi q(\mathbf{z}) }{ \pi q(\mathbf{z})  + (1-\pi) p(\mathbf{z})   },
\end{equation}
then the Bayes classifier is given as $g_{1/2}^{*}(\mathbf{Z})=\Ind{\{ \eta(\mathbf{Z}) > 1/2 \} }$, see e.g., \cite{ptpr}. It is the classifier with minimal classification error, designated the Bayes error $L_{\pi}^{(g_{1/2}^{*})}= \Prob(g_{1/2}^{*}(\mathbf{Z}) \neq \ell )$. Under $H_0$, this Bayes error will be $\min(\pi,1-\pi)$.

An interesting question is whether $g_{1/2}^{*}$ leads to a consistent test in our framework. We first define consistency for a \emph{hypothesis test}: Let $\Theta$ be the space of tuples of all distributions on $\R^d$, $\theta=(P,Q) \in \Theta$, $\Theta_0=\{(P,Q): P=Q \}$, $\Theta_1=\{(P,Q): P \neq Q \}$. Let $\delta: \X^N \to \{0,1 \}$ be a decision rule and $\phi(\theta):= \E_{\theta}[\delta]$. Following e.g., \cite{vaart_1998} we call a test consistent at level $\alpha$ (for $\Theta_1$), if $\limsup_{N} \sup_{\theta \in \Theta_0} \phi(\theta) \leq \alpha$ and for any $\theta \in \Theta_1$, $\liminf_{N} \phi(\theta)=1$. For theoretical purposes, we extend this definition also to $\delta$ that depend on the unknown $\theta$ itself, for instance via the densities of $P$ and $Q$ respectively. 

Under the assumption of equal class probabilities $\pi=1/2$ the Bayes error has the property that,
\begin{equation}
L^{(g_{1/2}^{*})}=1/2 (1-TV(P, Q)),
\label{eq:TVanderror}
\end{equation}
where $TV(P, Q)$ is the total variation distance between $P$, $Q$: $TV(P, Q)= 2\sup_{A} |P(A) - Q(A)|$, with the supremum taken over all Borel sets on $\R^d$. As $TV$ defines a metric on the space of all probability measures on $\R^d$, it holds that $P=Q \iff TV(P, Q)=0$.
%
Consequently, as soon as there is any difference in $P$ and $Q$, $TV(P, Q) > 0$ and $L^{(g_{1/2}^{*})} < 1/2$. Thus we would expect a test based on $g_{1/2}^{*}$ to be consistent. More generally, \cite{DPLBpublished} prove that if the classifier $\hat{g}$ is such that
\begin{align}\label{Ramdasassumption}
    \hat{L}^{(\hat{g})}_0 =  L_{0} + o_{\Prob}(1), \   \hat{L}^{(\hat{g})}_1 =  L_{1} + o_{\Prob}(1), \text{ for some $L_0,L_1 \in (0,1)$ with $L_0 + L_1 = 1-\varepsilon$, for any $\varepsilon > 0$},
\end{align}
then the decision rule in \eqref{Binomialtest} is consistent.

Unfortunately, this assumption doesn't hold for $g_{1/2}^*$, if $\pi \neq 1/2.$ In this case, simple counterexamples show that even when $P,Q$ are different, it might still be that $L^{(g_{1/2}^{*})}_0 +  L^{(g_{1/2}^{*})}_1=1$.

\begin{lemma} \label{consistencylemma0}
Take $\X \subset \R$ and $\pi \neq 1/2$. Then no decision rule of the form, $\delta(D_N)=\delta( g^*_{1/2}(D_N) ) $ 
is consistent.
\end{lemma}

Thus even though we allow the classifier $g^*_{1/2}$ to depend for each $(P,Q) \in \Theta_1$ on the densities $p$ of $P$ and $q$ of $Q$, we are not able to
construct a consistent test. The problem appears to be that the Bayes classifier minimizes the \emph{overall} classification loss, so that condition \eqref{Ramdasassumption} cannot hold. In doing so, it focuses too much on the overrepresented class.
Indeed, we might define the following alternative classifier: For given $P$, $Q$ let $g_{\pi}^*$ be the classifier that minimizes the error $L_{1/2}^{g}$, i.e. a classifier that solves the problem
\begin{equation}\label{newproblem}
\argmin \{ L_{1/2}^{g}:  g: \X \to \{0,1 \}  \text{ a classifier} \}.    
\end{equation}

It turns out a slight variation to the Bayes classifier solves this problem:

\begin{lemma} \label{consistencylemma2}
The classifier
\begin{equation}\label{gpi}
    g_{\pi}^*(\mathbf{z}) = \Ind \left\{ \eta(\mathbf{z}) > \pi  \right\},
\end{equation}
is a solution to \eqref{newproblem}. Moreover it holds that 
\begin{equation}\label{eq:TVanderror2}
    1 - TV(P,Q)    = L_0^{g_{\pi}^*} + L_1^{g_{\pi}^*},
\end{equation}
for any $\pi \in (0,1)$.
\end{lemma}

Thus for this classifier a generalization of \eqref{eq:TVanderror} holds for any $\pi \in (0,1)$. In particular, it now yields a consistent test:

\begin{corollary}\label{consistencycor}
The decision rule $\delta_B(g_{\pi}^*(D_N))$ in \eqref{Binomialtest} is consistent for any $\pi \in (0,1)$.
\end{corollary}

Since this theoretical classifier needs no training, the two testing approaches coincide with an evaluation of the classifier loss on the overall data $D_N$. While this analysis with theoretical classifiers is by no means sufficient for the much more complicated case of a classifier $\hat{g}$ trained on data, it suggests that adapting the ``cutoff'' in a given classifier might improve consistency issues. Indeed, we use the classifier
\[
\hat{g}(\mathbf{z})=\Ind\{\hat{\eta}(\mathbf{z}) > \hat{\pi}  \},
\]
where $\hat{\pi}$ is an estimate of the prior probability based on the \emph{training} data. As long as the later is used (as opposed to the test data), the tests above are still valid. 

\section{Tests based on U-Statistics} \label{U-stats-Test}

To avoid the splitting in training and test set, we introduced an OOB error-based test in Section \ref{finaltest}. In this section, we discuss a potential framework to analyse a version of such a test theoretically. For $N_{train} \leq N$, let again, $n_{0, train}= \sum_{i=1}^{N_{train}} \Ind\{\ell_i=0\}$ and $n_{1, train}= \sum_{i=1}^{N_{train}} \Ind\{\ell_i=1\}$. Let $D_{N_{train}}^{-i}$ denote the data set without observation $(\ell_i, \mathbf{Z}_i)$. Then we consider the class-wise OOB error based on $N_{train}$ observations:
\begin{align}\label{overalloob}
    h_{N_{train}}((\ell_1, \mathbf{Z}_{1}), \ldots, (\ell_{N_{train}}, \mathbf{Z}_{N_{train}}))&:= \frac{1}{2} \left( \frac{1}{n_{0, train}} \sum_{i: \ell_i=0} \Ind{ \{\hat{g}_{-i} (\mathbf{Z}_i) = 1 \}} + \frac{1}{n_{1, train}} \sum_{i: \ell_i=1} \Ind{ \{\hat{g}_{-i} (\mathbf{Z}_i) = 0 \}} \right) \nonumber \\
    &=\frac{1}{2} \sum_{i=1}^{N_{train}} \varepsilon_{i}^{oob},
\end{align}
where 
\[
\varepsilon_{i}^{oob}:=\Ind{ \{\hat{g}_{-i} (\mathbf{Z}_i) \neq \ell_i \}} \left( \frac{1-\ell_{i}}{n_{0, train}} + \frac{\ell_{i}}{n_{1, train}}  \right),
\]
for $\hat{g}_{-i}$ trained on $D_{N_{train}}^{-i}$. Also recall that $L_{j}^{\hat{g}}=\Prob(\hat{g}(\mathbf{Z}) = j| D_{N_{train}}, \ell \neq j)$ for $j \in \{0,1\}$ and $L_{1/2}^{(\hat{g})}=1/2(L_{0}^{(\hat{g})} + L_{1}^{(\hat{g})} )$.
We assume that the number of classifiers in the ensemble, $B \to \infty$, so that $\hat{g}(\mathbf{Z}) \to \E_{\nu}[\hat{g}_{\nu}(\mathbf{Z})]$, almost surely. We refer to the function $h_{N_{train}}$ as kernel of size $N_{train}$ and define the incomplete U-Statistics,
\begin{equation} \label{estimatedU}
    \hat{U}_{N,K}:=\frac{1}{K} \sum  h_{N_{train}}((\mathbf{Z}_{i_1},\ell_{i_1}),\ldots, (\mathbf{Z}_{i_{N_{train}}},\ell_{i_{N_{train}}})),
\end{equation}
where the sum is taken over $K$ randomly chosen subsets of size $N_{train}$ - see e.g., \cite{Ustat90}, \cite{epub17654},  \cite{RFuncertainty}, \cite{peng2019asymptotic}. We assume that $K$ goes to infinity as $N$ goes to infinity. Since we are only considering learners for which the $i$th sample point is not included, we may simply see $\hat{g}_{-i}$ as an infinite ensemble build on the dataset $D_{N_{train}}^{-i}$ only. Consequently, with the assumption of an infinite number of learners, the OOB error is ``almost'' unbiased for $\E[L_{1/2}^{(\hat{g})}]$.

\begin{lemma}\label{expectationlemma}
$\E[h_{N_{train}}((\ell_1, \mathbf{Z}_{N_{train}}), \ldots, (\ell_{N_{train}}, \mathbf{Z}_{N_{train}}))]=\E[L_{1/2}^{(\hat{g}_{-i})}]$.
\end{lemma}

Here, $\E[L_{1/2}^{(\hat{g}_{-i})}]$ refers to the expected value of the error based on the classifier trained on $N_{train}-1$ data points. As such, it does not depend on $i$. This is essentially the same result as in \cite{Luntz} in the case of the leave-one-out error. 

We are now able to show that $h_{N_{train}}$ in \eqref{overalloob} is a \emph{symmetric} function, unbiased for $\E[L_{1/2}^{(\hat{g}_{-i})}]$: 

\begin{lemma}\label{hntrainlemma}
$h_{N_{train}}$ is a valid kernel for the expectation $\E[L_{1/2}^{(\hat{g}_{-i})}]$.
\end{lemma}

Combining arguments from \cite{RFuncertainty} and \cite{wager2017estimation}, we obtain the conditions for asymptotic normality listed in Theorem \ref{asymptoticnormalxx}. Though both paper consider the asymptotic distribution of a Random Forest prediction at a fixed $\mathbf{z}$, the $U$-Statistics theory they develop can be used in our context as well. We also refer to \cite{peng2019asymptotic} and \cite{RomanoUstats}, who already refined the results of \cite{RFuncertainty} for asymptotic normality of a $U$-statistics with growing kernel size. \cite{peng2019asymptotic} in particular, derived a similar result to Theorem \ref{asymptoticnormalxx} independently from us. Let for random variables $\xi_1, \xi_2$, $\Var(\xi_1)$, $\mbox{Cov}(\xi_1,\xi_2)$ be the variance and covariance respectively and define for the following, for $c \in \{1,\ldots,N_{train} \}$,
\begin{align}
    \zeta_{c,N_{train}}&=\Var(\E[h_{N_{train}}((\mathbf{Z}_{1},\ell_1),\ldots, (\mathbf{Z}_{{N_{train}}},\ell_{{N_{train}}}))|(\mathbf{Z}_{1},\ell_1),\ldots, (\mathbf{Z}_{c},\ell_c)]).
\end{align}

In particular, $ \zeta_{1,N_{train}}$ and $ \zeta_{N_{train},N_{train}}$ will be of special interest. \cite{Ustat90} provides an immediate important result:

\begin{lemma} \label{zeta1result}
$N_{train} \zeta_{1,N_{train}} \leq \zeta_{N_{train},N_{train}}$
\end{lemma}

Lemma \ref{zeta1result}, which is actually true for any $U$-statistics, shows that, whenever the second moment of the kernel $h_{N_{train}}$ exists, $\zeta_{1,N_{train}}=O(N_{train}^{-1})$. Then


\begin{theorem}\label{asymptoticnormalxx}
Assume that for $N \to \infty$, $N_{train}=N_{train}(N) \to \infty$ and $K=K(N) \to \infty$,
\begin{align}
\lim_{N} \frac{K N_{train}^2}{N} \frac{\zeta_{1,N_{train}}}{{\zeta_{N_{train},N_{train}}}} &= 0, \label{zetacond}\\
\lim_{N} \frac{\sqrt{K} N_{train}}{N}&= 0. \label{Kocond}
\end{align}
Then,
\begin{align} \label{normality2}
\frac{\sqrt{K}(\hat{U}_{N,K} - \E[L_{1/2}^{(\hat{g}_{-1})}] ) }{\sqrt{    \zeta_{N_{train},N_{train}}}} \stackrel{D}{\to} N(0,1).
\end{align}
\end{theorem}

Condition \eqref{zetacond} is hard to control in general, but with Lemma \ref{zeta1result}, it can be seen that choosing
\begin{align}\label{Kocondd}
    \frac{K N_{train}}{N} \to 0,
\end{align}
is sufficient for both \eqref{zetacond} and \eqref{Kocond}. If $K=\log(N_{train})^{1+d}$, this corresponds to the condition $\log(N_{train})^{1+d} N_{train}/N \to 0$ required by \cite{wager2017estimation}. In the context of Random Forest, Theorem \ref{asymptoticnormalxx} essentially proves that the OOB error of a prediction function that is bounded, is asymptotically normal if the number of trees is ``high'' and if $K$ forests are trained on subsamples such that \eqref{zetacond} and \eqref{Kocond} are true. Since the OOB error with infinite learners is essentially the leave-one-out error in the context of cross-validation, this also means that a test of the cross-validation error could be derived under much weaker assumption as for instance in \cite{epub17654}. The key reason for the generality of the result, as was also realized by \cite{peng2019asymptotic}, is that $K$ should be chosen small relative to $N$. This introduces additional variance, such that conditions on $\zeta_{1,N_{train}}$ usually required in such results, see e.g., \cite{RomanoUstats}, can be replaced by \eqref{Kocondd}. This has an additional computational advantage, but it may come at the price of reduced power, as will be seen in Corollary \ref{asymptoticpower2}.

\begin{figure}
    \centering
      \scalebox{0.8}{
  \input{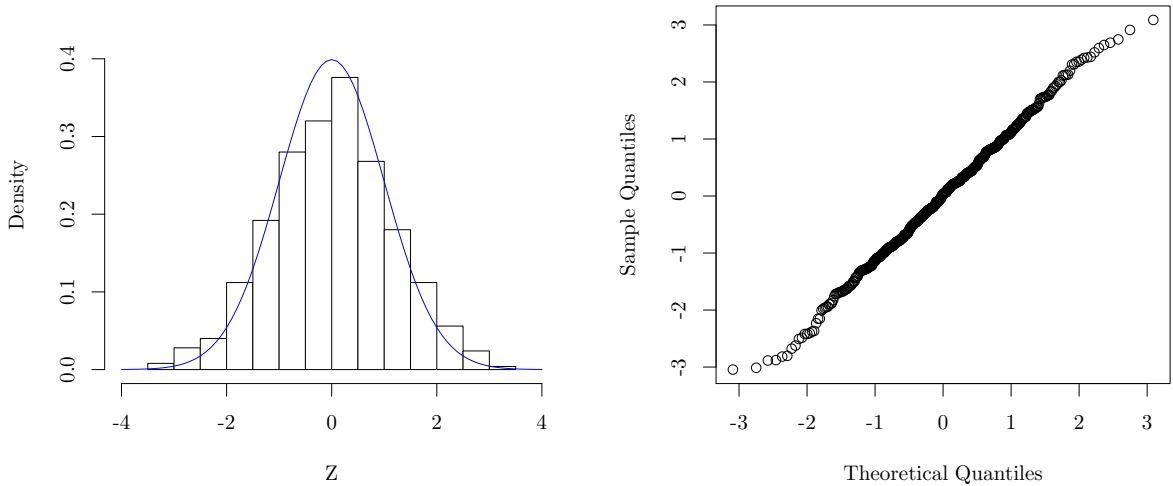}
  }
    \caption{Illustration of the asymptotic normality of the OOB error based test-statistic for the Random Forest classifier. In this example, $P=N(\mathbf{0}, I_{10 \times 10})$ and $Q=N(0.4/\sqrt{10} \cdot \mathbf{1}, I_{10 \times 10})$, an $N=6000$, $K=2\ceil{\log(N)}=17$ were chosen over 500 replications. }
    \label{asymptoticnormalityillustrationHA}
\end{figure}

\cite[Section 3]{RFuncertainty} also provide a consistent estimate for $\zeta_{c,N_{train}}$, denoted $\hat{\zeta}_{c,N_{train}}$, for any $c \in \{ 1,\ldots, N_{train}\}$. As its population counterpart, this estimator is also bounded by 1 for all $c$ and $N_{train}$ in our case. Thus if for a classifier \eqref{zetacond} and \eqref{Kocond} are true, the decision rule
\begin{equation}\label{test3}
 \delta(\hat{g}(D_{N}))=\Ind \left \{\frac{\sqrt{K} (\hat{U}_{N,K} - 1/2)}{\sqrt{\hat{\zeta}_{N_{train},N_{train}}}} < \Phi^{-1}(\alpha) \right\},
\end{equation}
constitutes a valid test. To illustrate Theorem \ref{asymptoticnormalxx}, Figure \ref{asymptoticnormalityillustrationHA} displays the simulated distribution of 
\begin{align}\label{teststatisticHA}
Z=\frac{\sqrt{K} (\hat{U}_{N,K} - \E[L_{1/2}^{(\hat{g}_{-i})}] )}{\sqrt{\hat{\zeta}_{N_{train},N_{train}}}},    
\end{align}
for $P=N(\boldsymbol{\mu}_1, I_{10 \times 10})$ and $Q=N(\boldsymbol{\mu}_2, I_{10 \times 10})$, with $\boldsymbol{\mu}_1=\mathbf{0}$ and $\boldsymbol{\mu}_2=0.4/\sqrt{10} \cdot \mathbf{1}$. We simulated $S=500$ replications using $N=6000$, $K=\ceil{2*\log(N)}=17$ and $N_{train}=\ceil{N/(K*\log(\log(N)))}=163$.

With this at hand, we can construct another test:

\begin{corollary} \label{asymptoticpower2}
Assume the conditions of Theorem \ref{asymptoticnormalxx} hold true and that $\hat{\zeta}_{N_{train},N_{train}}/\zeta_{N_{train},N_{train}} \stackrel{p}{\to} 1$. Then the decision rule in \eqref{test3} conserves the level asymptotically and has approximate power
\begin{align}\label{powerexpression}
    \Phi \left( \Phi^{-1}(\alpha) + \sqrt{\frac{K}{\zeta_{N_{train},N_{train}}}} (1/2 - \E[L_{1/2}^{(\hat{g}_{-i})}]) \right).
\end{align}
\end{corollary}

The test has thus power going to one, as soon as
\begin{align} \label{whatwewant2}
    \limsup_{N} \E[L_{1/2}^{(\hat{g}_{-i})}]  < 1/2.
\end{align}
Condition \eqref{whatwewant2} mirrors condition (A9) in \cite{DPLBpublished}, in that it asks for a better than chance prediction in expectation. Crucially, Corollary \ref{asymptoticpower2} also illustrates the downside of the weak assumptions used in Theorem \ref{asymptoticnormalxx}: The power is dependent on $\sqrt{K}$, as well as the accuracy of the trained classifier through $\E[L^{(\hat{g}_{-i})}]$. Since our theory requires that $K$ is of small order compared to $N$, we lose power, at least theoretically. In practice, it appears from simulations with Random Forest that $\zeta_{N_{train}, N_{train}}$ decreases to zero and roughly behaves like $1/N_{train}$. From the asymptotic power expression above, it can be seen that this would offset the small order $K$. Nonetheless, the test of Corollary \ref{asymptoticpower2} appears less powerful than the Binomial and hypoRF test. In the example of Figure \ref{asymptoticnormalityillustrationHA}, plugging the estimate of $\E[L_{1/2}^{(\hat{g}_{-i})}]$ obtained from the 500 repetitions into \eqref{powerexpression} and averaging, we obtain an expected power of 0.63. The actual power, i.e. the fraction of rejected tests over the 500 repetitions, is given as 0.61. The Binomial test with Random Forest on the other hand, reaches a power of 1. This illustrates that the test derived in this section, still lacks behind the test that uses sample-splitting. Nonetheless, modern $U$-statistics theory gives powerful theoretical tools to construct OOB-error based tests with tractable asymptotic power.




\section{Application} \label{Application}

In this section, we first describe the proposed significance threshold for the variable importance measure and apply the hypoRF test to simulated and real application cases. In the simulation section, we will compare the hypoRF to recent kernel-based tests by investigating the power of a selected scenario. A more extensive simulation study is given in \ref{simusec}. In Section \ref{sec:Real}, two real data sets from biology and finance are considered.

\subsection{Variable importance measure} \label{varimportance}

Variable importance measures in the context of Random Forest are practical tools introduced by \cite{Breiman2001}. As a by-product of the hypoRF test of Section \ref{finaltest}, we obtain a significance threshold for such a given variable importance measure: For each permutation, we record the maximum variable importance measure $I_{\sigma}$ over all variables, thus approximating the distribution of $I_{\sigma}$ under $H_0$. The estimated $1-\alpha$ quantile of this distribution will then be used as the significance threshold. Every variable with an importance measure above this threshold will be called significant. This should serve as an additional hint, in which components a rejection decision might originate from.
We will use in all instances the ``Gini'' importance measure or ``Mean Decrease Impurity'', see e.g., \cite[Section 5]{Biau2016}.


Obtaining $p$-values for the variable importance measure by permuting the response vector was developed much earlier in \cite{variableimportance1} and further developed in \cite{Janitza2018}. As we are not directly interested in $p$-values for each variable, our approach differs slightly and is more in the spirit of the Westfall-Young permutation approach, see e.g., \cite{westfallyoung}. Since we use a permutation approach already to define the decision rule of the hypoRF test, the significance threshold for the variable importance arises without any additional cost.

\begin{figure}
\centering
\scalebox{0.4}{
\input{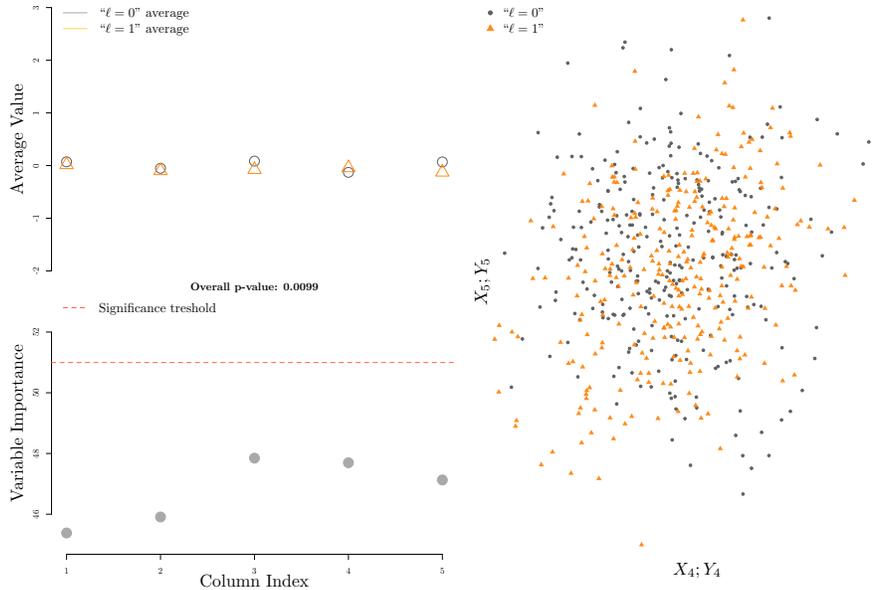}
}
\caption{\textbf{(Application)} We sampled $300$ observations from a $d=5$ dimensional multivariate normal, with no correlation between the marginals. Likewise $300$ observations were sampled from a multivariate normal, where the pairwise correlation between the columns is 0.4. The Random Forest used $500$ trees.}
\label{fig:intro2}
\end{figure}

Figure \ref{fig:intro} in Section \ref{motexamplesec} demonstrates that in this example the Random Forest is able to correctly identify the effect of the last two components. This appears remarkable, as there is only a change in dependence, but no marginal change. On the other hand, one could imagine a situation, where no significant variable may be identified, but the test overall still rejects. This is illustrated in Figure \ref{fig:intro2}. In this example, instead of endowing only the last two components with correlations, we introduced correlations of 0.4 between all variables when changing from $P$ to $Q$. Again the hypoRF test manages to differentiate between the two distributions. However this time, no significant variables can be identified. This seems sensible, as the source of change is divided equally between the different components in this example. Any situation could also be a mixture of the above extreme examples: There could be one or several significant variables, but the test still rejects, even after removing them. Section \ref{sec:Real} will show real-world examples in which some variables can be identified to be significant in the above sense.

\subsection{Simulation} \label{contamin}

In what follows, we will demonstrate the power of the proposed tests through simulation, and compare it with 3 kernel methods and a recently proposed Random Forest test based on the classification probability. To this end, we will use both the first version of the test, as described in Algorithm \ref{RFTestalg1} (``Binomial'' test), and the refined version in Algorithm \ref{permutationalg} (``hypoRF'' test). For the latter, as mentioned in Section \ref{finaltest}, we will use $K=100$ permutations. For the Binomial test described in Algorithm \ref{RFTestalg1} we decided to set $N_{train}=N_{test}$, as taking half of the data as training and the other half as test set seems to be a sensible solution a priori. To conduct our simulations we will use the R-package ``hypoRF'' developed by the authors, which consists of the ``hypoRF'' function including the two proposed tests. For each pair of samples, we run all tests and save the decisions. The estimated power is then the fraction of rejected among the $S$ tests.

The 3 kernel-based tests include the ``quadratic time MMD'' \citep{Twosampletest} using a permutation approach to approximate the $H_0$ distribution (``MMDboot''), its optimized version ``MMD-full'', as well as the ``ME'' test with optimized locations, ``ME-full'' \citep{NIPS2016_6148}. The original idea of the ``MMD-full'' was formulated in \cite{NIPS2012_4727}, however they subsequently used a linear version of the MMD. We instead use the approach of \cite{NIPS2016_6148}, which uses the optimization procedure of \cite{NIPS2012_4727} together with the quadratic MMD from \cite{Twosampletest}. 
A Python implementation of these methods is available from the link provided in \cite{NIPS2016_6148} (\url{https://github.com/wittawatj/interpretable-test}). Among these tests, it seems the MMDboot still is somewhat of a gold-standard, with newer methods such as presented in \cite{NIPS2012_4727}, \cite{NIPS2015_5685} and \cite{NIPS2016_6148}, more focused on developing more efficient versions of the test that are nearly as good. Nonetheless, the new methods often end up being surprisingly competitive or even better in some situations, as recently demonstrated in \cite{NIPS2016_6148}. Thus our choice to include MMD-full, ME-full as well. For all tests, we use a Gaussian kernel, which is a standard and reasonable choice if no a priori knowledge about the optimal kernel is available. The Gaussian kernel requires a bandwidth parameter $\sigma$, which is tuned in MMD-full and ME-full based on training data. For MMDboot we use the ``median heuristic'', as described in \cite[Section 8]{Twosampletest}, which takes $\sigma$ to be the median (Euclidean) distance between the elements in $(\mathbf{Z}_i)_{i=1}^{2n}$. 

Finally, we consider the method of \cite{Cal2020}, which is a test based on the classification probability of Random Forest. We would like to emphasize that their first publication on arXiv appeared more than 6 months after our first upload on arXiv. As such, we do not view them as a direct competitor. Nonetheless, it seems interesting to compare their performance to the one of hypoRF, as they use a permutation approach based on the \emph{in-sample} probability estimates.


We would like to stress that we did not use any tuning for the parameters of the RF-based tests, just as we did not use any tuning for MMDboot. As such, comparing the MMD/ME-full to the other methods might not be entirely fair. On the other hand, our chosen sample size might be too small for the optimized versions to work in full capacity. In particular, all optimized tests suffer from a similar drawback as our Binomial test: The tuning of the method takes up half of the available data. While \cite{NIPS2016_6148} find that ME-full outperforms the MMD, they only observe settings where the latter also uses half of the data to tune its kernel, as proposed in \cite{NIPS2012_4727}. In our terminology, they only compare ME-full to MMD-full, instead of MMDboot. It seems unclear a priori what happens if we instead employ the median heuristic for the MMD and let it use all of the available data, as in \cite{Twosampletest}. It should also be said that both optimization and testing of the ME-full scale linearly in $N$, making its performance below all the more impressive. On the other hand, the optimization depends on some hyperparameters common in gradient-based optimization, such as step size taken in the gradient step, the maximum number of iterations, etc. As this optimization is rather complicated for large $d$, some parameter choices sometimes lead to a longer runtime of the ME than the calculation-intensive hypoRF and CPT-RF. In general, it seems both runtime and performance of ME-full are in practice highly dependent on the chosen hyperparameters; we tried 3 different sets of parameters based on the code in \url{https://github.com/wittawatj/interpretable-test} with very different power results. The setting used in this simulation study is the exact same as used in their simulation study.

As discussed in \cite{DPLB:Powerdiscussion}, changing the parameters of our experiments (for instance the dimension $d$) should be done in a way that leaves the Kullback-Leibler (KL) Divergence constant. When varying the dimension $d$ we generally follow this suggestion, though in our case, this is not as imminent; whatever unconscious advantage we might give our testing procedure is also inherent in the competing methods. Finally, also note that, while our methods would be in principle applicable to arbitrary classifiers, we did not compare our proposed tests with tests based on other classifiers, such as those used in \cite{facebookguys}. Rather, we believe the choice of classifiers for binary classification is a more general problem and should be studied separately, as for example done extensively in \cite{RFsuperiority}. The only exception to this, is our use of an LDA classifier-based test for the example of a Gaussian mean-shift in \ref{subsec:gauss_dsmall}.

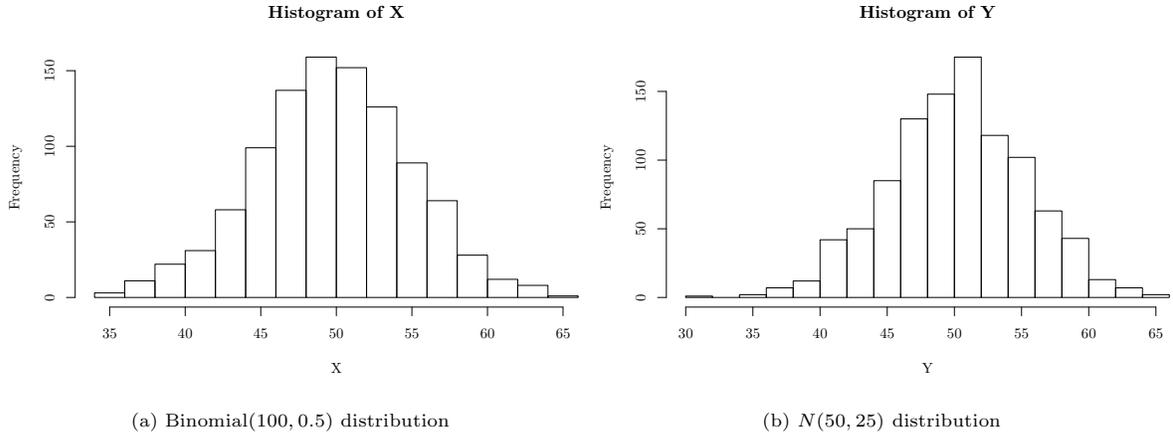
\begin{figure}
\centering
\begin{subfigure}[b]{0.475\textwidth}
  \centering
  \scalebox{0.55}{
  \input{4_contaminated_Binomial.tex}
  }
  \caption{$\text{Binomial}(100,0.5)$ distribution}
\end{subfigure}
\begin{subfigure}[b]{0.475\textwidth}
  \centering
  \scalebox{0.55}{
  \input{4_contaminated_Gaussian.tex}
  }
  \caption{$N(50,25)$ distribution}
\end{subfigure}
\caption{\textbf{(Contamination)} Illustration of the difference in marginals in the $c$ columns of $H_c$.}
\label{contaminationillustration}
\end{figure}

Where not differently stated, we use for the following experiments: $N=600$ observations, $300$ per class, $d=200$ dimensions, $K=100$ permutations and $600$ trees for the RF-based tests. In some examples, we additionally study a sparse case, where the intended change in distribution appears only in $c < d$ components. Throughout, notation such as
\[
P=\sum_{t=1}^T \omega_{t} N(\boldsymbol{\mu}_t, \Sigma_t),
\]
with $\omega_{t} \geq 0$, $\sum_{t=1}^T \omega_t=1$, $\boldsymbol{\mu}_t \in \R^d$, $\Sigma_t \in \R^{d\times d}$ means $P$ is a discrete mixture of $T$ $d$-valued Gaussians. Moreover, if $P_1, \ldots, P_d$ are distributions on $\R$, we will denote by
\[
P=\prod_{j=1}^d P_j,
\]
their product measure on $\R^d$. In other words, in this case, we simply take all the components of $\mathbf{X}$ to be independent.\\
The prime example which we present here in the main text is rather challenging. Let $P=N(\boldsymbol{\mu},\Sigma)$ with $\boldsymbol{\mu}$ set to $50 \cdot \mathbf{1}$ and $\Sigma=25 \cdot I_{d\times d}$. For the alternative, we consider the mixture
\[
Q=\lambda H_c + (1-\lambda) P,
\]
$\lambda \in [0,1]$, and $H_c$ some distribution on $\R^d$. This is a ``contamination'' of $P$ by $H_c$ with $\lambda$ determining the contamination strength. Here, we take $H_c$ to be another independent $(d-c)$-variate Gaussian together with $c$ components that are in turn independent $\text{Binomial}(100,0.5)$ distributed. We thereby choose parameters such that the Binomial components in $H_c$ have the same mean and variance as the Gaussian components and such that differentiating between Binomial and Gaussian is known to be difficult. Figure \ref{contaminationillustration} displays two realizations of a Gaussian and Binomial component respectively. We take $d=200$ and $c$ to be $10\%$ of 200, or $c=20$. 

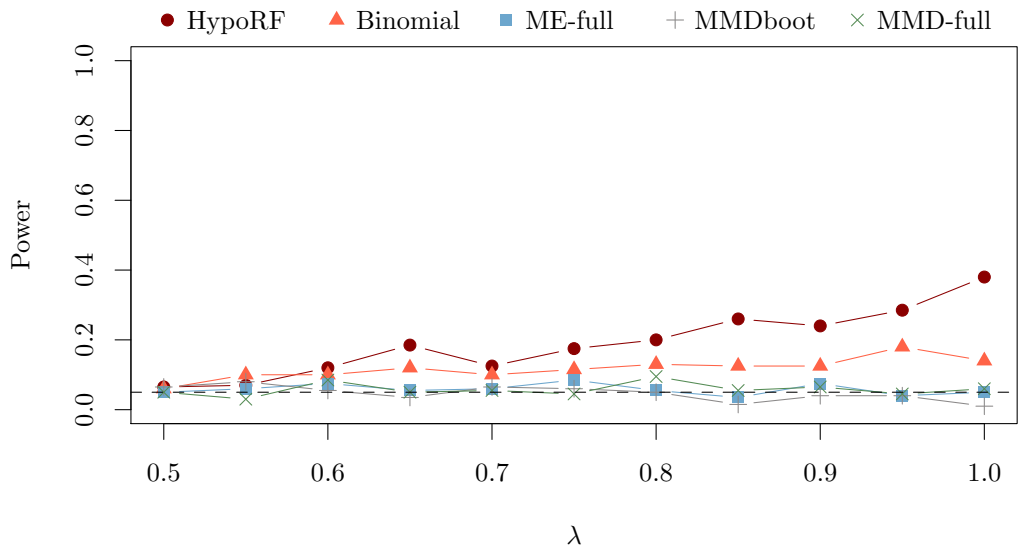
\begin{figure}
\centering
\scalebox{1}{
\input{4_plot_contamination_K100_Normapprox_F_with_Cai.tex}
}
\caption{\textbf{(Contamination)} A point in the figure represents a simulation of size $S=200$ for a specific test and a $\lambda \in (0.5,0.55,...,1)$. Each of the $S=200$ simulation runs we sampled $300$ observations from the contaminated distribution with $\lambda \in (0.5,0.55,...,1)$ and $c=20$. Likewise $300$ observations were sampled from $d=200$ independent standard normal distributions. The Random Forest used $600$ trees and a minimal node size to consider a random split of 4.}
\label{fig:test4K100}
\end{figure}

This problem is difficult; the Binomial and Gaussian components can hardly be differentiated by eye, the contamination level varies and the contamination is only in $c$ out of $d$ components actually detectable. Moreover, the combination of discrete and continuous components means the optimal kernel choice might not be clear, even with full information. Thus even for $300$ observations for each class, no test displays any power until we reach a contamination level of $0.5$. However, for higher contamination levels, Figure \ref{fig:test4K100} clearly displays the superiority of the RF-based tests: None of the kernel tests appear to significantly rise over the level of $5 \%$. On the other hand, the two proposed tests slowly grow from around $0.05$ to almost $0.4$ in the case of the hypoRF test. Interestingly, while relatively close at first, the difference in power between the Binomial test and the hypoRF grows and is starkest for $\lambda=1$, again demonstrating the benefit of using the OOB error as a test statistic. Although slightly worse than the hypoRF, the CPT-RF is also clearly beating the Binomial test, highlighting the benefit of using the permutation approach with (in-sample) classification probabilities.


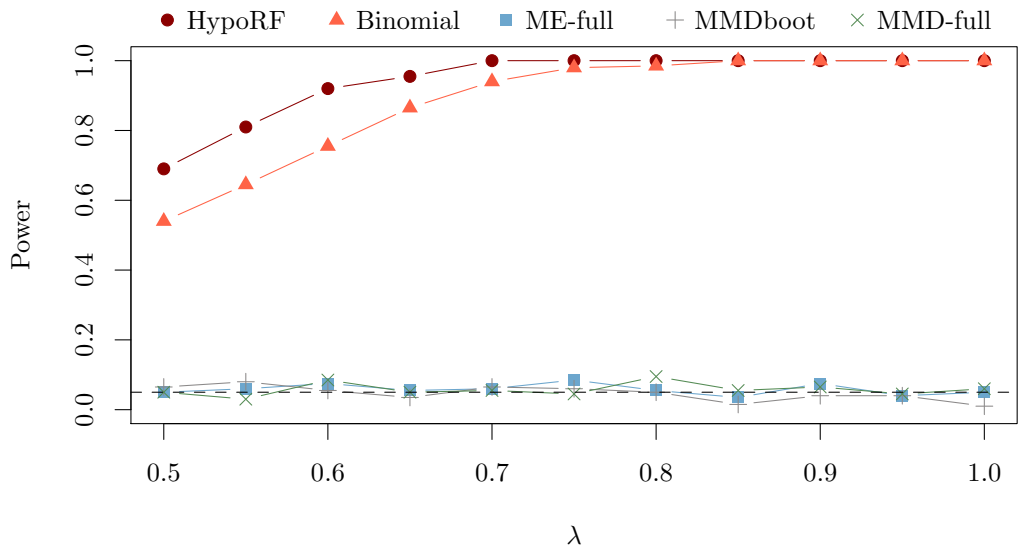
\begin{figure}
\centering
\scalebox{1}{
\input{4_plot_contamination_K100_dp_Normapprox_F_with_Cai.tex}
}
\caption{\textbf{(Contamination)} A point in the figure represents a simulation of size $S=200$ for a specific test and a $\lambda \in (0.5,0.55,...,1)$. Each of the $S=200$ simulation runs we sampled $300$ observations from the contaminated distribution with $\lambda \in (0.5,0.55,...,1)$ and $d=c$. Likewise $300$ observations were sampled from $d=200$ independent standard normal distributions. The Random Forest used $600$ trees and a minimal node size to consider a random split of 4.}
\label{fig:test5K100}
\end{figure}

Finally, we consider the case $d=c$, so that $H_c$ simply consists out of $d$ independent Binomial distributions. The result is displayed in Figure \ref{fig:test5K100} and all RF-based tests are now extremely strong, while the kernel tests fail to detect any signal.\\
More simulation examples can be found in \ref{simusec}.

\subsection{Real Data}
\label{sec:Real}

\begin{figure}
\centering
\scalebox{0.6}{
\input{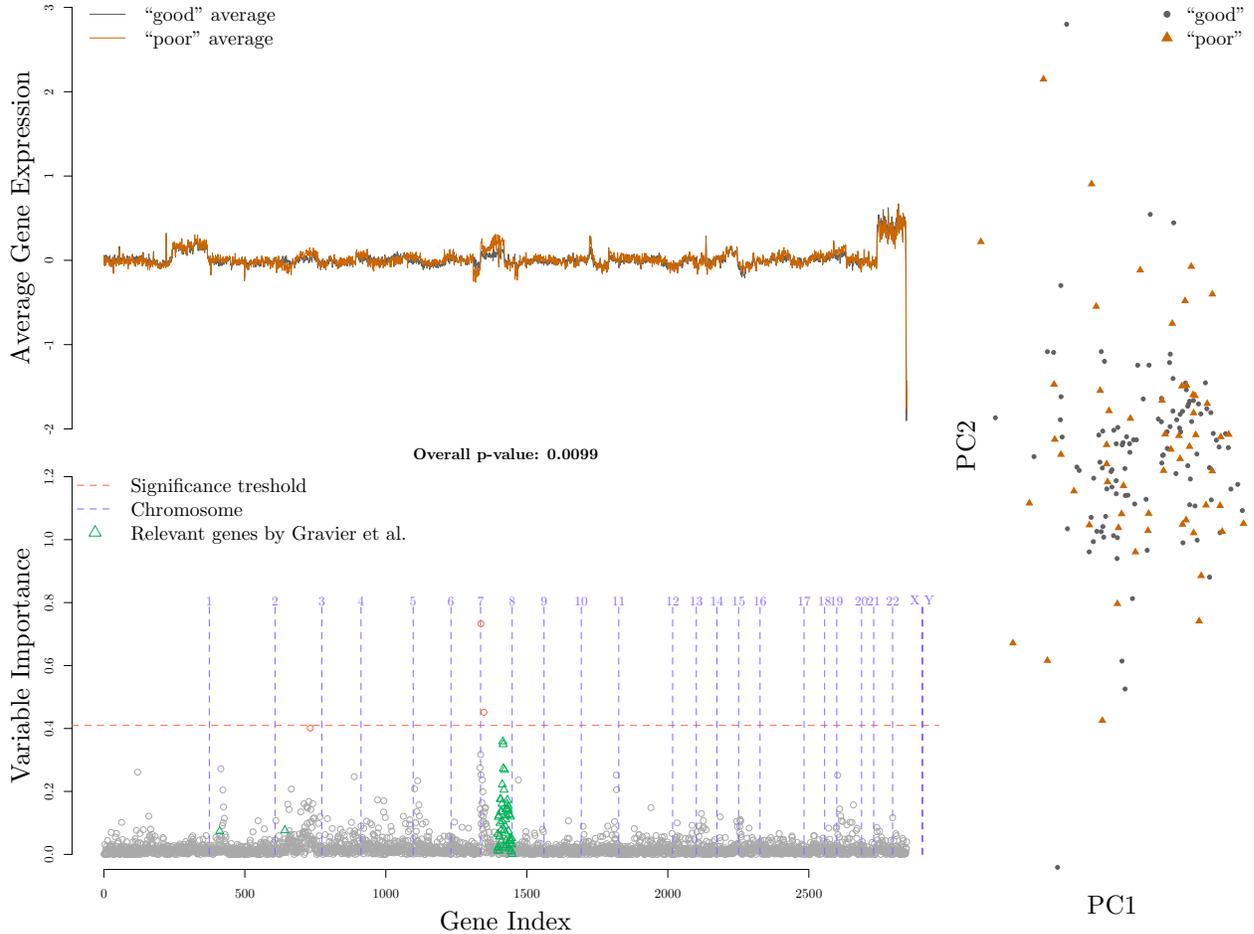}
}
\caption{\textbf{(Genes)} The variable importance (gene importance) combined with the average gene expression is illustrated. The test rejects the null hypothesis that the two groups ``good'' and ``poor'' come from the same distribution with a $p$-value of 0.0099. The 3 significant genes are ``8p23'', ``8p21'' and ``3q25'' (marked in red). The green triangles represent the important genes reported by \cite{Gravier2010}. Additionally, the plot of the first two principal components highlights the fact that there seem to be no obvious clusters. Note: only $15\%$ of the total variance is explained by the first 2 principal components. The Random Forest used $1000$ trees and a minimal node size to consider a random split of 4.}
\label{fig:genes}
\end{figure}

As a first application, we consider a high-dimensional microarray data set from \cite{githubdata}. The data set is about breast cancer, originally provided by \cite{Gravier2010}. They examined 168 patients with 2905 gene expressions, each over a five-year period. The 111 patients with no metastasis of small node-negative breast carcinoma after diagnosis were labeled ``good'', and the 57 patients with early metastasis were labeled ``poor''.\\
The application of the hypoRF to the two groups is summarized in Figure \ref{fig:genes}. The test detects a clear difference between the groups ``good'' and ``poor'' with ``8p23'', ``8p21'' and ``3q25'' being the most important (and significant) genes. There seems to be a high correlation between the genes that are located close to each other (especially within the same chromosome). This has the effect that the Random Forest takes a more or less arbitrary choice at a split point between those highly correlated genes. This in turn is reflected in the variable importance measure. For this reason, one should be careful when interpreting the variable importance measure on a gene level. It appears that chromosomes 8 and 3 play an important role in distinguishing the two groups. This finding is in line with \cite[Figure 2, p. 1129]{Gravier2010}.


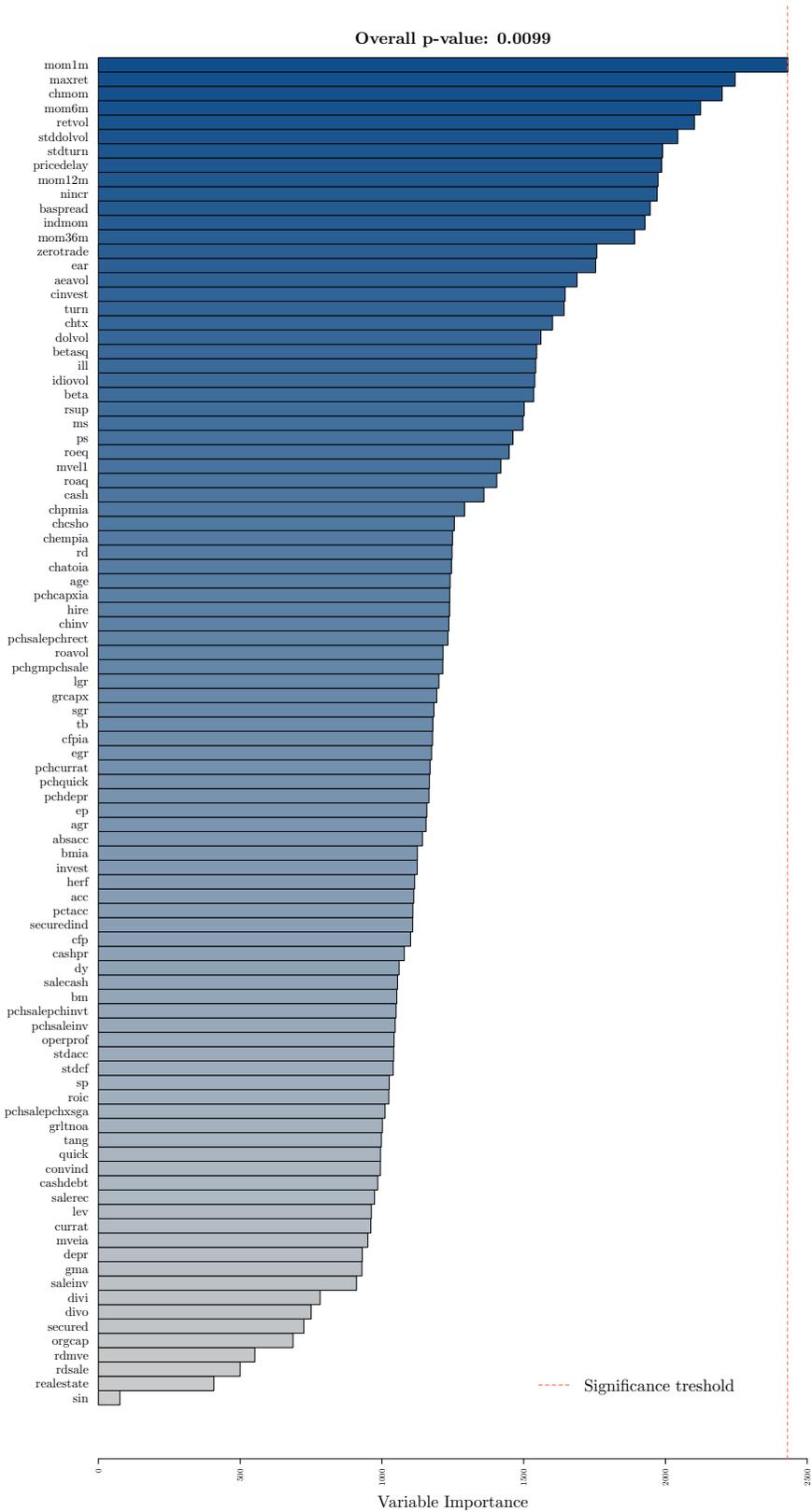
\begin{figure}
\centering
\scalebox{0.34}{
\input{riskfactors_plot_K100_ntree600_margin0}
}
\caption{\textbf{(Riskfactors)} The sorted variable importance of the 94 stock-specific characteristics are illustrated. More information on the 94 characteristics are listed in Tables \ref{table:factors1} and \ref{table:factors2} of \ref{factors}. The Test rejects with a $p$-value of almost zero. Nevertheless, the only significant characteristic is the 1-month momentum.}
\label{fig:riskfactors}
\end{figure}

In the second example, we are interested in the relative importance of financial risk factors (asset-specific characteristics). We claim that a financial risk factor has explanatory power if it contributes significantly to the classification of individual stock returns above or below the overall median. We use monthly stock return data from the Center for Research in Security Prices (CRSP). Our sample period starts in January 1977 and ends in December 2016, totaling 40 years. Additionally, we obtain the 94 stock-level predictive characteristics used by \cite{dacheng2020} from Dacheng Xiu's webpage - see, \url{http://dachxiu.chicagobooth.edu}. Between 1977 and 2016 we only use stocks for which we have a full return history. This leads to 501 stocks with 94 stock-specific characteristics. The group ``positive'' contains stocks and time points for which the return was above the overall median and vice versa for the ``negative'' group. The two groups are balanced and contain more than 120'000 observations each.\\
The application of the hpyoRF test on the two groups is summarized in Figure \ref{fig:riskfactors}. The ordering of the different risk factors is in line with the findings in \cite[Figure 5, p. 34]{dacheng2020}, 1-month momentum being the most important characteristic.\\
One could argue that stocks that are at time point $t$ close to the overall median are more or less randomly assigned to one of the two groups. Hence, a possible option is to only assign a stock and time point to a certain group if the return is above (below) a certain threshold, i.e., overall median $\pm \epsilon$. However, we observed that the result is very robust for different values of $\epsilon$.

\section{Discussion}

We discussed in this paper two easy to use and powerful tests based on Random Forest and empirically demonstrated their efficacy. We presented some consistency and power results and showed a way of adapting the Bayes classifier to obtain a consistent test. This adaptation consisted simply in changing the ``cutoff'' of the classifier. Especially the test based on the OOB statistics (hypoRF) proved to be powerful and additionally delivered a way to assess the significance of individual variables. This was demonstrated in applications using medical and financial data. 

After our first publication on arXiv, \cite{Cal2020} developed an approach based on a smooth transformation of the in-sample probabilities. Interestingly, experiments using their approach with OOB probability estimates, as a hybrid of their and our methodology, delivered very promising results. Investigating this further could lead to a further improve in power for RF-based tests.

\bibliography{RFTest}

\appendix

\section{Proofs}

\subsection{Proofs to Section \ref{framework}}

\begin{proposition}[Restatement of Proposition \ref{Prob1}] 
The decision rule in \eqref{Binomialtest} conserves the level asymptotically, i.e.
\[
\limsup_{N_{test} \to \infty} \Prob \left(\delta_{B}(\hat{g}(D_{N_{test}}))=1 \right) \leq \alpha,
\]
under $H_0:P=Q$.
\end{proposition}

\begin{proof}

Let $\mathcal{H}_{N}=\{D_{N_{train}},n_{1,test} \}$. Note that, $n_{1,test}$, $n_{0,test}$ contain the same probabilistic information, so it does not matter which we condition on. We first prove that,
\begin{align}\label{conddist}
    n_{j,test} \hat{L}_j^{(\hat{g})}|\mathcal{H}_{N}  ,   \sim \func{Bin}(n_{j,test}, L_j^{(\hat{g})}),
\end{align}
for $j \in \{0,1 \}$ and $\hat{L}_0^{(\hat{g})}$, $\hat{L}_1^{(\hat{g})}$ are conditionally independent given $D_{N_{train}}$, $n_{1,test}$. To prove \eqref{conddist} first note that by exchangeability (due to iid sampling),
\[
\sum_{i: \ell_i=j} \Ind\{\hat{g}(\mathbf{Z}_i) \neq \ell_i\} \stackrel{D}{=} \sum_{i=1}^{n_{j,test}} \Ind\{\hat{g}(\mathbf{Z}_i) \neq j\},
\]
$j \in \{0,1 \}$. Conditional on $\mathcal{H}_{N}$, the above is a sum of $n_{j,test}$ iid, elements $\Ind\{\hat{g}(\mathbf{Z}_i) \neq j\}$, with
\[
\Ind\{\hat{g}(\mathbf{Z}_i) \neq j\} | \mathcal{H}_{N} \sim \mbox{Bin}(1, \Prob(\hat{g}(\mathbf{Z}_i) \neq j|\mathcal{H}_{N})).
\]
Finally, since the event $\hat{g}(\mathbf{Z}_i) \neq j$ is independent of $n_{j,test}$,
\begin{align*}
    \Prob(\hat{g}(\mathbf{Z}_i) \neq j|\mathcal{H}_{N})&=\Prob(\hat{g}(\mathbf{Z}_i) \neq j|D_{N_{train}})\\
   &=\Prob(\hat{g}(\mathbf{Z}_i) \neq \ell_i|D_{N_{train}}, \ell_i=0)\\
    &= L_j^{(\hat{g})}.
\end{align*}

Let $\tilde{\sigma}_c^2:=L_{0}^{(\hat{g})} (1-L_{0}^{(\hat{g})}) + L_{1}^{(\hat{g})}(1-L_{1}^{(\hat{g})}) $ and recall that
\[
\hat{\sigma}_{c}^2=\frac{\hat{L}_{0}^{(\hat{g})} (1-\hat{L}_{0}^{(\hat{g})})}{n_{test,0}} + \frac{\hat{L}_{1}^{(\hat{g})} (1-\hat{L}_{1}^{(\hat{g})})}{n_{test,1}}.
\]
Moreover, set for all $N_{test}$:
\[
\epsilon_{N_{test}}:=\epsilon \cdot \frac{1}{N_{test}^{\nu}},
\]
for some $\epsilon > 0$ and $\nu \in (1/2,1)$. Note that we assume $N_{test} \to \infty$, while $N_{train}$ might also increase to infinity at any rate, or stay constant. Let for the following
\[
E:= \left \{ \frac{n_{1,test}}{N_{test}} \to \pi \right \}.
\]
Then $\Prob(E)=1$, as $\frac{n_{1,test}}{N_{test}} \to \pi$ a.s.

First assume for a realized sequence of $D_{N_{train}}$, $N_{test} \tilde{\sigma}_c^2 \to \infty$ holds. Then for a realized sequence of $n_{1,test}$, with the property that $n_{1,test}/N_{test} \to \pi$ (i.e. on $E$), it holds that 
\[
\limsup_{N_{test} \to \infty} \Prob \left(\delta_{B}(D_N)=1 | \mathcal{H}_N \right) \leq \Phi(\Phi^{-1}(\alpha))  = \alpha.
\]
Indeed, if $N_{test} L_{0}^{(\hat{g})} (1-L_{0}^{(\hat{g})}) \to \infty $ and $N_{test} L_{1}^{(\hat{g})}(1-L_{1}^{(\hat{g})}) \to \infty$, then conditional on $\mathcal{H}_N$,
\begin{align}\label{asymptoticnormalnicecase}
    \frac{\hat{L}^{(\hat{g})}_{1/2} - 1/2 }{\hat{\sigma}_{c}} \to N(0,1),
\end{align}
by the Lindeberg-Feller Central Limit Theorem. On the other hand assume $N_{test} L_{0}^{(\hat{g})} (1-L_{0}^{(\hat{g})}) \to \infty $ does not hold, but $N_{test} L_{1}^{(\hat{g})} (1-L_{1}^{(\hat{g})}) \to \infty$ is still true. The former holds if and only if $N_{test} L_{0}^{(\hat{g})}$ does not go to infinity (iff $L_{0}^{(\hat{g})} \to 0$) or $N_{test} (1- L_{0}^{(\hat{g})})$ does not go to infinity (iff $L_{0}^{(\hat{g})} \to 1$). Then we may write
\begin{align*}
    \frac{\hat{L}^{(\hat{g})}_{1/2} - 1/2 }{\hat{\sigma}_{c}} = \left( \frac{\sqrt{N_{test}} (\hat{L}^{(\hat{g})}_{0} - j) }{\sqrt{L_{1}^{(\hat{g})}(1-L_{1}^{(\hat{g})})}} +  \frac{\sqrt{N_{test}} (\hat{L}^{(\hat{g})}_{1} - (1-j)) }{\sqrt{L_{1}^{(\hat{g})}(1-L_{1}^{(\hat{g})})}}   \right) \frac{\sqrt{L_{1}^{(\hat{g})}(1-L_{1}^{(\hat{g})})}}{\sqrt{N_{test}}\hat{\sigma}_{c}},
\end{align*}
for $j \in \{0,1 \}$.
In this case, 
\[
\frac{\sqrt{L_{1}^{(\hat{g})}(1-L_{1}^{(\hat{g})})}}{\sqrt{N_{test}}\hat{\sigma}_{c}}  \stackrel{p}{\to} 1.
\]
Moreover, for all $\delta > 0$,
\begin{align*}
    \Prob \left(\frac{\sqrt{N_{test}}}{\sqrt{L_{1}^{(\hat{g})}(1-L_{1}^{(\hat{g})})}}  \Big| \hat{L}_{0}^{(\hat{g})} - L_{0}^{(\hat{g})}  | > \delta |\mathcal{H}_N  \right) &\leq \frac{N_{test} }{\delta^2 L_{1}^{(\hat{g})}(1-L_{1}^{(\hat{g})})} \Var(\hat{L}_{0}^{(\hat{g})} |\mathcal{H}_N) \\
    &=  \frac{N_{test} }{\delta^2 L_{1}^{(\hat{g})}(1-L_{1}^{(\hat{g})}) } \frac{L_{0}^{(\hat{g})}(1-L_{0}^{(\hat{g})})}{n_{0,test}} \\
    &\approx\frac{L_{0}^{(\hat{g})}(1-L_{0}^{(\hat{g})})}{ L_{1}^{(\hat{g})}(1-L_{1}^{(\hat{g})})},
\end{align*}
on $E$. Since $N_{test} L_{1}^{(\hat{g})}(1-L_{1}^{(\hat{g})}) \to \infty$ is still true, this means that 
\[
\frac{L_{0}^{(\hat{g})}(1-L_{0}^{(\hat{g})})}{L_{1}^{(\hat{g})}(1-L_{1}^{(\hat{g})})} = \frac{N_{test} L_{0}^{(\hat{g})}(1-L_{0}^{(\hat{g})})}{N_{test}L_{1}^{(\hat{g})}(1-L_{1}^{(\hat{g})})}  \to  0,
\]
on $E$ and thus,
\begin{align*}
\frac{\sqrt{N_{test}} (\hat{L}^{(\hat{g})}_{0} - j) }{\sqrt{L_{1}^{(\hat{g})}(1-L_{1}^{(\hat{g})})}} = \frac{\sqrt{N_{test}} (\hat{L}^{(\hat{g})}_{0} - L_{0}^{(\hat{g})}) }{\sqrt{L_{1}^{(\hat{g})}(1-L_{1}^{(\hat{g})})}} + \frac{\sqrt{N_{test}} (L_{0}^{(\hat{g})} - 0) }{\sqrt{ N_{test} L_{1}^{(\hat{g})}(1-L_{1}^{(\hat{g})})}} \stackrel{p}{\to} 0,
\end{align*}
and \eqref{asymptoticnormalnicecase} remains true. Finally note that $\epsilon_{N_{test}}$ is of too small order to make a difference in that case, since by the above $\sqrt{N_{test}} (\hat{L}^{(\hat{g})}_{1/2} - 1/2) = O_{\Prob}(1) $, while $\sqrt{N_{test}} \epsilon_{N_{test}} \to 0$.

Now assume that $N_{train}$, $D_{N_{train}}$ are such that $\liminf_{N_{test}} N_{test} \tilde{\sigma}_c^2 \to \infty$ does not hold. In this case, using again Markov's inequality,
\[
N_{test}(\hat{L}_{1/2} - 1/2) = O_{\Prob}(1),
\]
i.e. $ \lim_{M \to \infty} \limsup_{N_{test}} \Prob( N_{test}(\hat{L}_{1/2} - 1/2)  > M |  \mathcal{H}_{N} )=0$. Thus,
\begin{align*}
    \Prob \left(\delta_{B}(D_N)=1 | \mathcal{H}_{N} \right) &\leq  \Prob \left( N_{test}(\hat{L}^{(\hat{g})}_{1/2} - 1/2) > \epsilon \cdot N_{test}^{1-\nu} | \mathcal{H}_{N}  \right) \to 0,
\end{align*}
as $\epsilon \cdot N_{test}^{1-\nu} \to \infty$.

Thus we have shown that for a realized sequence of $D_{N_{train}},n_{1,test}$, with the property that $n_{1,test}/N_{test} \to \pi$, it holds that 
\[
\limsup_{N_{test} \to \infty} \Prob \left(\delta_{B}(D_N)=1 | \mathcal{H}_N \right) \leq  \alpha.
\]
On the other hand,
\begin{align*}
    &\limsup_{N_{test} \to \infty} \Prob \left(\delta_{B}(D_N)=1 \right)= \\
    &= \limsup_{N_{test} \to \infty} \E \left[ \Prob \left(\delta_{B}(D_N)=1| D_{N_{train}},n_{1,test}  \right) \Ind_E\right]\\
    & \leq  \E \left[ \limsup_{N_{test} \to \infty} \Prob \left(\delta_{B}(D_N)=1 | D_{N_{train}},n_{1,test}  \right) \Ind_E\right]\\
    &\leq \alpha.
\end{align*}
\end{proof}

\begin{lemma} [Restatement of Lemma \ref{consistencylemma0}]
Take $\X \subset \R$ and $\pi \neq 1/2$. Then no decision rule of the form, $\delta(D_N)=\delta( g^*_{1/2}(D_N) ) $ 
is consistent.
\end{lemma}

\begin{proof}
We first show that if $\pi \neq \frac{1}{2}$, one can construct $(P,Q) \in \Theta_1$ that the Bayes classifier is not able to differentiate. Consider $\pi > 1/2$,  $d=1$ and $Q$ being the uniform distribution on $(0,1)$, with density $q(z)=\Ind \{ z \in (0,1)  \}$. We write $q=\Ind{(0,1)}$ for short. $P$ is a mixture of $Q$ and another uniform on $R \subset (0,1)$, so that 
\[
p=  (1-\alpha)\Ind{(0,1)} + \alpha\frac{\Ind{R}}{|R|}. 
\]
Giving $Q$ a label of 1 and $P$ a label of 0 when observing $(1-\pi) P + \pi Q$, and taking $|R|=1/2$, the Bayes classifier is then given as $g_{1/2}^{*}(z)=\Ind \{\eta(z) >  1/2 \}$, where
\begin{align*}
    \eta(z):=\begin{cases}\pi /( \pi + (1- \pi)(1+\alpha) ), & \text{ if } z \in R\\
    \pi /(\pi + (1-\pi) (1-\alpha)), & \text{ if } z \notin R\end{cases}.
\end{align*}
Simple algebra shows that for any $\alpha <   \min(\pi/(1-\pi) - 1,1)$,  $\eta(z) > 1/2$ and thus $g_{1/2}^*(z)=1$ for all $z \in (0,1)$. 
In particular, $L_0^{(g_{1/2}^{*})}=1$ and $L_0^{(g_{1/2}^{*})}=0$ and both
$L_0^{(g_{1/2}^{*})} + L_1^{(g_{1/2}^{*})}=1$ and $L^{(g_{1/2}^{*})}=1-\pi=\min(\pi,1-\pi)$.

On the other hand, for any $\theta_0 \in \Theta_0$, simple evaluation of $\eta(z)$ shows that $g_{1/2}^*(z)=1$ for all $z$.
Consequently, for $\theta_1=(P,Q)$ in the above example and $\theta_0 \in \Theta_0$ arbitrary, it holds that 
\[
\E_{\theta_0}[ f( g^*_{1/2}(D_N) ) ] = \E_{\theta_1}[ f( g^*_{1/2}(D_N) ) ],
\]
for any bounded measurable function $f: \{0,1 \}^N \to \R$. In particular, since the test conserves the level by assumption, $\phi(\theta_1)=\phi(\theta_0)\leq \alpha$ and the test has no power.


\end{proof}

\begin{lemma}[Restatement of Lemma \ref{consistencylemma2}]
The classifier
\begin{equation}
    g_{\pi}^*(\mathbf{z}) = \Ind \left\{ \eta(\mathbf{z}) > \pi  \right\},
\end{equation}
is a solution to \eqref{newproblem}. Moreover it holds that 
\begin{equation}\label{eq:TVanderror2d}
    1 - TV(P,Q)    = L_0^{g_{\pi}^*} + L_1^{g_{\pi}^*},
\end{equation}
for any $\pi \in (0,1)$.
\end{lemma}

\begin{proof}
We show Relation \eqref{eq:TVanderror2d} for the classifier
\[
g^*(\mathbf{z}): = \Ind \left\{ \eta(\mathbf{z}) > \pi  \right\}.
\]
If this is true, it will immediately follows that $g^*=g_{\pi}^*$. Indeed, let $h_{\#}P$ be the push-forward measure of $P$ through a measurable function $h: \X \to \R$. Taking $h=g$, for an arbitrary classifier $g$, it holds that
\begin{align*}
    1-(L_0^{g_{\pi}^*} + L_1^{g_{\pi}^*} ) &= TV(P,Q)\\
    &\geq  P(g(\mathbf{X}) = 0 ) - Q(g(\mathbf{Y}) = 0 )\\
    &=\Prob(g(\mathbf{Z}) = 0 | \ell=0 ) - \Prob(g(\mathbf{Z}) = 0 | \ell=1 )  \\
    &=1-(L_0^{g} + L_1^{g} ),
\end{align*}
where the first inequality follows, because $\{\mathbf{x}: g(\mathbf{x}) = 0 \}$ and $\{\mathbf{y}: g(\mathbf{y}) = 0 \}$ are two Borel sets on $\X$. Consequently, it also holds for any classifier $g$ that 
\[
L_{1/2}^{g} = \frac{1}{2} (L_0^{g} + L_1^{g} ) \geq \frac{1}{2}  (L_0^{g_{\pi}^*} + L_1^{g_{\pi}^*} ) = L_{1/2}^{g_{\pi}^*},
\]
or $g^*=g_{\pi}^*$.

It remains to prove \eqref{eq:TVanderror2d} for $g^*$: It is well-known that (one of) the sets attaining the maximum in the definition of $TV(P,Q)$ is given by $A^*:=\{\mathbf{z}: q(\mathbf{z} ) \leq  p(\mathbf{z}) \}$. It is possible to rewrite $A^*$:
\begin{align*}
    A^*&=\left \{ \mathbf{z}: \frac{\pi q(\mathbf{z})}{ (1-\pi) p(\mathbf{z})+ \pi q(\mathbf{z})  } \leq \frac{\pi}{1-\pi} \frac{(1-\pi) p(\mathbf{z})}{ (1-\pi) p(\mathbf{z})+ \pi q(\mathbf{z}) }  \right \}\\
        &= \left \{ \mathbf{z}: \eta(\mathbf{z}) \leq \frac{\pi}{1-\pi} (1-\eta(\mathbf{z}))  \right\}\\
    &=  \{ \mathbf{z}:\eta(\mathbf{z}) \leq \pi \}.
\end{align*}
Thus
\begin{align*}
    TV(P,Q) = P(A^*) - Q(A^*) &= \Prob( \eta(\mathbf{z}) \leq  \pi | \ell=0 ) - \Prob( \eta(\mathbf{z}) \leq \pi | \ell=1 ) \\
    &= 1- \Prob( \eta(\mathbf{z}) >   \pi | \ell=0 )  - \Prob( \eta(\mathbf{z}) \leq \pi | \ell=1 )\\
    &=1 - (\Prob( \eta(\mathbf{z}) >   \pi | \ell=0 )  + \Prob( \eta(\mathbf{z}) \leq \pi | \ell=1 ))\\
    &= 1- (L_0^{g_{\pi}^*} + L_1^{g_{\pi}^*}).
\end{align*}


\end{proof}

\begin{corollary}[Restatement of Corollary \ref{consistencycor}]
The decision rule $\delta_B(g_{\pi}^*(D_N))$ in \eqref{Binomialtest} is consistent for any $\pi \in (0,1)$.
\end{corollary}

\begin{proof}

We restate here the decision rule in \eqref{Binomialtest} for completeness,
\begin{align*}
  \delta_{B}(g^*_{\pi}(D_{N})) = \Ind \left\{   \hat{L}_{1/2}^{(g^*_{\pi})} - 1/2  < \hat{\sigma}_c \Phi^{-1}(\alpha) + \epsilon_{N} \right\},
\end{align*}
since $N_{test}=N$.

First we show that the decision rule conserves the level, for $\epsilon_{N}=0$ for all $N$. Since, for any $P,Q$, $P=Q$, $\eta(z)=\pi$, $\hat{L}_0^{(g_{\pi}^*)}=1$ and $\hat{L}_1^{(g_{\pi}^*)}=0$ a.s., so that for all $\theta_0 \in \Theta_0$ and any sample size,
\begin{align*}
    \phi(\theta_0)=\Prob_{\theta_0}( \hat{L}_{1/2}^{(g_{\pi}^*)} < 1/2  )=0.
\end{align*}
Thus in particular $\sup_{\Theta_0} \phi(\theta_0) = 0 \leq \alpha$.

Assume $\theta \in \Theta_1$, so that $TV(P,Q) > 0$. We assume first that also $TV(P,Q) < 1$.
Since now the classifier itself does not need to be estimated, it holds that
\[
N_{j} \hat{L}_j^{(g_{\pi}^*)}  | N_{j}  \sim \func{Bin}(N_{j}, L_j^{(g_{\pi}^*)}),
\]
as proven in Proposition \ref{Prob1}. Since $1 > TV(P,Q) > 0$, $0 < L_0^{(g_{\pi}^*)} + L_1^{(g_{\pi}^*)} < 1$, so that $N L_j^{(g_{\pi}^*)}(1-L_j^{(g_{\pi}^*)}) \to \infty$ for $j=0$ or $j=1$. Conditional on any sequence of $N_0, N_1$, such that $N_0 \to \infty$ and $N_1 \to \infty$, as $N\to \infty$, 
\[
\sqrt{N_{0}} (\hat{L}_0^{(g_{\pi}^*)} - L_0^{(g_{\pi}^*)} )\stackrel{D}{\to} N(0, L_0^{(g_{\pi}^*)}(1-L_0^{(g_{\pi}^*)})) \text{ and } \sqrt{N_{1}} (\hat{L}_1^{(g_{\pi}^*)} - L_1^{(g_{\pi}^*)} )\stackrel{D}{\to} N(0, L_1^{(g_{\pi}^*)}(1-L_1^{(g_{\pi}^*)})),
\]
and since $\hat{L}_0^{(g_{\pi}^*)}$, $\hat{L}_1^{(g_{\pi}^*)}$ are conditionally independent, it holds that
\begin{align*}
\frac{ \hat{L}_{1/2}^{(g_{\pi}^*)}  - L_{1/2}^{(g_{\pi}^*)}  }{ 1/2 \sqrt{ \frac{\hat{L}_0^{(g_{\pi}^*)}(1-\hat{L}_0^{(g_{\pi}^*)})}{N_{0}} + \frac{\hat{L}_1^{(g_{\pi}^*)}(1-\hat{L}_1^{(g_{\pi}^*)})}{N_{1}} }  }=\frac{ \hat{L}_{1/2}^{(g_{\pi}^*)}  - L_{1/2}^{(g_{\pi}^*)}  }{ \hat{\sigma}_c  }  \stackrel{D}{\to} N(0,1),
\end{align*}
as in Proposition \ref{Prob1}. Consequently, 
\begin{align*}
    \Prob \left(  \frac{\hat{L}_{1/2}^{(g_{\pi}^*)}  - 1/2  }{ \hat{\sigma}_c } < \Phi^{-1}(\alpha)  \Big| N_0 \right)& =   \Prob \left(  \frac{\hat{L}_{1/2}^{(g_{\pi}^*)}  - L_{1/2}^{(g_{\pi}^*)}  }{ \hat{\sigma}_c } < \Phi^{-1}(\alpha) - \frac{
    L_{1/2}^{(g_{\pi}^*)  } - 1/2}{ \hat{\sigma}_c  }   \Big| N_0 \right).
\end{align*}
Now for any realized sequence of $N_0$, $N_1$ such that $N_0 \to \infty$ and $N_1 \to \infty$, as $N\to \infty$, this probability goes to 1, since $L_{1/2}^{(g_{\pi}^*)} - 1/2 < 0$ and $\hat{\sigma}_c=O(N_0^{-1/2}) \to 0$. Since $N_1/N \to \pi$, a.s., and $N_0=N-N_1$, this will be true for almost all sequences. Thus applying dominated convergence to the above conditional result, one sees that
\[
\Prob \left(  \frac{\hat{L}_{1/2}^{(g_{\pi}^*)}  - 1/2  }{ \hat{\sigma}_c } < \Phi^{-1}(\alpha)  \right) \to 1.
\]

If $TV(P,Q)=1$ on the other hand, $L_{1/2}^{(g_{\pi}^*)}=0$ and $\hat{\sigma}_c=0$ a.s. and trivially the rejection probability becomes
\[
\Prob( \hat{L}_{1/2}^{(g_{\pi}^*)}  < 1/2  )=1.
\]

\end{proof}

\subsection{Proofs to Section \ref{U-stats-Test}}

\begin{lemma}[Restatement of Lemma \ref{expectationlemma}]
$\E[h_{N_{train}}((\ell_1, \mathbf{Z}_{N_{train}}), \ldots, (\ell_{N_{train}}, \mathbf{Z}_{N_{train}}))]=\E[L_{1/2}^{(\hat{g}_{-i})}]$.
\end{lemma}

\begin{proof}

First we note that 
\[
\E[L_{1/2}^{\hat{g}_{-i}}]=\frac{1}{2} \left( \Prob(\hat{g}_{-i} (\mathbf{Z}_i) \neq \ell_i | \ell_i=1 )  +  \Prob(\hat{g}_{-i} (\mathbf{Z}_i) \neq \ell_i | \ell_i=0)  \right).
\]

Let $B(i) \leq B$ be the number of classifiers in the ensemble, not containing observation $i$. Since we assume that each classifier in the ensemble receives a bootstrapped version of $D_{N_{train}}$, there is a probability $p > 0$, that any given classifier $\hat{g}_{\nu_b}$ will not contain observation $i$. Since this bootstrapping is done independently for each classifier, we have that $B(i) \sim \mbox{Bin}(p, B)$. Thus as $B \to \infty$, also $B(i) \to \infty$ a.s. and thus $\hat{g}_{-i}(\mathbf{Z})=\E_{\nu}[\hat{g}_{\nu}(D_{N_{train}}^{-i})(\mathbf{Z})]$, or
\begin{align*}
\E[\varepsilon_{i}^{oob}]
&=\E[\Ind{ \{\hat{g}_{-i} (\mathbf{Z}_i) \neq \ell_i \}} \left( \frac{1-\ell_{i}}{n_{0, train}} + \frac{\ell_{i}}{n_{1, train}}  \right)  ]\\
&=\E \left[\Ind{ \{\hat{g}_{-i} (\mathbf{Z}_i) \neq \ell_i \}}  \frac{1-\ell_{i}}{n_{0, train}}   \right]  + \E \left[\Ind{ \{\hat{g}_{-i} (\mathbf{Z}_i) \neq \ell_i \}}  \frac{\ell_{i}}{n_{1, train}}    \right] .    
\end{align*}
Now, since $\ell_i=\Ind\{ \ell_i=1 \}$, it holds that
\begin{align*}
\E \left[\Ind{ \{\hat{g}_{-i} (\mathbf{Z}_i) \neq \ell_i \}}  \frac{\ell_{i}}{n_{1, train}}    \right]&=\E \left[\frac{1}{n_{1, train}} \cdot \Prob(\hat{g}_{-i} (\mathbf{Z}_i) \neq \ell_i , \ell_i=1 | n_{1, train} )  \right]\\
&=\E \left[\frac{\Prob(\ell_i=1| n_{1, train})}{n_{1, train}} \cdot \Prob(\hat{g}_{-i} (\mathbf{Z}_i) \neq \ell_i, | n_{1, train}, \ell_i=1 )   \right]\\
&=\E \left[\frac{\Prob(\ell_i=1| n_{1, train})}{n_{1, train}}    \right] \cdot \Prob(\hat{g}_{-i} (\mathbf{Z}_i) \neq \ell_i, |  \ell_i=1 ),
\end{align*}
since the event $\hat{g}_{-i} (\mathbf{Z}_i) \neq \ell_i$ is independent of $n_{1, train}$ given the event $\ell_i=1$. Finally,
\begin{align*}
\E \left[\frac{\Prob(\ell_i=1| n_{1, train})}{n_{1, train}}    \right]&= \frac{1}{N_{train}} \E \left[ \sum_{i=1}^{N_{train}} \frac{\Prob(\ell_i=1| n_{1, train})}{n_{1, train}}    \right] \\
&=\frac{1}{N_{train}} \E \left[  \E\left[\frac{1}{n_{1, train}}\sum_{i=1}^{N_{train}} \ell_i \Big| n_{1, train} \right]  \right]\\
&=\frac{1}{N_{train}}.
\end{align*}
Similarly, 
\begin{align*}
    \E \left[\Ind{ \{\hat{g}_{-i} (\mathbf{Z}_i) \neq \ell_i \}}  \frac{1-\ell_{i}}{n_{0, train}}   \right] = \frac{1}{N_{train}} \Prob(\hat{g}_{-i} (\mathbf{Z}_i) \neq \ell_i, |  \ell_i=0 ).
\end{align*}
Thus indeed,
\[
\E[h_{N_{train}}((\ell_1, \mathbf{Z}_{N_{train}}), \ldots, (\ell_{N_{train}}, \mathbf{Z}_{N_{train}}))]=N_{train} \E[\varepsilon_{1}^{oob} ]=\E[L_{1/2}^{\hat{g}_{-i}}].
\]

\end{proof}

\begin{lemma}
$h_{N_{train}}$ is a valid kernel for the expectation $\E[L_{1/2}^{(\hat{g}_{-i})}]$.
\end{lemma}

\begin{proof}
Unbiasedness was proven above. Symmetry follows, since for any two permutations $\sigma_1$, $\sigma_2$,
there exists $i,j$ such that $\sigma_{1}(j)=\sigma_{2}(i):=u$, and thus
\begin{align*}
\varepsilon_{\sigma_1(i)}^{oob}&=\E[\Ind{ \{g(\mathbf{Z}_{\sigma_1(i)}, D_{N_{train}}^{-\sigma_1(i)}, \theta) \neq \ell_{\sigma_1(i)} } \} \left( \frac{1-\ell_{\sigma_1(i)}}{n_{0, train}} + \frac{\ell_{\sigma_1(i)}}{n_{1, train}}  \right) | D_{N_{train}}^{\sigma_1}] \\
&= \E[\Ind{ \{g(\mathbf{Z}_{u}, D_{N_{train}}^{-u}, \theta) \neq \ell_{u} \} }  \left( \frac{1-\ell_{u}}{n_{0, train}} + \frac{\ell_{u}}{n_{1, train}}  \right)  | D_{N_{train}}^{\sigma_1} ] \\
&= \E[\Ind{ \{g(\mathbf{Z}_{u}, D_{N_{train}}^{-u}, \theta) \neq \ell_{u} \} } \left( \frac{1-\ell_{u}}{n_{0, train}} + \frac{\ell_{u}}{n_{1, train}}  \right)    | D_{N_{train}}^{\sigma_2}]  \\
&= \varepsilon_{\sigma_2(j)}^{oob},    
\end{align*}
where $D_{N_{train}}^{\sigma_s}= (\mathbf{Z}_{\sigma_s(1)},\ell_{\sigma_s(1)}),\ldots,(\mathbf{Z}_{\sigma_s(N_{train})},\ell_{\sigma_s(N_{train})})$, $s \in \{1,2 \}$. But that means the sum in \eqref{overalloob} does not change.
\end{proof}

We also need a well-known auxiliary result:

\begin{lemma}\label{subsequenceconvergencelemma}
Let $(\xi_N)_{N}$, $\xi$ be an arbitrary sequence of random variables. If every subsequence has a subsequence such that $\xi_{N(k(l))} \stackrel{D}{\to}  \xi$, then $\xi_{N} \stackrel{D}{\to}  \xi$. 
\end{lemma}

\begin{theorem}[Restatement of Theorem \ref{asymptoticnormalxx}]
Assume that for $N \to \infty$, $N_{train}=N_{train}(N) \to \infty$ and $K=K(N) \to \infty$,
\begin{align}
\lim_{N} \frac{K N_{train}^2}{N} \frac{\zeta_{1,N_{train}}}{{\zeta_{N_{train},N_{train}}}} &= 0, \label{zetacondapp}\\
\lim_{N} \frac{\sqrt{K} N_{train}}{N}&= 0. \label{Kocondapp}
\end{align}
Then,
\begin{align} \label{normality2app}
\frac{\sqrt{K}(\hat{U}_{N,K} - \E[L_{1/2}^{(\hat{g}_{-1})}] ) }{\sqrt{    \zeta_{N_{train},N_{train}}}} \stackrel{D}{\to} N(0,1).
\end{align}
\end{theorem}

\begin{proof}
Let for the following $\xi_i= (\mathbf{Z}_{i},\ell_{i})$ for brevity and consider the complete U-statistics
\begin{equation}\label{completeU}
\hat{U}_{N}:=\frac{1}{\binom{N}{N_{train}}} \sum  h_{N_{train}}(\xi_{i_1},\ldots, \xi_{i_{N_{train}}}), 
\end{equation}
where the sum is taken over all $\binom{N}{N_{train}}$ possible subsets of size $N_{train} \leq N$ from $\{1,\ldots, N\}$. From the ``H-Decomposition'', see e.g., \cite{Ustat90}, the variance of $ \hat{U}_{N}$ can be bounded as,
\begin{align*}
    \Var(  \hat{U}_{N}) 
    &\leq \frac{N_{train}^2}{N} \zeta_{1,N_{train}} +  \frac{N_{train}^2}{N^2} \Var(h)\\
    &\leq \frac{N_{train}^2}{N} \zeta_{1,N_{train}} + \frac{N_{train}^2}{N^2} \zeta_{N_{train},N_{train}} ,
\end{align*}
see also \cite[Lemma 7]{wager2017estimation}. Thus it holds for all $\varepsilon > 0$ that
\begin{align*}
    \Prob \left( \frac{\sqrt{K} |\hat{U}_{N} - \E[L_{1/2}^{\hat{g}_{-1}} ] |}{\sqrt{\zeta_{N_{train},N_{train}}} } > \varepsilon \right) &\leq \frac{K\Var(\hat{U}_{N})}{\varepsilon^2 \zeta_{N_{train},N_{train}}} \\
    &= \frac{1}{\varepsilon^2} \left( \frac{K N_{train}^2}{N} \frac{\zeta_{1,N_{train}}}{\zeta_{N_{train},N_{train}}} +  \frac{K N_{train}^2}{N^2} \right) \\
    & \to 0,
\end{align*}
by \eqref{zetacondapp} and \eqref{Kocondapp}.

We now use the idea of \cite[Lemma A]{Ustat90} to prove \eqref{normality2}: As in \cite{RFuncertainty}, we denote by $\mathcal{S}_{N,N_{train}}=\{S_j: j=1,\ldots, \binom{N}{N_{train}}  \}$ all possible subsamples of size $N_{train}$ sampled without replacement. Let $M_{N, N_{train}}=(M_{S_1}, \ldots, M_{S_{(N,N_{train})}})$ be the number of times each subsample appears when sampling $K$ times. Then $M_{N, N_{train}} | \xi_1, \xi_2, \ldots $ is multinomial distributed. Thus
\begin{align}\label{divideandconquer}
    \frac{\sqrt{K} \left( \hat{U}_{N,K} - \E[L^{\hat{g}_{-1}}_{1/2}] \right)}{ \sqrt{\zeta_{N_{train},N_{train}}}} & \stackrel{D}{=}  \sqrt{K}^{-1} \left( \sum_{i=1}^{N,N_{train}} M_{S_i} ( h_{N_{train}}( S_i)  - \E[L^{\hat{g}_{-1}}_{1/2}] )  \right)/ \sqrt{\zeta_{N_{train},N_{train}}} \nonumber \\
    &\stackrel{D}{=} \frac{1}{ \sqrt{\zeta_{N_{train},N_{train}}} \sqrt{K}} \left( \sum_{i=1}^{(N,N_{train})} \frac{K}{\binom{N}{N_{train}}} ( h_{N_{train}}( S_i)  - \E[L^{\hat{g}_{-1}}_{1/2}] )  \right) + \nonumber \\
    &\frac{1}{ \sqrt{\zeta_{N_{train},N_{train}}}\sqrt{K}} \left( \sum_{i=1}^{(N,N_{train})} (M_{S_i} - \frac{K}{\binom{N}{N_{train}}}) ( h_{N_{train}}( S_i)  - \E[L^{\hat{g}_{-1}}_{1/2}] )  \right) \nonumber \\
    &\stackrel{D}{=} \frac{\sqrt{K}(\hat{U}_{N} - \E[L^{\hat{g}_{-1}}_{1/2}])}{\sqrt{\zeta_{N_{train},N_{train}}} } + \nonumber \\
    &\sqrt{K}\left( \frac{1}{K} \sum_{i=1}^{(N,N_{train})} (M_{S_i} - \frac{K}{\binom{N}{N_{train}}}) \frac{( h_{N_{train}}( S_i)  - \E[L^{\hat{g}_{-1}}_{1/2}] )}{\sqrt{\zeta_{N_{train},N_{train}}}}  \right).
\end{align}

Let $a_i=(h_{N_{train}}( S_i)  - \E[L^{\hat{g}_{-1}}_{1/2}])/\sqrt{\zeta_{N_{train},N_{train}}}  $, as in \cite{Ustat90}. Then
\[
\hat{U}_{N,2}=\binom{N}{N_{train}}^{-1} \sum_{i=1}^{(N,N_{train})} a_i^2,
\]
is again a U-statistics with $\E[\hat{U}_{N,2}]=1$ and
\begin{align*}
    \Prob (|\hat{U}_{N,2} -1 |  > \varepsilon)&\leq \frac{1}{\varepsilon} \frac{N_{train}^2}{N} \Var(\E[( h_{N_{train}}( S_i)  - \E[L^{\hat{g}_{-1}}_{1/2}] )^2|\xi_1]) + \frac{N_{train}^2}{N^2}\\
    &= O\left(\frac{N_{train}}{N} \right)\\
    &=o(K),
\end{align*}
using Lemma \ref{zeta1result}. Thus, $\hat{U}_{N,2} \stackrel{p}{\to} 1$ and this will be true for any given subsequence as well. Similarly, 
\[
\binom{N}{N_{train}}^{-1} \sum_{i=1}^{(N,N_{train})} a_i \leq \frac{\sqrt{K}(\hat{U}_{N} - \E[L^{\hat{g}_{-1}}_{1/2}]) }{\sqrt{\zeta_{N_{train},N_{train}}}} \stackrel{p}{\to} 0.
\]
For each given subsequence we can thus choose a further subsequence such that $\hat{U}_{N,2} \stackrel{a.s.}{\to} 1$, as well as $\sqrt{K} (U - \E[L^{\hat{g}_{-1}}_{1/2}])/\sqrt{\zeta_{N_{train},N_{train}}}  \stackrel{a.s.}{\to} 0$. Then it follows from \eqref{divideandconquer} and the same characteristic function arguments as in \cite[Lemma A]{Ustat90} that,
\begin{align*}
    &\lim_{N \to \infty} \E[\exp\left( \iota t \sqrt{K} (\hat{U}_{N,K} - \E[L^{\hat{g}_{-1}}_{1/2}])/\zeta_{N_{train},N_{train}} \right)] = \\
    &\lim_{N \to \infty} \E\left[ \exp \left( \iota t \sqrt{K} (\hat{U}_{N} - \E[L^{\hat{g}_{-1}}_{1/2}]) /\zeta_{N_{train},N_{train}}\right) \right] \cdot \exp \left( -\frac{t^2}{2} \right)\\
    &=\exp \left( -\frac{t^2}{2} \right),
\end{align*}
where we suppressed the dependence on the chosen subsequence.
Thus the subsequence converges in distribution to $N(0,1)$ and by Lemma \ref{subsequenceconvergencelemma}, so does the overall sequence.

\end{proof}

\begin{corollary}[Restatement of Corollary \ref{asymptoticpower2}] 
Assume the conditions of Theorem \ref{asymptoticnormalxx} hold true and that $\hat{\zeta}_{N_{train},N_{train}}/\zeta_{N_{train},N_{train}} \stackrel{p}{\to} 1$. Then the decision rule in \eqref{test3} conserves the level asymptotically and has approximate power
\begin{align}\label{powerexpressionapp}
    \Phi \left( \Phi^{-1}(\alpha) + \sqrt{\frac{K}{\zeta_{N_{train},N_{train}}}} (1/2 - \E[L_{1/2}^{(\hat{g}_{-i})}]) \right).
\end{align}
\end{corollary}

\begin{proof}
From Theorem \ref{asymptoticnormalxx} and the assumption that $\hat{\zeta}_{N_{train},N_{train}}$ is a consistent estimator, it follows that
\begin{align*}
\frac{\sqrt{K} (\hat{U}_{N,K} - \E[L^{\hat{g}_{-i}}_{1/2}])}{\sqrt{\hat{\zeta}_{N_{train},N_{train}}}}
&\stackrel{D}{\to} N(0,1).
\end{align*}
In particular, under $H_0$, as $\E[L^{\hat{g}_{-i}}_{1/2}]=1/2$:
\[
\frac{\sqrt{K} (\hat{U}_{N,K} - 1/2)}{\sqrt{\hat{\zeta}_{N_{train},N_{train}}}} \stackrel{D}{\to} N(0,1),
\]
so that the decision rule \eqref{test3} attains the right level as $K \to \infty$. Moreover, under the alternative, for $t^*:=\Phi^{-1}(\alpha)$,
\begin{align*}
&\Prob\left(\frac{\sqrt{K} (\hat{U}_{N,K} - 1/2)}{\sqrt{\hat{\zeta}_{N_{train},N_{train}}}} < t^* \right)\\
&= \Prob \left(\frac{\sqrt{K} (\hat{U}_{N,K} - \E[L^{\hat{g}_{-i}}_{1/2}])}{\sqrt{\hat{\zeta}_{N_{train},N_{train}}}}  < t^* - \frac{\sqrt{K} (\E[L^{\hat{g}_{-i}}_{1/2}] - 1/2)}{\sqrt{\hat{\zeta}_{N_{train},N_{train}}}} \right)\\
&= \Phi \left( t^* + \frac{\sqrt{K} (1/2 - \E[L^{\hat{g}_{-i}}_{1/2}])}{\sqrt{\zeta_{N_{train},N_{train}}}} \right) + o_{\Prob}(1).
\end{align*}
\end{proof}

\section{Further Simulations}  \label{simusec}

Additional simulation examples can be found in the next three subsections.

\subsubsection{Gaussian Mean Shift} \label{subsec:gauss_dsmall}



The classical and most prominent example of two-sample testing is the detection of a mean-shifts between two Gaussians. That is, we assume $\Prob_X=N(\boldsymbol{\mu}_1,I_{d\times d})$ and $\Prob_Y=N(\boldsymbol{\mu}_2,I_{d\times d})$ so that the testing problem reduces to
\[
H_0: \boldsymbol{\mu}_1=\boldsymbol{\mu}_2 \ \ \text{vs} \ \ H_1:\boldsymbol{\mu}_1 \neq \boldsymbol{\mu}_2.
\]
We will implement this by simply taking $\boldsymbol{\mu}_2=\boldsymbol{\mu}_1 + (\delta/\sqrt{d}) \cdot \mathbf{1}$, for some $\delta \in \R$.

It appears clear that our test should not be the first to choose here. For $d$ much smaller than $n$, the optimal test would be given by Hotelling's test \citep{hotelling1931}. For $d$ approaching and even superseding $n$, the MMD with a Gaussian kernel, or an LDA classifier as in \cite{DPLBpublished}, might be the logical next choice. For this reason, we also included the LDA classifier in this example. For all the other examples, the simulated power of the LDA two-sample test is always no better than the level - as expected. Allowing the trees in the forest to grow fully, i.e., setting the minimum node size to a low number like 1, one observes a type of overfitting of the Random Forest. Thus we would expect our test to be beaten at least by MMDboot. Surprisingly this does not happen: As can be seen in Figure \ref{fig:1a}, all the RF-based tests display an impressive amount of power, where our hypoRF test is the strongest in all the provided mean shift scenarios. The Binomial test is even stronger than MMDboot and LDA, which seems surprising given the known strong performance of the MMD and LDA in this situation. The hypoRF test on the other hand towers above all others, together with MMD-full. In fact, the hypoRF and Binomial test almost appear to give respectively an upper and lower bound for the MMD-full in this example. Aside from the impressive power of our tests, it is also interesting to note the difference between MMD-full and MMDboot. While this seems not surprising, given that MMD-full is essentially the optimized version of MMDboot, we will see in subsequent examples that their power ranking is often reversed.

To make the example more interesting, one might ask what happens if the mean shift is not present in all of the $d$ components, but only in $c < d$ of them? This was noted to be a difficult problem in \cite{NIPS2015_5685}. We therefore study a ``sparse'' case $c=2$ ($1\%$ out of $d=200$) and a ``moderately sparse'' case $c=20$ ($10\%$ out of $d=200$), now considering $\boldsymbol{\mu}_2=\boldsymbol{\mu}_1 + (\delta/\sqrt{c}) \cdot \mathbf{1}$. Note that there is some advantage here, as we now scale $\delta$ only by a factor of $\sqrt{c} < \sqrt{d}$. Thus, if a test is able to detect the sparse changes well, it should display a higher power than before. Indeed as seen in Figure \ref{fig:1b}, the performance of the kernel tests are remarkably stable (given the randomness inherent in the simulation), when changing from $c=d=200$ to $c=20$ to $c=2$. On the other hand, the performance of the RF-based tests appear to increase. Thus the odds only shift in favor of our tests and the test of \cite{Cal2020}: For $c=20$ the optimized MMD, MMD-full, is still very competitive, though MMDboot, ME-full, and LDA fall further behind. While the hypoRF, the CPT-RF and the fully optimized MMD test reach a power of close to $1$, the remaining kernel tests and LDA stay below 0.7. The Binomial test, on the other hand, displays almost the same performance as MMD-full, ending with a power of a bit over 0.8. Its performance is amplified in the sparse case, in which the Binomial, CPT-RF and hypoRF test beat the other tests by a large margin. The power of both tests quickly increases from around 0.05 to 1, as $\delta$ passes from 0.2 to 1. While the performance of the Binomial test is impressive, the hypoRF test manages to pick up the nuanced changes even faster, at times almost doubling the power of the Binomial test. Though the price to pay for this is a much higher computational effort.

It should be said that both the sparse and moderately sparse case here are tailor-made for a RF-based classifier; not only are the changes only appearing in a few components, but they appear marginally and are thus easy to detect in the splitting process of the trees. Nonetheless, it seems surprising how strong the tests perform.
We will now turn to more complex examples, where changes in the marginals alone are not as easy, or even impossible to detect.

\begin{figure}
\centering
\scalebox{1}{
\input{1a_plot_gaussian_meanshift_Normapprox_F_LDA_with_Cai.tex}
}
\caption{\textbf{(Mean Shift)} A point in the figure represents a simulation of size $S=200$ for a specific test and a $\delta \in (0,0.0667,0.1334,0.2,\dots,1)$. Each of the $S=200$ simulation runs we sampled $300$ observations from a $d=200$ dimensional multivariate normal distribution with a mean shift of $\frac{\delta}{\sqrt{d}}$ and likewise $n=300$ observations from $d=200$ independent standard normal distributions. The Random Forest used $600$ trees and a minimal node size to consider a random split of 4.}
\label{fig:1a}
\end{figure}
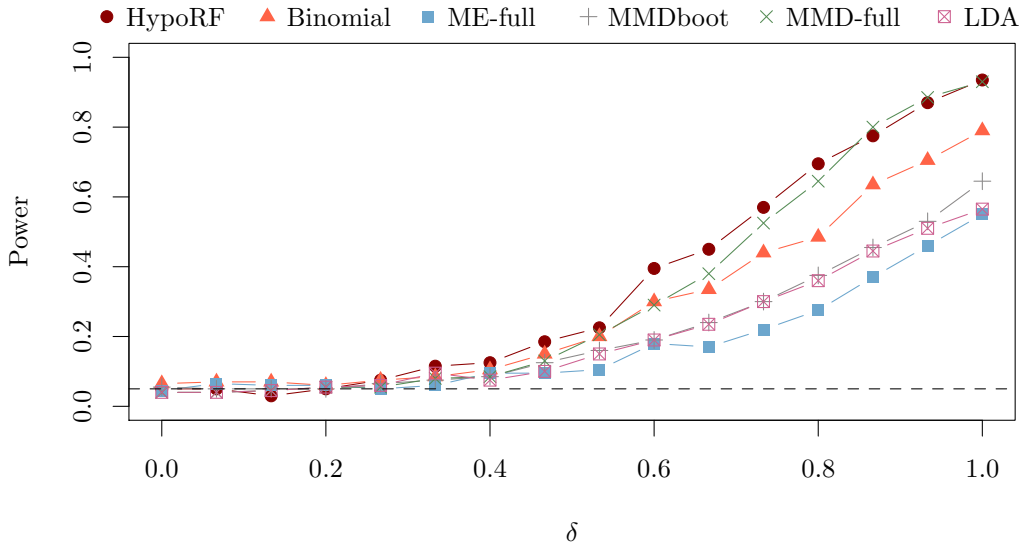

\begin{figure}
\centering
\begin{subfigure}{1\textwidth}
  \centering
  \scalebox{0.9}{
  \input{1b_plot_gaussian_meanshift_d20_Normapprox_F_LDA_with_Cai.tex}
  }
  \caption{$c=20$, moderately sparse case.}
\end{subfigure}%

\begin{subfigure}{1\textwidth}
  \centering
  \scalebox{0.9}{
  \input{1b_plot_gaussian_meanshift_d2_Normapprox_F_LDA_with_Cai.tex}
  }
  \caption{$c=2$ sparse case.}
\end{subfigure}
\caption{\textbf{(Mean Shift)} A point in the figures represents a simulation of size $S=200$ for a specific test and a $\delta \in (0,0.125,0.25,...,1)$. Each of the $S=200$ simulation runs we sampled $n=300$ observations from a $d=200$ dimensional multivariate Gaussian distribution, where $c$ columns have a shift in mean of $\frac{\delta}{\sqrt{c}}$ and likewise $n=300$ observations from $d=200$ independent standard normal distributions. The Random Forest used $600$ trees and a minimal node size to consider a random split of 4.}
\label{fig:1b}
\end{figure}



\subsubsection{Changing the Dependency Structure}

The previous example focused only on cases where the changes in distribution can be observed marginally. For these examples, it would in principle be enough to compare the marginal distributions to detect the difference between $Q$ and $P$. An interesting class of problems arises when we instead leave the marginal distribution unchanged but change the \emph{dependency structure} when moving from $P$ to $Q$. We will hereafter study two examples; the first one concerning a simple change from a multivariate Gaussian with independent components to one with nonzero correlation. The second one again takes $P$ to have independent Gaussian components, but induces a more complex dependence structure on $Q$, via a $t$-copula. Thus for what follows, we set $P= N(0, I_{d\times d})$.

First, consider $Q=N(0, \Sigma)$, where $\Sigma$ is some positive definite correlation matrix. As for any $d$ there are potentially $d(d-1)/2$ unique correlation coefficients in this matrix, the number of possible specifications is enormous even for small $d$. For simplicity, we only consider a single correlation number $\rho$, which we either use (I) in all $d(d-1)/2$ or (II) in only $c < d(d-1)/2$ cases. 

Figure \ref{fig:3a_gaussian_dep} displays the result of case (I). Now the superiority of our hypoRF test is challenged, though it manages to at least hold its own against MMD-full and ME-full. The roles of MMD-full and MMD are also reversed, the latter now displaying a much higher power, that in fact dwarfs the power of all other tests. MMD-full displays together with the Binomial test the smallest amount of power, both apparently suffering from the decrease in sample size. ME-full on the other hand, which suffers the same drawback, manages to put up a very strong performance, on par with the hypoRF. This is all the more impressive, keeping in mind that the ME is a test that scales linearly in $N$. Case (II) can be seen in Figure \ref{fig:3b_gaussian_dep}. Again the resulting ``sparsity'' is beneficial for our test, with the hypoRF now being on par with the powerful MMD test, and with ME-full only slightly above the Binomial test.

\begin{figure}
\centering
\scalebox{1}{
\input{3a_plot_gaussian_dep_K100_Normapprox_F_with_Cai.tex}
}
\caption{\textbf{(Dependency)} A point in the figure represents a simulation of size $S=200$ for a specific test and a $\rho \in (0,0.01,0.02...,0.15)$. Each of the $S=200$ simulation runs we sampled $300$ observations from a $d=60$ dimensional multivariate normal distribution with $\rho \in (0,0.01,0.02...,0.15)$, representing $Q$. Likewise $300$ observations were sampled from a $d=60$ dimensional multivariate normal distribution using $\rho = 0$, representing $P$. The Random Forest used $600$ trees and a minimal node size to consider a random split of 4.}
\label{fig:3a_gaussian_dep}
\end{figure}
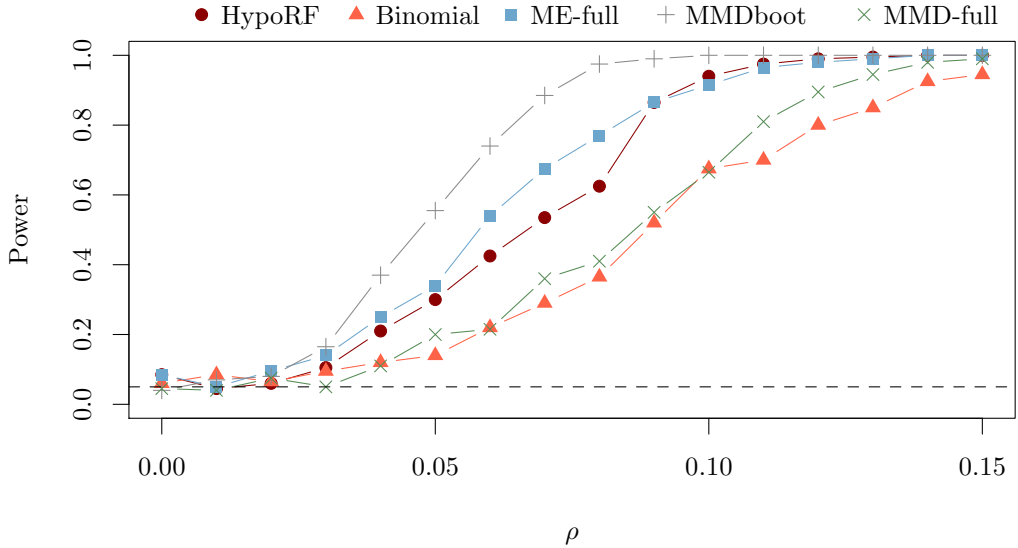

\begin{figure}
\centering
\scalebox{1}{
\input{3b_plot_gaussian_dep_K100_Normapprox_F_with_Cai.tex}
}
\caption{\textbf{(Dependency)} A point in the figure represents a simulation of size $S=200$ for a specific test and a $\rho \in (0,0.025,0.05...,0.375)$. Each of the $S=200$ simulation runs we sampled $300$ observations from a $d=10$ dimensional multivariate normal distribution with $c=4$ values in the correlation matrix equal to $\rho \in (0,0.025,0.05...,0.375)$, representing $Q$. Likewise $300$ observations were sampled from a multivariate normal distribution using $\rho = 0$, representing $P$. The Random Forest used $600$ trees and a minimal node size to consider a random split of 4.}
\label{fig:3b_gaussian_dep}
\end{figure}
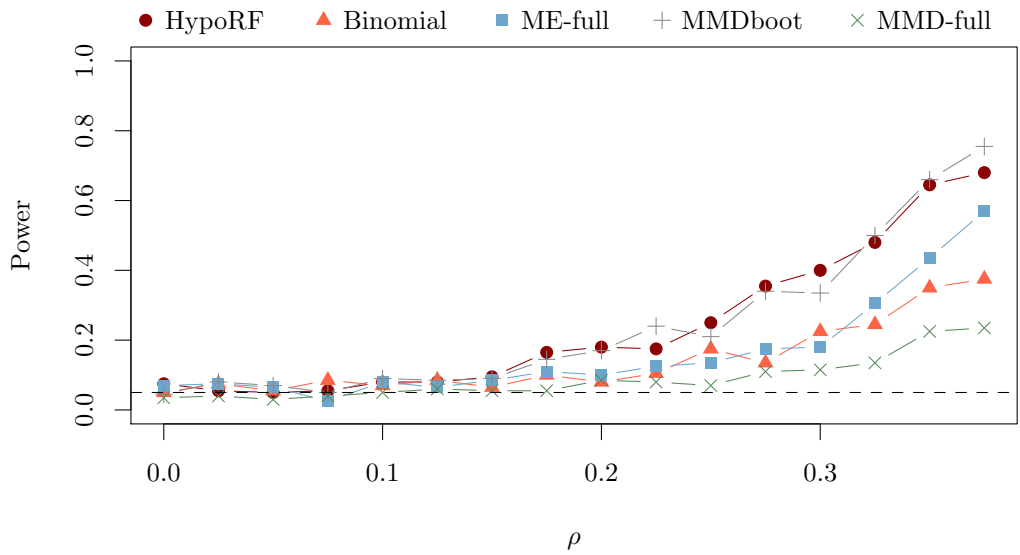

In the second example, we study a change in dependence, which is more interesting than the simple change of the covariance matrix. In particular, $Q$ is now given by a distribution that has standard Gaussian marginals bound together by a $t$-copula, see e.g., \cite{DeMc:05} or \cite[Chapter 5]{McFrEm:15}. While the density and cdf of the resulting distribution $Q$ are relatively complicated, it is simple and insightful to simulate from this distribution, as described in \cite{DeMc:05}: Let $x \mapsto t_{v}(x)$ denote the cdf of a univariate $t$-distribution with $\nu$ degrees of freedom, and $T_{\nu}(R)$ the multivariate $t$-distribution with dispersion matrix $R$ and $\nu$ degrees of freedom. We first simulate from a multivariate $t$-distribution with dispersion matrix $R$ and degrees of freedom $\nu$, to obtain $\mathbf{T} \sim T_{\nu}(R)$. 
In the second step, simply set $\mathbf{Y}:= \left(\Phi^{-1} (t_{v}(T_1)), \ldots, \Phi^{-1} (t_{v}(T_p)) \right)^T$. We denote $Q= T_{\Phi}(\nu, R)$. What kind of dependency structure does $\mathbf{Y}$ have? It is well known that $\mathbf{T} \sim t_{\nu}(R)$ has 
\[
\mathbf{T} \stackrel{D}{=} G^{-1/2} \mathbf{N},
\]
with $\mathbf{N} \sim N(0, R)$ and $G \sim \func{Gamma}(\nu/2, \nu/2)$ independent of $\mathbf{N}$. As such, the dependence induced in $\mathbf{T}$, and therefore in $Q$, is dictated through the mutual latent random variable $G$. It persists, even if $R=I_{d\times d}$ and induces more complex dependencies than mere correlation. These dependencies are moreover stronger, the smaller $\nu$, though this effect is hard to quantify. One reason this dependency structure is particularly interesting in our case is that it spans more than two columns, contrary to correlation which is an inherent bivariate property. We again study the case (I) with all $d$ components tied together by the $t$-copula, and (II) only the first $c=20< d$ components having a $t$-copula dependency, while the remaining $d-c=180$ columns are again independent $N(0,1)$.

\begin{figure}
\centering
\scalebox{1}{
\input{3a_plot_copula_K100_Normapprox_F_with_Cai.tex}
}
\caption{\textbf{(Dependency)} A point in the figure represents a simulation of size $S=200$ for a specific test and a $v \in (1,1.5,...,8)$. Each of the $S=200$ simulation runs we sampled $300$ observations from the Student-t Copula with $R=I_{d\times d}$, $v \in (1,1.5,...,8)$ and $d=60$ standard normally distributed margins and likewise $300$ observations from the multivariate normal. The Random Forest used $600$ trees and a minimal node size to consider a random split of 4.}
\label{fig:3a_copula}
\end{figure}
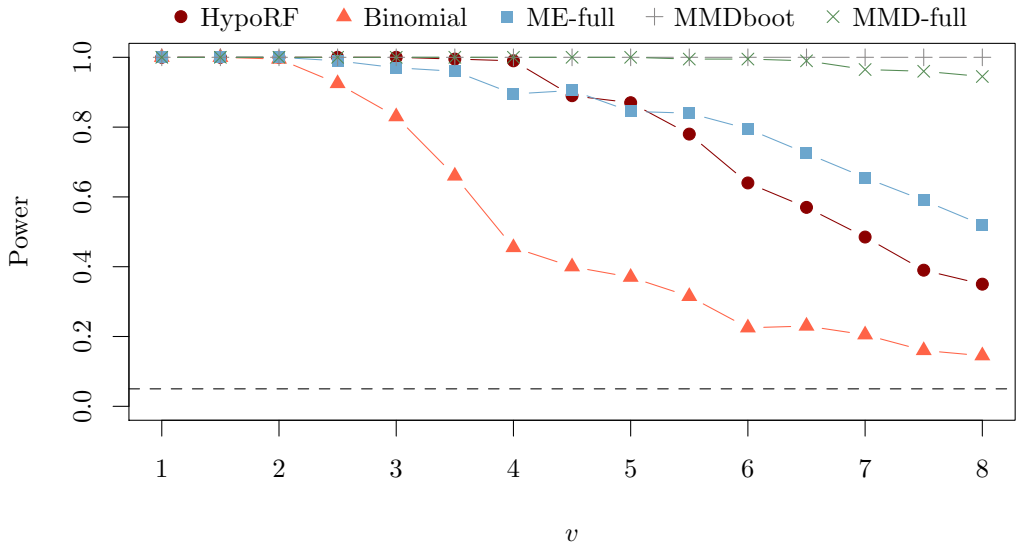

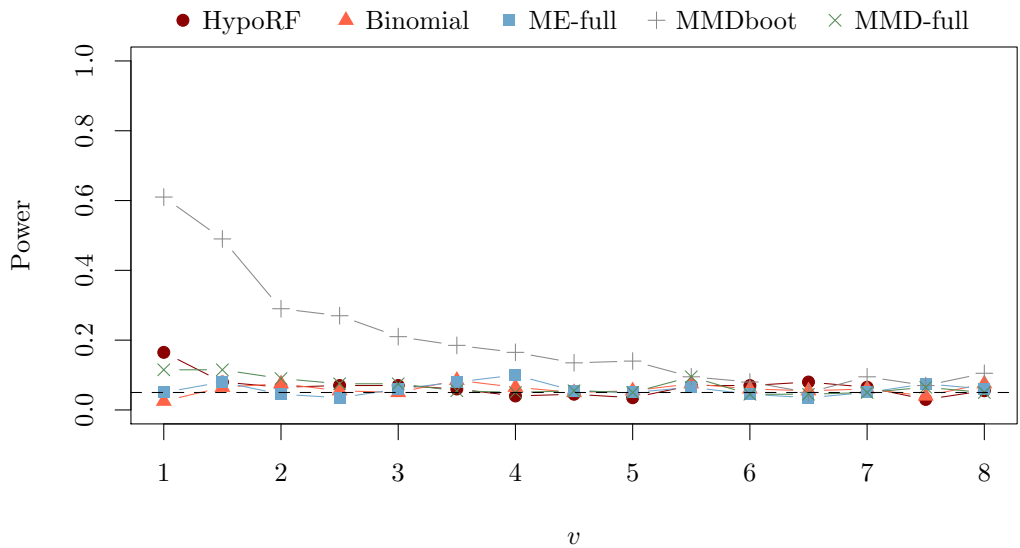
\begin{figure}
\centering
\scalebox{1}{
\input{3b_plot_copula_K100_Normapprox_F_with_Cai.tex}
}
\caption{\textbf{(Dependency)} A point in the figure represents a simulation of size $S=200$ for a specific test and a $v \in (1,1.5,...,8)$. Each of the $S=200$ simulation runs we sampled $300$ observations from a $d-c=180$ dimensional multivariate Gaussian distribution and a $d=20$ dimensional Student-t Copula with $R=I_{d\times d}$, $v \in (1,1.5,...,8)$ and standard normally distributed margins, representing $Q$. Likewise $300$ observations were sampled from a multivariate normal distribution, representing $P$. The Random Forest used $600$ trees and a minimal node size to consider a random split of 4.}
\label{fig:3b_copula}
\end{figure}

The results for case (I) are shown in Figure \ref{fig:3a_copula}. Now our tests, together with ME-full cannot compete with CPT-RF, MMD and MMD-full. However for the ME-full, this very much depends again on the hyperparameters chosen, for some settings ME-full was as good as MMD-full. Though there appears to be no clear way how to determine this. Both MMD-based tests manage to stay at almost one, even for $\nu=8$, which seems to be an extremely impressive feat. The CPT-RF test falls behind the two MMD-based tests, but has still an impressively high power, compared to our hypoRF test. Our best test, on the other hand, loses power quickly for $\nu > 4$, while the Binomial test does so even for $\nu > 2$. The results for case (II) shown in Figure \ref{fig:3b_copula}, are similarly insightful. Given the difficulty of this problem, it is not surprising that almost all of the tests fail to have any power for $\nu > 3$. The exception is once again the MMD, performing incredibly strong up to $\nu=5$. The performance of MMDboot is not only interesting in that it beats our tests, but also in how it beats all other kernel approaches in the same way. In particular, MMD-full stands no chance, which again is likely, in part, due to the reduced sample size the MMDboot has available for testing.
Though hard to generalize, it appears from this analysis that a complex, rather weak dependence, is a job best done by the plain MMDboot.

\subsubsection{Multivariate Blob} \label{Blobsection}

A well-known difficult example is the ``Gaussian Blob'', an example where ``the main data variation does not reflect the difference between between $P$ and $Q$'' \citep{NIPS2012_4727}, see e.g., \cite{NIPS2012_4727} and \cite{NIPS2016_6148}. We study here the following generalization of this idea: Let $T \in \N$, $\boldsymbol{\mu}=\left( \boldsymbol{\mu}_t \right)_{t=1}^T$, $\boldsymbol{\mu}_t \in \R^d$, and $\boldsymbol{\Sigma}=\left( \Sigma_t \right)_{t=1}^T$, with $\Sigma_t$ a positive definite $d \times d$ matrix. We consider the mixture
\[
N( \boldsymbol{\mu}, \boldsymbol{\Sigma}):= \sum_{t=1}^T \frac{1}{T} N(\boldsymbol{\mu}_t, \Sigma_t).
\]
For $\boldsymbol{\mu}$, we will always use a baseline vector of size $d$, $w$ say, and include in $\boldsymbol{\mu}$ all possible enumerations of choosing $d$ elements from $w \in \R^d$ with replacement. This gives a total number of $T=c^d$ possibilities and each $\boldsymbol{\mu}_t \in \R^d$ is one possible such enumeration. For example, if $c=d=2$ and $w=(1,2)$ then we may set $\boldsymbol{\mu}_1=(1,1)$, $\boldsymbol{\mu}_2=(2,2)$, $\boldsymbol{\mu}_3=(1,2)$, $\boldsymbol{\mu}_4=(2,1)$. We will refer to each element of this mixture as a ``Blob'' and study two experiments where we change the covariance matrices $\Sigma_t$ of the blobs when changing from $P$ to $Q$, i.e.,
\[
P=N( \boldsymbol{\mu}, \boldsymbol{\Sigma}_X), \ \ Q=N(\boldsymbol{\mu}, \boldsymbol{\Sigma}_Y).
\]

Obviously it quickly gets infeasible to simulate from $N(  \boldsymbol{\mu}, \boldsymbol{\Sigma})$, as with increasing $d$ the number of blobs explodes. Though, as shown below, this difficulty can be circumvented when $\Sigma_t$ is diagonal for all $t$. The example also considerably worsens the curse of dimensionality, as even for small $d$ the numbers of observations in each Blob is likely to be very small.
Thus for $300$ observations, we have a rather difficult example at hand.

We will subsequently study two experiments. The first one takes $w=\left( 1,2,3 \right)$, $\Sigma_{1,X}=\Sigma_{2,X}=\ldots =\Sigma_{t,X}=I_{d\times d}$ and $\Sigma_{1,Y}=\Sigma_{2,Y}=\ldots =\Sigma_{t,Y}=\Sigma$ to be a correlation matrix with nonzero elements on the off-diagonal. In particular, we generate $\Sigma$ randomly at the beginning of the $S$ trials for a given $d$, such that (1) it is a positive definite correlation matrix and (2) it has a ratio of minimal to maximal eigenvalue of at most $1-1/\sqrt{d}$. For $d=2$, this corresponds to the original Blob example as in \cite{NIPS2012_4727}, albeit with a less strict bound on the eigenvalue ratio. The resulting distribution for $d=1$ and $d=2$ is plotted in Figure \ref{figblobillustration1}. 

Table \ref{originalblobresults} displays the result of the experiment with our usual set-up and a variation of $d=2,3$ and the number of blobs being $2^d$ and $3^d$.
Very surprisingly our hypoRF test is the only one displaying notable power throughout the example. MMD and MMD-full are not able to detect any difference between the distribution with this sample size. Interestingly, the ME which we would have expected to work well in this example is also only at the level. However, this again depends on the specification chosen for the hyperparameters of the optimization. For another parametrization, we obtained a power of 0.116 for $d=2$, $blobs=2^2$ and $0.082$ for $d=2$ and $blobs=3^2$, all other values being on the level.
 
\begin{table}
    \centering
\begin{tabular}{l*{8}{c}}
N   & d & Blobs & ME-full  & MMD & MMD-full  & Binomial & hypoRF \\
\hline
600 & 2 & $2^2$ & 0.056   & 0.054 & 0.072 & 0.204 & 0.306  \\
600 & 2  & $3^2$ & 0.064 & 0.048 &  0.070 & 0.070 &  0.190  \\
600 & 3 &  $2^3$ & 0.052  & 0.040 & 0.060 & 0.088 & 0.116  \\
600 & 3 & $3^3$ &  0.056  & 0.060 &  0.060 & 0.064 &  0.084  \\
\end{tabular}
    \caption{\textbf{(Blob)} Power for different $N$, $d$ and number of Blobs. Each power was calculated with a simulation of size $S=500$ for a specific test.}
    \label{originalblobresults}
\end{table}

The second experiment takes $w=\left( -5,0, 5 \right)$ and for all $t$, $\Sigma_{t,X}$, $\Sigma_{t,Y}$ to be diagonal and generated similarly to $\boldsymbol{\mu}$. That is, we take $\Sigma_{t,X}=\mbox{diag}(\sigma_{t,X}^2)$, where each $\sigma_{t,X}$ is a vector including $d$ draws with replacement from a base vector $v_X \in \R^d$, and analogously with $\Sigma_{t,Y}$. In this case, it is possible to rewrite $P$ and $Q$, as
\[
P= \prod_{j=1}^d P_X \text{ and } Q= \prod_{j=1}^d P_Y,
\]
with
\[
P_X=\frac{1}{3} N(w_1, v^2_{1,X}) + \frac{1}{3} N(w_2, v^2_{2,X}) + \frac{1}{3} N(w_3, v^2_{3,X}),
\]
and
\[
P_Y=\frac{1}{3} N(w_1, v^2_{1,Y}) + \frac{1}{3} N(w_2, v^2_{2,Y}) + \frac{1}{3} N(w_3, v^2_{3,Y}).
\]
As such, it is feasible to simulate from $P$ and $Q$, even for large $d$, by simply simulating $d$ times from $P_X$ and $P_Y$. We consider $w=\left( -5,0,5 \right)$ and the standard deviations 
\begin{align*}
\left( v_{1,X},v_{2,X},v_{3,X}\right)&= \left(1,1,1 \right),\\
\left( v_{1,Y},v_{2,Y},v_{3,Y}\right)&= \left(1,2,1 \right).
\end{align*}
The change between the distributions is subtle even in notation; only the standard deviation of the middle mixture component is changed from 1 to 2. This has the effect that the middle component gets spread out more, causing it to melt into the other two. The resulting distribution for $d=1$ and $d=2$ is plotted in Figure \ref{figblobillustration2}. Unsurprisingly, $P$ looks quite similar as in Figure \ref{figblobillustration1}. The marginal plots ($d=1$) appear to be very different, though this is only an effect of having centers $(-5,0,5)$ instead of $(1,2,3)$. On the other hand, while not clearly visible, it can be seen that the different blobs of $Q$ display different behavior in variance; every Blob in positions $(2,1)$, $(2,2)$, $(2,3)$, $(1,2)$, $(3,2)$ on the $3 \times 3$ grid has its variance increased.

The results of the simulations are seen in Figure \ref{newblobresults}. The Binomial, CPT-RF and hypoRF test display a power quickly increasing with dimensions, regardless of the decreasing number of observations in each Blob. This also holds true, to a smaller degree, for the ME-full, which due to its location optimization appears to be able to adapt to the problem structure. However, its power considerably lacks behind the RF-based tests. In contrast, the behavior of the MMD-based tests quickly deteriorates as the number of samples per Blob decreases. Indeed from a kernel perspective, all points have more or less the same distance from each other, whether they are coming from $P$ or $Q$. Thus the extreme power of the MMD to detect ``joint'' changes in the structure of the data (i.e., dependency changes) cements its downfall here, as it is unable to detect the marginal difference.

This example might appear rather strange; it has a flavor of a mathematical counterexample, simple or even nonsensical on the outset, but proving an important point: While the differences between $P$ and $Q$ are obvious to the naked eye if only
one marginal each is plotted with a histogram, the example manages to completely fool the kernel tests (under a Gaussian kernel at least).
As such it is not only a demonstration of the merits of our test but also a way of fooling very general kernel tests. It might be interesting to find real-world applications, where such data structure is likely.

\begin{figure}
\centering
\begin{subfigure}[b]{0.475\textwidth}
  \centering
  \scalebox{0.55}{
  \input{6_Bloboriginal_histX_p1.tex}
  }
\end{subfigure}
\begin{subfigure}[b]{0.475\textwidth}
  \centering
  \scalebox{0.55}{
  \input{6_Bloboriginal_histY_p1.tex}
  }
\end{subfigure}
\begin{subfigure}[b]{0.475\textwidth}
  \centering
  \scalebox{0.55}{
  \input{6_Bloboriginal_plotX_p2.tex}
  }
  \caption{$P$}
\end{subfigure}
\quad
\begin{subfigure}[b]{0.475\textwidth}
  \centering
  \scalebox{0.55}{
  \input{6_Bloboriginal_plotY_p2.tex}
  }
  \caption{$Q$}
\end{subfigure}
\caption{\textbf{(Blob)} Illustration of the original Blob example. Below: Illustration for $d=2$. Above: First marginals of $P$ and $Q$ respectively.}
\label{figblobillustration1}
\end{figure}

\begin{figure}
\centering
\begin{subfigure}[b]{0.475\textwidth}
  \centering
  \scalebox{0.55}{
  \input{5_Blob_histX_p1.tex}
  }
\end{subfigure}
\begin{subfigure}[b]{0.475\textwidth}
  \centering
  \scalebox{0.55}{
  \input{5_Blob_histY_p1.tex}
  }
\end{subfigure}
\begin{subfigure}[b]{0.475\textwidth}
  \centering
  \scalebox{0.55}{
  \input{5_Blob_plotX_p2.tex}
  }
  \caption{$P$}
\end{subfigure}
\quad
\begin{subfigure}[b]{0.475\textwidth}
  \centering
  \scalebox{0.55}{
  \input{5_Blob_plotY_p2.tex}
  }
  \caption{$Q$}
\end{subfigure}
\caption{\textbf{(Blob)} Illustration of the second Blob example. Below: Illustration for $d=2$. Above: First marginals of $P$ and $Q$ respectively.}
\label{figblobillustration2}
\end{figure}

\begin{figure}
\centering
  \scalebox{1}{
  \input{5_plot_Blob_K100_small_Normapprox_F_with_Cai.tex}
  }

\caption{\textbf{(Blob)} A point in the figure represents a simulation of size $S=200$ for a specific test and a $d \in (2,4,6,8,10,20,40,80,120,200)$. Each of the $S=200$ simulation runs we sampled $300$ observations from $N( \boldsymbol{\mu}, \boldsymbol{\Sigma}_X)$ and likewise $300$ observations from $N( \boldsymbol{\mu}, \boldsymbol{\Sigma}_Y)$. The Random Forest used $600$ trees and a minimal node size to consider a random split of 4.}
\label{newblobresults}
\end{figure}
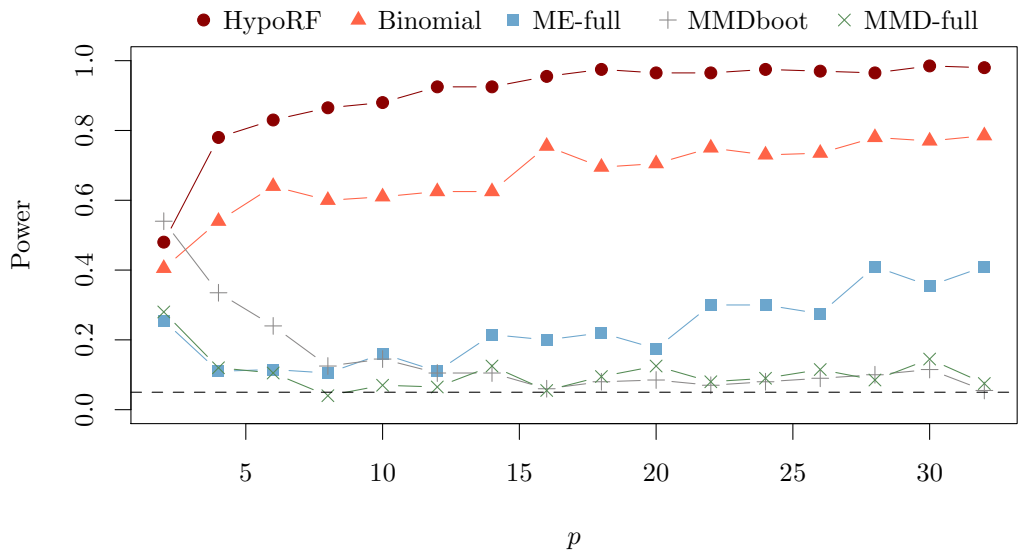


\section{Financial Riskfactors}  \label{factors}
\begin{table}
\begin{small}
\begin{center}
\captionsetup{type=table}
\begin{align*}
\resizebox{12cm}{!}{$\displaystyle
\begin{array}[t]{lllll}\toprule
\mbox{No.}& \mbox{Acronym}& \mbox{Firm Characteristic} & \mbox{Frequency} & \mbox{Literature} \\
\hline 
1 & \mbox{absacc}   &\mbox{Absolute accruals}   &\mbox{Annual}  &\mbox{\cite{bandyopadhyay2010accrual}}\\
2 & \mbox{acc}   &\mbox{Working capital accruals}   &\mbox{Annual}  &\mbox{\cite{sloan1996stock}}\\
3 & \mbox{aeavol}   &\mbox{Abnormal earnings announcement volume}   &\mbox{Quarterly}  &\mbox{\cite{lerman2008high}}\\
4 & \mbox{age}   &\mbox{Years since first Compustat coverage}   &\mbox{Annual}  &\mbox{\cite{jiang:2005}}\\
5 & \mbox{agr}   &\mbox{Asset growth}   &\mbox{Annual}  &\mbox{\cite{cooper2008asset}}\\
6 & \mbox{baspread}   &\mbox{Bid-ask spread}   &\mbox{Monthly}  &\mbox{\cite{amihud:1989}}\\
7 & \mbox{beta}   &\mbox{Beta}   &\mbox{Monthly}  &\mbox{\cite{fama:1973}}\\
8 & \mbox{betasq}   &\mbox{Beta squared}   &\mbox{Monthly}  &\mbox{\cite{fama:1973}}\\
9 & \mbox{bm}   &\mbox{Book-to-market}   &\mbox{Annual}  &\mbox{\cite{rosenberg1985persuasive}}\\
10 & \mbox{bmia}   &\mbox{Industry-adjusted book-to-market}   &\mbox{Annual}  &\mbox{\cite{asness:2000}}\\
11 & \mbox{cash}   &\mbox{Cash holdings}   &\mbox{Quarterly}  &\mbox{\cite{palazzo:2012}}\\
12 & \mbox{cashdebt}   &\mbox{Cash flow to debt}   &\mbox{Annual}  &\mbox{\cite{ou:1989}}\\
13 & \mbox{cashpr}   &\mbox{Cash productivity}   &\mbox{Annual}  &\mbox{\cite{chandrashekar2009productivity}}\\
14 & \mbox{cfp}   &\mbox{Cash flow to price ratio}   &\mbox{Annual}  &\mbox{\cite{desai:2004}}\\
15 & \mbox{cfpia}   &\mbox{Industry-adjusted cash flow to price ratio}   &\mbox{Annual}  &\mbox{\cite{asness:2000}}\\
16 & \mbox{chatoia}   &\mbox{Industry-adjusted change in asset turnover}   &\mbox{Annual}  &\mbox{\cite{soliman2008use}}\\
17 & \mbox{chcsho}   &\mbox{Change in shares outstanding}   &\mbox{Annual}  &\mbox{\cite{pontiff2008share}}\\
18 & \mbox{chempia}   &\mbox{Industry-adjusted change in employees}   &\mbox{Annual}  &\mbox{\cite{asness:2000}}\\
19 & \mbox{chinv}   &\mbox{Change in inventory}   &\mbox{Annual}  &\mbox{\cite{thomas2002inventory}}\\
20 & \mbox{chmom}   &\mbox{Change in 6-month momentum}   &\mbox{Monthly}  &\mbox{\cite{gettleman2006acceleration}}\\
21 & \mbox{chpmia}   &\mbox{Industry-adjusted change in profit margin}   &\mbox{Annual}  &\mbox{\cite{soliman2008use}}\\
22 & \mbox{chtx}   &\mbox{Change in tax expense}   &\mbox{Quarterly}  &\mbox{\cite{thomas2011tax}}\\
23 & \mbox{cinvest}   &\mbox{Corporate investment}   &\mbox{Quarterly}  &\mbox{\cite{titman2004capital}}\\
24 & \mbox{convind}   &\mbox{Convertible debt indicator}   &\mbox{Annual}  &\mbox{\cite{valta:2016}}\\
25 & \mbox{currat}   &\mbox{Current ratio}   &\mbox{Annual}  &\mbox{\cite{ou:1989}}\\
26 & \mbox{depr}   &\mbox{Depreciation / PP\&E}   &\mbox{Annual}  &\mbox{\cite{holthausen:1992}}\\
27 & \mbox{divi}   &\mbox{Dividend initiation}   &\mbox{Annual}  &\mbox{\cite{michaely:1995}}\\
28 & \mbox{divo}   &\mbox{Dividend omission}   &\mbox{Annual}  &\mbox{\cite{michaely:1995}}\\
29 & \mbox{dolvol}   &\mbox{Dollar trading volume}   &\mbox{Monthly}  &\mbox{\cite{chordia2001trading}}\\
30 & \mbox{dy}   &\mbox{Dividend to price}   &\mbox{Annual}  &\mbox{\cite{litzenberger:1982}}\\
31 & \mbox{ear}   &\mbox{Earnings announcement return}   &\mbox{Quarterly}  &\mbox{\cite{kishore:2008}}\\
32 & \mbox{egr}   &\mbox{Growth in common shareholder equity}   &\mbox{Annual}  &\mbox{\cite{richardson2005accrual}}\\
33 & \mbox{ep}   &\mbox{Earnings to price}   &\mbox{Annual}  &\mbox{\cite{basu:1977}}\\
34 & \mbox{gma}   &\mbox{Gross profitability}   &\mbox{Annual}  &\mbox{\cite{novy:2013}}\\
35 & \mbox{grcapx}   &\mbox{Growth in capital expenditures}   &\mbox{Annual}  &\mbox{\cite{anderson:2006}}\\
36 & \mbox{grltnoa}   &\mbox{Growth in long term net operating assets}   &\mbox{Annual}  &\mbox{\cite{fairfield:2003}}\\
37 & \mbox{herf}   &\mbox{Industry sales concentration}   &\mbox{Annual}  &\mbox{\cite{hou:2006}}\\
38 & \mbox{hire}   &\mbox{Employee growth rate}   &\mbox{Annual}  &\mbox{\cite{belo:2014}}\\
39 & \mbox{idiovol}   &\mbox{Idiosyncratic return volatility}   &\mbox{Monthly}  &\mbox{\cite{ali:2003}}\\
40 & \mbox{ill}   &\mbox{Illiquidity}   &\mbox{Monthly}  &\mbox{\cite{amihud:2002}}\\
41 & \mbox{indmom}   &\mbox{Industry momentum}   &\mbox{Monthly}  &\mbox{\cite{moskowitz:1999}}\\
42 & \mbox{invest}   &\mbox{Capital expenditures and inventory}   &\mbox{Annual}  &\mbox{\cite{chen:2010}}\\
43 & \mbox{lev}   &\mbox{Leverage}   &\mbox{Annual}  &\mbox{\cite{bhandari1988debt}}\\
44 & \mbox{lgr}   &\mbox{Growth in long-term debt}   &\mbox{Annual}  &\mbox{\cite{richardson2005accrual}}\\
45 & \mbox{maxret}   &\mbox{Maximum daily return}   &\mbox{Monthly}  &\mbox{\cite{bali2011maxing}}\\
46 & \mbox{mom12m}   &\mbox{12-month momentum}   &\mbox{Monthly}  &\mbox{\cite{jegadeesh:titman:1993}}\\
47 & \mbox{mom1m}   &\mbox{1-month momentum}   &\mbox{Monthly}  &\mbox{\cite{jegadeesh:titman:1993}}\\
48 & \mbox{mom36m}   &\mbox{36-month momentum}   &\mbox{Monthly}  &\mbox{\cite{jegadeesh:titman:1993}}\\
49 & \mbox{mom6m}   &\mbox{6-month momentum}   &\mbox{Monthly}  &\mbox{\cite{jegadeesh:titman:1993}}\\
50 & \mbox{ms}   &\mbox{Financial statement score}   &\mbox{Quarterly}  &\mbox{\cite{mohanram:2005}}\\
\bottomrule
\end{array}
$}
\end{align*}

\captionof{table}{\textbf{(Riskfactors)} This table lists the 94 financial characteristics we use in Section \ref{sec:Real}. We obtain the characteristics used by \cite{dacheng2020} from Dacheng Xiu's webpage; see \url{http://dachxiu.chicagobooth.edu}. Note that the data is collected in \cite{green:2017}.}
\label{table:factors1}
\end{center}
\end{small}
\end{table}

\begin{table}
\begin{small}
\begin{center}
\begin{align*}
\resizebox{12cm}{!}{$\displaystyle
\begin{array}{lllll}\toprule
\mbox{No.}& \mbox{Acronym}& \mbox{Firm Characteristic} & \mbox{Frequency} & \mbox{Literature} \\
\hline 
51 & \mbox{mvel1}   &\mbox{Size}   &\mbox{Monthly}  &\mbox{\cite{banz1981relationship}}\\
52 & \mbox{mveia}   &\mbox{Industry-adjusted size}   &\mbox{Annual}  &\mbox{\cite{asness:2000}}\\
53 & \mbox{nincr}   &\mbox{Number of earnings increases}   &\mbox{Quarterly}  &\mbox{\cite{barth:1999}}\\
54 & \mbox{operprof}   &\mbox{Operating profitability}   &\mbox{Annual}  &\mbox{\cite{fama:french:2015}}\\
55 & \mbox{orgcap}   &\mbox{Organizational capital}   &\mbox{Annual}  &\mbox{\cite{eisfeldt:2013}}\\
56 & \mbox{pchcapxia}   &\mbox{Industry adjusted change in capital exp.}   &\mbox{Annual}  &\mbox{\cite{abarbanell:1998}}\\
57 & \mbox{pchcurrat}   &\mbox{Change in current ratio}   &\mbox{Annual}  &\mbox{\cite{ou:1989}}\\
58 & \mbox{pchdepr}   &\mbox{Change in depreciation}   &\mbox{Annual}  &\mbox{\cite{holthausen:1992}}\\
59 & \mbox{pchgmpchsale}   &\mbox{Change in gross margin - change in sales}   &\mbox{Annual}  &\mbox{\cite{abarbanell:1998}}\\
60 & \mbox{pchquick}   &\mbox{Change in quick ratio}   &\mbox{Annual}  &\mbox{\cite{ou:1989}}\\
61 & \mbox{pchsalepchinvt}   &\mbox{Change in sales - change in inventory}   &\mbox{Annual}  &\mbox{\cite{abarbanell:1998}}\\
62 & \mbox{pchsalepchrect}   &\mbox{Change in sales -  change in A/R}   &\mbox{Annual}  &\mbox{\cite{abarbanell:1998}}\\
63 & \mbox{pchsalepchxsga}   &\mbox{Change in sales - change in SG\&A}   &\mbox{Annual}  &\mbox{\cite{abarbanell:1998}}\\
64 & \mbox{ppchsaleinv}   &\mbox{Change sales-to-inventory}   &\mbox{Annual}  &\mbox{\cite{ou:1989}}\\
65 & \mbox{pctacc}   &\mbox{Percent accruals}   &\mbox{Annual}  &\mbox{\cite{hafzalla2011percent}}\\
66 & \mbox{pricedelay}   &\mbox{Price delay}   &\mbox{Monthly}  &\mbox{\cite{hou:2005}}\\
67 & \mbox{ps}   &\mbox{Financial statements score}   &\mbox{Annual}  &\mbox{\cite{piotroski2000value}}\\
68 & \mbox{quick}   &\mbox{Quick ratio}   &\mbox{Annual}  &\mbox{\cite{ou:1989}}\\
69 & \mbox{rd}   &\mbox{R\&D increase}   &\mbox{Annual}  &\mbox{\cite{eberhart2004examination}}\\
70 & \mbox{rdmve}   &\mbox{R\&D to market capitalization}   &\mbox{Annual}  &\mbox{\cite{guo:2006}}\\
71 & \mbox{rdsale}   &\mbox{R\&D to sales}   &\mbox{Annual}  &\mbox{\cite{guo:2006}}\\
72 & \mbox{realestate}   &\mbox{Real estate holdings}   &\mbox{Annual}  &\mbox{\cite{tuzel:2010}}\\
73 & \mbox{retvol}   &\mbox{Return volatility}   &\mbox{Monthly}  &\mbox{\cite{ang2006cross}}\\
74 & \mbox{roaq}   &\mbox{Return on assets}   &\mbox{Quarterly}  &\mbox{\cite{balakrishnan2010post}}\\
75 & \mbox{roavol}   &\mbox{Earnings volatility}   &\mbox{Quarterly}  &\mbox{\cite{francis:2004}}\\
76 & \mbox{roeq}   &\mbox{Return on equity}   &\mbox{Quarterly}  &\mbox{\cite{hou2015digesting}}\\
77 & \mbox{roic}   &\mbox{Return on invested capital}   &\mbox{Annual}  &\mbox{\cite{brown:2007}}\\
78 & \mbox{rsup}   &\mbox{Revenue surprise}   &\mbox{Quarterly}  &\mbox{\cite{kama:2009}}\\
79 & \mbox{salecash}   &\mbox{Sales to cash}   &\mbox{Annual}  &\mbox{\cite{ou:1989}}\\
80 & \mbox{saleinv}   &\mbox{Sales to inventory}   &\mbox{Annual}  &\mbox{\cite{ou:1989}}\\
81 & \mbox{salerec}   &\mbox{Sales to receivables}   &\mbox{Annual}  &\mbox{\cite{ou:1989}}\\
82 & \mbox{secured}   &\mbox{Secured debt}   &\mbox{Annual}  &\mbox{\cite{valta:2016}}\\
83 & \mbox{securedind}   &\mbox{Secured debt indicator}   &\mbox{Annual}  &\mbox{\cite{valta:2016}}\\
84 & \mbox{sgr}   &\mbox{Sales growth}   &\mbox{Annual}  &\mbox{\cite{lakonishok1994contrarian}}\\
85 & \mbox{sin}   &\mbox{Sin stocks}   &\mbox{Annual}  &\mbox{\cite{hong:2009}}\\
86 & \mbox{sp}   &\mbox{Sales to price}   &\mbox{Annual}  &\mbox{\cite{barbee:1996}}\\
87 & \mbox{stddolvol}   &\mbox{Volatility of liquidity (dollar trading volume)}   &\mbox{Monthly}  &\mbox{\cite{chordia2001trading}}\\
88 & \mbox{stdturn}   &\mbox{Volatility of liquidity (share turnover)}   &\mbox{Monthly}  &\mbox{\cite{chordia2001trading}}\\
89 & \mbox{stdacc}   &\mbox{Accrual volatility}   &\mbox{Quarterly}  &\mbox{\cite{bandyopadhyay2010accrual}}\\
90 & \mbox{stdcf}   &\mbox{Cash flow volatility}   &\mbox{Quarterly}  &\mbox{\cite{huang2009cross}}\\
91 & \mbox{tang}   &\mbox{Debt capacity/firm tangibility}   &\mbox{Annual}  &\mbox{\cite{almeida:2007}}\\
92 & \mbox{tb}   &\mbox{Tax income to book income}   &\mbox{Annual}  &\mbox{\cite{lev:2004}}\\
93 & \mbox{turn}   &\mbox{Share turnover}   &\mbox{Monthly}  &\mbox{\cite{datar1998liquidity}}\\
94 & \mbox{zerotrade}   &\mbox{Zero trading days}   &\mbox{Monthly}  &\mbox{\cite{liu:2006}}\\
\bottomrule
\end{array}
$}
\end{align*}
\captionof{table}{\textbf{(Riskfactors)} Table \ref{table:factors1} continued.}
\label{table:factors2}
\end{center}
\end{small}
\end{table}

\end{document}

%% file: 4_contaminated_Binomial.tex
\begin{tikzpicture}[x=1pt,y=1pt]
\definecolor{fillColor}{RGB}{255,255,255}
\path[use as bounding box,fill=fillColor,fill opacity=0.00] (0,0) rectangle (433.62,289.08);
\begin{scope}
\path[clip] (  0.00,  0.00) rectangle (433.62,289.08);
\definecolor{drawColor}{RGB}{0,0,0}

\node[text=drawColor,anchor=base,inner sep=0pt, outer sep=0pt, scale=  1.20] at (228.81,260.29) {\bfseries Histogram of X};

\node[text=drawColor,anchor=base,inner sep=0pt, outer sep=0pt, scale=  1.00] at (228.81, 15.60) {X};

\node[text=drawColor,rotate= 90.00,anchor=base,inner sep=0pt, outer sep=0pt, scale=  1.00] at ( 10.80,150.54) {Frequency};
\end{scope}
\begin{scope}
\path[clip] (  0.00,  0.00) rectangle (433.62,289.08);
\definecolor{drawColor}{RGB}{0,0,0}

\path[draw=drawColor,line width= 0.4pt,line join=round,line cap=round] ( 72.90, 61.20) -- (384.72, 61.20);

\path[draw=drawColor,line width= 0.4pt,line join=round,line cap=round] ( 72.90, 61.20) -- ( 72.90, 55.20);

\path[draw=drawColor,line width= 0.4pt,line join=round,line cap=round] (124.87, 61.20) -- (124.87, 55.20);

\path[draw=drawColor,line width= 0.4pt,line join=round,line cap=round] (176.84, 61.20) -- (176.84, 55.20);

\path[draw=drawColor,line width= 0.4pt,line join=round,line cap=round] (228.81, 61.20) -- (228.81, 55.20);

\path[draw=drawColor,line width= 0.4pt,line join=round,line cap=round] (280.78, 61.20) -- (280.78, 55.20);

\path[draw=drawColor,line width= 0.4pt,line join=round,line cap=round] (332.75, 61.20) -- (332.75, 55.20);

\path[draw=drawColor,line width= 0.4pt,line join=round,line cap=round] (384.72, 61.20) -- (384.72, 55.20);

\node[text=drawColor,anchor=base,inner sep=0pt, outer sep=0pt, scale=  1.00] at ( 72.90, 39.60) {35};

\node[text=drawColor,anchor=base,inner sep=0pt, outer sep=0pt, scale=  1.00] at (124.87, 39.60) {40};

\node[text=drawColor,anchor=base,inner sep=0pt, outer sep=0pt, scale=  1.00] at (176.84, 39.60) {45};

\node[text=drawColor,anchor=base,inner sep=0pt, outer sep=0pt, scale=  1.00] at (228.81, 39.60) {50};

\node[text=drawColor,anchor=base,inner sep=0pt, outer sep=0pt, scale=  1.00] at (280.78, 39.60) {55};

\node[text=drawColor,anchor=base,inner sep=0pt, outer sep=0pt, scale=  1.00] at (332.75, 39.60) {60};

\node[text=drawColor,anchor=base,inner sep=0pt, outer sep=0pt, scale=  1.00] at (384.72, 39.60) {65};

\path[draw=drawColor,line width= 0.4pt,line join=round,line cap=round] ( 49.20, 67.82) -- ( 49.20,223.90);

\path[draw=drawColor,line width= 0.4pt,line join=round,line cap=round] ( 49.20, 67.82) -- ( 43.20, 67.82);

\path[draw=drawColor,line width= 0.4pt,line join=round,line cap=round] ( 49.20,119.84) -- ( 43.20,119.84);

\path[draw=drawColor,line width= 0.4pt,line join=round,line cap=round] ( 49.20,171.87) -- ( 43.20,171.87);

\path[draw=drawColor,line width= 0.4pt,line join=round,line cap=round] ( 49.20,223.90) -- ( 43.20,223.90);

\node[text=drawColor,rotate= 90.00,anchor=base,inner sep=0pt, outer sep=0pt, scale=  1.00] at ( 34.80, 67.82) {0};

\node[text=drawColor,rotate= 90.00,anchor=base,inner sep=0pt, outer sep=0pt, scale=  1.00] at ( 34.80,119.84) {50};

\node[text=drawColor,rotate= 90.00,anchor=base,inner sep=0pt, outer sep=0pt, scale=  1.00] at ( 34.80,171.87) {100};

\node[text=drawColor,rotate= 90.00,anchor=base,inner sep=0pt, outer sep=0pt, scale=  1.00] at ( 34.80,223.90) {150};
\end{scope}
\begin{scope}
\path[clip] ( 49.20, 61.20) rectangle (408.42,239.88);
\definecolor{drawColor}{RGB}{0,0,0}

\path[draw=drawColor,line width= 0.4pt,line join=round,line cap=round] ( 62.50, 67.82) rectangle ( 83.29, 70.94);

\path[draw=drawColor,line width= 0.4pt,line join=round,line cap=round] ( 83.29, 67.82) rectangle (104.08, 79.26);

\path[draw=drawColor,line width= 0.4pt,line join=round,line cap=round] (104.08, 67.82) rectangle (124.87, 90.71);

\path[draw=drawColor,line width= 0.4pt,line join=round,line cap=round] (124.87, 67.82) rectangle (145.66,100.07);

\path[draw=drawColor,line width= 0.4pt,line join=round,line cap=round] (145.66, 67.82) rectangle (166.45,128.17);

\path[draw=drawColor,line width= 0.4pt,line join=round,line cap=round] (166.45, 67.82) rectangle (187.23,170.83);

\path[draw=drawColor,line width= 0.4pt,line join=round,line cap=round] (187.23, 67.82) rectangle (208.02,210.37);

\path[draw=drawColor,line width= 0.4pt,line join=round,line cap=round] (208.02, 67.82) rectangle (228.81,233.26);

\path[draw=drawColor,line width= 0.4pt,line join=round,line cap=round] (228.81, 67.82) rectangle (249.60,225.98);

\path[draw=drawColor,line width= 0.4pt,line join=round,line cap=round] (249.60, 67.82) rectangle (270.39,198.92);

\path[draw=drawColor,line width= 0.4pt,line join=round,line cap=round] (270.39, 67.82) rectangle (291.17,160.43);

\path[draw=drawColor,line width= 0.4pt,line join=round,line cap=round] (291.17, 67.82) rectangle (311.96,134.41);

\path[draw=drawColor,line width= 0.4pt,line join=round,line cap=round] (311.96, 67.82) rectangle (332.75, 96.95);

\path[draw=drawColor,line width= 0.4pt,line join=round,line cap=round] (332.75, 67.82) rectangle (353.54, 80.30);

\path[draw=drawColor,line width= 0.4pt,line join=round,line cap=round] (353.54, 67.82) rectangle (374.33, 76.14);

\path[draw=drawColor,line width= 0.4pt,line join=round,line cap=round] (374.33, 67.82) rectangle (395.12, 68.86);
\end{scope}
\end{tikzpicture}

%% file: 4_contaminated_Gaussian.tex
\begin{tikzpicture}[x=1pt,y=1pt]
\definecolor{fillColor}{RGB}{255,255,255}
\path[use as bounding box,fill=fillColor,fill opacity=0.00] (0,0) rectangle (433.62,289.08);
\begin{scope}
\path[clip] (  0.00,  0.00) rectangle (433.62,289.08);
\definecolor{drawColor}{RGB}{0,0,0}

\node[text=drawColor,anchor=base,inner sep=0pt, outer sep=0pt, scale=  1.20] at (228.81,260.29) {\bfseries Histogram of Y};

\node[text=drawColor,anchor=base,inner sep=0pt, outer sep=0pt, scale=  1.00] at (228.81, 15.60) {Y};

\node[text=drawColor,rotate= 90.00,anchor=base,inner sep=0pt, outer sep=0pt, scale=  1.00] at ( 10.80,150.54) {Frequency};
\end{scope}
\begin{scope}
\path[clip] (  0.00,  0.00) rectangle (433.62,289.08);
\definecolor{drawColor}{RGB}{0,0,0}

\path[draw=drawColor,line width= 0.4pt,line join=round,line cap=round] ( 62.50, 61.20) -- (385.88, 61.20);

\path[draw=drawColor,line width= 0.4pt,line join=round,line cap=round] ( 62.50, 61.20) -- ( 62.50, 55.20);

\path[draw=drawColor,line width= 0.4pt,line join=round,line cap=round] (108.70, 61.20) -- (108.70, 55.20);

\path[draw=drawColor,line width= 0.4pt,line join=round,line cap=round] (154.90, 61.20) -- (154.90, 55.20);

\path[draw=drawColor,line width= 0.4pt,line join=round,line cap=round] (201.09, 61.20) -- (201.09, 55.20);

\path[draw=drawColor,line width= 0.4pt,line join=round,line cap=round] (247.29, 61.20) -- (247.29, 55.20);

\path[draw=drawColor,line width= 0.4pt,line join=round,line cap=round] (293.48, 61.20) -- (293.48, 55.20);

\path[draw=drawColor,line width= 0.4pt,line join=round,line cap=round] (339.68, 61.20) -- (339.68, 55.20);

\path[draw=drawColor,line width= 0.4pt,line join=round,line cap=round] (385.88, 61.20) -- (385.88, 55.20);

\node[text=drawColor,anchor=base,inner sep=0pt, outer sep=0pt, scale=  1.00] at ( 62.50, 39.60) {30};

\node[text=drawColor,anchor=base,inner sep=0pt, outer sep=0pt, scale=  1.00] at (108.70, 39.60) {35};

\node[text=drawColor,anchor=base,inner sep=0pt, outer sep=0pt, scale=  1.00] at (154.90, 39.60) {40};

\node[text=drawColor,anchor=base,inner sep=0pt, outer sep=0pt, scale=  1.00] at (201.09, 39.60) {45};

\node[text=drawColor,anchor=base,inner sep=0pt, outer sep=0pt, scale=  1.00] at (247.29, 39.60) {50};

\node[text=drawColor,anchor=base,inner sep=0pt, outer sep=0pt, scale=  1.00] at (293.48, 39.60) {55};

\node[text=drawColor,anchor=base,inner sep=0pt, outer sep=0pt, scale=  1.00] at (339.68, 39.60) {60};

\node[text=drawColor,anchor=base,inner sep=0pt, outer sep=0pt, scale=  1.00] at (385.88, 39.60) {65};

\path[draw=drawColor,line width= 0.4pt,line join=round,line cap=round] ( 49.20, 67.82) -- ( 49.20,209.63);

\path[draw=drawColor,line width= 0.4pt,line join=round,line cap=round] ( 49.20, 67.82) -- ( 43.20, 67.82);

\path[draw=drawColor,line width= 0.4pt,line join=round,line cap=round] ( 49.20,115.09) -- ( 43.20,115.09);

\path[draw=drawColor,line width= 0.4pt,line join=round,line cap=round] ( 49.20,162.36) -- ( 43.20,162.36);

\path[draw=drawColor,line width= 0.4pt,line join=round,line cap=round] ( 49.20,209.63) -- ( 43.20,209.63);

\node[text=drawColor,rotate= 90.00,anchor=base,inner sep=0pt, outer sep=0pt, scale=  1.00] at ( 34.80, 67.82) {0};

\node[text=drawColor,rotate= 90.00,anchor=base,inner sep=0pt, outer sep=0pt, scale=  1.00] at ( 34.80,115.09) {50};

\node[text=drawColor,rotate= 90.00,anchor=base,inner sep=0pt, outer sep=0pt, scale=  1.00] at ( 34.80,162.36) {100};

\node[text=drawColor,rotate= 90.00,anchor=base,inner sep=0pt, outer sep=0pt, scale=  1.00] at ( 34.80,209.63) {150};
\end{scope}
\begin{scope}
\path[clip] ( 49.20, 61.20) rectangle (408.42,239.88);
\definecolor{drawColor}{RGB}{0,0,0}

\path[draw=drawColor,line width= 0.4pt,line join=round,line cap=round] ( 62.50, 67.82) rectangle ( 80.98, 68.76);

\path[draw=drawColor,line width= 0.4pt,line join=round,line cap=round] ( 80.98, 67.82) rectangle ( 99.46, 67.82);

\path[draw=drawColor,line width= 0.4pt,line join=round,line cap=round] ( 99.46, 67.82) rectangle (117.94, 69.71);

\path[draw=drawColor,line width= 0.4pt,line join=round,line cap=round] (117.94, 67.82) rectangle (136.42, 74.44);

\path[draw=drawColor,line width= 0.4pt,line join=round,line cap=round] (136.42, 67.82) rectangle (154.90, 79.16);

\path[draw=drawColor,line width= 0.4pt,line join=round,line cap=round] (154.90, 67.82) rectangle (173.37,107.52);

\path[draw=drawColor,line width= 0.4pt,line join=round,line cap=round] (173.37, 67.82) rectangle (191.85,115.09);

\path[draw=drawColor,line width= 0.4pt,line join=round,line cap=round] (191.85, 67.82) rectangle (210.33,148.18);

\path[draw=drawColor,line width= 0.4pt,line join=round,line cap=round] (210.33, 67.82) rectangle (228.81,190.72);

\path[draw=drawColor,line width= 0.4pt,line join=round,line cap=round] (228.81, 67.82) rectangle (247.29,207.74);

\path[draw=drawColor,line width= 0.4pt,line join=round,line cap=round] (247.29, 67.82) rectangle (265.77,233.26);

\path[draw=drawColor,line width= 0.4pt,line join=round,line cap=round] (265.77, 67.82) rectangle (284.25,179.37);

\path[draw=drawColor,line width= 0.4pt,line join=round,line cap=round] (284.25, 67.82) rectangle (302.72,164.25);

\path[draw=drawColor,line width= 0.4pt,line join=round,line cap=round] (302.72, 67.82) rectangle (321.20,127.38);

\path[draw=drawColor,line width= 0.4pt,line join=round,line cap=round] (321.20, 67.82) rectangle (339.68,108.47);

\path[draw=drawColor,line width= 0.4pt,line join=round,line cap=round] (339.68, 67.82) rectangle (358.16, 80.11);

\path[draw=drawColor,line width= 0.4pt,line join=round,line cap=round] (358.16, 67.82) rectangle (376.64, 74.44);

\path[draw=drawColor,line width= 0.4pt,line join=round,line cap=round] (376.64, 67.82) rectangle (395.12, 69.71);
\end{scope}
\end{tikzpicture}

%% file: 4_plot_contamination_K100_Normapprox_F_with_Cai.tex
\begin{tikzpicture}[x=1pt,y=1pt]
\definecolor{fillColor}{RGB}{255,255,255}
\path[use as bounding box,fill=fillColor,fill opacity=0.00] (0,0) rectangle (505.89,289.08);
\begin{scope}
\path[clip] (  0.00,  0.00) rectangle (505.89,289.08);
\definecolor{drawColor}{RGB}{139,0,0}

\path[draw=drawColor,line width= 0.4pt,line join=round,line cap=round] ( 68.51, 78.72) -- ( 89.80, 79.25);

\path[draw=drawColor,line width= 0.4pt,line join=round,line cap=round] (101.62, 80.85) -- (123.26, 86.22);

\path[draw=drawColor,line width= 0.4pt,line join=round,line cap=round] (134.80, 89.52) -- (156.66, 96.58);

\path[draw=drawColor,line width= 0.4pt,line join=round,line cap=round] (168.12, 96.71) -- (189.91, 90.21);

\path[draw=drawColor,line width= 0.4pt,line join=round,line cap=round] (201.48, 89.95) -- (223.12, 95.32);

\path[draw=drawColor,line width= 0.4pt,line join=round,line cap=round] (234.90, 97.51) -- (256.28,100.17);

\path[draw=drawColor,line width= 0.4pt,line join=round,line cap=round] (267.98,102.62) -- (289.77,109.12);

\path[draw=drawColor,line width= 0.4pt,line join=round,line cap=round] (301.49,110.24) -- (322.83,108.12);

\path[draw=drawColor,line width= 0.4pt,line join=round,line cap=round] (334.66,108.83) -- (356.23,113.66);

\path[draw=drawColor,line width= 0.4pt,line join=round,line cap=round] (367.52,117.53) -- (389.95,128.12);
\definecolor{fillColor}{RGB}{139,0,0}

\path[draw=drawColor,line width= 0.4pt,line join=round,line cap=round,fill=fillColor] ( 62.51, 78.57) circle (  2.25);

\path[draw=drawColor,line width= 0.4pt,line join=round,line cap=round,fill=fillColor] ( 95.80, 79.40) circle (  2.25);

\path[draw=drawColor,line width= 0.4pt,line join=round,line cap=round,fill=fillColor] (129.09, 87.67) circle (  2.25);

\path[draw=drawColor,line width= 0.4pt,line join=round,line cap=round,fill=fillColor] (162.37, 98.43) circle (  2.25);

\path[draw=drawColor,line width= 0.4pt,line join=round,line cap=round,fill=fillColor] (195.66, 88.50) circle (  2.25);

\path[draw=drawColor,line width= 0.4pt,line join=round,line cap=round,fill=fillColor] (228.95, 96.77) circle (  2.25);

\path[draw=drawColor,line width= 0.4pt,line join=round,line cap=round,fill=fillColor] (262.23,100.91) circle (  2.25);

\path[draw=drawColor,line width= 0.4pt,line join=round,line cap=round,fill=fillColor] (295.52,110.83) circle (  2.25);

\path[draw=drawColor,line width= 0.4pt,line join=round,line cap=round,fill=fillColor] (328.80,107.52) circle (  2.25);

\path[draw=drawColor,line width= 0.4pt,line join=round,line cap=round,fill=fillColor] (362.09,114.97) circle (  2.25);

\path[draw=drawColor,line width= 0.4pt,line join=round,line cap=round,fill=fillColor] (395.38,130.69) circle (  2.25);
\end{scope}
\begin{scope}
\path[clip] (  0.00,  0.00) rectangle (505.89,289.08);
\definecolor{drawColor}{RGB}{0,0,0}

\path[draw=drawColor,line width= 0.4pt,line join=round,line cap=round] ( 62.51, 61.20) -- (395.38, 61.20);

\path[draw=drawColor,line width= 0.4pt,line join=round,line cap=round] ( 62.51, 61.20) -- ( 62.51, 55.20);

\path[draw=drawColor,line width= 0.4pt,line join=round,line cap=round] (129.09, 61.20) -- (129.09, 55.20);

\path[draw=drawColor,line width= 0.4pt,line join=round,line cap=round] (195.66, 61.20) -- (195.66, 55.20);

\path[draw=drawColor,line width= 0.4pt,line join=round,line cap=round] (262.23, 61.20) -- (262.23, 55.20);

\path[draw=drawColor,line width= 0.4pt,line join=round,line cap=round] (328.80, 61.20) -- (328.80, 55.20);

\path[draw=drawColor,line width= 0.4pt,line join=round,line cap=round] (395.38, 61.20) -- (395.38, 55.20);

\node[text=drawColor,anchor=base,inner sep=0pt, outer sep=0pt, scale=  1.00] at ( 62.51, 39.60) {0.5};

\node[text=drawColor,anchor=base,inner sep=0pt, outer sep=0pt, scale=  1.00] at (129.09, 39.60) {0.6};

\node[text=drawColor,anchor=base,inner sep=0pt, outer sep=0pt, scale=  1.00] at (195.66, 39.60) {0.7};

\node[text=drawColor,anchor=base,inner sep=0pt, outer sep=0pt, scale=  1.00] at (262.23, 39.60) {0.8};

\node[text=drawColor,anchor=base,inner sep=0pt, outer sep=0pt, scale=  1.00] at (328.80, 39.60) {0.9};

\node[text=drawColor,anchor=base,inner sep=0pt, outer sep=0pt, scale=  1.00] at (395.38, 39.60) {1.0};

\path[draw=drawColor,line width= 0.4pt,line join=round,line cap=round] ( 49.20, 67.82) -- ( 49.20,233.26);

\path[draw=drawColor,line width= 0.4pt,line join=round,line cap=round] ( 49.20, 67.82) -- ( 43.20, 67.82);

\path[draw=drawColor,line width= 0.4pt,line join=round,line cap=round] ( 49.20,100.91) -- ( 43.20,100.91);

\path[draw=drawColor,line width= 0.4pt,line join=round,line cap=round] ( 49.20,134.00) -- ( 43.20,134.00);

\path[draw=drawColor,line width= 0.4pt,line join=round,line cap=round] ( 49.20,167.08) -- ( 43.20,167.08);

\path[draw=drawColor,line width= 0.4pt,line join=round,line cap=round] ( 49.20,200.17) -- ( 43.20,200.17);

\path[draw=drawColor,line width= 0.4pt,line join=round,line cap=round] ( 49.20,233.26) -- ( 43.20,233.26);

\node[text=drawColor,rotate= 90.00,anchor=base,inner sep=0pt, outer sep=0pt, scale=  1.00] at ( 34.80, 67.82) {0.0};

\node[text=drawColor,rotate= 90.00,anchor=base,inner sep=0pt, outer sep=0pt, scale=  1.00] at ( 34.80,100.91) {0.2};

\node[text=drawColor,rotate= 90.00,anchor=base,inner sep=0pt, outer sep=0pt, scale=  1.00] at ( 34.80,134.00) {0.4};

\node[text=drawColor,rotate= 90.00,anchor=base,inner sep=0pt, outer sep=0pt, scale=  1.00] at ( 34.80,167.08) {0.6};

\node[text=drawColor,rotate= 90.00,anchor=base,inner sep=0pt, outer sep=0pt, scale=  1.00] at ( 34.80,200.17) {0.8};

\node[text=drawColor,rotate= 90.00,anchor=base,inner sep=0pt, outer sep=0pt, scale=  1.00] at ( 34.80,233.26) {1.0};

\path[draw=drawColor,line width= 0.4pt,line join=round,line cap=round] ( 49.20, 61.20) --
	(408.69, 61.20) --
	(408.69,239.88) --
	( 49.20,239.88) --
	( 49.20, 61.20);
\end{scope}
\begin{scope}
\path[clip] (  0.00,  0.00) rectangle (505.89,289.08);
\definecolor{drawColor}{RGB}{0,0,0}

\node[text=drawColor,anchor=base,inner sep=0pt, outer sep=0pt, scale=  1.00] at (228.95, 15.60) {$\lambda$};

\node[text=drawColor,rotate= 90.00,anchor=base,inner sep=0pt, outer sep=0pt, scale=  1.00] at ( 10.80,150.54) {Power};
\definecolor{drawColor}{RGB}{255,99,71}

\path[draw=drawColor,line width= 0.4pt,line join=round,line cap=round] ( 68.40, 78.91) -- ( 89.92, 83.19);

\path[draw=drawColor,line width= 0.4pt,line join=round,line cap=round] (101.80, 84.36) -- (123.09, 84.36);

\path[draw=drawColor,line width= 0.4pt,line join=round,line cap=round] (135.06, 84.96) -- (156.40, 87.08);

\path[draw=drawColor,line width= 0.4pt,line join=round,line cap=round] (168.34, 87.08) -- (189.69, 84.96);

\path[draw=drawColor,line width= 0.4pt,line join=round,line cap=round] (201.64, 84.81) -- (222.96, 86.40);

\path[draw=drawColor,line width= 0.4pt,line join=round,line cap=round] (234.93, 87.29) -- (256.25, 88.88);

\path[draw=drawColor,line width= 0.4pt,line join=round,line cap=round] (268.23, 89.18) -- (289.52, 88.65);

\path[draw=drawColor,line width= 0.4pt,line join=round,line cap=round] (301.52, 88.50) -- (322.80, 88.50);

\path[draw=drawColor,line width= 0.4pt,line join=round,line cap=round] (334.59, 90.08) -- (356.30, 96.02);

\path[draw=drawColor,line width= 0.4pt,line join=round,line cap=round] (367.97, 96.43) -- (389.49, 92.15);
\definecolor{fillColor}{RGB}{255,99,71}

\path[fill=fillColor] ( 62.51, 81.24) --
	( 65.54, 75.99) --
	( 59.48, 75.99) --
	cycle;

\path[fill=fillColor] ( 95.80, 87.86) --
	( 98.83, 82.61) --
	( 92.77, 82.61) --
	cycle;

\path[fill=fillColor] (129.09, 87.86) --
	(132.12, 82.61) --
	(126.06, 82.61) --
	cycle;

\path[fill=fillColor] (162.37, 91.17) --
	(165.40, 85.92) --
	(159.34, 85.92) --
	cycle;

\path[fill=fillColor] (195.66, 87.86) --
	(198.69, 82.61) --
	(192.63, 82.61) --
	cycle;

\path[fill=fillColor] (228.95, 90.34) --
	(231.98, 85.09) --
	(225.91, 85.09) --
	cycle;

\path[fill=fillColor] (262.23, 92.82) --
	(265.26, 87.58) --
	(259.20, 87.58) --
	cycle;

\path[fill=fillColor] (295.52, 92.00) --
	(298.55, 86.75) --
	(292.49, 86.75) --
	cycle;

\path[fill=fillColor] (328.80, 92.00) --
	(331.83, 86.75) --
	(325.77, 86.75) --
	cycle;

\path[fill=fillColor] (362.09,101.10) --
	(365.12, 95.85) --
	(359.06, 95.85) --
	cycle;

\path[fill=fillColor] (395.38, 94.48) --
	(398.41, 89.23) --
	(392.35, 89.23) --
	cycle;
\definecolor{drawColor}{RGB}{108,166,205}

\path[draw=drawColor,line width= 0.4pt,line join=round,line cap=round] ( 68.51, 76.39) -- ( 89.81, 77.45);

\path[draw=drawColor,line width= 0.4pt,line join=round,line cap=round] (101.78, 78.19) -- (123.10, 79.78);

\path[draw=drawColor,line width= 0.4pt,line join=round,line cap=round] (135.06, 79.63) -- (156.40, 77.51);

\path[draw=drawColor,line width= 0.4pt,line join=round,line cap=round] (168.37, 77.07) -- (189.66, 77.60);

\path[draw=drawColor,line width= 0.4pt,line join=round,line cap=round] (201.61, 78.48) -- (222.99, 81.14);

\path[draw=drawColor,line width= 0.4pt,line join=round,line cap=round] (234.88, 81.00) -- (256.30, 77.80);

\path[draw=drawColor,line width= 0.4pt,line join=round,line cap=round] (268.20, 76.32) -- (289.55, 74.20);

\path[draw=drawColor,line width= 0.4pt,line join=round,line cap=round] (301.40, 74.78) -- (322.92, 79.06);

\path[draw=drawColor,line width= 0.4pt,line join=round,line cap=round] (334.71, 79.20) -- (356.18, 75.46);

\path[draw=drawColor,line width= 0.4pt,line join=round,line cap=round] (368.08, 74.73) -- (389.38, 75.79);
\definecolor{fillColor}{RGB}{108,166,205}

\path[fill=fillColor] ( 60.26, 73.84) --
	( 64.76, 73.84) --
	( 64.76, 78.34) --
	( 60.26, 78.34) --
	cycle;

\path[fill=fillColor] ( 93.55, 75.49) --
	( 98.05, 75.49) --
	( 98.05, 79.99) --
	( 93.55, 79.99) --
	cycle;

\path[fill=fillColor] (126.84, 77.98) --
	(131.34, 77.98) --
	(131.34, 82.48) --
	(126.84, 82.48) --
	cycle;

\path[fill=fillColor] (160.12, 74.67) --
	(164.62, 74.67) --
	(164.62, 79.17) --
	(160.12, 79.17) --
	cycle;

\path[fill=fillColor] (193.41, 75.49) --
	(197.91, 75.49) --
	(197.91, 79.99) --
	(193.41, 79.99) --
	cycle;

\path[fill=fillColor] (226.70, 79.63) --
	(231.20, 79.63) --
	(231.20, 84.13) --
	(226.70, 84.13) --
	cycle;

\path[fill=fillColor] (259.98, 74.67) --
	(264.48, 74.67) --
	(264.48, 79.17) --
	(259.98, 79.17) --
	cycle;

\path[fill=fillColor] (293.27, 71.36) --
	(297.77, 71.36) --
	(297.77, 75.86) --
	(293.27, 75.86) --
	cycle;

\path[fill=fillColor] (326.55, 77.98) --
	(331.05, 77.98) --
	(331.05, 82.48) --
	(326.55, 82.48) --
	cycle;

\path[fill=fillColor] (359.84, 72.19) --
	(364.34, 72.19) --
	(364.34, 76.69) --
	(359.84, 76.69) --
	cycle;

\path[fill=fillColor] (393.13, 73.84) --
	(397.63, 73.84) --
	(397.63, 78.34) --
	(393.13, 78.34) --
	cycle;
\definecolor{drawColor}{RGB}{139,137,137}

\path[draw=drawColor,line width= 0.4pt,line join=round,line cap=round] ( 68.50, 79.02) -- ( 89.82, 80.61);

\path[draw=drawColor,line width= 0.4pt,line join=round,line cap=round] (101.75, 80.31) -- (123.13, 77.66);

\path[draw=drawColor,line width= 0.4pt,line join=round,line cap=round] (135.06, 76.32) -- (156.40, 74.20);

\path[draw=drawColor,line width= 0.4pt,line join=round,line cap=round] (168.31, 74.49) -- (189.72, 77.69);

\path[draw=drawColor,line width= 0.4pt,line join=round,line cap=round] (201.66, 78.42) -- (222.95, 77.89);

\path[draw=drawColor,line width= 0.4pt,line join=round,line cap=round] (234.94, 77.45) -- (256.24, 76.39);

\path[draw=drawColor,line width= 0.4pt,line join=round,line cap=round] (268.14, 75.06) -- (289.61, 71.33);

\path[draw=drawColor,line width= 0.4pt,line join=round,line cap=round] (301.47, 71.04) -- (322.85, 73.70);

\path[draw=drawColor,line width= 0.4pt,line join=round,line cap=round] (334.80, 74.44) -- (356.09, 74.44);

\path[draw=drawColor,line width= 0.4pt,line join=round,line cap=round] (368.02, 73.55) -- (389.44, 70.36);

\path[draw=drawColor,line width= 0.4pt,line join=round,line cap=round] ( 59.33, 78.57) -- ( 65.70, 78.57);

\path[draw=drawColor,line width= 0.4pt,line join=round,line cap=round] ( 62.51, 75.39) -- ( 62.51, 81.75);

\path[draw=drawColor,line width= 0.4pt,line join=round,line cap=round] ( 92.62, 81.05) -- ( 98.98, 81.05);

\path[draw=drawColor,line width= 0.4pt,line join=round,line cap=round] ( 95.80, 77.87) -- ( 95.80, 84.24);

\path[draw=drawColor,line width= 0.4pt,line join=round,line cap=round] (125.90, 76.92) -- (132.27, 76.92);

\path[draw=drawColor,line width= 0.4pt,line join=round,line cap=round] (129.09, 73.74) -- (129.09, 80.10);

\path[draw=drawColor,line width= 0.4pt,line join=round,line cap=round] (159.19, 73.61) -- (165.55, 73.61);

\path[draw=drawColor,line width= 0.4pt,line join=round,line cap=round] (162.37, 70.43) -- (162.37, 76.79);

\path[draw=drawColor,line width= 0.4pt,line join=round,line cap=round] (192.48, 78.57) -- (198.84, 78.57);

\path[draw=drawColor,line width= 0.4pt,line join=round,line cap=round] (195.66, 75.39) -- (195.66, 81.75);

\path[draw=drawColor,line width= 0.4pt,line join=round,line cap=round] (225.76, 77.74) -- (232.13, 77.74);

\path[draw=drawColor,line width= 0.4pt,line join=round,line cap=round] (228.95, 74.56) -- (228.95, 80.93);

\path[draw=drawColor,line width= 0.4pt,line join=round,line cap=round] (259.05, 76.09) -- (265.41, 76.09);

\path[draw=drawColor,line width= 0.4pt,line join=round,line cap=round] (262.23, 72.91) -- (262.23, 79.27);

\path[draw=drawColor,line width= 0.4pt,line join=round,line cap=round] (292.34, 70.30) -- (298.70, 70.30);

\path[draw=drawColor,line width= 0.4pt,line join=round,line cap=round] (295.52, 67.12) -- (295.52, 73.48);

\path[draw=drawColor,line width= 0.4pt,line join=round,line cap=round] (325.62, 74.44) -- (331.99, 74.44);

\path[draw=drawColor,line width= 0.4pt,line join=round,line cap=round] (328.80, 71.25) -- (328.80, 77.62);

\path[draw=drawColor,line width= 0.4pt,line join=round,line cap=round] (358.91, 74.44) -- (365.27, 74.44);

\path[draw=drawColor,line width= 0.4pt,line join=round,line cap=round] (362.09, 71.25) -- (362.09, 77.62);

\path[draw=drawColor,line width= 0.4pt,line join=round,line cap=round] (392.19, 69.47) -- (398.56, 69.47);

\path[draw=drawColor,line width= 0.4pt,line join=round,line cap=round] (395.38, 66.29) -- (395.38, 72.65);
\definecolor{drawColor}{RGB}{84,139,84}

\path[draw=drawColor,line width= 0.4pt,line join=round,line cap=round] ( 68.49, 75.50) -- ( 89.83, 73.37);

\path[draw=drawColor,line width= 0.4pt,line join=round,line cap=round] (101.59, 74.36) -- (123.30, 80.30);

\path[draw=drawColor,line width= 0.4pt,line join=round,line cap=round] (135.00, 80.85) -- (156.46, 77.12);

\path[draw=drawColor,line width= 0.4pt,line join=round,line cap=round] (168.37, 76.24) -- (189.66, 76.77);

\path[draw=drawColor,line width= 0.4pt,line join=round,line cap=round] (201.65, 76.62) -- (222.95, 75.56);

\path[draw=drawColor,line width= 0.4pt,line join=round,line cap=round] (234.77, 76.71) -- (256.41, 82.09);

\path[draw=drawColor,line width= 0.4pt,line join=round,line cap=round] (268.12, 82.37) -- (289.63, 78.09);

\path[draw=drawColor,line width= 0.4pt,line join=round,line cap=round] (301.51, 77.22) -- (322.81, 78.27);

\path[draw=drawColor,line width= 0.4pt,line join=round,line cap=round] (334.77, 77.98) -- (356.12, 75.86);

\path[draw=drawColor,line width= 0.4pt,line join=round,line cap=round] (368.07, 75.71) -- (389.39, 77.30);

\path[draw=drawColor,line width= 0.4pt,line join=round,line cap=round] ( 60.26, 73.84) -- ( 64.76, 78.34);

\path[draw=drawColor,line width= 0.4pt,line join=round,line cap=round] ( 60.26, 78.34) -- ( 64.76, 73.84);

\path[draw=drawColor,line width= 0.4pt,line join=round,line cap=round] ( 93.55, 70.53) -- ( 98.05, 75.03);

\path[draw=drawColor,line width= 0.4pt,line join=round,line cap=round] ( 93.55, 75.03) -- ( 98.05, 70.53);

\path[draw=drawColor,line width= 0.4pt,line join=round,line cap=round] (126.84, 79.63) -- (131.34, 84.13);

\path[draw=drawColor,line width= 0.4pt,line join=round,line cap=round] (126.84, 84.13) -- (131.34, 79.63);

\path[draw=drawColor,line width= 0.4pt,line join=round,line cap=round] (160.12, 73.84) -- (164.62, 78.34);

\path[draw=drawColor,line width= 0.4pt,line join=round,line cap=round] (160.12, 78.34) -- (164.62, 73.84);

\path[draw=drawColor,line width= 0.4pt,line join=round,line cap=round] (193.41, 74.67) -- (197.91, 79.17);

\path[draw=drawColor,line width= 0.4pt,line join=round,line cap=round] (193.41, 79.17) -- (197.91, 74.67);

\path[draw=drawColor,line width= 0.4pt,line join=round,line cap=round] (226.70, 73.01) -- (231.20, 77.51);

\path[draw=drawColor,line width= 0.4pt,line join=round,line cap=round] (226.70, 77.51) -- (231.20, 73.01);

\path[draw=drawColor,line width= 0.4pt,line join=round,line cap=round] (259.98, 81.28) -- (264.48, 85.78);

\path[draw=drawColor,line width= 0.4pt,line join=round,line cap=round] (259.98, 85.78) -- (264.48, 81.28);

\path[draw=drawColor,line width= 0.4pt,line join=round,line cap=round] (293.27, 74.67) -- (297.77, 79.17);

\path[draw=drawColor,line width= 0.4pt,line join=round,line cap=round] (293.27, 79.17) -- (297.77, 74.67);

\path[draw=drawColor,line width= 0.4pt,line join=round,line cap=round] (326.55, 76.32) -- (331.05, 80.82);

\path[draw=drawColor,line width= 0.4pt,line join=round,line cap=round] (326.55, 80.82) -- (331.05, 76.32);

\path[draw=drawColor,line width= 0.4pt,line join=round,line cap=round] (359.84, 73.01) -- (364.34, 77.51);

\path[draw=drawColor,line width= 0.4pt,line join=round,line cap=round] (359.84, 77.51) -- (364.34, 73.01);

\path[draw=drawColor,line width= 0.4pt,line join=round,line cap=round] (393.13, 75.49) -- (397.63, 79.99);

\path[draw=drawColor,line width= 0.4pt,line join=round,line cap=round] (393.13, 79.99) -- (397.63, 75.49);

\path[draw=drawColor,line width= 0.4pt,line join=round,line cap=round] ( 68.49, 75.50) -- ( 89.83, 73.37);

\path[draw=drawColor,line width= 0.4pt,line join=round,line cap=round] (101.59, 74.36) -- (123.30, 80.30);

\path[draw=drawColor,line width= 0.4pt,line join=round,line cap=round] (135.00, 80.85) -- (156.46, 77.12);

\path[draw=drawColor,line width= 0.4pt,line join=round,line cap=round] (168.37, 76.24) -- (189.66, 76.77);

\path[draw=drawColor,line width= 0.4pt,line join=round,line cap=round] (201.65, 76.62) -- (222.95, 75.56);

\path[draw=drawColor,line width= 0.4pt,line join=round,line cap=round] (234.77, 76.71) -- (256.41, 82.09);

\path[draw=drawColor,line width= 0.4pt,line join=round,line cap=round] (268.12, 82.37) -- (289.63, 78.09);

\path[draw=drawColor,line width= 0.4pt,line join=round,line cap=round] (301.51, 77.22) -- (322.81, 78.27);

\path[draw=drawColor,line width= 0.4pt,line join=round,line cap=round] (334.77, 77.98) -- (356.12, 75.86);

\path[draw=drawColor,line width= 0.4pt,line join=round,line cap=round] (368.07, 75.71) -- (389.39, 77.30);

\path[draw=drawColor,line width= 0.4pt,line join=round,line cap=round] ( 60.26, 73.84) -- ( 64.76, 78.34);

\path[draw=drawColor,line width= 0.4pt,line join=round,line cap=round] ( 60.26, 78.34) -- ( 64.76, 73.84);

\path[draw=drawColor,line width= 0.4pt,line join=round,line cap=round] ( 93.55, 70.53) -- ( 98.05, 75.03);

\path[draw=drawColor,line width= 0.4pt,line join=round,line cap=round] ( 93.55, 75.03) -- ( 98.05, 70.53);

\path[draw=drawColor,line width= 0.4pt,line join=round,line cap=round] (126.84, 79.63) -- (131.34, 84.13);

\path[draw=drawColor,line width= 0.4pt,line join=round,line cap=round] (126.84, 84.13) -- (131.34, 79.63);

\path[draw=drawColor,line width= 0.4pt,line join=round,line cap=round] (160.12, 73.84) -- (164.62, 78.34);

\path[draw=drawColor,line width= 0.4pt,line join=round,line cap=round] (160.12, 78.34) -- (164.62, 73.84);

\path[draw=drawColor,line width= 0.4pt,line join=round,line cap=round] (193.41, 74.67) -- (197.91, 79.17);

\path[draw=drawColor,line width= 0.4pt,line join=round,line cap=round] (193.41, 79.17) -- (197.91, 74.67);

\path[draw=drawColor,line width= 0.4pt,line join=round,line cap=round] (226.70, 73.01) -- (231.20, 77.51);

\path[draw=drawColor,line width= 0.4pt,line join=round,line cap=round] (226.70, 77.51) -- (231.20, 73.01);

\path[draw=drawColor,line width= 0.4pt,line join=round,line cap=round] (259.98, 81.28) -- (264.48, 85.78);

\path[draw=drawColor,line width= 0.4pt,line join=round,line cap=round] (259.98, 85.78) -- (264.48, 81.28);

\path[draw=drawColor,line width= 0.4pt,line join=round,line cap=round] (293.27, 74.67) -- (297.77, 79.17);

\path[draw=drawColor,line width= 0.4pt,line join=round,line cap=round] (293.27, 79.17) -- (297.77, 74.67);

\path[draw=drawColor,line width= 0.4pt,line join=round,line cap=round] (326.55, 76.32) -- (331.05, 80.82);

\path[draw=drawColor,line width= 0.4pt,line join=round,line cap=round] (326.55, 80.82) -- (331.05, 76.32);

\path[draw=drawColor,line width= 0.4pt,line join=round,line cap=round] (359.84, 73.01) -- (364.34, 77.51);

\path[draw=drawColor,line width= 0.4pt,line join=round,line cap=round] (359.84, 77.51) -- (364.34, 73.01);

\path[draw=drawColor,line width= 0.4pt,line join=round,line cap=round] (393.13, 75.49) -- (397.63, 79.99);

\path[draw=drawColor,line width= 0.4pt,line join=round,line cap=round] (393.13, 79.99) -- (397.63, 75.49);
\definecolor{drawColor}{RGB}{205,96,144}

\path[draw=drawColor,line width= 0.4pt,line join=round,line cap=round] ( 68.51, 82.03) -- ( 89.80, 82.56);

\path[draw=drawColor,line width= 0.4pt,line join=round,line cap=round] (101.79, 83.01) -- (123.09, 84.06);

\path[draw=drawColor,line width= 0.4pt,line join=round,line cap=round] (135.06, 84.96) -- (156.40, 87.08);

\path[draw=drawColor,line width= 0.4pt,line join=round,line cap=round] (168.33, 86.93) -- (189.70, 84.27);

\path[draw=drawColor,line width= 0.4pt,line join=round,line cap=round] (201.23, 85.75) -- (223.37, 94.55);

\path[draw=drawColor,line width= 0.4pt,line join=round,line cap=round] (234.94, 97.07) -- (256.24, 98.13);

\path[draw=drawColor,line width= 0.4pt,line join=round,line cap=round] (268.21, 97.98) -- (289.53, 96.39);

\path[draw=drawColor,line width= 0.4pt,line join=round,line cap=round] (301.27, 97.66) -- (323.05,104.16);

\path[draw=drawColor,line width= 0.4pt,line join=round,line cap=round] (334.69,107.04) -- (356.20,111.32);

\path[draw=drawColor,line width= 0.4pt,line join=round,line cap=round] (368.08,112.79) -- (389.38,113.84);

\path[draw=drawColor,line width= 0.4pt,line join=round,line cap=round] ( 60.26, 79.63) rectangle ( 64.76, 84.13);

\path[draw=drawColor,line width= 0.4pt,line join=round,line cap=round] ( 60.26, 79.63) -- ( 64.76, 84.13);

\path[draw=drawColor,line width= 0.4pt,line join=round,line cap=round] ( 60.26, 84.13) -- ( 64.76, 79.63);

\path[draw=drawColor,line width= 0.4pt,line join=round,line cap=round] ( 93.55, 80.46) rectangle ( 98.05, 84.96);

\path[draw=drawColor,line width= 0.4pt,line join=round,line cap=round] ( 93.55, 80.46) -- ( 98.05, 84.96);

\path[draw=drawColor,line width= 0.4pt,line join=round,line cap=round] ( 93.55, 84.96) -- ( 98.05, 80.46);

\path[draw=drawColor,line width= 0.4pt,line join=round,line cap=round] (126.84, 82.11) rectangle (131.34, 86.61);

\path[draw=drawColor,line width= 0.4pt,line join=round,line cap=round] (126.84, 82.11) -- (131.34, 86.61);

\path[draw=drawColor,line width= 0.4pt,line join=round,line cap=round] (126.84, 86.61) -- (131.34, 82.11);

\path[draw=drawColor,line width= 0.4pt,line join=round,line cap=round] (160.12, 85.42) rectangle (164.62, 89.92);

\path[draw=drawColor,line width= 0.4pt,line join=round,line cap=round] (160.12, 85.42) -- (164.62, 89.92);

\path[draw=drawColor,line width= 0.4pt,line join=round,line cap=round] (160.12, 89.92) -- (164.62, 85.42);

\path[draw=drawColor,line width= 0.4pt,line join=round,line cap=round] (193.41, 81.28) rectangle (197.91, 85.78);

\path[draw=drawColor,line width= 0.4pt,line join=round,line cap=round] (193.41, 81.28) -- (197.91, 85.78);

\path[draw=drawColor,line width= 0.4pt,line join=round,line cap=round] (193.41, 85.78) -- (197.91, 81.28);

\path[draw=drawColor,line width= 0.4pt,line join=round,line cap=round] (226.70, 94.52) rectangle (231.20, 99.02);

\path[draw=drawColor,line width= 0.4pt,line join=round,line cap=round] (226.70, 94.52) -- (231.20, 99.02);

\path[draw=drawColor,line width= 0.4pt,line join=round,line cap=round] (226.70, 99.02) -- (231.20, 94.52);

\path[draw=drawColor,line width= 0.4pt,line join=round,line cap=round] (259.98, 96.18) rectangle (264.48,100.68);

\path[draw=drawColor,line width= 0.4pt,line join=round,line cap=round] (259.98, 96.18) -- (264.48,100.68);

\path[draw=drawColor,line width= 0.4pt,line join=round,line cap=round] (259.98,100.68) -- (264.48, 96.18);

\path[draw=drawColor,line width= 0.4pt,line join=round,line cap=round] (293.27, 93.69) rectangle (297.77, 98.19);

\path[draw=drawColor,line width= 0.4pt,line join=round,line cap=round] (293.27, 93.69) -- (297.77, 98.19);

\path[draw=drawColor,line width= 0.4pt,line join=round,line cap=round] (293.27, 98.19) -- (297.77, 93.69);

\path[draw=drawColor,line width= 0.4pt,line join=round,line cap=round] (326.55,103.62) rectangle (331.05,108.12);

\path[draw=drawColor,line width= 0.4pt,line join=round,line cap=round] (326.55,103.62) -- (331.05,108.12);

\path[draw=drawColor,line width= 0.4pt,line join=round,line cap=round] (326.55,108.12) -- (331.05,103.62);

\path[draw=drawColor,line width= 0.4pt,line join=round,line cap=round] (359.84,110.24) rectangle (364.34,114.74);

\path[draw=drawColor,line width= 0.4pt,line join=round,line cap=round] (359.84,110.24) -- (364.34,114.74);

\path[draw=drawColor,line width= 0.4pt,line join=round,line cap=round] (359.84,114.74) -- (364.34,110.24);

\path[draw=drawColor,line width= 0.4pt,line join=round,line cap=round] (393.13,111.89) rectangle (397.63,116.39);

\path[draw=drawColor,line width= 0.4pt,line join=round,line cap=round] (393.13,111.89) -- (397.63,116.39);

\path[draw=drawColor,line width= 0.4pt,line join=round,line cap=round] (393.13,116.39) -- (397.63,111.89);
\definecolor{drawColor}{RGB}{0,0,0}

\path[draw=drawColor,line width= 0.4pt,dash pattern=on 4pt off 4pt ,line join=round,line cap=round] ( 61.18, 76.09) --
	( 64.57, 76.09) --
	( 67.96, 76.09) --
	( 71.35, 76.09) --
	( 74.74, 76.09) --
	( 78.13, 76.09) --
	( 81.52, 76.09) --
	( 84.91, 76.09) --
	( 88.30, 76.09) --
	( 91.69, 76.09) --
	( 95.07, 76.09) --
	( 98.46, 76.09) --
	(101.85, 76.09) --
	(105.24, 76.09) --
	(108.63, 76.09) --
	(112.02, 76.09) --
	(115.41, 76.09) --
	(118.80, 76.09) --
	(122.19, 76.09) --
	(125.58, 76.09) --
	(128.97, 76.09) --
	(132.35, 76.09) --
	(135.74, 76.09) --
	(139.13, 76.09) --
	(142.52, 76.09) --
	(145.91, 76.09) --
	(149.30, 76.09) --
	(152.69, 76.09) --
	(156.08, 76.09) --
	(159.47, 76.09) --
	(162.86, 76.09) --
	(166.25, 76.09) --
	(169.64, 76.09) --
	(173.02, 76.09) --
	(176.41, 76.09) --
	(179.80, 76.09) --
	(183.19, 76.09) --
	(186.58, 76.09) --
	(189.97, 76.09) --
	(193.36, 76.09) --
	(196.75, 76.09) --
	(200.14, 76.09) --
	(203.53, 76.09) --
	(206.92, 76.09) --
	(210.30, 76.09) --
	(213.69, 76.09) --
	(217.08, 76.09) --
	(220.47, 76.09) --
	(223.86, 76.09) --
	(227.25, 76.09) --
	(230.64, 76.09) --
	(234.03, 76.09) --
	(237.42, 76.09) --
	(240.81, 76.09) --
	(244.20, 76.09) --
	(247.59, 76.09) --
	(250.97, 76.09) --
	(254.36, 76.09) --
	(257.75, 76.09) --
	(261.14, 76.09) --
	(264.53, 76.09) --
	(267.92, 76.09) --
	(271.31, 76.09) --
	(274.70, 76.09) --
	(278.09, 76.09) --
	(281.48, 76.09) --
	(284.87, 76.09) --
	(288.25, 76.09) --
	(291.64, 76.09) --
	(295.03, 76.09) --
	(298.42, 76.09) --
	(301.81, 76.09) --
	(305.20, 76.09) --
	(308.59, 76.09) --
	(311.98, 76.09) --
	(315.37, 76.09) --
	(318.76, 76.09) --
	(322.15, 76.09) --
	(325.54, 76.09) --
	(328.92, 76.09) --
	(332.31, 76.09) --
	(335.70, 76.09) --
	(339.09, 76.09) --
	(342.48, 76.09) --
	(345.87, 76.09) --
	(349.26, 76.09) --
	(352.65, 76.09) --
	(356.04, 76.09) --
	(359.43, 76.09) --
	(362.82, 76.09) --
	(366.20, 76.09) --
	(369.59, 76.09) --
	(372.98, 76.09) --
	(376.37, 76.09) --
	(379.76, 76.09) --
	(383.15, 76.09) --
	(386.54, 76.09) --
	(389.93, 76.09) --
	(393.32, 76.09) --
	(396.71, 76.09);
\definecolor{drawColor}{RGB}{139,0,0}
\definecolor{fillColor}{RGB}{139,0,0}

\path[draw=drawColor,line width= 0.4pt,line join=round,line cap=round,fill=fillColor] (421.11,227.88) circle (  2.25);
\definecolor{fillColor}{RGB}{255,99,71}

\path[fill=fillColor] (421.11,219.38) --
	(424.15,214.13) --
	(418.08,214.13) --
	cycle;
\definecolor{fillColor}{RGB}{108,166,205}

\path[fill=fillColor] (418.86,201.63) --
	(423.36,201.63) --
	(423.36,206.13) --
	(418.86,206.13) --
	cycle;
\definecolor{drawColor}{RGB}{139,137,137}

\path[draw=drawColor,line width= 0.4pt,line join=round,line cap=round] (417.93,191.88) -- (424.30,191.88);

\path[draw=drawColor,line width= 0.4pt,line join=round,line cap=round] (421.11,188.70) -- (421.11,195.06);
\definecolor{drawColor}{RGB}{84,139,84}

\path[draw=drawColor,line width= 0.4pt,line join=round,line cap=round] (418.86,177.63) -- (423.36,182.13);

\path[draw=drawColor,line width= 0.4pt,line join=round,line cap=round] (418.86,182.13) -- (423.36,177.63);
\definecolor{drawColor}{RGB}{205,96,144}

\path[draw=drawColor,line width= 0.4pt,line join=round,line cap=round] (418.86,165.63) rectangle (423.36,170.13);

\path[draw=drawColor,line width= 0.4pt,line join=round,line cap=round] (418.86,165.63) -- (423.36,170.13);

\path[draw=drawColor,line width= 0.4pt,line join=round,line cap=round] (418.86,170.13) -- (423.36,165.63);
\definecolor{drawColor}{RGB}{0,0,0}

\node[text=drawColor,anchor=base west,inner sep=0pt, outer sep=0pt, scale=  1.00] at (430.11,224.44) {HypoRF};

\node[text=drawColor,anchor=base west,inner sep=0pt, outer sep=0pt, scale=  1.00] at (430.11,212.44) {Binomial};

\node[text=drawColor,anchor=base west,inner sep=0pt, outer sep=0pt, scale=  1.00] at (430.11,200.44) {ME-full};

\node[text=drawColor,anchor=base west,inner sep=0pt, outer sep=0pt, scale=  1.00] at (430.11,188.44) {MMDboot};

\node[text=drawColor,anchor=base west,inner sep=0pt, outer sep=0pt, scale=  1.00] at (430.11,176.44) {MMD-full};

\node[text=drawColor,anchor=base west,inner sep=0pt, outer sep=0pt, scale=  1.00] at (430.11,164.44) {CPT-RF};
\end{scope}
\end{tikzpicture}

%% file: 4_plot_contamination_K100_dp_Normapprox_F_with_Cai.tex
\begin{tikzpicture}[x=1pt,y=1pt]
\definecolor{fillColor}{RGB}{255,255,255}
\path[use as bounding box,fill=fillColor,fill opacity=0.00] (0,0) rectangle (505.89,289.08);
\begin{scope}
\path[clip] (  0.00,  0.00) rectangle (505.89,289.08);
\definecolor{drawColor}{RGB}{139,0,0}

\path[draw=drawColor,line width= 0.4pt,line join=round,line cap=round] ( 67.67,185.05) -- ( 90.65,198.75);

\path[draw=drawColor,line width= 0.4pt,line join=round,line cap=round] (101.07,204.71) -- (123.82,217.15);

\path[draw=drawColor,line width= 0.4pt,line join=round,line cap=round] (135.00,221.06) -- (156.46,224.79);

\path[draw=drawColor,line width= 0.4pt,line join=round,line cap=round] (168.23,227.13) -- (189.80,231.95);

\path[draw=drawColor,line width= 0.4pt,line join=round,line cap=round] (201.66,233.26) -- (222.94,233.26);

\path[draw=drawColor,line width= 0.4pt,line join=round,line cap=round] (234.95,233.26) -- (256.23,233.26);

\path[draw=drawColor,line width= 0.4pt,line join=round,line cap=round] (268.23,233.26) -- (289.52,233.26);

\path[draw=drawColor,line width= 0.4pt,line join=round,line cap=round] (301.52,233.26) -- (322.80,233.26);

\path[draw=drawColor,line width= 0.4pt,line join=round,line cap=round] (334.80,233.26) -- (356.09,233.26);

\path[draw=drawColor,line width= 0.4pt,line join=round,line cap=round] (368.09,233.26) -- (389.38,233.26);
\definecolor{fillColor}{RGB}{139,0,0}

\path[draw=drawColor,line width= 0.4pt,line join=round,line cap=round,fill=fillColor] ( 62.51,181.97) circle (  2.25);

\path[draw=drawColor,line width= 0.4pt,line join=round,line cap=round,fill=fillColor] ( 95.80,201.83) circle (  2.25);

\path[draw=drawColor,line width= 0.4pt,line join=round,line cap=round,fill=fillColor] (129.09,220.03) circle (  2.25);

\path[draw=drawColor,line width= 0.4pt,line join=round,line cap=round,fill=fillColor] (162.37,225.82) circle (  2.25);

\path[draw=drawColor,line width= 0.4pt,line join=round,line cap=round,fill=fillColor] (195.66,233.26) circle (  2.25);

\path[draw=drawColor,line width= 0.4pt,line join=round,line cap=round,fill=fillColor] (228.95,233.26) circle (  2.25);

\path[draw=drawColor,line width= 0.4pt,line join=round,line cap=round,fill=fillColor] (262.23,233.26) circle (  2.25);

\path[draw=drawColor,line width= 0.4pt,line join=round,line cap=round,fill=fillColor] (295.52,233.26) circle (  2.25);

\path[draw=drawColor,line width= 0.4pt,line join=round,line cap=round,fill=fillColor] (328.80,233.26) circle (  2.25);

\path[draw=drawColor,line width= 0.4pt,line join=round,line cap=round,fill=fillColor] (362.09,233.26) circle (  2.25);

\path[draw=drawColor,line width= 0.4pt,line join=round,line cap=round,fill=fillColor] (395.38,233.26) circle (  2.25);
\end{scope}
\begin{scope}
\path[clip] (  0.00,  0.00) rectangle (505.89,289.08);
\definecolor{drawColor}{RGB}{0,0,0}

\path[draw=drawColor,line width= 0.4pt,line join=round,line cap=round] ( 62.51, 61.20) -- (395.38, 61.20);

\path[draw=drawColor,line width= 0.4pt,line join=round,line cap=round] ( 62.51, 61.20) -- ( 62.51, 55.20);

\path[draw=drawColor,line width= 0.4pt,line join=round,line cap=round] (129.09, 61.20) -- (129.09, 55.20);

\path[draw=drawColor,line width= 0.4pt,line join=round,line cap=round] (195.66, 61.20) -- (195.66, 55.20);

\path[draw=drawColor,line width= 0.4pt,line join=round,line cap=round] (262.23, 61.20) -- (262.23, 55.20);

\path[draw=drawColor,line width= 0.4pt,line join=round,line cap=round] (328.80, 61.20) -- (328.80, 55.20);

\path[draw=drawColor,line width= 0.4pt,line join=round,line cap=round] (395.38, 61.20) -- (395.38, 55.20);

\node[text=drawColor,anchor=base,inner sep=0pt, outer sep=0pt, scale=  1.00] at ( 62.51, 39.60) {0.5};

\node[text=drawColor,anchor=base,inner sep=0pt, outer sep=0pt, scale=  1.00] at (129.09, 39.60) {0.6};

\node[text=drawColor,anchor=base,inner sep=0pt, outer sep=0pt, scale=  1.00] at (195.66, 39.60) {0.7};

\node[text=drawColor,anchor=base,inner sep=0pt, outer sep=0pt, scale=  1.00] at (262.23, 39.60) {0.8};

\node[text=drawColor,anchor=base,inner sep=0pt, outer sep=0pt, scale=  1.00] at (328.80, 39.60) {0.9};

\node[text=drawColor,anchor=base,inner sep=0pt, outer sep=0pt, scale=  1.00] at (395.38, 39.60) {1.0};

\path[draw=drawColor,line width= 0.4pt,line join=round,line cap=round] ( 49.20, 67.82) -- ( 49.20,233.26);

\path[draw=drawColor,line width= 0.4pt,line join=round,line cap=round] ( 49.20, 67.82) -- ( 43.20, 67.82);

\path[draw=drawColor,line width= 0.4pt,line join=round,line cap=round] ( 49.20,100.91) -- ( 43.20,100.91);

\path[draw=drawColor,line width= 0.4pt,line join=round,line cap=round] ( 49.20,134.00) -- ( 43.20,134.00);

\path[draw=drawColor,line width= 0.4pt,line join=round,line cap=round] ( 49.20,167.08) -- ( 43.20,167.08);

\path[draw=drawColor,line width= 0.4pt,line join=round,line cap=round] ( 49.20,200.17) -- ( 43.20,200.17);

\path[draw=drawColor,line width= 0.4pt,line join=round,line cap=round] ( 49.20,233.26) -- ( 43.20,233.26);

\node[text=drawColor,rotate= 90.00,anchor=base,inner sep=0pt, outer sep=0pt, scale=  1.00] at ( 34.80, 67.82) {0.0};

\node[text=drawColor,rotate= 90.00,anchor=base,inner sep=0pt, outer sep=0pt, scale=  1.00] at ( 34.80,100.91) {0.2};

\node[text=drawColor,rotate= 90.00,anchor=base,inner sep=0pt, outer sep=0pt, scale=  1.00] at ( 34.80,134.00) {0.4};

\node[text=drawColor,rotate= 90.00,anchor=base,inner sep=0pt, outer sep=0pt, scale=  1.00] at ( 34.80,167.08) {0.6};

\node[text=drawColor,rotate= 90.00,anchor=base,inner sep=0pt, outer sep=0pt, scale=  1.00] at ( 34.80,200.17) {0.8};

\node[text=drawColor,rotate= 90.00,anchor=base,inner sep=0pt, outer sep=0pt, scale=  1.00] at ( 34.80,233.26) {1.0};

\path[draw=drawColor,line width= 0.4pt,line join=round,line cap=round] ( 49.20, 61.20) --
	(408.69, 61.20) --
	(408.69,239.88) --
	( 49.20,239.88) --
	( 49.20, 61.20);
\end{scope}
\begin{scope}
\path[clip] (  0.00,  0.00) rectangle (505.89,289.08);
\definecolor{drawColor}{RGB}{0,0,0}

\node[text=drawColor,anchor=base,inner sep=0pt, outer sep=0pt, scale=  1.00] at (228.95, 15.60) {$\lambda$};

\node[text=drawColor,rotate= 90.00,anchor=base,inner sep=0pt, outer sep=0pt, scale=  1.00] at ( 10.80,150.54) {Power};
\definecolor{drawColor}{RGB}{255,99,71}

\path[draw=drawColor,line width= 0.4pt,line join=round,line cap=round] ( 67.83,159.93) -- ( 90.48,171.75);

\path[draw=drawColor,line width= 0.4pt,line join=round,line cap=round] (101.07,177.41) -- (123.82,189.85);

\path[draw=drawColor,line width= 0.4pt,line join=round,line cap=round] (134.35,195.61) -- (157.11,208.05);

\path[draw=drawColor,line width= 0.4pt,line join=round,line cap=round] (167.99,213.02) -- (190.04,221.24);

\path[draw=drawColor,line width= 0.4pt,line join=round,line cap=round] (201.54,224.51) -- (223.06,228.78);

\path[draw=drawColor,line width= 0.4pt,line join=round,line cap=round] (234.94,230.10) -- (256.23,230.63);

\path[draw=drawColor,line width= 0.4pt,line join=round,line cap=round] (268.21,231.23) -- (289.53,232.82);

\path[draw=drawColor,line width= 0.4pt,line join=round,line cap=round] (301.52,233.26) -- (322.80,233.26);

\path[draw=drawColor,line width= 0.4pt,line join=round,line cap=round] (334.80,233.26) -- (356.09,233.26);

\path[draw=drawColor,line width= 0.4pt,line join=round,line cap=round] (368.09,233.26) -- (389.38,233.26);
\definecolor{fillColor}{RGB}{255,99,71}

\path[fill=fillColor] ( 62.51,160.66) --
	( 65.54,155.41) --
	( 59.48,155.41) --
	cycle;

\path[fill=fillColor] ( 95.80,178.03) --
	( 98.83,172.78) --
	( 92.77,172.78) --
	cycle;

\path[fill=fillColor] (129.09,196.23) --
	(132.12,190.98) --
	(126.06,190.98) --
	cycle;

\path[fill=fillColor] (162.37,214.43) --
	(165.40,209.18) --
	(159.34,209.18) --
	cycle;

\path[fill=fillColor] (195.66,226.83) --
	(198.69,221.59) --
	(192.63,221.59) --
	cycle;

\path[fill=fillColor] (228.95,233.45) --
	(231.98,228.20) --
	(225.91,228.20) --
	cycle;

\path[fill=fillColor] (262.23,234.28) --
	(265.26,229.03) --
	(259.20,229.03) --
	cycle;

\path[fill=fillColor] (295.52,236.76) --
	(298.55,231.51) --
	(292.49,231.51) --
	cycle;

\path[fill=fillColor] (328.80,236.76) --
	(331.83,231.51) --
	(325.77,231.51) --
	cycle;

\path[fill=fillColor] (362.09,236.76) --
	(365.12,231.51) --
	(359.06,231.51) --
	cycle;

\path[fill=fillColor] (395.38,236.76) --
	(398.41,231.51) --
	(392.35,231.51) --
	cycle;
\definecolor{drawColor}{RGB}{108,166,205}

\path[draw=drawColor,line width= 0.4pt,line join=round,line cap=round] ( 68.51, 76.39) -- ( 89.81, 77.45);

\path[draw=drawColor,line width= 0.4pt,line join=round,line cap=round] (101.78, 78.19) -- (123.10, 79.78);

\path[draw=drawColor,line width= 0.4pt,line join=round,line cap=round] (135.06, 79.63) -- (156.40, 77.51);

\path[draw=drawColor,line width= 0.4pt,line join=round,line cap=round] (168.37, 77.07) -- (189.66, 77.60);

\path[draw=drawColor,line width= 0.4pt,line join=round,line cap=round] (201.61, 78.48) -- (222.99, 81.14);

\path[draw=drawColor,line width= 0.4pt,line join=round,line cap=round] (234.88, 81.00) -- (256.30, 77.80);

\path[draw=drawColor,line width= 0.4pt,line join=round,line cap=round] (268.20, 76.32) -- (289.55, 74.20);

\path[draw=drawColor,line width= 0.4pt,line join=round,line cap=round] (301.40, 74.78) -- (322.92, 79.06);

\path[draw=drawColor,line width= 0.4pt,line join=round,line cap=round] (334.71, 79.20) -- (356.18, 75.46);

\path[draw=drawColor,line width= 0.4pt,line join=round,line cap=round] (368.08, 74.73) -- (389.38, 75.79);
\definecolor{fillColor}{RGB}{108,166,205}

\path[fill=fillColor] ( 60.26, 73.84) --
	( 64.76, 73.84) --
	( 64.76, 78.34) --
	( 60.26, 78.34) --
	cycle;

\path[fill=fillColor] ( 93.55, 75.49) --
	( 98.05, 75.49) --
	( 98.05, 79.99) --
	( 93.55, 79.99) --
	cycle;

\path[fill=fillColor] (126.84, 77.98) --
	(131.34, 77.98) --
	(131.34, 82.48) --
	(126.84, 82.48) --
	cycle;

\path[fill=fillColor] (160.12, 74.67) --
	(164.62, 74.67) --
	(164.62, 79.17) --
	(160.12, 79.17) --
	cycle;

\path[fill=fillColor] (193.41, 75.49) --
	(197.91, 75.49) --
	(197.91, 79.99) --
	(193.41, 79.99) --
	cycle;

\path[fill=fillColor] (226.70, 79.63) --
	(231.20, 79.63) --
	(231.20, 84.13) --
	(226.70, 84.13) --
	cycle;

\path[fill=fillColor] (259.98, 74.67) --
	(264.48, 74.67) --
	(264.48, 79.17) --
	(259.98, 79.17) --
	cycle;

\path[fill=fillColor] (293.27, 71.36) --
	(297.77, 71.36) --
	(297.77, 75.86) --
	(293.27, 75.86) --
	cycle;

\path[fill=fillColor] (326.55, 77.98) --
	(331.05, 77.98) --
	(331.05, 82.48) --
	(326.55, 82.48) --
	cycle;

\path[fill=fillColor] (359.84, 72.19) --
	(364.34, 72.19) --
	(364.34, 76.69) --
	(359.84, 76.69) --
	cycle;

\path[fill=fillColor] (393.13, 73.84) --
	(397.63, 73.84) --
	(397.63, 78.34) --
	(393.13, 78.34) --
	cycle;
\definecolor{drawColor}{RGB}{139,137,137}

\path[draw=drawColor,line width= 0.4pt,line join=round,line cap=round] ( 68.50, 79.02) -- ( 89.82, 80.61);

\path[draw=drawColor,line width= 0.4pt,line join=round,line cap=round] (101.75, 80.31) -- (123.13, 77.66);

\path[draw=drawColor,line width= 0.4pt,line join=round,line cap=round] (135.06, 76.32) -- (156.40, 74.20);

\path[draw=drawColor,line width= 0.4pt,line join=round,line cap=round] (168.31, 74.49) -- (189.72, 77.69);

\path[draw=drawColor,line width= 0.4pt,line join=round,line cap=round] (201.66, 78.42) -- (222.95, 77.89);

\path[draw=drawColor,line width= 0.4pt,line join=round,line cap=round] (234.94, 77.45) -- (256.24, 76.39);

\path[draw=drawColor,line width= 0.4pt,line join=round,line cap=round] (268.14, 75.06) -- (289.61, 71.33);

\path[draw=drawColor,line width= 0.4pt,line join=round,line cap=round] (301.47, 71.04) -- (322.85, 73.70);

\path[draw=drawColor,line width= 0.4pt,line join=round,line cap=round] (334.80, 74.44) -- (356.09, 74.44);

\path[draw=drawColor,line width= 0.4pt,line join=round,line cap=round] (368.02, 73.55) -- (389.44, 70.36);

\path[draw=drawColor,line width= 0.4pt,line join=round,line cap=round] ( 59.33, 78.57) -- ( 65.70, 78.57);

\path[draw=drawColor,line width= 0.4pt,line join=round,line cap=round] ( 62.51, 75.39) -- ( 62.51, 81.75);

\path[draw=drawColor,line width= 0.4pt,line join=round,line cap=round] ( 92.62, 81.05) -- ( 98.98, 81.05);

\path[draw=drawColor,line width= 0.4pt,line join=round,line cap=round] ( 95.80, 77.87) -- ( 95.80, 84.24);

\path[draw=drawColor,line width= 0.4pt,line join=round,line cap=round] (125.90, 76.92) -- (132.27, 76.92);

\path[draw=drawColor,line width= 0.4pt,line join=round,line cap=round] (129.09, 73.74) -- (129.09, 80.10);

\path[draw=drawColor,line width= 0.4pt,line join=round,line cap=round] (159.19, 73.61) -- (165.55, 73.61);

\path[draw=drawColor,line width= 0.4pt,line join=round,line cap=round] (162.37, 70.43) -- (162.37, 76.79);

\path[draw=drawColor,line width= 0.4pt,line join=round,line cap=round] (192.48, 78.57) -- (198.84, 78.57);

\path[draw=drawColor,line width= 0.4pt,line join=round,line cap=round] (195.66, 75.39) -- (195.66, 81.75);

\path[draw=drawColor,line width= 0.4pt,line join=round,line cap=round] (225.76, 77.74) -- (232.13, 77.74);

\path[draw=drawColor,line width= 0.4pt,line join=round,line cap=round] (228.95, 74.56) -- (228.95, 80.93);

\path[draw=drawColor,line width= 0.4pt,line join=round,line cap=round] (259.05, 76.09) -- (265.41, 76.09);

\path[draw=drawColor,line width= 0.4pt,line join=round,line cap=round] (262.23, 72.91) -- (262.23, 79.27);

\path[draw=drawColor,line width= 0.4pt,line join=round,line cap=round] (292.34, 70.30) -- (298.70, 70.30);

\path[draw=drawColor,line width= 0.4pt,line join=round,line cap=round] (295.52, 67.12) -- (295.52, 73.48);

\path[draw=drawColor,line width= 0.4pt,line join=round,line cap=round] (325.62, 74.44) -- (331.99, 74.44);

\path[draw=drawColor,line width= 0.4pt,line join=round,line cap=round] (328.80, 71.25) -- (328.80, 77.62);

\path[draw=drawColor,line width= 0.4pt,line join=round,line cap=round] (358.91, 74.44) -- (365.27, 74.44);

\path[draw=drawColor,line width= 0.4pt,line join=round,line cap=round] (362.09, 71.25) -- (362.09, 77.62);

\path[draw=drawColor,line width= 0.4pt,line join=round,line cap=round] (392.19, 69.47) -- (398.56, 69.47);

\path[draw=drawColor,line width= 0.4pt,line join=round,line cap=round] (395.38, 66.29) -- (395.38, 72.65);
\definecolor{drawColor}{RGB}{84,139,84}

\path[draw=drawColor,line width= 0.4pt,line join=round,line cap=round] ( 68.49, 75.50) -- ( 89.83, 73.37);

\path[draw=drawColor,line width= 0.4pt,line join=round,line cap=round] (101.59, 74.36) -- (123.30, 80.30);

\path[draw=drawColor,line width= 0.4pt,line join=round,line cap=round] (135.00, 80.85) -- (156.46, 77.12);

\path[draw=drawColor,line width= 0.4pt,line join=round,line cap=round] (168.37, 76.24) -- (189.66, 76.77);

\path[draw=drawColor,line width= 0.4pt,line join=round,line cap=round] (201.65, 76.62) -- (222.95, 75.56);

\path[draw=drawColor,line width= 0.4pt,line join=round,line cap=round] (234.77, 76.71) -- (256.41, 82.09);

\path[draw=drawColor,line width= 0.4pt,line join=round,line cap=round] (268.12, 82.37) -- (289.63, 78.09);

\path[draw=drawColor,line width= 0.4pt,line join=round,line cap=round] (301.51, 77.22) -- (322.81, 78.27);

\path[draw=drawColor,line width= 0.4pt,line join=round,line cap=round] (334.77, 77.98) -- (356.12, 75.86);

\path[draw=drawColor,line width= 0.4pt,line join=round,line cap=round] (368.07, 75.71) -- (389.39, 77.30);

\path[draw=drawColor,line width= 0.4pt,line join=round,line cap=round] ( 60.26, 73.84) -- ( 64.76, 78.34);

\path[draw=drawColor,line width= 0.4pt,line join=round,line cap=round] ( 60.26, 78.34) -- ( 64.76, 73.84);

\path[draw=drawColor,line width= 0.4pt,line join=round,line cap=round] ( 93.55, 70.53) -- ( 98.05, 75.03);

\path[draw=drawColor,line width= 0.4pt,line join=round,line cap=round] ( 93.55, 75.03) -- ( 98.05, 70.53);

\path[draw=drawColor,line width= 0.4pt,line join=round,line cap=round] (126.84, 79.63) -- (131.34, 84.13);

\path[draw=drawColor,line width= 0.4pt,line join=round,line cap=round] (126.84, 84.13) -- (131.34, 79.63);

\path[draw=drawColor,line width= 0.4pt,line join=round,line cap=round] (160.12, 73.84) -- (164.62, 78.34);

\path[draw=drawColor,line width= 0.4pt,line join=round,line cap=round] (160.12, 78.34) -- (164.62, 73.84);

\path[draw=drawColor,line width= 0.4pt,line join=round,line cap=round] (193.41, 74.67) -- (197.91, 79.17);

\path[draw=drawColor,line width= 0.4pt,line join=round,line cap=round] (193.41, 79.17) -- (197.91, 74.67);

\path[draw=drawColor,line width= 0.4pt,line join=round,line cap=round] (226.70, 73.01) -- (231.20, 77.51);

\path[draw=drawColor,line width= 0.4pt,line join=round,line cap=round] (226.70, 77.51) -- (231.20, 73.01);

\path[draw=drawColor,line width= 0.4pt,line join=round,line cap=round] (259.98, 81.28) -- (264.48, 85.78);

\path[draw=drawColor,line width= 0.4pt,line join=round,line cap=round] (259.98, 85.78) -- (264.48, 81.28);

\path[draw=drawColor,line width= 0.4pt,line join=round,line cap=round] (293.27, 74.67) -- (297.77, 79.17);

\path[draw=drawColor,line width= 0.4pt,line join=round,line cap=round] (293.27, 79.17) -- (297.77, 74.67);

\path[draw=drawColor,line width= 0.4pt,line join=round,line cap=round] (326.55, 76.32) -- (331.05, 80.82);

\path[draw=drawColor,line width= 0.4pt,line join=round,line cap=round] (326.55, 80.82) -- (331.05, 76.32);

\path[draw=drawColor,line width= 0.4pt,line join=round,line cap=round] (359.84, 73.01) -- (364.34, 77.51);

\path[draw=drawColor,line width= 0.4pt,line join=round,line cap=round] (359.84, 77.51) -- (364.34, 73.01);

\path[draw=drawColor,line width= 0.4pt,line join=round,line cap=round] (393.13, 75.49) -- (397.63, 79.99);

\path[draw=drawColor,line width= 0.4pt,line join=round,line cap=round] (393.13, 79.99) -- (397.63, 75.49);

\path[draw=drawColor,line width= 0.4pt,line join=round,line cap=round] ( 68.49, 75.50) -- ( 89.83, 73.37);

\path[draw=drawColor,line width= 0.4pt,line join=round,line cap=round] (101.59, 74.36) -- (123.30, 80.30);

\path[draw=drawColor,line width= 0.4pt,line join=round,line cap=round] (135.00, 80.85) -- (156.46, 77.12);

\path[draw=drawColor,line width= 0.4pt,line join=round,line cap=round] (168.37, 76.24) -- (189.66, 76.77);

\path[draw=drawColor,line width= 0.4pt,line join=round,line cap=round] (201.65, 76.62) -- (222.95, 75.56);

\path[draw=drawColor,line width= 0.4pt,line join=round,line cap=round] (234.77, 76.71) -- (256.41, 82.09);

\path[draw=drawColor,line width= 0.4pt,line join=round,line cap=round] (268.12, 82.37) -- (289.63, 78.09);

\path[draw=drawColor,line width= 0.4pt,line join=round,line cap=round] (301.51, 77.22) -- (322.81, 78.27);

\path[draw=drawColor,line width= 0.4pt,line join=round,line cap=round] (334.77, 77.98) -- (356.12, 75.86);

\path[draw=drawColor,line width= 0.4pt,line join=round,line cap=round] (368.07, 75.71) -- (389.39, 77.30);

\path[draw=drawColor,line width= 0.4pt,line join=round,line cap=round] ( 60.26, 73.84) -- ( 64.76, 78.34);

\path[draw=drawColor,line width= 0.4pt,line join=round,line cap=round] ( 60.26, 78.34) -- ( 64.76, 73.84);

\path[draw=drawColor,line width= 0.4pt,line join=round,line cap=round] ( 93.55, 70.53) -- ( 98.05, 75.03);

\path[draw=drawColor,line width= 0.4pt,line join=round,line cap=round] ( 93.55, 75.03) -- ( 98.05, 70.53);

\path[draw=drawColor,line width= 0.4pt,line join=round,line cap=round] (126.84, 79.63) -- (131.34, 84.13);

\path[draw=drawColor,line width= 0.4pt,line join=round,line cap=round] (126.84, 84.13) -- (131.34, 79.63);

\path[draw=drawColor,line width= 0.4pt,line join=round,line cap=round] (160.12, 73.84) -- (164.62, 78.34);

\path[draw=drawColor,line width= 0.4pt,line join=round,line cap=round] (160.12, 78.34) -- (164.62, 73.84);

\path[draw=drawColor,line width= 0.4pt,line join=round,line cap=round] (193.41, 74.67) -- (197.91, 79.17);

\path[draw=drawColor,line width= 0.4pt,line join=round,line cap=round] (193.41, 79.17) -- (197.91, 74.67);

\path[draw=drawColor,line width= 0.4pt,line join=round,line cap=round] (226.70, 73.01) -- (231.20, 77.51);

\path[draw=drawColor,line width= 0.4pt,line join=round,line cap=round] (226.70, 77.51) -- (231.20, 73.01);

\path[draw=drawColor,line width= 0.4pt,line join=round,line cap=round] (259.98, 81.28) -- (264.48, 85.78);

\path[draw=drawColor,line width= 0.4pt,line join=round,line cap=round] (259.98, 85.78) -- (264.48, 81.28);

\path[draw=drawColor,line width= 0.4pt,line join=round,line cap=round] (293.27, 74.67) -- (297.77, 79.17);

\path[draw=drawColor,line width= 0.4pt,line join=round,line cap=round] (293.27, 79.17) -- (297.77, 74.67);

\path[draw=drawColor,line width= 0.4pt,line join=round,line cap=round] (326.55, 76.32) -- (331.05, 80.82);

\path[draw=drawColor,line width= 0.4pt,line join=round,line cap=round] (326.55, 80.82) -- (331.05, 76.32);

\path[draw=drawColor,line width= 0.4pt,line join=round,line cap=round] (359.84, 73.01) -- (364.34, 77.51);

\path[draw=drawColor,line width= 0.4pt,line join=round,line cap=round] (359.84, 77.51) -- (364.34, 73.01);

\path[draw=drawColor,line width= 0.4pt,line join=round,line cap=round] (393.13, 75.49) -- (397.63, 79.99);

\path[draw=drawColor,line width= 0.4pt,line join=round,line cap=round] (393.13, 79.99) -- (397.63, 75.49);
\definecolor{drawColor}{RGB}{205,96,144}

\path[draw=drawColor,line width= 0.4pt,line join=round,line cap=round] ( 67.83,200.47) -- ( 90.48,212.29);

\path[draw=drawColor,line width= 0.4pt,line join=round,line cap=round] (101.55,216.78) -- (123.34,223.28);

\path[draw=drawColor,line width= 0.4pt,line join=round,line cap=round] (134.97,226.16) -- (156.49,230.44);

\path[draw=drawColor,line width= 0.4pt,line join=round,line cap=round] (168.37,231.76) -- (189.66,232.29);

\path[draw=drawColor,line width= 0.4pt,line join=round,line cap=round] (201.66,232.58) -- (222.95,233.11);

\path[draw=drawColor,line width= 0.4pt,line join=round,line cap=round] (234.95,233.26) -- (256.23,233.26);

\path[draw=drawColor,line width= 0.4pt,line join=round,line cap=round] (268.23,233.26) -- (289.52,233.26);

\path[draw=drawColor,line width= 0.4pt,line join=round,line cap=round] (301.52,233.26) -- (322.80,233.26);

\path[draw=drawColor,line width= 0.4pt,line join=round,line cap=round] (334.80,233.26) -- (356.09,233.26);

\path[draw=drawColor,line width= 0.4pt,line join=round,line cap=round] (368.09,233.26) -- (389.38,233.26);

\path[draw=drawColor,line width= 0.4pt,line join=round,line cap=round] ( 60.26,195.44) rectangle ( 64.76,199.94);

\path[draw=drawColor,line width= 0.4pt,line join=round,line cap=round] ( 60.26,195.44) -- ( 64.76,199.94);

\path[draw=drawColor,line width= 0.4pt,line join=round,line cap=round] ( 60.26,199.94) -- ( 64.76,195.44);

\path[draw=drawColor,line width= 0.4pt,line join=round,line cap=round] ( 93.55,212.81) rectangle ( 98.05,217.31);

\path[draw=drawColor,line width= 0.4pt,line join=round,line cap=round] ( 93.55,212.81) -- ( 98.05,217.31);

\path[draw=drawColor,line width= 0.4pt,line join=round,line cap=round] ( 93.55,217.31) -- ( 98.05,212.81);

\path[draw=drawColor,line width= 0.4pt,line join=round,line cap=round] (126.84,222.74) rectangle (131.34,227.24);

\path[draw=drawColor,line width= 0.4pt,line join=round,line cap=round] (126.84,222.74) -- (131.34,227.24);

\path[draw=drawColor,line width= 0.4pt,line join=round,line cap=round] (126.84,227.24) -- (131.34,222.74);

\path[draw=drawColor,line width= 0.4pt,line join=round,line cap=round] (160.12,229.36) rectangle (164.62,233.86);

\path[draw=drawColor,line width= 0.4pt,line join=round,line cap=round] (160.12,229.36) -- (164.62,233.86);

\path[draw=drawColor,line width= 0.4pt,line join=round,line cap=round] (160.12,233.86) -- (164.62,229.36);

\path[draw=drawColor,line width= 0.4pt,line join=round,line cap=round] (193.41,230.18) rectangle (197.91,234.68);

\path[draw=drawColor,line width= 0.4pt,line join=round,line cap=round] (193.41,230.18) -- (197.91,234.68);

\path[draw=drawColor,line width= 0.4pt,line join=round,line cap=round] (193.41,234.68) -- (197.91,230.18);

\path[draw=drawColor,line width= 0.4pt,line join=round,line cap=round] (226.70,231.01) rectangle (231.20,235.51);

\path[draw=drawColor,line width= 0.4pt,line join=round,line cap=round] (226.70,231.01) -- (231.20,235.51);

\path[draw=drawColor,line width= 0.4pt,line join=round,line cap=round] (226.70,235.51) -- (231.20,231.01);

\path[draw=drawColor,line width= 0.4pt,line join=round,line cap=round] (259.98,231.01) rectangle (264.48,235.51);

\path[draw=drawColor,line width= 0.4pt,line join=round,line cap=round] (259.98,231.01) -- (264.48,235.51);

\path[draw=drawColor,line width= 0.4pt,line join=round,line cap=round] (259.98,235.51) -- (264.48,231.01);

\path[draw=drawColor,line width= 0.4pt,line join=round,line cap=round] (293.27,231.01) rectangle (297.77,235.51);

\path[draw=drawColor,line width= 0.4pt,line join=round,line cap=round] (293.27,231.01) -- (297.77,235.51);

\path[draw=drawColor,line width= 0.4pt,line join=round,line cap=round] (293.27,235.51) -- (297.77,231.01);

\path[draw=drawColor,line width= 0.4pt,line join=round,line cap=round] (326.55,231.01) rectangle (331.05,235.51);

\path[draw=drawColor,line width= 0.4pt,line join=round,line cap=round] (326.55,231.01) -- (331.05,235.51);

\path[draw=drawColor,line width= 0.4pt,line join=round,line cap=round] (326.55,235.51) -- (331.05,231.01);

\path[draw=drawColor,line width= 0.4pt,line join=round,line cap=round] (359.84,231.01) rectangle (364.34,235.51);

\path[draw=drawColor,line width= 0.4pt,line join=round,line cap=round] (359.84,231.01) -- (364.34,235.51);

\path[draw=drawColor,line width= 0.4pt,line join=round,line cap=round] (359.84,235.51) -- (364.34,231.01);

\path[draw=drawColor,line width= 0.4pt,line join=round,line cap=round] (393.13,231.01) rectangle (397.63,235.51);

\path[draw=drawColor,line width= 0.4pt,line join=round,line cap=round] (393.13,231.01) -- (397.63,235.51);

\path[draw=drawColor,line width= 0.4pt,line join=round,line cap=round] (393.13,235.51) -- (397.63,231.01);
\definecolor{drawColor}{RGB}{0,0,0}

\path[draw=drawColor,line width= 0.4pt,dash pattern=on 4pt off 4pt ,line join=round,line cap=round] ( 61.18, 76.09) --
	( 64.57, 76.09) --
	( 67.96, 76.09) --
	( 71.35, 76.09) --
	( 74.74, 76.09) --
	( 78.13, 76.09) --
	( 81.52, 76.09) --
	( 84.91, 76.09) --
	( 88.30, 76.09) --
	( 91.69, 76.09) --
	( 95.07, 76.09) --
	( 98.46, 76.09) --
	(101.85, 76.09) --
	(105.24, 76.09) --
	(108.63, 76.09) --
	(112.02, 76.09) --
	(115.41, 76.09) --
	(118.80, 76.09) --
	(122.19, 76.09) --
	(125.58, 76.09) --
	(128.97, 76.09) --
	(132.35, 76.09) --
	(135.74, 76.09) --
	(139.13, 76.09) --
	(142.52, 76.09) --
	(145.91, 76.09) --
	(149.30, 76.09) --
	(152.69, 76.09) --
	(156.08, 76.09) --
	(159.47, 76.09) --
	(162.86, 76.09) --
	(166.25, 76.09) --
	(169.64, 76.09) --
	(173.02, 76.09) --
	(176.41, 76.09) --
	(179.80, 76.09) --
	(183.19, 76.09) --
	(186.58, 76.09) --
	(189.97, 76.09) --
	(193.36, 76.09) --
	(196.75, 76.09) --
	(200.14, 76.09) --
	(203.53, 76.09) --
	(206.92, 76.09) --
	(210.30, 76.09) --
	(213.69, 76.09) --
	(217.08, 76.09) --
	(220.47, 76.09) --
	(223.86, 76.09) --
	(227.25, 76.09) --
	(230.64, 76.09) --
	(234.03, 76.09) --
	(237.42, 76.09) --
	(240.81, 76.09) --
	(244.20, 76.09) --
	(247.59, 76.09) --
	(250.97, 76.09) --
	(254.36, 76.09) --
	(257.75, 76.09) --
	(261.14, 76.09) --
	(264.53, 76.09) --
	(267.92, 76.09) --
	(271.31, 76.09) --
	(274.70, 76.09) --
	(278.09, 76.09) --
	(281.48, 76.09) --
	(284.87, 76.09) --
	(288.25, 76.09) --
	(291.64, 76.09) --
	(295.03, 76.09) --
	(298.42, 76.09) --
	(301.81, 76.09) --
	(305.20, 76.09) --
	(308.59, 76.09) --
	(311.98, 76.09) --
	(315.37, 76.09) --
	(318.76, 76.09) --
	(322.15, 76.09) --
	(325.54, 76.09) --
	(328.92, 76.09) --
	(332.31, 76.09) --
	(335.70, 76.09) --
	(339.09, 76.09) --
	(342.48, 76.09) --
	(345.87, 76.09) --
	(349.26, 76.09) --
	(352.65, 76.09) --
	(356.04, 76.09) --
	(359.43, 76.09) --
	(362.82, 76.09) --
	(366.20, 76.09) --
	(369.59, 76.09) --
	(372.98, 76.09) --
	(376.37, 76.09) --
	(379.76, 76.09) --
	(383.15, 76.09) --
	(386.54, 76.09) --
	(389.93, 76.09) --
	(393.32, 76.09) --
	(396.71, 76.09);
\definecolor{drawColor}{RGB}{139,0,0}
\definecolor{fillColor}{RGB}{139,0,0}

\path[draw=drawColor,line width= 0.4pt,line join=round,line cap=round,fill=fillColor] (421.11,227.88) circle (  2.25);
\definecolor{fillColor}{RGB}{255,99,71}

\path[fill=fillColor] (421.11,219.38) --
	(424.15,214.13) --
	(418.08,214.13) --
	cycle;
\definecolor{fillColor}{RGB}{108,166,205}

\path[fill=fillColor] (418.86,201.63) --
	(423.36,201.63) --
	(423.36,206.13) --
	(418.86,206.13) --
	cycle;
\definecolor{drawColor}{RGB}{139,137,137}

\path[draw=drawColor,line width= 0.4pt,line join=round,line cap=round] (417.93,191.88) -- (424.30,191.88);

\path[draw=drawColor,line width= 0.4pt,line join=round,line cap=round] (421.11,188.70) -- (421.11,195.06);
\definecolor{drawColor}{RGB}{84,139,84}

\path[draw=drawColor,line width= 0.4pt,line join=round,line cap=round] (418.86,177.63) -- (423.36,182.13);

\path[draw=drawColor,line width= 0.4pt,line join=round,line cap=round] (418.86,182.13) -- (423.36,177.63);
\definecolor{drawColor}{RGB}{205,96,144}

\path[draw=drawColor,line width= 0.4pt,line join=round,line cap=round] (418.86,165.63) rectangle (423.36,170.13);

\path[draw=drawColor,line width= 0.4pt,line join=round,line cap=round] (418.86,165.63) -- (423.36,170.13);

\path[draw=drawColor,line width= 0.4pt,line join=round,line cap=round] (418.86,170.13) -- (423.36,165.63);
\definecolor{drawColor}{RGB}{0,0,0}

\node[text=drawColor,anchor=base west,inner sep=0pt, outer sep=0pt, scale=  1.00] at (430.11,224.44) {HypoRF};

\node[text=drawColor,anchor=base west,inner sep=0pt, outer sep=0pt, scale=  1.00] at (430.11,212.44) {Binomial};

\node[text=drawColor,anchor=base west,inner sep=0pt, outer sep=0pt, scale=  1.00] at (430.11,200.44) {ME-full};

\node[text=drawColor,anchor=base west,inner sep=0pt, outer sep=0pt, scale=  1.00] at (430.11,188.44) {MMDboot};

\node[text=drawColor,anchor=base west,inner sep=0pt, outer sep=0pt, scale=  1.00] at (430.11,176.44) {MMD-full};

\node[text=drawColor,anchor=base west,inner sep=0pt, outer sep=0pt, scale=  1.00] at (430.11,164.44) {CPT-RF};
\end{scope}
\end{tikzpicture}

%% file: riskfactors_plot_K100_ntree600_margin0.tex
\begin{tikzpicture}[x=1pt,y=1pt]
\definecolor{fillColor}{RGB}{255,255,255}
\path[use as bounding box,fill=fillColor,fill opacity=0.00] (0,0) rectangle (1011.78,1806.75);
\begin{scope}
\path[clip] (  0.00,  0.00) rectangle (1011.78,1806.75);
\definecolor{drawColor}{RGB}{0,0,0}
\definecolor{fillColor}{gray}{0.80}

\path[draw=drawColor,line width= 0.4pt,line join=round,line cap=round,fill=fillColor] (120.00,124.25) rectangle (145.52,141.34);
\definecolor{fillColor}{RGB}{201,202,203}

\path[draw=drawColor,line width= 0.4pt,line join=round,line cap=round,fill=fillColor] (120.00,141.34) rectangle (257.43,158.43);
\definecolor{fillColor}{RGB}{199,201,202}

\path[draw=drawColor,line width= 0.4pt,line join=round,line cap=round,fill=fillColor] (120.00,158.43) rectangle (288.69,175.51);
\definecolor{fillColor}{RGB}{197,199,201}

\path[draw=drawColor,line width= 0.4pt,line join=round,line cap=round,fill=fillColor] (120.00,175.51) rectangle (306.24,192.60);
\definecolor{fillColor}{RGB}{195,198,201}

\path[draw=drawColor,line width= 0.4pt,line join=round,line cap=round,fill=fillColor] (120.00,192.60) rectangle (351.63,209.69);
\definecolor{fillColor}{RGB}{193,197,200}

\path[draw=drawColor,line width= 0.4pt,line join=round,line cap=round,fill=fillColor] (120.00,209.69) rectangle (364.57,226.78);
\definecolor{fillColor}{RGB}{191,195,199}

\path[draw=drawColor,line width= 0.4pt,line join=round,line cap=round,fill=fillColor] (120.00,226.78) rectangle (373.07,243.86);
\definecolor{fillColor}{RGB}{189,194,199}

\path[draw=drawColor,line width= 0.4pt,line join=round,line cap=round,fill=fillColor] (120.00,243.86) rectangle (383.86,260.95);
\definecolor{fillColor}{RGB}{187,193,198}

\path[draw=drawColor,line width= 0.4pt,line join=round,line cap=round,fill=fillColor] (120.00,260.95) rectangle (427.17,278.04);
\definecolor{fillColor}{RGB}{185,191,197}

\path[draw=drawColor,line width= 0.4pt,line join=round,line cap=round,fill=fillColor] (120.00,278.04) rectangle (433.68,295.13);
\definecolor{fillColor}{RGB}{183,190,197}

\path[draw=drawColor,line width= 0.4pt,line join=round,line cap=round,fill=fillColor] (120.00,295.13) rectangle (434.07,312.22);
\definecolor{fillColor}{RGB}{181,189,196}

\path[draw=drawColor,line width= 0.4pt,line join=round,line cap=round,fill=fillColor] (120.00,312.22) rectangle (440.67,329.30);
\definecolor{fillColor}{RGB}{179,187,195}

\path[draw=drawColor,line width= 0.4pt,line join=round,line cap=round,fill=fillColor] (120.00,329.30) rectangle (444.35,346.39);
\definecolor{fillColor}{RGB}{177,186,194}

\path[draw=drawColor,line width= 0.4pt,line join=round,line cap=round,fill=fillColor] (120.00,346.39) rectangle (444.86,363.48);
\definecolor{fillColor}{RGB}{175,185,194}

\path[draw=drawColor,line width= 0.4pt,line join=round,line cap=round,fill=fillColor] (120.00,363.48) rectangle (448.69,380.57);
\definecolor{fillColor}{RGB}{173,183,193}

\path[draw=drawColor,line width= 0.4pt,line join=round,line cap=round,fill=fillColor] (120.00,380.57) rectangle (452.54,397.65);
\definecolor{fillColor}{RGB}{171,182,192}

\path[draw=drawColor,line width= 0.4pt,line join=round,line cap=round,fill=fillColor] (120.00,397.65) rectangle (455.71,414.74);
\definecolor{fillColor}{RGB}{169,180,192}

\path[draw=drawColor,line width= 0.4pt,line join=round,line cap=round,fill=fillColor] (120.00,414.74) rectangle (455.82,431.83);
\definecolor{fillColor}{RGB}{167,179,191}

\path[draw=drawColor,line width= 0.4pt,line join=round,line cap=round,fill=fillColor] (120.00,431.83) rectangle (456.87,448.92);
\definecolor{fillColor}{RGB}{165,178,190}

\path[draw=drawColor,line width= 0.4pt,line join=round,line cap=round,fill=fillColor] (120.00,448.92) rectangle (458.09,466.01);
\definecolor{fillColor}{RGB}{163,176,190}

\path[draw=drawColor,line width= 0.4pt,line join=round,line cap=round,fill=fillColor] (120.00,466.01) rectangle (461.21,483.09);
\definecolor{fillColor}{RGB}{161,175,189}

\path[draw=drawColor,line width= 0.4pt,line join=round,line cap=round,fill=fillColor] (120.00,483.09) rectangle (465.70,500.18);
\definecolor{fillColor}{RGB}{159,174,188}

\path[draw=drawColor,line width= 0.4pt,line join=round,line cap=round,fill=fillColor] (120.00,500.18) rectangle (466.20,517.27);
\definecolor{fillColor}{RGB}{157,172,187}

\path[draw=drawColor,line width= 0.4pt,line join=round,line cap=round,fill=fillColor] (120.00,517.27) rectangle (470.94,534.36);
\definecolor{fillColor}{RGB}{155,171,187}

\path[draw=drawColor,line width= 0.4pt,line join=round,line cap=round,fill=fillColor] (120.00,534.36) rectangle (471.52,551.44);
\definecolor{fillColor}{RGB}{153,170,186}

\path[draw=drawColor,line width= 0.4pt,line join=round,line cap=round,fill=fillColor] (120.00,551.44) rectangle (471.98,568.53);
\definecolor{fillColor}{RGB}{151,168,185}

\path[draw=drawColor,line width= 0.4pt,line join=round,line cap=round,fill=fillColor] (120.00,568.53) rectangle (473.28,585.62);
\definecolor{fillColor}{RGB}{149,167,185}

\path[draw=drawColor,line width= 0.4pt,line join=round,line cap=round,fill=fillColor] (120.00,585.62) rectangle (474.38,602.71);
\definecolor{fillColor}{RGB}{147,166,184}

\path[draw=drawColor,line width= 0.4pt,line join=round,line cap=round,fill=fillColor] (120.00,602.71) rectangle (475.23,619.80);
\definecolor{fillColor}{RGB}{145,164,183}

\path[draw=drawColor,line width= 0.4pt,line join=round,line cap=round,fill=fillColor] (120.00,619.80) rectangle (476.07,636.88);
\definecolor{fillColor}{RGB}{143,163,183}

\path[draw=drawColor,line width= 0.4pt,line join=round,line cap=round,fill=fillColor] (120.00,636.88) rectangle (477.96,653.97);
\definecolor{fillColor}{RGB}{141,162,182}

\path[draw=drawColor,line width= 0.4pt,line join=round,line cap=round,fill=fillColor] (120.00,653.97) rectangle (484.14,671.06);
\definecolor{fillColor}{RGB}{139,160,181}

\path[draw=drawColor,line width= 0.4pt,line join=round,line cap=round,fill=fillColor] (120.00,671.06) rectangle (491.59,688.15);
\definecolor{fillColor}{RGB}{137,159,180}

\path[draw=drawColor,line width= 0.4pt,line join=round,line cap=round,fill=fillColor] (120.00,688.15) rectangle (494.16,705.23);
\definecolor{fillColor}{RGB}{135,157,180}

\path[draw=drawColor,line width= 0.4pt,line join=round,line cap=round,fill=fillColor] (120.00,705.23) rectangle (494.35,722.32);
\definecolor{fillColor}{RGB}{133,156,179}

\path[draw=drawColor,line width= 0.4pt,line join=round,line cap=round,fill=fillColor] (120.00,722.32) rectangle (495.42,739.41);
\definecolor{fillColor}{RGB}{131,155,178}

\path[draw=drawColor,line width= 0.4pt,line join=round,line cap=round,fill=fillColor] (120.00,739.41) rectangle (496.46,756.50);
\definecolor{fillColor}{RGB}{129,153,178}

\path[draw=drawColor,line width= 0.4pt,line join=round,line cap=round,fill=fillColor] (120.00,756.50) rectangle (499.56,773.59);
\definecolor{fillColor}{RGB}{127,152,177}

\path[draw=drawColor,line width= 0.4pt,line join=round,line cap=round,fill=fillColor] (120.00,773.59) rectangle (499.74,790.67);
\definecolor{fillColor}{RGB}{125,151,176}

\path[draw=drawColor,line width= 0.4pt,line join=round,line cap=round,fill=fillColor] (120.00,790.67) rectangle (505.77,807.76);
\definecolor{fillColor}{RGB}{123,149,176}

\path[draw=drawColor,line width= 0.4pt,line join=round,line cap=round,fill=fillColor] (120.00,807.76) rectangle (510.06,824.85);
\definecolor{fillColor}{RGB}{121,148,175}

\path[draw=drawColor,line width= 0.4pt,line join=round,line cap=round,fill=fillColor] (120.00,824.85) rectangle (511.00,841.94);
\definecolor{fillColor}{RGB}{119,147,174}

\path[draw=drawColor,line width= 0.4pt,line join=round,line cap=round,fill=fillColor] (120.00,841.94) rectangle (513.48,859.02);
\definecolor{fillColor}{RGB}{117,145,173}

\path[draw=drawColor,line width= 0.4pt,line join=round,line cap=round,fill=fillColor] (120.00,859.02) rectangle (514.03,876.11);
\definecolor{fillColor}{RGB}{115,144,173}

\path[draw=drawColor,line width= 0.4pt,line join=round,line cap=round,fill=fillColor] (120.00,876.11) rectangle (514.75,893.20);
\definecolor{fillColor}{RGB}{113,143,172}

\path[draw=drawColor,line width= 0.4pt,line join=round,line cap=round,fill=fillColor] (120.00,893.20) rectangle (516.66,910.29);
\definecolor{fillColor}{RGB}{111,141,171}

\path[draw=drawColor,line width= 0.4pt,line join=round,line cap=round,fill=fillColor] (120.00,910.29) rectangle (517.79,927.38);
\definecolor{fillColor}{RGB}{108,140,171}

\path[draw=drawColor,line width= 0.4pt,line join=round,line cap=round,fill=fillColor] (120.00,927.38) rectangle (518.08,944.46);
\definecolor{fillColor}{RGB}{106,138,170}

\path[draw=drawColor,line width= 0.4pt,line join=round,line cap=round,fill=fillColor] (120.00,944.46) rectangle (519.40,961.55);
\definecolor{fillColor}{RGB}{104,137,169}

\path[draw=drawColor,line width= 0.4pt,line join=round,line cap=round,fill=fillColor] (120.00,961.55) rectangle (522.86,978.64);
\definecolor{fillColor}{RGB}{102,136,169}

\path[draw=drawColor,line width= 0.4pt,line join=round,line cap=round,fill=fillColor] (120.00,978.64) rectangle (525.40,995.73);
\definecolor{fillColor}{RGB}{100,134,168}

\path[draw=drawColor,line width= 0.4pt,line join=round,line cap=round,fill=fillColor] (120.00,995.73) rectangle (529.98,1012.81);
\definecolor{fillColor}{RGB}{98,133,167}

\path[draw=drawColor,line width= 0.4pt,line join=round,line cap=round,fill=fillColor] (120.00,1012.81) rectangle (530.19,1029.90);
\definecolor{fillColor}{RGB}{96,132,166}

\path[draw=drawColor,line width= 0.4pt,line join=round,line cap=round,fill=fillColor] (120.00,1029.90) rectangle (536.00,1046.99);
\definecolor{fillColor}{RGB}{94,130,166}

\path[draw=drawColor,line width= 0.4pt,line join=round,line cap=round,fill=fillColor] (120.00,1046.99) rectangle (537.15,1064.08);
\definecolor{fillColor}{RGB}{92,129,165}

\path[draw=drawColor,line width= 0.4pt,line join=round,line cap=round,fill=fillColor] (120.00,1064.08) rectangle (537.96,1081.16);
\definecolor{fillColor}{RGB}{90,128,164}

\path[draw=drawColor,line width= 0.4pt,line join=round,line cap=round,fill=fillColor] (120.00,1081.16) rectangle (538.35,1098.25);
\definecolor{fillColor}{RGB}{88,126,164}

\path[draw=drawColor,line width= 0.4pt,line join=round,line cap=round,fill=fillColor] (120.00,1098.25) rectangle (538.71,1115.34);
\definecolor{fillColor}{RGB}{86,125,163}

\path[draw=drawColor,line width= 0.4pt,line join=round,line cap=round,fill=fillColor] (120.00,1115.34) rectangle (540.23,1132.43);
\definecolor{fillColor}{RGB}{84,124,162}

\path[draw=drawColor,line width= 0.4pt,line join=round,line cap=round,fill=fillColor] (120.00,1132.43) rectangle (540.98,1149.52);
\definecolor{fillColor}{RGB}{82,122,162}

\path[draw=drawColor,line width= 0.4pt,line join=round,line cap=round,fill=fillColor] (120.00,1149.52) rectangle (541.52,1166.60);
\definecolor{fillColor}{RGB}{80,121,161}

\path[draw=drawColor,line width= 0.4pt,line join=round,line cap=round,fill=fillColor] (120.00,1166.60) rectangle (543.67,1183.69);
\definecolor{fillColor}{RGB}{78,120,160}

\path[draw=drawColor,line width= 0.4pt,line join=round,line cap=round,fill=fillColor] (120.00,1183.69) rectangle (555.84,1200.78);
\definecolor{fillColor}{RGB}{76,118,159}

\path[draw=drawColor,line width= 0.4pt,line join=round,line cap=round,fill=fillColor] (120.00,1200.78) rectangle (578.90,1217.87);
\definecolor{fillColor}{RGB}{74,117,159}

\path[draw=drawColor,line width= 0.4pt,line join=round,line cap=round,fill=fillColor] (120.00,1217.87) rectangle (594.25,1234.95);
\definecolor{fillColor}{RGB}{72,115,158}

\path[draw=drawColor,line width= 0.4pt,line join=round,line cap=round,fill=fillColor] (120.00,1234.95) rectangle (598.97,1252.04);
\definecolor{fillColor}{RGB}{70,114,157}

\path[draw=drawColor,line width= 0.4pt,line join=round,line cap=round,fill=fillColor] (120.00,1252.04) rectangle (608.83,1269.13);
\definecolor{fillColor}{RGB}{68,113,157}

\path[draw=drawColor,line width= 0.4pt,line join=round,line cap=round,fill=fillColor] (120.00,1269.13) rectangle (613.39,1286.22);
\definecolor{fillColor}{RGB}{66,111,156}

\path[draw=drawColor,line width= 0.4pt,line join=round,line cap=round,fill=fillColor] (120.00,1286.22) rectangle (625.39,1303.31);
\definecolor{fillColor}{RGB}{64,110,155}

\path[draw=drawColor,line width= 0.4pt,line join=round,line cap=round,fill=fillColor] (120.00,1303.31) rectangle (626.96,1320.39);
\definecolor{fillColor}{RGB}{62,109,155}

\path[draw=drawColor,line width= 0.4pt,line join=round,line cap=round,fill=fillColor] (120.00,1320.39) rectangle (638.27,1337.48);
\definecolor{fillColor}{RGB}{60,107,154}

\path[draw=drawColor,line width= 0.4pt,line join=round,line cap=round,fill=fillColor] (120.00,1337.48) rectangle (639.53,1354.57);
\definecolor{fillColor}{RGB}{58,106,153}

\path[draw=drawColor,line width= 0.4pt,line join=round,line cap=round,fill=fillColor] (120.00,1354.57) rectangle (640.67,1371.66);
\definecolor{fillColor}{RGB}{56,105,152}

\path[draw=drawColor,line width= 0.4pt,line join=round,line cap=round,fill=fillColor] (120.00,1371.66) rectangle (641.77,1388.74);
\definecolor{fillColor}{RGB}{54,103,152}

\path[draw=drawColor,line width= 0.4pt,line join=round,line cap=round,fill=fillColor] (120.00,1388.74) rectangle (646.76,1405.83);
\definecolor{fillColor}{RGB}{52,102,151}

\path[draw=drawColor,line width= 0.4pt,line join=round,line cap=round,fill=fillColor] (120.00,1405.83) rectangle (660.68,1422.92);
\definecolor{fillColor}{RGB}{50,101,150}

\path[draw=drawColor,line width= 0.4pt,line join=round,line cap=round,fill=fillColor] (120.00,1422.92) rectangle (674.43,1440.01);
\definecolor{fillColor}{RGB}{48,99,150}

\path[draw=drawColor,line width= 0.4pt,line join=round,line cap=round,fill=fillColor] (120.00,1440.01) rectangle (675.65,1457.10);
\definecolor{fillColor}{RGB}{46,98,149}

\path[draw=drawColor,line width= 0.4pt,line join=round,line cap=round,fill=fillColor] (120.00,1457.10) rectangle (689.71,1474.18);
\definecolor{fillColor}{RGB}{44,96,148}

\path[draw=drawColor,line width= 0.4pt,line join=round,line cap=round,fill=fillColor] (120.00,1474.18) rectangle (712.02,1491.27);
\definecolor{fillColor}{RGB}{42,95,148}

\path[draw=drawColor,line width= 0.4pt,line join=round,line cap=round,fill=fillColor] (120.00,1491.27) rectangle (713.43,1508.36);
\definecolor{fillColor}{RGB}{40,94,147}

\path[draw=drawColor,line width= 0.4pt,line join=round,line cap=round,fill=fillColor] (120.00,1508.36) rectangle (758.28,1525.45);
\definecolor{fillColor}{RGB}{38,92,146}

\path[draw=drawColor,line width= 0.4pt,line join=round,line cap=round,fill=fillColor] (120.00,1525.45) rectangle (770.66,1542.53);
\definecolor{fillColor}{RGB}{36,91,145}

\path[draw=drawColor,line width= 0.4pt,line join=round,line cap=round,fill=fillColor] (120.00,1542.53) rectangle (776.91,1559.62);
\definecolor{fillColor}{RGB}{34,90,145}

\path[draw=drawColor,line width= 0.4pt,line join=round,line cap=round,fill=fillColor] (120.00,1559.62) rectangle (785.15,1576.71);
\definecolor{fillColor}{RGB}{32,88,144}

\path[draw=drawColor,line width= 0.4pt,line join=round,line cap=round,fill=fillColor] (120.00,1576.71) rectangle (786.34,1593.80);
\definecolor{fillColor}{RGB}{30,87,143}

\path[draw=drawColor,line width= 0.4pt,line join=round,line cap=round,fill=fillColor] (120.00,1593.80) rectangle (790.39,1610.89);
\definecolor{fillColor}{RGB}{28,86,143}

\path[draw=drawColor,line width= 0.4pt,line join=round,line cap=round,fill=fillColor] (120.00,1610.89) rectangle (791.50,1627.97);
\definecolor{fillColor}{RGB}{26,84,142}

\path[draw=drawColor,line width= 0.4pt,line join=round,line cap=round,fill=fillColor] (120.00,1627.97) rectangle (809.82,1645.06);
\definecolor{fillColor}{RGB}{24,83,141}

\path[draw=drawColor,line width= 0.4pt,line join=round,line cap=round,fill=fillColor] (120.00,1645.06) rectangle (829.70,1662.15);
\definecolor{fillColor}{RGB}{22,82,141}

\path[draw=drawColor,line width= 0.4pt,line join=round,line cap=round,fill=fillColor] (120.00,1662.15) rectangle (836.82,1679.24);
\definecolor{fillColor}{RGB}{20,80,140}

\path[draw=drawColor,line width= 0.4pt,line join=round,line cap=round,fill=fillColor] (120.00,1679.24) rectangle (862.61,1696.32);
\definecolor{fillColor}{RGB}{18,79,139}

\path[draw=drawColor,line width= 0.4pt,line join=round,line cap=round,fill=fillColor] (120.00,1696.32) rectangle (878.04,1713.41);
\definecolor{fillColor}{RGB}{16,78,139}

\path[draw=drawColor,line width= 0.4pt,line join=round,line cap=round,fill=fillColor] (120.00,1713.41) rectangle (940.83,1730.50);
\end{scope}
\begin{scope}
\path[clip] (  0.00,  0.00) rectangle (1011.78,1806.75);
\definecolor{drawColor}{RGB}{0,0,0}

\node[text=drawColor,anchor=base east,inner sep=0pt, outer sep=0pt, scale=  1.50] at (108.00,127.63) {sin};

\node[text=drawColor,anchor=base east,inner sep=0pt, outer sep=0pt, scale=  1.50] at (108.00,144.72) {realestate};

\node[text=drawColor,anchor=base east,inner sep=0pt, outer sep=0pt, scale=  1.50] at (108.00,161.80) {rdsale};

\node[text=drawColor,anchor=base east,inner sep=0pt, outer sep=0pt, scale=  1.50] at (108.00,178.89) {rdmve};

\node[text=drawColor,anchor=base east,inner sep=0pt, outer sep=0pt, scale=  1.50] at (108.00,195.98) {orgcap};

\node[text=drawColor,anchor=base east,inner sep=0pt, outer sep=0pt, scale=  1.50] at (108.00,213.07) {secured};

\node[text=drawColor,anchor=base east,inner sep=0pt, outer sep=0pt, scale=  1.50] at (108.00,230.15) {divo};

\node[text=drawColor,anchor=base east,inner sep=0pt, outer sep=0pt, scale=  1.50] at (108.00,247.24) {divi};

\node[text=drawColor,anchor=base east,inner sep=0pt, outer sep=0pt, scale=  1.50] at (108.00,264.33) {saleinv};

\node[text=drawColor,anchor=base east,inner sep=0pt, outer sep=0pt, scale=  1.50] at (108.00,281.42) {gma};

\node[text=drawColor,anchor=base east,inner sep=0pt, outer sep=0pt, scale=  1.50] at (108.00,298.50) {depr};

\node[text=drawColor,anchor=base east,inner sep=0pt, outer sep=0pt, scale=  1.50] at (108.00,315.59) {mveia};

\node[text=drawColor,anchor=base east,inner sep=0pt, outer sep=0pt, scale=  1.50] at (108.00,332.68) {currat};

\node[text=drawColor,anchor=base east,inner sep=0pt, outer sep=0pt, scale=  1.50] at (108.00,349.77) {lev};

\node[text=drawColor,anchor=base east,inner sep=0pt, outer sep=0pt, scale=  1.50] at (108.00,366.86) {salerec};

\node[text=drawColor,anchor=base east,inner sep=0pt, outer sep=0pt, scale=  1.50] at (108.00,383.94) {cashdebt};

\node[text=drawColor,anchor=base east,inner sep=0pt, outer sep=0pt, scale=  1.50] at (108.00,401.03) {convind};

\node[text=drawColor,anchor=base east,inner sep=0pt, outer sep=0pt, scale=  1.50] at (108.00,418.12) {quick};

\node[text=drawColor,anchor=base east,inner sep=0pt, outer sep=0pt, scale=  1.50] at (108.00,435.21) {tang};

\node[text=drawColor,anchor=base east,inner sep=0pt, outer sep=0pt, scale=  1.50] at (108.00,452.29) {grltnoa};

\node[text=drawColor,anchor=base east,inner sep=0pt, outer sep=0pt, scale=  1.50] at (108.00,469.38) {pchsalepchxsga};

\node[text=drawColor,anchor=base east,inner sep=0pt, outer sep=0pt, scale=  1.50] at (108.00,486.47) {roic};

\node[text=drawColor,anchor=base east,inner sep=0pt, outer sep=0pt, scale=  1.50] at (108.00,503.56) {sp};

\node[text=drawColor,anchor=base east,inner sep=0pt, outer sep=0pt, scale=  1.50] at (108.00,520.65) {stdcf};

\node[text=drawColor,anchor=base east,inner sep=0pt, outer sep=0pt, scale=  1.50] at (108.00,537.73) {stdacc};

\node[text=drawColor,anchor=base east,inner sep=0pt, outer sep=0pt, scale=  1.50] at (108.00,554.82) {operprof};

\node[text=drawColor,anchor=base east,inner sep=0pt, outer sep=0pt, scale=  1.50] at (108.00,571.91) {pchsaleinv};

\node[text=drawColor,anchor=base east,inner sep=0pt, outer sep=0pt, scale=  1.50] at (108.00,589.00) {pchsalepchinvt};

\node[text=drawColor,anchor=base east,inner sep=0pt, outer sep=0pt, scale=  1.50] at (108.00,606.08) {bm};

\node[text=drawColor,anchor=base east,inner sep=0pt, outer sep=0pt, scale=  1.50] at (108.00,623.17) {salecash};

\node[text=drawColor,anchor=base east,inner sep=0pt, outer sep=0pt, scale=  1.50] at (108.00,640.26) {dy};

\node[text=drawColor,anchor=base east,inner sep=0pt, outer sep=0pt, scale=  1.50] at (108.00,657.35) {cashpr};

\node[text=drawColor,anchor=base east,inner sep=0pt, outer sep=0pt, scale=  1.50] at (108.00,674.44) {cfp};

\node[text=drawColor,anchor=base east,inner sep=0pt, outer sep=0pt, scale=  1.50] at (108.00,691.52) {securedind};

\node[text=drawColor,anchor=base east,inner sep=0pt, outer sep=0pt, scale=  1.50] at (108.00,708.61) {pctacc};

\node[text=drawColor,anchor=base east,inner sep=0pt, outer sep=0pt, scale=  1.50] at (108.00,725.70) {acc};

\node[text=drawColor,anchor=base east,inner sep=0pt, outer sep=0pt, scale=  1.50] at (108.00,742.79) {herf};

\node[text=drawColor,anchor=base east,inner sep=0pt, outer sep=0pt, scale=  1.50] at (108.00,759.87) {invest};

\node[text=drawColor,anchor=base east,inner sep=0pt, outer sep=0pt, scale=  1.50] at (108.00,776.96) {bmia};

\node[text=drawColor,anchor=base east,inner sep=0pt, outer sep=0pt, scale=  1.50] at (108.00,794.05) {absacc};

\node[text=drawColor,anchor=base east,inner sep=0pt, outer sep=0pt, scale=  1.50] at (108.00,811.14) {agr};

\node[text=drawColor,anchor=base east,inner sep=0pt, outer sep=0pt, scale=  1.50] at (108.00,828.23) {ep};

\node[text=drawColor,anchor=base east,inner sep=0pt, outer sep=0pt, scale=  1.50] at (108.00,845.31) {pchdepr};

\node[text=drawColor,anchor=base east,inner sep=0pt, outer sep=0pt, scale=  1.50] at (108.00,862.40) {pchquick};

\node[text=drawColor,anchor=base east,inner sep=0pt, outer sep=0pt, scale=  1.50] at (108.00,879.49) {pchcurrat};

\node[text=drawColor,anchor=base east,inner sep=0pt, outer sep=0pt, scale=  1.50] at (108.00,896.58) {egr};

\node[text=drawColor,anchor=base east,inner sep=0pt, outer sep=0pt, scale=  1.50] at (108.00,913.66) {cfpia};

\node[text=drawColor,anchor=base east,inner sep=0pt, outer sep=0pt, scale=  1.50] at (108.00,930.75) {tb};

\node[text=drawColor,anchor=base east,inner sep=0pt, outer sep=0pt, scale=  1.50] at (108.00,947.84) {sgr};

\node[text=drawColor,anchor=base east,inner sep=0pt, outer sep=0pt, scale=  1.50] at (108.00,964.93) {grcapx};

\node[text=drawColor,anchor=base east,inner sep=0pt, outer sep=0pt, scale=  1.50] at (108.00,982.02) {lgr};

\node[text=drawColor,anchor=base east,inner sep=0pt, outer sep=0pt, scale=  1.50] at (108.00,999.10) {pchgmpchsale};

\node[text=drawColor,anchor=base east,inner sep=0pt, outer sep=0pt, scale=  1.50] at (108.00,1016.19) {roavol};

\node[text=drawColor,anchor=base east,inner sep=0pt, outer sep=0pt, scale=  1.50] at (108.00,1033.28) {pchsalepchrect};

\node[text=drawColor,anchor=base east,inner sep=0pt, outer sep=0pt, scale=  1.50] at (108.00,1050.37) {chinv};

\node[text=drawColor,anchor=base east,inner sep=0pt, outer sep=0pt, scale=  1.50] at (108.00,1067.45) {hire};

\node[text=drawColor,anchor=base east,inner sep=0pt, outer sep=0pt, scale=  1.50] at (108.00,1084.54) {pchcapxia};

\node[text=drawColor,anchor=base east,inner sep=0pt, outer sep=0pt, scale=  1.50] at (108.00,1101.63) {age};

\node[text=drawColor,anchor=base east,inner sep=0pt, outer sep=0pt, scale=  1.50] at (108.00,1118.72) {chatoia};

\node[text=drawColor,anchor=base east,inner sep=0pt, outer sep=0pt, scale=  1.50] at (108.00,1135.81) {rd};

\node[text=drawColor,anchor=base east,inner sep=0pt, outer sep=0pt, scale=  1.50] at (108.00,1152.89) {chempia};

\node[text=drawColor,anchor=base east,inner sep=0pt, outer sep=0pt, scale=  1.50] at (108.00,1169.98) {chcsho};

\node[text=drawColor,anchor=base east,inner sep=0pt, outer sep=0pt, scale=  1.50] at (108.00,1187.07) {chpmia};

\node[text=drawColor,anchor=base east,inner sep=0pt, outer sep=0pt, scale=  1.50] at (108.00,1204.16) {cash};

\node[text=drawColor,anchor=base east,inner sep=0pt, outer sep=0pt, scale=  1.50] at (108.00,1221.24) {roaq};

\node[text=drawColor,anchor=base east,inner sep=0pt, outer sep=0pt, scale=  1.50] at (108.00,1238.33) {mvel1};

\node[text=drawColor,anchor=base east,inner sep=0pt, outer sep=0pt, scale=  1.50] at (108.00,1255.42) {roeq};

\node[text=drawColor,anchor=base east,inner sep=0pt, outer sep=0pt, scale=  1.50] at (108.00,1272.51) {ps};

\node[text=drawColor,anchor=base east,inner sep=0pt, outer sep=0pt, scale=  1.50] at (108.00,1289.60) {ms};

\node[text=drawColor,anchor=base east,inner sep=0pt, outer sep=0pt, scale=  1.50] at (108.00,1306.68) {rsup};

\node[text=drawColor,anchor=base east,inner sep=0pt, outer sep=0pt, scale=  1.50] at (108.00,1323.77) {beta};

\node[text=drawColor,anchor=base east,inner sep=0pt, outer sep=0pt, scale=  1.50] at (108.00,1340.86) {idiovol};

\node[text=drawColor,anchor=base east,inner sep=0pt, outer sep=0pt, scale=  1.50] at (108.00,1357.95) {ill};

\node[text=drawColor,anchor=base east,inner sep=0pt, outer sep=0pt, scale=  1.50] at (108.00,1375.03) {betasq};

\node[text=drawColor,anchor=base east,inner sep=0pt, outer sep=0pt, scale=  1.50] at (108.00,1392.12) {dolvol};

\node[text=drawColor,anchor=base east,inner sep=0pt, outer sep=0pt, scale=  1.50] at (108.00,1409.21) {chtx};

\node[text=drawColor,anchor=base east,inner sep=0pt, outer sep=0pt, scale=  1.50] at (108.00,1426.30) {turn};

\node[text=drawColor,anchor=base east,inner sep=0pt, outer sep=0pt, scale=  1.50] at (108.00,1443.39) {cinvest};

\node[text=drawColor,anchor=base east,inner sep=0pt, outer sep=0pt, scale=  1.50] at (108.00,1460.47) {aeavol};

\node[text=drawColor,anchor=base east,inner sep=0pt, outer sep=0pt, scale=  1.50] at (108.00,1477.56) {ear};

\node[text=drawColor,anchor=base east,inner sep=0pt, outer sep=0pt, scale=  1.50] at (108.00,1494.65) {zerotrade};

\node[text=drawColor,anchor=base east,inner sep=0pt, outer sep=0pt, scale=  1.50] at (108.00,1511.74) {mom36m};

\node[text=drawColor,anchor=base east,inner sep=0pt, outer sep=0pt, scale=  1.50] at (108.00,1528.82) {indmom};

\node[text=drawColor,anchor=base east,inner sep=0pt, outer sep=0pt, scale=  1.50] at (108.00,1545.91) {baspread};

\node[text=drawColor,anchor=base east,inner sep=0pt, outer sep=0pt, scale=  1.50] at (108.00,1563.00) {nincr};

\node[text=drawColor,anchor=base east,inner sep=0pt, outer sep=0pt, scale=  1.50] at (108.00,1580.09) {mom12m};

\node[text=drawColor,anchor=base east,inner sep=0pt, outer sep=0pt, scale=  1.50] at (108.00,1597.18) {pricedelay};

\node[text=drawColor,anchor=base east,inner sep=0pt, outer sep=0pt, scale=  1.50] at (108.00,1614.26) {stdturn};

\node[text=drawColor,anchor=base east,inner sep=0pt, outer sep=0pt, scale=  1.50] at (108.00,1631.35) {stddolvol};

\node[text=drawColor,anchor=base east,inner sep=0pt, outer sep=0pt, scale=  1.50] at (108.00,1648.44) {retvol};

\node[text=drawColor,anchor=base east,inner sep=0pt, outer sep=0pt, scale=  1.50] at (108.00,1665.53) {mom6m};

\node[text=drawColor,anchor=base east,inner sep=0pt, outer sep=0pt, scale=  1.50] at (108.00,1682.61) {chmom};

\node[text=drawColor,anchor=base east,inner sep=0pt, outer sep=0pt, scale=  1.50] at (108.00,1699.70) {maxret};

\node[text=drawColor,anchor=base east,inner sep=0pt, outer sep=0pt, scale=  1.50] at (108.00,1716.79) {mom1m};

\path[draw=drawColor,line width= 0.4pt,line join=round,line cap=round] (120.00, 60.00) -- (963.78, 60.00);

\path[draw=drawColor,line width= 0.4pt,line join=round,line cap=round] (120.00, 60.00) -- (120.00, 54.00);

\path[draw=drawColor,line width= 0.4pt,line join=round,line cap=round] (288.76, 60.00) -- (288.76, 54.00);

\path[draw=drawColor,line width= 0.4pt,line join=round,line cap=round] (457.51, 60.00) -- (457.51, 54.00);

\path[draw=drawColor,line width= 0.4pt,line join=round,line cap=round] (626.27, 60.00) -- (626.27, 54.00);

\path[draw=drawColor,line width= 0.4pt,line join=round,line cap=round] (795.02, 60.00) -- (795.02, 54.00);

\path[draw=drawColor,line width= 0.4pt,line join=round,line cap=round] (963.78, 60.00) -- (963.78, 54.00);

\node[text=drawColor,rotate= 90.00,anchor=base east,inner sep=0pt, outer sep=0pt, scale=  1.00] at (123.44, 48.00) {0};

\node[text=drawColor,rotate= 90.00,anchor=base east,inner sep=0pt, outer sep=0pt, scale=  1.00] at (292.20, 48.00) {500};

\node[text=drawColor,rotate= 90.00,anchor=base east,inner sep=0pt, outer sep=0pt, scale=  1.00] at (460.96, 48.00) {1000};

\node[text=drawColor,rotate= 90.00,anchor=base east,inner sep=0pt, outer sep=0pt, scale=  1.00] at (629.71, 48.00) {1500};

\node[text=drawColor,rotate= 90.00,anchor=base east,inner sep=0pt, outer sep=0pt, scale=  1.00] at (798.47, 48.00) {2000};

\node[text=drawColor,rotate= 90.00,anchor=base east,inner sep=0pt, outer sep=0pt, scale=  1.00] at (967.22, 48.00) {2500};
\end{scope}
\begin{scope}
\path[clip] (  0.00,  0.00) rectangle (1011.78,1806.75);
\definecolor{drawColor}{RGB}{0,0,0}

\node[text=drawColor,anchor=base,inner sep=0pt, outer sep=0pt, scale=  2.00] at (541.89,1746.75) {\bfseries Overall p-value: 0.0099};
\end{scope}
\begin{scope}
\path[clip] (120.00, 60.00) rectangle (963.78,1794.75);
\definecolor{drawColor}{RGB}{255,99,71}

\path[draw=drawColor,line width= 0.4pt,dash pattern=on 4pt off 4pt ,line join=round,line cap=round] (940.40, 60.00) -- (940.40,1794.75);

\path[draw=drawColor,line width= 0.4pt,dash pattern=on 4pt off 4pt ,line join=round,line cap=round] (644.27,148.10) -- (680.27,148.10);
\definecolor{drawColor}{RGB}{0,0,0}

\node[text=drawColor,anchor=base west,inner sep=0pt, outer sep=0pt, scale=  2.00] at (698.27,141.21) {Significance treshold};
\end{scope}
\begin{scope}
\path[clip] (  0.00,  0.00) rectangle (1011.78,1806.75);
\definecolor{drawColor}{RGB}{0,0,0}

\node[text=drawColor,anchor=base,inner sep=0pt, outer sep=0pt, scale=  2.00] at (541.89,  2.40) {Variable Importance};
\end{scope}
\end{tikzpicture}

%% file: 1a_plot_gaussian_meanshift_Normapprox_F_LDA_with_Cai.tex
\begin{tikzpicture}[x=1pt,y=1pt]
\definecolor{fillColor}{RGB}{255,255,255}
\path[use as bounding box,fill=fillColor,fill opacity=0.00] (0,0) rectangle (505.89,289.08);
\begin{scope}
\path[clip] (  0.00,  0.00) rectangle (505.89,289.08);
\definecolor{drawColor}{RGB}{139,0,0}

\path[draw=drawColor,line width= 0.4pt,line join=round,line cap=round] ( 68.51, 76.09) -- ( 78.71, 76.09);

\path[draw=drawColor,line width= 0.4pt,line join=round,line cap=round] ( 90.64, 75.21) -- (100.96, 73.67);

\path[draw=drawColor,line width= 0.4pt,line join=round,line cap=round] (112.83, 73.67) -- (123.15, 75.21);

\path[draw=drawColor,line width= 0.4pt,line join=round,line cap=round] (134.99, 77.19) -- (145.38, 79.13);

\path[draw=drawColor,line width= 0.4pt,line join=round,line cap=round] (157.03, 81.94) -- (167.72, 85.13);

\path[draw=drawColor,line width= 0.4pt,line join=round,line cap=round] (179.45, 87.29) -- (189.68, 88.05);

\path[draw=drawColor,line width= 0.4pt,line join=round,line cap=round] (201.14, 90.95) -- (212.37, 95.97);

\path[draw=drawColor,line width= 0.4pt,line join=round,line cap=round] (223.60,100.14) -- (234.29,103.33);

\path[draw=drawColor,line width= 0.4pt,line join=round,line cap=round] (243.76,109.75) -- (258.51,128.46);

\path[draw=drawColor,line width= 0.4pt,line join=round,line cap=round] (267.78,135.44) -- (278.87,139.99);

\path[draw=drawColor,line width= 0.4pt,line join=round,line cap=round] (288.89,146.27) -- (302.14,158.12);

\path[draw=drawColor,line width= 0.4pt,line join=round,line cap=round] (311.00,166.21) -- (324.41,178.71);

\path[draw=drawColor,line width= 0.4pt,line join=round,line cap=round] (333.96,185.88) -- (345.84,192.96);

\path[draw=drawColor,line width= 0.4pt,line join=round,line cap=round] (355.89,199.51) -- (368.29,208.29);

\path[draw=drawColor,line width= 0.4pt,line join=round,line cap=round] (378.58,214.37) -- (389.98,219.89);
\definecolor{fillColor}{RGB}{139,0,0}

\path[draw=drawColor,line width= 0.4pt,line join=round,line cap=round,fill=fillColor] ( 62.51, 76.09) circle (  2.25);

\path[draw=drawColor,line width= 0.4pt,line join=round,line cap=round,fill=fillColor] ( 84.71, 76.09) circle (  2.25);

\path[draw=drawColor,line width= 0.4pt,line join=round,line cap=round,fill=fillColor] (106.90, 72.78) circle (  2.25);

\path[draw=drawColor,line width= 0.4pt,line join=round,line cap=round,fill=fillColor] (129.09, 76.09) circle (  2.25);

\path[draw=drawColor,line width= 0.4pt,line join=round,line cap=round,fill=fillColor] (151.28, 80.23) circle (  2.25);

\path[draw=drawColor,line width= 0.4pt,line join=round,line cap=round,fill=fillColor] (173.47, 86.84) circle (  2.25);

\path[draw=drawColor,line width= 0.4pt,line join=round,line cap=round,fill=fillColor] (195.66, 88.50) circle (  2.25);

\path[draw=drawColor,line width= 0.4pt,line join=round,line cap=round,fill=fillColor] (217.85, 98.43) circle (  2.25);

\path[draw=drawColor,line width= 0.4pt,line join=round,line cap=round,fill=fillColor] (240.04,105.04) circle (  2.25);

\path[draw=drawColor,line width= 0.4pt,line join=round,line cap=round,fill=fillColor] (262.23,133.17) circle (  2.25);

\path[draw=drawColor,line width= 0.4pt,line join=round,line cap=round,fill=fillColor] (284.42,142.27) circle (  2.25);

\path[draw=drawColor,line width= 0.4pt,line join=round,line cap=round,fill=fillColor] (306.61,162.12) circle (  2.25);

\path[draw=drawColor,line width= 0.4pt,line join=round,line cap=round,fill=fillColor] (328.80,182.80) circle (  2.25);

\path[draw=drawColor,line width= 0.4pt,line join=round,line cap=round,fill=fillColor] (350.99,196.04) circle (  2.25);

\path[draw=drawColor,line width= 0.4pt,line join=round,line cap=round,fill=fillColor] (373.18,211.75) circle (  2.25);

\path[draw=drawColor,line width= 0.4pt,line join=round,line cap=round,fill=fillColor] (395.38,222.51) circle (  2.25);
\end{scope}
\begin{scope}
\path[clip] (  0.00,  0.00) rectangle (505.89,289.08);
\definecolor{drawColor}{RGB}{0,0,0}

\path[draw=drawColor,line width= 0.4pt,line join=round,line cap=round] ( 62.51, 61.20) -- (395.38, 61.20);

\path[draw=drawColor,line width= 0.4pt,line join=round,line cap=round] ( 62.51, 61.20) -- ( 62.51, 55.20);

\path[draw=drawColor,line width= 0.4pt,line join=round,line cap=round] (129.09, 61.20) -- (129.09, 55.20);

\path[draw=drawColor,line width= 0.4pt,line join=round,line cap=round] (195.66, 61.20) -- (195.66, 55.20);

\path[draw=drawColor,line width= 0.4pt,line join=round,line cap=round] (262.23, 61.20) -- (262.23, 55.20);

\path[draw=drawColor,line width= 0.4pt,line join=round,line cap=round] (328.80, 61.20) -- (328.80, 55.20);

\path[draw=drawColor,line width= 0.4pt,line join=round,line cap=round] (395.38, 61.20) -- (395.38, 55.20);

\node[text=drawColor,anchor=base,inner sep=0pt, outer sep=0pt, scale=  1.00] at ( 62.51, 39.60) {0.0};

\node[text=drawColor,anchor=base,inner sep=0pt, outer sep=0pt, scale=  1.00] at (129.09, 39.60) {0.2};

\node[text=drawColor,anchor=base,inner sep=0pt, outer sep=0pt, scale=  1.00] at (195.66, 39.60) {0.4};

\node[text=drawColor,anchor=base,inner sep=0pt, outer sep=0pt, scale=  1.00] at (262.23, 39.60) {0.6};

\node[text=drawColor,anchor=base,inner sep=0pt, outer sep=0pt, scale=  1.00] at (328.80, 39.60) {0.8};

\node[text=drawColor,anchor=base,inner sep=0pt, outer sep=0pt, scale=  1.00] at (395.38, 39.60) {1.0};

\path[draw=drawColor,line width= 0.4pt,line join=round,line cap=round] ( 49.20, 67.82) -- ( 49.20,233.26);

\path[draw=drawColor,line width= 0.4pt,line join=round,line cap=round] ( 49.20, 67.82) -- ( 43.20, 67.82);

\path[draw=drawColor,line width= 0.4pt,line join=round,line cap=round] ( 49.20,100.91) -- ( 43.20,100.91);

\path[draw=drawColor,line width= 0.4pt,line join=round,line cap=round] ( 49.20,134.00) -- ( 43.20,134.00);

\path[draw=drawColor,line width= 0.4pt,line join=round,line cap=round] ( 49.20,167.08) -- ( 43.20,167.08);

\path[draw=drawColor,line width= 0.4pt,line join=round,line cap=round] ( 49.20,200.17) -- ( 43.20,200.17);

\path[draw=drawColor,line width= 0.4pt,line join=round,line cap=round] ( 49.20,233.26) -- ( 43.20,233.26);

\node[text=drawColor,rotate= 90.00,anchor=base,inner sep=0pt, outer sep=0pt, scale=  1.00] at ( 34.80, 67.82) {0.0};

\node[text=drawColor,rotate= 90.00,anchor=base,inner sep=0pt, outer sep=0pt, scale=  1.00] at ( 34.80,100.91) {0.2};

\node[text=drawColor,rotate= 90.00,anchor=base,inner sep=0pt, outer sep=0pt, scale=  1.00] at ( 34.80,134.00) {0.4};

\node[text=drawColor,rotate= 90.00,anchor=base,inner sep=0pt, outer sep=0pt, scale=  1.00] at ( 34.80,167.08) {0.6};

\node[text=drawColor,rotate= 90.00,anchor=base,inner sep=0pt, outer sep=0pt, scale=  1.00] at ( 34.80,200.17) {0.8};

\node[text=drawColor,rotate= 90.00,anchor=base,inner sep=0pt, outer sep=0pt, scale=  1.00] at ( 34.80,233.26) {1.0};

\path[draw=drawColor,line width= 0.4pt,line join=round,line cap=round] ( 49.20, 61.20) --
	(408.69, 61.20) --
	(408.69,239.88) --
	( 49.20,239.88) --
	( 49.20, 61.20);
\end{scope}
\begin{scope}
\path[clip] (  0.00,  0.00) rectangle (505.89,289.08);
\definecolor{drawColor}{RGB}{0,0,0}

\node[text=drawColor,anchor=base,inner sep=0pt, outer sep=0pt, scale=  1.00] at (228.94, 15.60) {$\delta$};

\node[text=drawColor,rotate= 90.00,anchor=base,inner sep=0pt, outer sep=0pt, scale=  1.00] at ( 10.80,150.54) {Power};
\definecolor{drawColor}{RGB}{255,99,71}

\path[draw=drawColor,line width= 0.4pt,line join=round,line cap=round] ( 68.51, 78.80) -- ( 78.71, 79.18);

\path[draw=drawColor,line width= 0.4pt,line join=round,line cap=round] ( 90.71, 79.40) -- (100.90, 79.40);

\path[draw=drawColor,line width= 0.4pt,line join=round,line cap=round] (112.88, 78.95) -- (123.10, 78.19);

\path[draw=drawColor,line width= 0.4pt,line join=round,line cap=round] (135.05, 78.41) -- (145.31, 79.56);

\path[draw=drawColor,line width= 0.4pt,line join=round,line cap=round] (157.26, 80.67) -- (167.48, 81.43);

\path[draw=drawColor,line width= 0.4pt,line join=round,line cap=round] (179.40, 82.77) -- (189.72, 84.30);

\path[draw=drawColor,line width= 0.4pt,line join=round,line cap=round] (201.35, 87.10) -- (212.16, 90.73);

\path[draw=drawColor,line width= 0.4pt,line join=round,line cap=round] (223.47, 94.73) -- (234.42, 98.81);

\path[draw=drawColor,line width= 0.4pt,line join=round,line cap=round] (244.85,104.49) -- (257.42,113.86);

\path[draw=drawColor,line width= 0.4pt,line join=round,line cap=round] (268.04,118.97) -- (278.62,121.73);

\path[draw=drawColor,line width= 0.4pt,line join=round,line cap=round] (289.15,126.94) -- (301.89,136.91);

\path[draw=drawColor,line width= 0.4pt,line join=round,line cap=round] (312.30,142.52) -- (323.11,146.15);

\path[draw=drawColor,line width= 0.4pt,line join=round,line cap=round] (332.80,152.53) -- (346.99,168.40);

\path[draw=drawColor,line width= 0.4pt,line join=round,line cap=round] (356.31,175.65) -- (367.87,181.68);

\path[draw=drawColor,line width= 0.4pt,line join=round,line cap=round] (378.25,187.67) -- (390.31,195.31);
\definecolor{fillColor}{RGB}{255,99,71}

\path[fill=fillColor] ( 62.51, 82.07) --
	( 65.54, 76.82) --
	( 59.48, 76.82) --
	cycle;

\path[fill=fillColor] ( 84.71, 82.90) --
	( 87.74, 77.65) --
	( 81.67, 77.65) --
	cycle;

\path[fill=fillColor] (106.90, 82.90) --
	(109.93, 77.65) --
	(103.87, 77.65) --
	cycle;

\path[fill=fillColor] (129.09, 81.24) --
	(132.12, 75.99) --
	(126.06, 75.99) --
	cycle;

\path[fill=fillColor] (151.28, 83.73) --
	(154.31, 78.48) --
	(148.25, 78.48) --
	cycle;

\path[fill=fillColor] (173.47, 85.38) --
	(176.50, 80.13) --
	(170.44, 80.13) --
	cycle;

\path[fill=fillColor] (195.66, 88.69) --
	(198.69, 83.44) --
	(192.63, 83.44) --
	cycle;

\path[fill=fillColor] (217.85, 96.13) --
	(220.88, 90.88) --
	(214.82, 90.88) --
	cycle;

\path[fill=fillColor] (240.04,104.41) --
	(243.07, 99.16) --
	(237.01, 99.16) --
	cycle;

\path[fill=fillColor] (262.23,120.95) --
	(265.26,115.70) --
	(259.20,115.70) --
	cycle;

\path[fill=fillColor] (284.42,126.74) --
	(287.45,121.49) --
	(281.39,121.49) --
	cycle;

\path[fill=fillColor] (306.61,144.11) --
	(309.64,138.86) --
	(303.58,138.86) --
	cycle;

\path[fill=fillColor] (328.80,151.56) --
	(331.83,146.31) --
	(325.77,146.31) --
	cycle;

\path[fill=fillColor] (350.99,176.37) --
	(354.02,171.13) --
	(347.96,171.13) --
	cycle;

\path[fill=fillColor] (373.18,187.96) --
	(376.22,182.71) --
	(370.15,182.71) --
	cycle;

\path[fill=fillColor] (395.38,202.02) --
	(398.41,196.77) --
	(392.35,196.77) --
	cycle;
\definecolor{drawColor}{RGB}{108,166,205}

\path[draw=drawColor,line width= 0.4pt,line join=round,line cap=round] ( 68.45, 76.15) -- ( 78.77, 77.69);

\path[draw=drawColor,line width= 0.4pt,line join=round,line cap=round] ( 90.70, 78.35) -- (100.90, 77.97);

\path[draw=drawColor,line width= 0.4pt,line join=round,line cap=round] (112.90, 77.74) -- (123.09, 77.74);

\path[draw=drawColor,line width= 0.4pt,line join=round,line cap=round] (135.07, 77.30) -- (145.29, 76.54);

\path[draw=drawColor,line width= 0.4pt,line join=round,line cap=round] (157.26, 76.54) -- (167.48, 77.30);

\path[draw=drawColor,line width= 0.4pt,line join=round,line cap=round] (179.27, 79.26) -- (189.85, 82.02);

\path[draw=drawColor,line width= 0.4pt,line join=round,line cap=round] (201.66, 83.53) -- (211.85, 83.53);

\path[draw=drawColor,line width= 0.4pt,line join=round,line cap=round] (223.83, 83.98) -- (234.06, 84.74);

\path[draw=drawColor,line width= 0.4pt,line join=round,line cap=round] (245.28, 88.12) -- (256.99, 94.67);

\path[draw=drawColor,line width= 0.4pt,line join=round,line cap=round] (268.21, 97.15) -- (278.44, 96.39);

\path[draw=drawColor,line width= 0.4pt,line join=round,line cap=round] (290.04, 98.04) -- (300.99,102.12);

\path[draw=drawColor,line width= 0.4pt,line join=round,line cap=round] (312.16,106.49) -- (323.25,111.04);

\path[draw=drawColor,line width= 0.4pt,line join=round,line cap=round] (333.70,116.78) -- (346.10,125.56);

\path[draw=drawColor,line width= 0.4pt,line join=round,line cap=round] (355.98,132.38) -- (368.20,140.58);

\path[draw=drawColor,line width= 0.4pt,line join=round,line cap=round] (378.17,147.27) -- (390.39,155.47);
\definecolor{fillColor}{RGB}{108,166,205}

\path[fill=fillColor] ( 60.26, 73.01) --
	( 64.76, 73.01) --
	( 64.76, 77.51) --
	( 60.26, 77.51) --
	cycle;

\path[fill=fillColor] ( 82.46, 76.32) --
	( 86.96, 76.32) --
	( 86.96, 80.82) --
	( 82.46, 80.82) --
	cycle;

\path[fill=fillColor] (104.65, 75.49) --
	(109.15, 75.49) --
	(109.15, 79.99) --
	(104.65, 79.99) --
	cycle;

\path[fill=fillColor] (126.84, 75.49) --
	(131.34, 75.49) --
	(131.34, 79.99) --
	(126.84, 79.99) --
	cycle;

\path[fill=fillColor] (149.03, 73.84) --
	(153.53, 73.84) --
	(153.53, 78.34) --
	(149.03, 78.34) --
	cycle;

\path[fill=fillColor] (171.22, 75.49) --
	(175.72, 75.49) --
	(175.72, 79.99) --
	(171.22, 79.99) --
	cycle;

\path[fill=fillColor] (193.41, 81.28) --
	(197.91, 81.28) --
	(197.91, 85.78) --
	(193.41, 85.78) --
	cycle;

\path[fill=fillColor] (215.60, 81.28) --
	(220.10, 81.28) --
	(220.10, 85.78) --
	(215.60, 85.78) --
	cycle;

\path[fill=fillColor] (237.79, 82.94) --
	(242.29, 82.94) --
	(242.29, 87.44) --
	(237.79, 87.44) --
	cycle;

\path[fill=fillColor] (259.98, 95.35) --
	(264.48, 95.35) --
	(264.48, 99.85) --
	(259.98, 99.85) --
	cycle;

\path[fill=fillColor] (282.17, 93.69) --
	(286.67, 93.69) --
	(286.67, 98.19) --
	(282.17, 98.19) --
	cycle;

\path[fill=fillColor] (304.36,101.97) --
	(308.86,101.97) --
	(308.86,106.47) --
	(304.36,106.47) --
	cycle;

\path[fill=fillColor] (326.55,111.07) --
	(331.05,111.07) --
	(331.05,115.57) --
	(326.55,115.57) --
	cycle;

\path[fill=fillColor] (348.74,126.78) --
	(353.24,126.78) --
	(353.24,131.28) --
	(348.74,131.28) --
	cycle;

\path[fill=fillColor] (370.93,141.67) --
	(375.43,141.67) --
	(375.43,146.17) --
	(370.93,146.17) --
	cycle;

\path[fill=fillColor] (393.13,156.56) --
	(397.63,156.56) --
	(397.63,161.06) --
	(393.13,161.06) --
	cycle;
\definecolor{drawColor}{RGB}{139,137,137}

\path[draw=drawColor,line width= 0.4pt,line join=round,line cap=round] ( 68.51, 76.69) -- ( 78.71, 76.31);

\path[draw=drawColor,line width= 0.4pt,line join=round,line cap=round] ( 90.71, 76.09) -- (100.90, 76.09);

\path[draw=drawColor,line width= 0.4pt,line join=round,line cap=round] (112.90, 76.09) -- (123.09, 76.09);

\path[draw=drawColor,line width= 0.4pt,line join=round,line cap=round] (135.05, 76.76) -- (145.31, 77.90);

\path[draw=drawColor,line width= 0.4pt,line join=round,line cap=round] (157.26, 79.02) -- (167.48, 79.78);

\path[draw=drawColor,line width= 0.4pt,line join=round,line cap=round] (179.45, 80.67) -- (189.68, 81.43);

\path[draw=drawColor,line width= 0.4pt,line join=round,line cap=round] (201.41, 83.60) -- (212.10, 86.78);

\path[draw=drawColor,line width= 0.4pt,line join=round,line cap=round] (223.66, 90.01) -- (234.23, 92.77);

\path[draw=drawColor,line width= 0.4pt,line join=round,line cap=round] (245.90, 95.60) -- (256.38, 97.94);

\path[draw=drawColor,line width= 0.4pt,line join=round,line cap=round] (267.85,101.35) -- (278.80,105.43);

\path[draw=drawColor,line width= 0.4pt,line join=round,line cap=round] (289.90,109.97) -- (301.14,115.00);

\path[draw=drawColor,line width= 0.4pt,line join=round,line cap=round] (311.85,120.38) -- (323.57,126.93);

\path[draw=drawColor,line width= 0.4pt,line join=round,line cap=round] (333.96,132.93) -- (345.84,140.02);

\path[draw=drawColor,line width= 0.4pt,line join=round,line cap=round] (356.23,146.02) -- (367.95,152.58);

\path[draw=drawColor,line width= 0.4pt,line join=round,line cap=round] (377.74,159.41) -- (390.82,170.62);

\path[draw=drawColor,line width= 0.4pt,line join=round,line cap=round] ( 59.33, 76.92) -- ( 65.70, 76.92);

\path[draw=drawColor,line width= 0.4pt,line join=round,line cap=round] ( 62.51, 73.74) -- ( 62.51, 80.10);

\path[draw=drawColor,line width= 0.4pt,line join=round,line cap=round] ( 81.52, 76.09) -- ( 87.89, 76.09);

\path[draw=drawColor,line width= 0.4pt,line join=round,line cap=round] ( 84.71, 72.91) -- ( 84.71, 79.27);

\path[draw=drawColor,line width= 0.4pt,line join=round,line cap=round] (103.71, 76.09) -- (110.08, 76.09);

\path[draw=drawColor,line width= 0.4pt,line join=round,line cap=round] (106.90, 72.91) -- (106.90, 79.27);

\path[draw=drawColor,line width= 0.4pt,line join=round,line cap=round] (125.90, 76.09) -- (132.27, 76.09);

\path[draw=drawColor,line width= 0.4pt,line join=round,line cap=round] (129.09, 72.91) -- (129.09, 79.27);

\path[draw=drawColor,line width= 0.4pt,line join=round,line cap=round] (148.10, 78.57) -- (154.46, 78.57);

\path[draw=drawColor,line width= 0.4pt,line join=round,line cap=round] (151.28, 75.39) -- (151.28, 81.75);

\path[draw=drawColor,line width= 0.4pt,line join=round,line cap=round] (170.29, 80.23) -- (176.65, 80.23);

\path[draw=drawColor,line width= 0.4pt,line join=round,line cap=round] (173.47, 77.04) -- (173.47, 83.41);

\path[draw=drawColor,line width= 0.4pt,line join=round,line cap=round] (192.48, 81.88) -- (198.84, 81.88);

\path[draw=drawColor,line width= 0.4pt,line join=round,line cap=round] (195.66, 78.70) -- (195.66, 85.06);

\path[draw=drawColor,line width= 0.4pt,line join=round,line cap=round] (214.67, 88.50) -- (221.03, 88.50);

\path[draw=drawColor,line width= 0.4pt,line join=round,line cap=round] (217.85, 85.32) -- (217.85, 91.68);

\path[draw=drawColor,line width= 0.4pt,line join=round,line cap=round] (236.86, 94.29) -- (243.22, 94.29);

\path[draw=drawColor,line width= 0.4pt,line join=round,line cap=round] (240.04, 91.11) -- (240.04, 97.47);

\path[draw=drawColor,line width= 0.4pt,line join=round,line cap=round] (259.05, 99.25) -- (265.41, 99.25);

\path[draw=drawColor,line width= 0.4pt,line join=round,line cap=round] (262.23, 96.07) -- (262.23,102.43);

\path[draw=drawColor,line width= 0.4pt,line join=round,line cap=round] (281.24,107.52) -- (287.60,107.52);

\path[draw=drawColor,line width= 0.4pt,line join=round,line cap=round] (284.42,104.34) -- (284.42,110.71);

\path[draw=drawColor,line width= 0.4pt,line join=round,line cap=round] (303.43,117.45) -- (309.79,117.45);

\path[draw=drawColor,line width= 0.4pt,line join=round,line cap=round] (306.61,114.27) -- (306.61,120.63);

\path[draw=drawColor,line width= 0.4pt,line join=round,line cap=round] (325.62,129.86) -- (331.99,129.86);

\path[draw=drawColor,line width= 0.4pt,line join=round,line cap=round] (328.80,126.68) -- (328.80,133.04);

\path[draw=drawColor,line width= 0.4pt,line join=round,line cap=round] (347.81,143.10) -- (354.18,143.10);

\path[draw=drawColor,line width= 0.4pt,line join=round,line cap=round] (350.99,139.91) -- (350.99,146.28);

\path[draw=drawColor,line width= 0.4pt,line join=round,line cap=round] (370.00,155.50) -- (376.37,155.50);

\path[draw=drawColor,line width= 0.4pt,line join=round,line cap=round] (373.18,152.32) -- (373.18,158.69);

\path[draw=drawColor,line width= 0.4pt,line join=round,line cap=round] (392.19,174.53) -- (398.56,174.53);

\path[draw=drawColor,line width= 0.4pt,line join=round,line cap=round] (395.38,171.35) -- (395.38,177.71);
\definecolor{drawColor}{RGB}{84,139,84}

\path[draw=drawColor,line width= 0.4pt,line join=round,line cap=round] ( 68.51, 74.44) -- ( 78.71, 74.44);

\path[draw=drawColor,line width= 0.4pt,line join=round,line cap=round] ( 90.70, 74.66) -- (100.90, 75.04);

\path[draw=drawColor,line width= 0.4pt,line join=round,line cap=round] (112.88, 75.71) -- (123.10, 76.47);

\path[draw=drawColor,line width= 0.4pt,line join=round,line cap=round] (135.09, 76.92) -- (145.28, 76.92);

\path[draw=drawColor,line width= 0.4pt,line join=round,line cap=round] (157.18, 78.02) -- (167.57, 79.95);

\path[draw=drawColor,line width= 0.4pt,line join=round,line cap=round] (179.46, 81.28) -- (189.66, 81.66);

\path[draw=drawColor,line width= 0.4pt,line join=round,line cap=round] (201.35, 83.79) -- (212.16, 87.42);

\path[draw=drawColor,line width= 0.4pt,line join=round,line cap=round] (223.09, 92.25) -- (234.80, 98.81);

\path[draw=drawColor,line width= 0.4pt,line join=round,line cap=round] (245.11,104.95) -- (257.16,112.58);

\path[draw=drawColor,line width= 0.4pt,line join=round,line cap=round] (267.21,119.14) -- (279.44,127.34);

\path[draw=drawColor,line width= 0.4pt,line join=round,line cap=round] (288.50,135.09) -- (302.54,150.27);

\path[draw=drawColor,line width= 0.4pt,line join=round,line cap=round] (311.08,158.68) -- (324.33,170.53);

\path[draw=drawColor,line width= 0.4pt,line join=round,line cap=round] (332.73,179.07) -- (347.07,195.64);

\path[draw=drawColor,line width= 0.4pt,line join=round,line cap=round] (356.06,203.39) -- (368.12,211.02);

\path[draw=drawColor,line width= 0.4pt,line join=round,line cap=round] (378.87,216.14) -- (389.69,219.77);

\path[draw=drawColor,line width= 0.4pt,line join=round,line cap=round] ( 60.26, 72.19) -- ( 64.76, 76.69);

\path[draw=drawColor,line width= 0.4pt,line join=round,line cap=round] ( 60.26, 76.69) -- ( 64.76, 72.19);

\path[draw=drawColor,line width= 0.4pt,line join=round,line cap=round] ( 82.46, 72.19) -- ( 86.96, 76.69);

\path[draw=drawColor,line width= 0.4pt,line join=round,line cap=round] ( 82.46, 76.69) -- ( 86.96, 72.19);

\path[draw=drawColor,line width= 0.4pt,line join=round,line cap=round] (104.65, 73.01) -- (109.15, 77.51);

\path[draw=drawColor,line width= 0.4pt,line join=round,line cap=round] (104.65, 77.51) -- (109.15, 73.01);

\path[draw=drawColor,line width= 0.4pt,line join=round,line cap=round] (126.84, 74.67) -- (131.34, 79.17);

\path[draw=drawColor,line width= 0.4pt,line join=round,line cap=round] (126.84, 79.17) -- (131.34, 74.67);

\path[draw=drawColor,line width= 0.4pt,line join=round,line cap=round] (149.03, 74.67) -- (153.53, 79.17);

\path[draw=drawColor,line width= 0.4pt,line join=round,line cap=round] (149.03, 79.17) -- (153.53, 74.67);

\path[draw=drawColor,line width= 0.4pt,line join=round,line cap=round] (171.22, 78.80) -- (175.72, 83.30);

\path[draw=drawColor,line width= 0.4pt,line join=round,line cap=round] (171.22, 83.30) -- (175.72, 78.80);

\path[draw=drawColor,line width= 0.4pt,line join=round,line cap=round] (193.41, 79.63) -- (197.91, 84.13);

\path[draw=drawColor,line width= 0.4pt,line join=round,line cap=round] (193.41, 84.13) -- (197.91, 79.63);

\path[draw=drawColor,line width= 0.4pt,line join=round,line cap=round] (215.60, 87.08) -- (220.10, 91.58);

\path[draw=drawColor,line width= 0.4pt,line join=round,line cap=round] (215.60, 91.58) -- (220.10, 87.08);

\path[draw=drawColor,line width= 0.4pt,line join=round,line cap=round] (237.79, 99.48) -- (242.29,103.98);

\path[draw=drawColor,line width= 0.4pt,line join=round,line cap=round] (237.79,103.98) -- (242.29, 99.48);

\path[draw=drawColor,line width= 0.4pt,line join=round,line cap=round] (259.98,113.55) -- (264.48,118.05);

\path[draw=drawColor,line width= 0.4pt,line join=round,line cap=round] (259.98,118.05) -- (264.48,113.55);

\path[draw=drawColor,line width= 0.4pt,line join=round,line cap=round] (282.17,128.44) -- (286.67,132.94);

\path[draw=drawColor,line width= 0.4pt,line join=round,line cap=round] (282.17,132.94) -- (286.67,128.44);

\path[draw=drawColor,line width= 0.4pt,line join=round,line cap=round] (304.36,152.43) -- (308.86,156.93);

\path[draw=drawColor,line width= 0.4pt,line join=round,line cap=round] (304.36,156.93) -- (308.86,152.43);

\path[draw=drawColor,line width= 0.4pt,line join=round,line cap=round] (326.55,172.28) -- (331.05,176.78);

\path[draw=drawColor,line width= 0.4pt,line join=round,line cap=round] (326.55,176.78) -- (331.05,172.28);

\path[draw=drawColor,line width= 0.4pt,line join=round,line cap=round] (348.74,197.92) -- (353.24,202.42);

\path[draw=drawColor,line width= 0.4pt,line join=round,line cap=round] (348.74,202.42) -- (353.24,197.92);

\path[draw=drawColor,line width= 0.4pt,line join=round,line cap=round] (370.93,211.99) -- (375.43,216.49);

\path[draw=drawColor,line width= 0.4pt,line join=round,line cap=round] (370.93,216.49) -- (375.43,211.99);

\path[draw=drawColor,line width= 0.4pt,line join=round,line cap=round] (393.13,219.43) -- (397.63,223.93);

\path[draw=drawColor,line width= 0.4pt,line join=round,line cap=round] (393.13,223.93) -- (397.63,219.43);
\definecolor{drawColor}{RGB}{238,180,34}

\path[draw=drawColor,line width= 0.4pt,line join=round,line cap=round] ( 68.51, 74.44) -- ( 78.71, 74.44);

\path[draw=drawColor,line width= 0.4pt,line join=round,line cap=round] ( 90.70, 74.66) -- (100.90, 75.04);

\path[draw=drawColor,line width= 0.4pt,line join=round,line cap=round] (112.88, 75.71) -- (123.10, 76.47);

\path[draw=drawColor,line width= 0.4pt,line join=round,line cap=round] (135.08, 77.14) -- (145.28, 77.52);

\path[draw=drawColor,line width= 0.4pt,line join=round,line cap=round] (157.08, 79.26) -- (167.66, 82.02);

\path[draw=drawColor,line width= 0.4pt,line join=round,line cap=round] (179.40, 82.65) -- (189.72, 81.11);

\path[draw=drawColor,line width= 0.4pt,line join=round,line cap=round] (201.56, 81.33) -- (211.95, 83.26);

\path[draw=drawColor,line width= 0.4pt,line join=round,line cap=round] (223.47, 86.46) -- (234.42, 90.54);

\path[draw=drawColor,line width= 0.4pt,line join=round,line cap=round] (245.79, 94.35) -- (256.48, 97.54);

\path[draw=drawColor,line width= 0.4pt,line join=round,line cap=round] (267.92,101.16) -- (278.73,104.79);

\path[draw=drawColor,line width= 0.4pt,line join=round,line cap=round] (289.82,109.31) -- (301.21,114.83);

\path[draw=drawColor,line width= 0.4pt,line join=round,line cap=round] (312.09,119.90) -- (323.33,124.93);

\path[draw=drawColor,line width= 0.4pt,line join=round,line cap=round] (333.87,130.59) -- (345.93,138.23);

\path[draw=drawColor,line width= 0.4pt,line join=round,line cap=round] (356.39,144.06) -- (367.79,149.58);

\path[draw=drawColor,line width= 0.4pt,line join=round,line cap=round] (378.74,154.47) -- (389.82,159.02);

\path[draw=drawColor,line width= 0.4pt,line join=round,line cap=round] ( 62.51, 74.44) circle (  2.25);

\path[draw=drawColor,line width= 0.4pt,line join=round,line cap=round] ( 60.26, 72.19) -- ( 64.76, 76.69);

\path[draw=drawColor,line width= 0.4pt,line join=round,line cap=round] ( 60.26, 76.69) -- ( 64.76, 72.19);

\path[draw=drawColor,line width= 0.4pt,line join=round,line cap=round] ( 84.71, 74.44) circle (  2.25);

\path[draw=drawColor,line width= 0.4pt,line join=round,line cap=round] ( 82.46, 72.19) -- ( 86.96, 76.69);

\path[draw=drawColor,line width= 0.4pt,line join=round,line cap=round] ( 82.46, 76.69) -- ( 86.96, 72.19);

\path[draw=drawColor,line width= 0.4pt,line join=round,line cap=round] (106.90, 75.26) circle (  2.25);

\path[draw=drawColor,line width= 0.4pt,line join=round,line cap=round] (104.65, 73.01) -- (109.15, 77.51);

\path[draw=drawColor,line width= 0.4pt,line join=round,line cap=round] (104.65, 77.51) -- (109.15, 73.01);

\path[draw=drawColor,line width= 0.4pt,line join=round,line cap=round] (129.09, 76.92) circle (  2.25);

\path[draw=drawColor,line width= 0.4pt,line join=round,line cap=round] (126.84, 74.67) -- (131.34, 79.17);

\path[draw=drawColor,line width= 0.4pt,line join=round,line cap=round] (126.84, 79.17) -- (131.34, 74.67);

\path[draw=drawColor,line width= 0.4pt,line join=round,line cap=round] (151.28, 77.74) circle (  2.25);

\path[draw=drawColor,line width= 0.4pt,line join=round,line cap=round] (149.03, 75.49) -- (153.53, 79.99);

\path[draw=drawColor,line width= 0.4pt,line join=round,line cap=round] (149.03, 79.99) -- (153.53, 75.49);

\path[draw=drawColor,line width= 0.4pt,line join=round,line cap=round] (173.47, 83.53) circle (  2.25);

\path[draw=drawColor,line width= 0.4pt,line join=round,line cap=round] (171.22, 81.28) -- (175.72, 85.78);

\path[draw=drawColor,line width= 0.4pt,line join=round,line cap=round] (171.22, 85.78) -- (175.72, 81.28);

\path[draw=drawColor,line width= 0.4pt,line join=round,line cap=round] (195.66, 80.23) circle (  2.25);

\path[draw=drawColor,line width= 0.4pt,line join=round,line cap=round] (193.41, 77.98) -- (197.91, 82.48);

\path[draw=drawColor,line width= 0.4pt,line join=round,line cap=round] (193.41, 82.48) -- (197.91, 77.98);

\path[draw=drawColor,line width= 0.4pt,line join=round,line cap=round] (217.85, 84.36) circle (  2.25);

\path[draw=drawColor,line width= 0.4pt,line join=round,line cap=round] (215.60, 82.11) -- (220.10, 86.61);

\path[draw=drawColor,line width= 0.4pt,line join=round,line cap=round] (215.60, 86.61) -- (220.10, 82.11);

\path[draw=drawColor,line width= 0.4pt,line join=round,line cap=round] (240.04, 92.63) circle (  2.25);

\path[draw=drawColor,line width= 0.4pt,line join=round,line cap=round] (237.79, 90.38) -- (242.29, 94.88);

\path[draw=drawColor,line width= 0.4pt,line join=round,line cap=round] (237.79, 94.88) -- (242.29, 90.38);

\path[draw=drawColor,line width= 0.4pt,line join=round,line cap=round] (262.23, 99.25) circle (  2.25);

\path[draw=drawColor,line width= 0.4pt,line join=round,line cap=round] (259.98, 97.00) -- (264.48,101.50);

\path[draw=drawColor,line width= 0.4pt,line join=round,line cap=round] (259.98,101.50) -- (264.48, 97.00);

\path[draw=drawColor,line width= 0.4pt,line join=round,line cap=round] (284.42,106.70) circle (  2.25);

\path[draw=drawColor,line width= 0.4pt,line join=round,line cap=round] (282.17,104.45) -- (286.67,108.95);

\path[draw=drawColor,line width= 0.4pt,line join=round,line cap=round] (282.17,108.95) -- (286.67,104.45);

\path[draw=drawColor,line width= 0.4pt,line join=round,line cap=round] (306.61,117.45) circle (  2.25);

\path[draw=drawColor,line width= 0.4pt,line join=round,line cap=round] (304.36,115.20) -- (308.86,119.70);

\path[draw=drawColor,line width= 0.4pt,line join=round,line cap=round] (304.36,119.70) -- (308.86,115.20);

\path[draw=drawColor,line width= 0.4pt,line join=round,line cap=round] (328.80,127.38) circle (  2.25);

\path[draw=drawColor,line width= 0.4pt,line join=round,line cap=round] (326.55,125.13) -- (331.05,129.63);

\path[draw=drawColor,line width= 0.4pt,line join=round,line cap=round] (326.55,129.63) -- (331.05,125.13);

\path[draw=drawColor,line width= 0.4pt,line join=round,line cap=round] (350.99,141.44) circle (  2.25);

\path[draw=drawColor,line width= 0.4pt,line join=round,line cap=round] (348.74,139.19) -- (353.24,143.69);

\path[draw=drawColor,line width= 0.4pt,line join=round,line cap=round] (348.74,143.69) -- (353.24,139.19);

\path[draw=drawColor,line width= 0.4pt,line join=round,line cap=round] (373.18,152.19) circle (  2.25);

\path[draw=drawColor,line width= 0.4pt,line join=round,line cap=round] (370.93,149.94) -- (375.43,154.44);

\path[draw=drawColor,line width= 0.4pt,line join=round,line cap=round] (370.93,154.44) -- (375.43,149.94);

\path[draw=drawColor,line width= 0.4pt,line join=round,line cap=round] (395.38,161.29) circle (  2.25);

\path[draw=drawColor,line width= 0.4pt,line join=round,line cap=round] (393.13,159.04) -- (397.63,163.54);

\path[draw=drawColor,line width= 0.4pt,line join=round,line cap=round] (393.13,163.54) -- (397.63,159.04);
\definecolor{drawColor}{RGB}{205,96,144}

\path[draw=drawColor,line width= 0.4pt,line join=round,line cap=round] ( 68.51, 76.09) -- ( 78.71, 76.09);

\path[draw=drawColor,line width= 0.4pt,line join=round,line cap=round] ( 90.67, 75.42) -- (100.93, 74.28);

\path[draw=drawColor,line width= 0.4pt,line join=round,line cap=round] (112.52, 75.70) -- (123.46, 79.78);

\path[draw=drawColor,line width= 0.4pt,line join=round,line cap=round] (135.08, 81.66) -- (145.28, 81.28);

\path[draw=drawColor,line width= 0.4pt,line join=round,line cap=round] (157.27, 80.83) -- (167.47, 80.45);

\path[draw=drawColor,line width= 0.4pt,line join=round,line cap=round] (179.47, 80.23) -- (189.66, 80.23);

\path[draw=drawColor,line width= 0.4pt,line join=round,line cap=round] (201.35, 82.13) -- (212.16, 85.76);

\path[draw=drawColor,line width= 0.4pt,line join=round,line cap=round] (223.00, 90.74) -- (234.89, 97.83);

\path[draw=drawColor,line width= 0.4pt,line join=round,line cap=round] (245.36,103.68) -- (256.91,109.71);

\path[draw=drawColor,line width= 0.4pt,line join=round,line cap=round] (267.78,114.76) -- (278.87,119.31);

\path[draw=drawColor,line width= 0.4pt,line join=round,line cap=round] (288.50,125.99) -- (302.54,141.17);

\path[draw=drawColor,line width= 0.4pt,line join=round,line cap=round] (310.92,149.75) -- (324.49,162.91);

\path[draw=drawColor,line width= 0.4pt,line join=round,line cap=round] (333.27,171.09) -- (346.52,182.94);

\path[draw=drawColor,line width= 0.4pt,line join=round,line cap=round] (356.68,188.85) -- (367.50,192.47);

\path[draw=drawColor,line width= 0.4pt,line join=round,line cap=round] (378.08,197.85) -- (390.48,206.63);

\path[draw=drawColor,line width= 0.4pt,line join=round,line cap=round] ( 60.26, 73.84) rectangle ( 64.76, 78.34);

\path[draw=drawColor,line width= 0.4pt,line join=round,line cap=round] ( 60.26, 73.84) -- ( 64.76, 78.34);

\path[draw=drawColor,line width= 0.4pt,line join=round,line cap=round] ( 60.26, 78.34) -- ( 64.76, 73.84);

\path[draw=drawColor,line width= 0.4pt,line join=round,line cap=round] ( 82.46, 73.84) rectangle ( 86.96, 78.34);

\path[draw=drawColor,line width= 0.4pt,line join=round,line cap=round] ( 82.46, 73.84) -- ( 86.96, 78.34);

\path[draw=drawColor,line width= 0.4pt,line join=round,line cap=round] ( 82.46, 78.34) -- ( 86.96, 73.84);

\path[draw=drawColor,line width= 0.4pt,line join=round,line cap=round] (104.65, 71.36) rectangle (109.15, 75.86);

\path[draw=drawColor,line width= 0.4pt,line join=round,line cap=round] (104.65, 71.36) -- (109.15, 75.86);

\path[draw=drawColor,line width= 0.4pt,line join=round,line cap=round] (104.65, 75.86) -- (109.15, 71.36);

\path[draw=drawColor,line width= 0.4pt,line join=round,line cap=round] (126.84, 79.63) rectangle (131.34, 84.13);

\path[draw=drawColor,line width= 0.4pt,line join=round,line cap=round] (126.84, 79.63) -- (131.34, 84.13);

\path[draw=drawColor,line width= 0.4pt,line join=round,line cap=round] (126.84, 84.13) -- (131.34, 79.63);

\path[draw=drawColor,line width= 0.4pt,line join=round,line cap=round] (149.03, 78.80) rectangle (153.53, 83.30);

\path[draw=drawColor,line width= 0.4pt,line join=round,line cap=round] (149.03, 78.80) -- (153.53, 83.30);

\path[draw=drawColor,line width= 0.4pt,line join=round,line cap=round] (149.03, 83.30) -- (153.53, 78.80);

\path[draw=drawColor,line width= 0.4pt,line join=round,line cap=round] (171.22, 77.98) rectangle (175.72, 82.48);

\path[draw=drawColor,line width= 0.4pt,line join=round,line cap=round] (171.22, 77.98) -- (175.72, 82.48);

\path[draw=drawColor,line width= 0.4pt,line join=round,line cap=round] (171.22, 82.48) -- (175.72, 77.98);

\path[draw=drawColor,line width= 0.4pt,line join=round,line cap=round] (193.41, 77.98) rectangle (197.91, 82.48);

\path[draw=drawColor,line width= 0.4pt,line join=round,line cap=round] (193.41, 77.98) -- (197.91, 82.48);

\path[draw=drawColor,line width= 0.4pt,line join=round,line cap=round] (193.41, 82.48) -- (197.91, 77.98);

\path[draw=drawColor,line width= 0.4pt,line join=round,line cap=round] (215.60, 85.42) rectangle (220.10, 89.92);

\path[draw=drawColor,line width= 0.4pt,line join=round,line cap=round] (215.60, 85.42) -- (220.10, 89.92);

\path[draw=drawColor,line width= 0.4pt,line join=round,line cap=round] (215.60, 89.92) -- (220.10, 85.42);

\path[draw=drawColor,line width= 0.4pt,line join=round,line cap=round] (237.79, 98.66) rectangle (242.29,103.16);

\path[draw=drawColor,line width= 0.4pt,line join=round,line cap=round] (237.79, 98.66) -- (242.29,103.16);

\path[draw=drawColor,line width= 0.4pt,line join=round,line cap=round] (237.79,103.16) -- (242.29, 98.66);

\path[draw=drawColor,line width= 0.4pt,line join=round,line cap=round] (259.98,110.24) rectangle (264.48,114.74);

\path[draw=drawColor,line width= 0.4pt,line join=round,line cap=round] (259.98,110.24) -- (264.48,114.74);

\path[draw=drawColor,line width= 0.4pt,line join=round,line cap=round] (259.98,114.74) -- (264.48,110.24);

\path[draw=drawColor,line width= 0.4pt,line join=round,line cap=round] (282.17,119.34) rectangle (286.67,123.84);

\path[draw=drawColor,line width= 0.4pt,line join=round,line cap=round] (282.17,119.34) -- (286.67,123.84);

\path[draw=drawColor,line width= 0.4pt,line join=round,line cap=round] (282.17,123.84) -- (286.67,119.34);

\path[draw=drawColor,line width= 0.4pt,line join=round,line cap=round] (304.36,143.33) rectangle (308.86,147.83);

\path[draw=drawColor,line width= 0.4pt,line join=round,line cap=round] (304.36,143.33) -- (308.86,147.83);

\path[draw=drawColor,line width= 0.4pt,line join=round,line cap=round] (304.36,147.83) -- (308.86,143.33);

\path[draw=drawColor,line width= 0.4pt,line join=round,line cap=round] (326.55,164.83) rectangle (331.05,169.33);

\path[draw=drawColor,line width= 0.4pt,line join=round,line cap=round] (326.55,164.83) -- (331.05,169.33);

\path[draw=drawColor,line width= 0.4pt,line join=round,line cap=round] (326.55,169.33) -- (331.05,164.83);

\path[draw=drawColor,line width= 0.4pt,line join=round,line cap=round] (348.74,184.69) rectangle (353.24,189.19);

\path[draw=drawColor,line width= 0.4pt,line join=round,line cap=round] (348.74,184.69) -- (353.24,189.19);

\path[draw=drawColor,line width= 0.4pt,line join=round,line cap=round] (348.74,189.19) -- (353.24,184.69);

\path[draw=drawColor,line width= 0.4pt,line join=round,line cap=round] (370.93,192.13) rectangle (375.43,196.63);

\path[draw=drawColor,line width= 0.4pt,line join=round,line cap=round] (370.93,192.13) -- (375.43,196.63);

\path[draw=drawColor,line width= 0.4pt,line join=round,line cap=round] (370.93,196.63) -- (375.43,192.13);

\path[draw=drawColor,line width= 0.4pt,line join=round,line cap=round] (393.13,207.85) rectangle (397.63,212.35);

\path[draw=drawColor,line width= 0.4pt,line join=round,line cap=round] (393.13,207.85) -- (397.63,212.35);

\path[draw=drawColor,line width= 0.4pt,line join=round,line cap=round] (393.13,212.35) -- (397.63,207.85);
\definecolor{drawColor}{RGB}{0,0,0}

\path[draw=drawColor,line width= 0.4pt,dash pattern=on 4pt off 4pt ,line join=round,line cap=round] ( 55.86, 76.09) --
	( 59.35, 76.09) --
	( 62.85, 76.09) --
	( 66.35, 76.09) --
	( 69.84, 76.09) --
	( 73.34, 76.09) --
	( 76.84, 76.09) --
	( 80.33, 76.09) --
	( 83.83, 76.09) --
	( 87.33, 76.09) --
	( 90.82, 76.09) --
	( 94.32, 76.09) --
	( 97.82, 76.09) --
	(101.31, 76.09) --
	(104.81, 76.09) --
	(108.31, 76.09) --
	(111.80, 76.09) --
	(115.30, 76.09) --
	(118.80, 76.09) --
	(122.29, 76.09) --
	(125.79, 76.09) --
	(129.29, 76.09) --
	(132.79, 76.09) --
	(136.28, 76.09) --
	(139.78, 76.09) --
	(143.28, 76.09) --
	(146.77, 76.09) --
	(150.27, 76.09) --
	(153.77, 76.09) --
	(157.26, 76.09) --
	(160.76, 76.09) --
	(164.26, 76.09) --
	(167.75, 76.09) --
	(171.25, 76.09) --
	(174.75, 76.09) --
	(178.24, 76.09) --
	(181.74, 76.09) --
	(185.24, 76.09) --
	(188.73, 76.09) --
	(192.23, 76.09) --
	(195.73, 76.09) --
	(199.22, 76.09) --
	(202.72, 76.09) --
	(206.22, 76.09) --
	(209.71, 76.09) --
	(213.21, 76.09) --
	(216.71, 76.09) --
	(220.20, 76.09) --
	(223.70, 76.09) --
	(227.20, 76.09) --
	(230.69, 76.09) --
	(234.19, 76.09) --
	(237.69, 76.09) --
	(241.18, 76.09) --
	(244.68, 76.09) --
	(248.18, 76.09) --
	(251.67, 76.09) --
	(255.17, 76.09) --
	(258.67, 76.09) --
	(262.16, 76.09) --
	(265.66, 76.09) --
	(269.16, 76.09) --
	(272.65, 76.09) --
	(276.15, 76.09) --
	(279.65, 76.09) --
	(283.14, 76.09) --
	(286.64, 76.09) --
	(290.14, 76.09) --
	(293.63, 76.09) --
	(297.13, 76.09) --
	(300.63, 76.09) --
	(304.12, 76.09) --
	(307.62, 76.09) --
	(311.12, 76.09) --
	(314.61, 76.09) --
	(318.11, 76.09) --
	(321.61, 76.09) --
	(325.10, 76.09) --
	(328.60, 76.09) --
	(332.10, 76.09) --
	(335.60, 76.09) --
	(339.09, 76.09) --
	(342.59, 76.09) --
	(346.09, 76.09) --
	(349.58, 76.09) --
	(353.08, 76.09) --
	(356.58, 76.09) --
	(360.07, 76.09) --
	(363.57, 76.09) --
	(367.07, 76.09) --
	(370.56, 76.09) --
	(374.06, 76.09) --
	(377.56, 76.09) --
	(381.05, 76.09) --
	(384.55, 76.09) --
	(388.05, 76.09) --
	(391.54, 76.09) --
	(395.04, 76.09) --
	(398.54, 76.09) --
	(402.03, 76.09);
\definecolor{drawColor}{RGB}{139,0,0}
\definecolor{fillColor}{RGB}{139,0,0}

\path[draw=drawColor,line width= 0.4pt,line join=round,line cap=round,fill=fillColor] (421.11,227.88) circle (  2.25);
\definecolor{fillColor}{RGB}{255,99,71}

\path[fill=fillColor] (421.11,219.38) --
	(424.15,214.13) --
	(418.08,214.13) --
	cycle;
\definecolor{fillColor}{RGB}{108,166,205}

\path[fill=fillColor] (418.86,201.63) --
	(423.36,201.63) --
	(423.36,206.13) --
	(418.86,206.13) --
	cycle;
\definecolor{drawColor}{RGB}{139,137,137}

\path[draw=drawColor,line width= 0.4pt,line join=round,line cap=round] (417.93,191.88) -- (424.30,191.88);

\path[draw=drawColor,line width= 0.4pt,line join=round,line cap=round] (421.11,188.70) -- (421.11,195.06);
\definecolor{drawColor}{RGB}{84,139,84}

\path[draw=drawColor,line width= 0.4pt,line join=round,line cap=round] (418.86,177.63) -- (423.36,182.13);

\path[draw=drawColor,line width= 0.4pt,line join=round,line cap=round] (418.86,182.13) -- (423.36,177.63);
\definecolor{drawColor}{RGB}{238,180,34}

\path[draw=drawColor,line width= 0.4pt,line join=round,line cap=round] (421.11,167.88) circle (  2.25);

\path[draw=drawColor,line width= 0.4pt,line join=round,line cap=round] (418.86,165.63) -- (423.36,170.13);

\path[draw=drawColor,line width= 0.4pt,line join=round,line cap=round] (418.86,170.13) -- (423.36,165.63);
\definecolor{drawColor}{RGB}{205,96,144}

\path[draw=drawColor,line width= 0.4pt,line join=round,line cap=round] (418.86,153.63) rectangle (423.36,158.13);

\path[draw=drawColor,line width= 0.4pt,line join=round,line cap=round] (418.86,153.63) -- (423.36,158.13);

\path[draw=drawColor,line width= 0.4pt,line join=round,line cap=round] (418.86,158.13) -- (423.36,153.63);
\definecolor{drawColor}{RGB}{0,0,0}

\node[text=drawColor,anchor=base west,inner sep=0pt, outer sep=0pt, scale=  1.00] at (430.11,224.44) {HypoRF};

\node[text=drawColor,anchor=base west,inner sep=0pt, outer sep=0pt, scale=  1.00] at (430.11,212.44) {Binomial};

\node[text=drawColor,anchor=base west,inner sep=0pt, outer sep=0pt, scale=  1.00] at (430.11,200.44) {ME-full};

\node[text=drawColor,anchor=base west,inner sep=0pt, outer sep=0pt, scale=  1.00] at (430.11,188.44) {MMDboot};

\node[text=drawColor,anchor=base west,inner sep=0pt, outer sep=0pt, scale=  1.00] at (430.11,176.44) {MMD-full};

\node[text=drawColor,anchor=base west,inner sep=0pt, outer sep=0pt, scale=  1.00] at (430.11,164.44) {LDA};

\node[text=drawColor,anchor=base west,inner sep=0pt, outer sep=0pt, scale=  1.00] at (430.11,152.44) {CPT-RF};
\end{scope}
\end{tikzpicture}

%% file: 1b_plot_gaussian_meanshift_d20_Normapprox_F_LDA_with_Cai.tex
\begin{tikzpicture}[x=1pt,y=1pt]
\definecolor{fillColor}{RGB}{255,255,255}
\path[use as bounding box,fill=fillColor,fill opacity=0.00] (0,0) rectangle (505.89,289.08);
\begin{scope}
\path[clip] (  0.00,  0.00) rectangle (505.89,289.08);
\definecolor{drawColor}{RGB}{139,0,0}

\path[draw=drawColor,line width= 0.4pt,line join=round,line cap=round] ( 68.51, 75.87) -- ( 78.71, 75.49);

\path[draw=drawColor,line width= 0.4pt,line join=round,line cap=round] ( 90.69, 75.71) -- (100.91, 76.47);

\path[draw=drawColor,line width= 0.4pt,line join=round,line cap=round] (112.52, 79.01) -- (123.46, 83.09);

\path[draw=drawColor,line width= 0.4pt,line join=round,line cap=round] (134.99, 84.09) -- (145.38, 82.15);

\path[draw=drawColor,line width= 0.4pt,line join=round,line cap=round] (157.21, 81.94) -- (167.53, 83.48);

\path[draw=drawColor,line width= 0.4pt,line join=round,line cap=round] (179.22, 86.08) -- (189.91, 89.27);

\path[draw=drawColor,line width= 0.4pt,line join=round,line cap=round] (200.30, 94.78) -- (213.21,105.37);

\path[draw=drawColor,line width= 0.4pt,line join=round,line cap=round] (223.00,112.25) -- (234.89,119.34);

\path[draw=drawColor,line width= 0.4pt,line join=round,line cap=round] (245.19,125.49) -- (257.08,132.58);

\path[draw=drawColor,line width= 0.4pt,line join=round,line cap=round] (266.23,140.12) -- (280.42,155.99);

\path[draw=drawColor,line width= 0.4pt,line join=round,line cap=round] (288.50,164.87) -- (302.54,180.05);

\path[draw=drawColor,line width= 0.4pt,line join=round,line cap=round] (311.08,188.46) -- (324.33,200.31);

\path[draw=drawColor,line width= 0.4pt,line join=round,line cap=round] (333.87,207.52) -- (345.93,215.16);

\path[draw=drawColor,line width= 0.4pt,line join=round,line cap=round] (356.74,220.09) -- (367.44,223.28);

\path[draw=drawColor,line width= 0.4pt,line join=round,line cap=round] (378.99,226.50) -- (389.57,229.27);
\definecolor{fillColor}{RGB}{139,0,0}

\path[draw=drawColor,line width= 0.4pt,line join=round,line cap=round,fill=fillColor] ( 62.51, 76.09) circle (  2.25);

\path[draw=drawColor,line width= 0.4pt,line join=round,line cap=round,fill=fillColor] ( 84.71, 75.26) circle (  2.25);

\path[draw=drawColor,line width= 0.4pt,line join=round,line cap=round,fill=fillColor] (106.90, 76.92) circle (  2.25);

\path[draw=drawColor,line width= 0.4pt,line join=round,line cap=round,fill=fillColor] (129.09, 85.19) circle (  2.25);

\path[draw=drawColor,line width= 0.4pt,line join=round,line cap=round,fill=fillColor] (151.28, 81.05) circle (  2.25);

\path[draw=drawColor,line width= 0.4pt,line join=round,line cap=round,fill=fillColor] (173.47, 84.36) circle (  2.25);

\path[draw=drawColor,line width= 0.4pt,line join=round,line cap=round,fill=fillColor] (195.66, 90.98) circle (  2.25);

\path[draw=drawColor,line width= 0.4pt,line join=round,line cap=round,fill=fillColor] (217.85,109.18) circle (  2.25);

\path[draw=drawColor,line width= 0.4pt,line join=round,line cap=round,fill=fillColor] (240.04,122.41) circle (  2.25);

\path[draw=drawColor,line width= 0.4pt,line join=round,line cap=round,fill=fillColor] (262.23,135.65) circle (  2.25);

\path[draw=drawColor,line width= 0.4pt,line join=round,line cap=round,fill=fillColor] (284.42,160.47) circle (  2.25);

\path[draw=drawColor,line width= 0.4pt,line join=round,line cap=round,fill=fillColor] (306.61,184.46) circle (  2.25);

\path[draw=drawColor,line width= 0.4pt,line join=round,line cap=round,fill=fillColor] (328.80,204.31) circle (  2.25);

\path[draw=drawColor,line width= 0.4pt,line join=round,line cap=round,fill=fillColor] (350.99,218.37) circle (  2.25);

\path[draw=drawColor,line width= 0.4pt,line join=round,line cap=round,fill=fillColor] (373.18,224.99) circle (  2.25);

\path[draw=drawColor,line width= 0.4pt,line join=round,line cap=round,fill=fillColor] (395.38,230.78) circle (  2.25);
\end{scope}
\begin{scope}
\path[clip] (  0.00,  0.00) rectangle (505.89,289.08);
\definecolor{drawColor}{RGB}{0,0,0}

\path[draw=drawColor,line width= 0.4pt,line join=round,line cap=round] ( 62.51, 61.20) -- (395.38, 61.20);

\path[draw=drawColor,line width= 0.4pt,line join=round,line cap=round] ( 62.51, 61.20) -- ( 62.51, 55.20);

\path[draw=drawColor,line width= 0.4pt,line join=round,line cap=round] (129.09, 61.20) -- (129.09, 55.20);

\path[draw=drawColor,line width= 0.4pt,line join=round,line cap=round] (195.66, 61.20) -- (195.66, 55.20);

\path[draw=drawColor,line width= 0.4pt,line join=round,line cap=round] (262.23, 61.20) -- (262.23, 55.20);

\path[draw=drawColor,line width= 0.4pt,line join=round,line cap=round] (328.80, 61.20) -- (328.80, 55.20);

\path[draw=drawColor,line width= 0.4pt,line join=round,line cap=round] (395.38, 61.20) -- (395.38, 55.20);

\node[text=drawColor,anchor=base,inner sep=0pt, outer sep=0pt, scale=  1.00] at ( 62.51, 39.60) {0.0};

\node[text=drawColor,anchor=base,inner sep=0pt, outer sep=0pt, scale=  1.00] at (129.09, 39.60) {0.2};

\node[text=drawColor,anchor=base,inner sep=0pt, outer sep=0pt, scale=  1.00] at (195.66, 39.60) {0.4};

\node[text=drawColor,anchor=base,inner sep=0pt, outer sep=0pt, scale=  1.00] at (262.23, 39.60) {0.6};

\node[text=drawColor,anchor=base,inner sep=0pt, outer sep=0pt, scale=  1.00] at (328.80, 39.60) {0.8};

\node[text=drawColor,anchor=base,inner sep=0pt, outer sep=0pt, scale=  1.00] at (395.38, 39.60) {1.0};

\path[draw=drawColor,line width= 0.4pt,line join=round,line cap=round] ( 49.20, 67.82) -- ( 49.20,233.26);

\path[draw=drawColor,line width= 0.4pt,line join=round,line cap=round] ( 49.20, 67.82) -- ( 43.20, 67.82);

\path[draw=drawColor,line width= 0.4pt,line join=round,line cap=round] ( 49.20,100.91) -- ( 43.20,100.91);

\path[draw=drawColor,line width= 0.4pt,line join=round,line cap=round] ( 49.20,134.00) -- ( 43.20,134.00);

\path[draw=drawColor,line width= 0.4pt,line join=round,line cap=round] ( 49.20,167.08) -- ( 43.20,167.08);

\path[draw=drawColor,line width= 0.4pt,line join=round,line cap=round] ( 49.20,200.17) -- ( 43.20,200.17);

\path[draw=drawColor,line width= 0.4pt,line join=round,line cap=round] ( 49.20,233.26) -- ( 43.20,233.26);

\node[text=drawColor,rotate= 90.00,anchor=base,inner sep=0pt, outer sep=0pt, scale=  1.00] at ( 34.80, 67.82) {0.0};

\node[text=drawColor,rotate= 90.00,anchor=base,inner sep=0pt, outer sep=0pt, scale=  1.00] at ( 34.80,100.91) {0.2};

\node[text=drawColor,rotate= 90.00,anchor=base,inner sep=0pt, outer sep=0pt, scale=  1.00] at ( 34.80,134.00) {0.4};

\node[text=drawColor,rotate= 90.00,anchor=base,inner sep=0pt, outer sep=0pt, scale=  1.00] at ( 34.80,167.08) {0.6};

\node[text=drawColor,rotate= 90.00,anchor=base,inner sep=0pt, outer sep=0pt, scale=  1.00] at ( 34.80,200.17) {0.8};

\node[text=drawColor,rotate= 90.00,anchor=base,inner sep=0pt, outer sep=0pt, scale=  1.00] at ( 34.80,233.26) {1.0};

\path[draw=drawColor,line width= 0.4pt,line join=round,line cap=round] ( 49.20, 61.20) --
	(408.69, 61.20) --
	(408.69,239.88) --
	( 49.20,239.88) --
	( 49.20, 61.20);
\end{scope}
\begin{scope}
\path[clip] (  0.00,  0.00) rectangle (505.89,289.08);
\definecolor{drawColor}{RGB}{0,0,0}

\node[text=drawColor,anchor=base,inner sep=0pt, outer sep=0pt, scale=  1.00] at (228.94, 15.60) {$\delta$};

\node[text=drawColor,rotate= 90.00,anchor=base,inner sep=0pt, outer sep=0pt, scale=  1.00] at ( 10.80,150.54) {Power};
\definecolor{drawColor}{RGB}{255,99,71}

\path[draw=drawColor,line width= 0.4pt,line join=round,line cap=round] ( 68.45, 77.69) -- ( 78.77, 76.15);

\path[draw=drawColor,line width= 0.4pt,line join=round,line cap=round] ( 90.71, 75.26) -- (100.90, 75.26);

\path[draw=drawColor,line width= 0.4pt,line join=round,line cap=round] (112.90, 75.26) -- (123.09, 75.26);

\path[draw=drawColor,line width= 0.4pt,line join=round,line cap=round] (134.99, 76.36) -- (145.38, 78.30);

\path[draw=drawColor,line width= 0.4pt,line join=round,line cap=round] (157.24, 80.07) -- (167.51, 81.21);

\path[draw=drawColor,line width= 0.4pt,line join=round,line cap=round] (179.22, 83.60) -- (189.91, 86.78);

\path[draw=drawColor,line width= 0.4pt,line join=round,line cap=round] (201.28, 90.59) -- (212.23, 94.67);

\path[draw=drawColor,line width= 0.4pt,line join=round,line cap=round] (223.33, 99.22) -- (234.56,104.25);

\path[draw=drawColor,line width= 0.4pt,line join=round,line cap=round] (245.02,110.04) -- (257.25,118.24);

\path[draw=drawColor,line width= 0.4pt,line join=round,line cap=round] (267.55,124.36) -- (279.10,130.39);

\path[draw=drawColor,line width= 0.4pt,line join=round,line cap=round] (288.81,137.26) -- (302.22,149.76);

\path[draw=drawColor,line width= 0.4pt,line join=round,line cap=round] (311.25,157.65) -- (324.16,168.24);

\path[draw=drawColor,line width= 0.4pt,line join=round,line cap=round] (333.11,176.22) -- (346.69,189.38);

\path[draw=drawColor,line width= 0.4pt,line join=round,line cap=round] (355.89,197.02) -- (368.29,205.80);

\path[draw=drawColor,line width= 0.4pt,line join=round,line cap=round] (378.81,211.37) -- (389.75,215.45);
\definecolor{fillColor}{RGB}{255,99,71}

\path[fill=fillColor] ( 62.51, 82.07) --
	( 65.54, 76.82) --
	( 59.48, 76.82) --
	cycle;

\path[fill=fillColor] ( 84.71, 78.76) --
	( 87.74, 73.51) --
	( 81.67, 73.51) --
	cycle;

\path[fill=fillColor] (106.90, 78.76) --
	(109.93, 73.51) --
	(103.87, 73.51) --
	cycle;

\path[fill=fillColor] (129.09, 78.76) --
	(132.12, 73.51) --
	(126.06, 73.51) --
	cycle;

\path[fill=fillColor] (151.28, 82.90) --
	(154.31, 77.65) --
	(148.25, 77.65) --
	cycle;

\path[fill=fillColor] (173.47, 85.38) --
	(176.50, 80.13) --
	(170.44, 80.13) --
	cycle;

\path[fill=fillColor] (195.66, 92.00) --
	(198.69, 86.75) --
	(192.63, 86.75) --
	cycle;

\path[fill=fillColor] (217.85,100.27) --
	(220.88, 95.02) --
	(214.82, 95.02) --
	cycle;

\path[fill=fillColor] (240.04,110.20) --
	(243.07,104.95) --
	(237.01,104.95) --
	cycle;

\path[fill=fillColor] (262.23,125.09) --
	(265.26,119.84) --
	(259.20,119.84) --
	cycle;

\path[fill=fillColor] (284.42,136.67) --
	(287.45,131.42) --
	(281.39,131.42) --
	cycle;

\path[fill=fillColor] (306.61,157.35) --
	(309.64,152.10) --
	(303.58,152.10) --
	cycle;

\path[fill=fillColor] (328.80,175.55) --
	(331.83,170.30) --
	(325.77,170.30) --
	cycle;

\path[fill=fillColor] (350.99,197.05) --
	(354.02,191.81) --
	(347.96,191.81) --
	cycle;

\path[fill=fillColor] (373.18,212.77) --
	(376.22,207.52) --
	(370.15,207.52) --
	cycle;

\path[fill=fillColor] (395.38,221.04) --
	(398.41,215.80) --
	(392.35,215.80) --
	cycle;
\definecolor{drawColor}{RGB}{108,166,205}

\path[draw=drawColor,line width= 0.4pt,line join=round,line cap=round] ( 68.51, 75.49) -- ( 78.71, 75.87);

\path[draw=drawColor,line width= 0.4pt,line join=round,line cap=round] ( 90.67, 75.42) -- (100.93, 74.28);

\path[draw=drawColor,line width= 0.4pt,line join=round,line cap=round] (112.83, 74.49) -- (123.15, 76.03);

\path[draw=drawColor,line width= 0.4pt,line join=round,line cap=round] (135.02, 77.80) -- (145.34, 79.34);

\path[draw=drawColor,line width= 0.4pt,line join=round,line cap=round] (157.21, 79.34) -- (167.53, 77.80);

\path[draw=drawColor,line width= 0.4pt,line join=round,line cap=round] (179.27, 78.43) -- (189.85, 81.19);

\path[draw=drawColor,line width= 0.4pt,line join=round,line cap=round] (201.65, 82.93) -- (211.85, 83.31);

\path[draw=drawColor,line width= 0.4pt,line join=round,line cap=round] (223.85, 83.76) -- (234.04, 84.14);

\path[draw=drawColor,line width= 0.4pt,line join=round,line cap=round] (245.66, 86.46) -- (256.61, 90.54);

\path[draw=drawColor,line width= 0.4pt,line join=round,line cap=round] (268.21, 93.08) -- (278.44, 93.84);

\path[draw=drawColor,line width= 0.4pt,line join=round,line cap=round] (289.82, 96.91) -- (301.21,102.43);

\path[draw=drawColor,line width= 0.4pt,line join=round,line cap=round] (311.59,108.39) -- (323.82,116.59);

\path[draw=drawColor,line width= 0.4pt,line join=round,line cap=round] (333.96,123.01) -- (345.84,130.09);

\path[draw=drawColor,line width= 0.4pt,line join=round,line cap=round] (356.47,135.62) -- (367.71,140.64);

\path[draw=drawColor,line width= 0.4pt,line join=round,line cap=round] (378.08,146.56) -- (390.48,155.34);
\definecolor{fillColor}{RGB}{108,166,205}

\path[fill=fillColor] ( 60.26, 73.01) --
	( 64.76, 73.01) --
	( 64.76, 77.51) --
	( 60.26, 77.51) --
	cycle;

\path[fill=fillColor] ( 82.46, 73.84) --
	( 86.96, 73.84) --
	( 86.96, 78.34) --
	( 82.46, 78.34) --
	cycle;

\path[fill=fillColor] (104.65, 71.36) --
	(109.15, 71.36) --
	(109.15, 75.86) --
	(104.65, 75.86) --
	cycle;

\path[fill=fillColor] (126.84, 74.67) --
	(131.34, 74.67) --
	(131.34, 79.17) --
	(126.84, 79.17) --
	cycle;

\path[fill=fillColor] (149.03, 77.98) --
	(153.53, 77.98) --
	(153.53, 82.48) --
	(149.03, 82.48) --
	cycle;

\path[fill=fillColor] (171.22, 74.67) --
	(175.72, 74.67) --
	(175.72, 79.17) --
	(171.22, 79.17) --
	cycle;

\path[fill=fillColor] (193.41, 80.46) --
	(197.91, 80.46) --
	(197.91, 84.96) --
	(193.41, 84.96) --
	cycle;

\path[fill=fillColor] (215.60, 81.28) --
	(220.10, 81.28) --
	(220.10, 85.78) --
	(215.60, 85.78) --
	cycle;

\path[fill=fillColor] (237.79, 82.11) --
	(242.29, 82.11) --
	(242.29, 86.61) --
	(237.79, 86.61) --
	cycle;

\path[fill=fillColor] (259.98, 90.38) --
	(264.48, 90.38) --
	(264.48, 94.88) --
	(259.98, 94.88) --
	cycle;

\path[fill=fillColor] (282.17, 92.04) --
	(286.67, 92.04) --
	(286.67, 96.54) --
	(282.17, 96.54) --
	cycle;

\path[fill=fillColor] (304.36,102.79) --
	(308.86,102.79) --
	(308.86,107.29) --
	(304.36,107.29) --
	cycle;

\path[fill=fillColor] (326.55,117.68) --
	(331.05,117.68) --
	(331.05,122.18) --
	(326.55,122.18) --
	cycle;

\path[fill=fillColor] (348.74,130.92) --
	(353.24,130.92) --
	(353.24,135.42) --
	(348.74,135.42) --
	cycle;

\path[fill=fillColor] (370.93,140.85) --
	(375.43,140.85) --
	(375.43,145.35) --
	(370.93,145.35) --
	cycle;

\path[fill=fillColor] (393.13,156.56) --
	(397.63,156.56) --
	(397.63,161.06) --
	(393.13,161.06) --
	cycle;
\definecolor{drawColor}{RGB}{139,137,137}

\path[draw=drawColor,line width= 0.4pt,line join=round,line cap=round] ( 68.51, 77.14) -- ( 78.71, 77.52);

\path[draw=drawColor,line width= 0.4pt,line join=round,line cap=round] ( 90.71, 77.74) -- (100.90, 77.74);

\path[draw=drawColor,line width= 0.4pt,line join=round,line cap=round] (112.89, 77.97) -- (123.09, 78.35);

\path[draw=drawColor,line width= 0.4pt,line join=round,line cap=round] (135.08, 78.80) -- (145.28, 79.18);

\path[draw=drawColor,line width= 0.4pt,line join=round,line cap=round] (157.24, 80.07) -- (167.51, 81.21);

\path[draw=drawColor,line width= 0.4pt,line join=round,line cap=round] (179.37, 82.98) -- (189.76, 84.92);

\path[draw=drawColor,line width= 0.4pt,line join=round,line cap=round] (201.41, 87.73) -- (212.10, 90.92);

\path[draw=drawColor,line width= 0.4pt,line join=round,line cap=round] (223.70, 93.94) -- (234.19, 96.29);

\path[draw=drawColor,line width= 0.4pt,line join=round,line cap=round] (245.73, 99.51) -- (256.54,103.13);

\path[draw=drawColor,line width= 0.4pt,line join=round,line cap=round] (267.98,106.76) -- (278.67,109.95);

\path[draw=drawColor,line width= 0.4pt,line join=round,line cap=round] (290.04,113.76) -- (300.99,117.84);

\path[draw=drawColor,line width= 0.4pt,line join=round,line cap=round] (311.51,123.40) -- (323.91,132.18);

\path[draw=drawColor,line width= 0.4pt,line join=round,line cap=round] (334.28,138.10) -- (345.52,143.13);

\path[draw=drawColor,line width= 0.4pt,line join=round,line cap=round] (356.15,148.65) -- (368.03,155.74);

\path[draw=drawColor,line width= 0.4pt,line join=round,line cap=round] (378.66,161.26) -- (389.90,166.29);

\path[draw=drawColor,line width= 0.4pt,line join=round,line cap=round] ( 59.33, 76.92) -- ( 65.70, 76.92);

\path[draw=drawColor,line width= 0.4pt,line join=round,line cap=round] ( 62.51, 73.74) -- ( 62.51, 80.10);

\path[draw=drawColor,line width= 0.4pt,line join=round,line cap=round] ( 81.52, 77.74) -- ( 87.89, 77.74);

\path[draw=drawColor,line width= 0.4pt,line join=round,line cap=round] ( 84.71, 74.56) -- ( 84.71, 80.93);

\path[draw=drawColor,line width= 0.4pt,line join=round,line cap=round] (103.71, 77.74) -- (110.08, 77.74);

\path[draw=drawColor,line width= 0.4pt,line join=round,line cap=round] (106.90, 74.56) -- (106.90, 80.93);

\path[draw=drawColor,line width= 0.4pt,line join=round,line cap=round] (125.90, 78.57) -- (132.27, 78.57);

\path[draw=drawColor,line width= 0.4pt,line join=round,line cap=round] (129.09, 75.39) -- (129.09, 81.75);

\path[draw=drawColor,line width= 0.4pt,line join=round,line cap=round] (148.10, 79.40) -- (154.46, 79.40);

\path[draw=drawColor,line width= 0.4pt,line join=round,line cap=round] (151.28, 76.22) -- (151.28, 82.58);

\path[draw=drawColor,line width= 0.4pt,line join=round,line cap=round] (170.29, 81.88) -- (176.65, 81.88);

\path[draw=drawColor,line width= 0.4pt,line join=round,line cap=round] (173.47, 78.70) -- (173.47, 85.06);

\path[draw=drawColor,line width= 0.4pt,line join=round,line cap=round] (192.48, 86.02) -- (198.84, 86.02);

\path[draw=drawColor,line width= 0.4pt,line join=round,line cap=round] (195.66, 82.83) -- (195.66, 89.20);

\path[draw=drawColor,line width= 0.4pt,line join=round,line cap=round] (214.67, 92.63) -- (221.03, 92.63);

\path[draw=drawColor,line width= 0.4pt,line join=round,line cap=round] (217.85, 89.45) -- (217.85, 95.82);

\path[draw=drawColor,line width= 0.4pt,line join=round,line cap=round] (236.86, 97.60) -- (243.22, 97.60);

\path[draw=drawColor,line width= 0.4pt,line join=round,line cap=round] (240.04, 94.42) -- (240.04,100.78);

\path[draw=drawColor,line width= 0.4pt,line join=round,line cap=round] (259.05,105.04) -- (265.41,105.04);

\path[draw=drawColor,line width= 0.4pt,line join=round,line cap=round] (262.23,101.86) -- (262.23,108.22);

\path[draw=drawColor,line width= 0.4pt,line join=round,line cap=round] (281.24,111.66) -- (287.60,111.66);

\path[draw=drawColor,line width= 0.4pt,line join=round,line cap=round] (284.42,108.48) -- (284.42,114.84);

\path[draw=drawColor,line width= 0.4pt,line join=round,line cap=round] (303.43,119.93) -- (309.79,119.93);

\path[draw=drawColor,line width= 0.4pt,line join=round,line cap=round] (306.61,116.75) -- (306.61,123.11);

\path[draw=drawColor,line width= 0.4pt,line join=round,line cap=round] (325.62,135.65) -- (331.99,135.65);

\path[draw=drawColor,line width= 0.4pt,line join=round,line cap=round] (328.80,132.47) -- (328.80,138.83);

\path[draw=drawColor,line width= 0.4pt,line join=round,line cap=round] (347.81,145.58) -- (354.18,145.58);

\path[draw=drawColor,line width= 0.4pt,line join=round,line cap=round] (350.99,142.39) -- (350.99,148.76);

\path[draw=drawColor,line width= 0.4pt,line join=round,line cap=round] (370.00,158.81) -- (376.37,158.81);

\path[draw=drawColor,line width= 0.4pt,line join=round,line cap=round] (373.18,155.63) -- (373.18,161.99);

\path[draw=drawColor,line width= 0.4pt,line join=round,line cap=round] (392.19,168.74) -- (398.56,168.74);

\path[draw=drawColor,line width= 0.4pt,line join=round,line cap=round] (395.38,165.56) -- (395.38,171.92);
\definecolor{drawColor}{RGB}{84,139,84}

\path[draw=drawColor,line width= 0.4pt,line join=round,line cap=round] ( 68.51, 74.44) -- ( 78.71, 74.44);

\path[draw=drawColor,line width= 0.4pt,line join=round,line cap=round] ( 90.71, 74.44) -- (100.90, 74.44);

\path[draw=drawColor,line width= 0.4pt,line join=round,line cap=round] (112.89, 74.66) -- (123.09, 75.04);

\path[draw=drawColor,line width= 0.4pt,line join=round,line cap=round] (135.07, 75.71) -- (145.29, 76.47);

\path[draw=drawColor,line width= 0.4pt,line join=round,line cap=round] (157.21, 77.80) -- (167.53, 79.34);

\path[draw=drawColor,line width= 0.4pt,line join=round,line cap=round] (179.37, 81.33) -- (189.76, 83.26);

\path[draw=drawColor,line width= 0.4pt,line join=round,line cap=round] (201.35, 86.27) -- (212.16, 89.90);

\path[draw=drawColor,line width= 0.4pt,line join=round,line cap=round] (223.09, 94.74) -- (234.80,101.29);

\path[draw=drawColor,line width= 0.4pt,line join=round,line cap=round] (245.36,106.99) -- (256.91,113.02);

\path[draw=drawColor,line width= 0.4pt,line join=round,line cap=round] (266.96,119.50) -- (279.70,129.47);

\path[draw=drawColor,line width= 0.4pt,line join=round,line cap=round] (288.73,137.34) -- (302.30,150.50);

\path[draw=drawColor,line width= 0.4pt,line join=round,line cap=round] (310.20,159.49) -- (325.22,179.64);

\path[draw=drawColor,line width= 0.4pt,line join=round,line cap=round] (333.19,188.55) -- (346.60,201.05);

\path[draw=drawColor,line width= 0.4pt,line join=round,line cap=round] (356.62,207.23) -- (367.56,211.31);

\path[draw=drawColor,line width= 0.4pt,line join=round,line cap=round] (378.50,216.18) -- (390.06,222.21);

\path[draw=drawColor,line width= 0.4pt,line join=round,line cap=round] ( 60.26, 72.19) -- ( 64.76, 76.69);

\path[draw=drawColor,line width= 0.4pt,line join=round,line cap=round] ( 60.26, 76.69) -- ( 64.76, 72.19);

\path[draw=drawColor,line width= 0.4pt,line join=round,line cap=round] ( 82.46, 72.19) -- ( 86.96, 76.69);

\path[draw=drawColor,line width= 0.4pt,line join=round,line cap=round] ( 82.46, 76.69) -- ( 86.96, 72.19);

\path[draw=drawColor,line width= 0.4pt,line join=round,line cap=round] (104.65, 72.19) -- (109.15, 76.69);

\path[draw=drawColor,line width= 0.4pt,line join=round,line cap=round] (104.65, 76.69) -- (109.15, 72.19);

\path[draw=drawColor,line width= 0.4pt,line join=round,line cap=round] (126.84, 73.01) -- (131.34, 77.51);

\path[draw=drawColor,line width= 0.4pt,line join=round,line cap=round] (126.84, 77.51) -- (131.34, 73.01);

\path[draw=drawColor,line width= 0.4pt,line join=round,line cap=round] (149.03, 74.67) -- (153.53, 79.17);

\path[draw=drawColor,line width= 0.4pt,line join=round,line cap=round] (149.03, 79.17) -- (153.53, 74.67);

\path[draw=drawColor,line width= 0.4pt,line join=round,line cap=round] (171.22, 77.98) -- (175.72, 82.48);

\path[draw=drawColor,line width= 0.4pt,line join=round,line cap=round] (171.22, 82.48) -- (175.72, 77.98);

\path[draw=drawColor,line width= 0.4pt,line join=round,line cap=round] (193.41, 82.11) -- (197.91, 86.61);

\path[draw=drawColor,line width= 0.4pt,line join=round,line cap=round] (193.41, 86.61) -- (197.91, 82.11);

\path[draw=drawColor,line width= 0.4pt,line join=round,line cap=round] (215.60, 89.56) -- (220.10, 94.06);

\path[draw=drawColor,line width= 0.4pt,line join=round,line cap=round] (215.60, 94.06) -- (220.10, 89.56);

\path[draw=drawColor,line width= 0.4pt,line join=round,line cap=round] (237.79,101.97) -- (242.29,106.47);

\path[draw=drawColor,line width= 0.4pt,line join=round,line cap=round] (237.79,106.47) -- (242.29,101.97);

\path[draw=drawColor,line width= 0.4pt,line join=round,line cap=round] (259.98,113.55) -- (264.48,118.05);

\path[draw=drawColor,line width= 0.4pt,line join=round,line cap=round] (259.98,118.05) -- (264.48,113.55);

\path[draw=drawColor,line width= 0.4pt,line join=round,line cap=round] (282.17,130.92) -- (286.67,135.42);

\path[draw=drawColor,line width= 0.4pt,line join=round,line cap=round] (282.17,135.42) -- (286.67,130.92);

\path[draw=drawColor,line width= 0.4pt,line join=round,line cap=round] (304.36,152.43) -- (308.86,156.93);

\path[draw=drawColor,line width= 0.4pt,line join=round,line cap=round] (304.36,156.93) -- (308.86,152.43);

\path[draw=drawColor,line width= 0.4pt,line join=round,line cap=round] (326.55,182.21) -- (331.05,186.71);

\path[draw=drawColor,line width= 0.4pt,line join=round,line cap=round] (326.55,186.71) -- (331.05,182.21);

\path[draw=drawColor,line width= 0.4pt,line join=round,line cap=round] (348.74,202.89) -- (353.24,207.39);

\path[draw=drawColor,line width= 0.4pt,line join=round,line cap=round] (348.74,207.39) -- (353.24,202.89);

\path[draw=drawColor,line width= 0.4pt,line join=round,line cap=round] (370.93,211.16) -- (375.43,215.66);

\path[draw=drawColor,line width= 0.4pt,line join=round,line cap=round] (370.93,215.66) -- (375.43,211.16);

\path[draw=drawColor,line width= 0.4pt,line join=round,line cap=round] (393.13,222.74) -- (397.63,227.24);

\path[draw=drawColor,line width= 0.4pt,line join=round,line cap=round] (393.13,227.24) -- (397.63,222.74);
\definecolor{drawColor}{RGB}{238,180,34}

\path[draw=drawColor,line width= 0.4pt,line join=round,line cap=round] ( 68.51, 74.44) -- ( 78.71, 74.44);

\path[draw=drawColor,line width= 0.4pt,line join=round,line cap=round] ( 90.69, 74.88) -- (100.91, 75.64);

\path[draw=drawColor,line width= 0.4pt,line join=round,line cap=round] (112.86, 76.76) -- (123.12, 77.90);

\path[draw=drawColor,line width= 0.4pt,line join=round,line cap=round] (135.08, 78.35) -- (145.28, 77.97);

\path[draw=drawColor,line width= 0.4pt,line join=round,line cap=round] (157.26, 78.19) -- (167.48, 78.95);

\path[draw=drawColor,line width= 0.4pt,line join=round,line cap=round] (179.45, 79.84) -- (189.68, 80.61);

\path[draw=drawColor,line width= 0.4pt,line join=round,line cap=round] (201.46, 82.57) -- (212.04, 85.33);

\path[draw=drawColor,line width= 0.4pt,line join=round,line cap=round] (223.47, 88.94) -- (234.42, 93.02);

\path[draw=drawColor,line width= 0.4pt,line join=round,line cap=round] (245.97, 96.00) -- (256.30, 97.54);

\path[draw=drawColor,line width= 0.4pt,line join=round,line cap=round] (267.78,100.70) -- (278.87,105.25);

\path[draw=drawColor,line width= 0.4pt,line join=round,line cap=round] (288.98,111.43) -- (302.06,122.65);

\path[draw=drawColor,line width= 0.4pt,line join=round,line cap=round] (312.36,128.27) -- (323.05,131.45);

\path[draw=drawColor,line width= 0.4pt,line join=round,line cap=round] (334.12,135.94) -- (345.67,141.97);

\path[draw=drawColor,line width= 0.4pt,line join=round,line cap=round] (355.47,148.75) -- (368.71,160.60);

\path[draw=drawColor,line width= 0.4pt,line join=round,line cap=round] (379.08,165.70) -- (389.48,167.64);

\path[draw=drawColor,line width= 0.4pt,line join=round,line cap=round] ( 62.51, 74.44) circle (  2.25);

\path[draw=drawColor,line width= 0.4pt,line join=round,line cap=round] ( 60.26, 72.19) -- ( 64.76, 76.69);

\path[draw=drawColor,line width= 0.4pt,line join=round,line cap=round] ( 60.26, 76.69) -- ( 64.76, 72.19);

\path[draw=drawColor,line width= 0.4pt,line join=round,line cap=round] ( 84.71, 74.44) circle (  2.25);

\path[draw=drawColor,line width= 0.4pt,line join=round,line cap=round] ( 82.46, 72.19) -- ( 86.96, 76.69);

\path[draw=drawColor,line width= 0.4pt,line join=round,line cap=round] ( 82.46, 76.69) -- ( 86.96, 72.19);

\path[draw=drawColor,line width= 0.4pt,line join=round,line cap=round] (106.90, 76.09) circle (  2.25);

\path[draw=drawColor,line width= 0.4pt,line join=round,line cap=round] (104.65, 73.84) -- (109.15, 78.34);

\path[draw=drawColor,line width= 0.4pt,line join=round,line cap=round] (104.65, 78.34) -- (109.15, 73.84);

\path[draw=drawColor,line width= 0.4pt,line join=round,line cap=round] (129.09, 78.57) circle (  2.25);

\path[draw=drawColor,line width= 0.4pt,line join=round,line cap=round] (126.84, 76.32) -- (131.34, 80.82);

\path[draw=drawColor,line width= 0.4pt,line join=round,line cap=round] (126.84, 80.82) -- (131.34, 76.32);

\path[draw=drawColor,line width= 0.4pt,line join=round,line cap=round] (151.28, 77.74) circle (  2.25);

\path[draw=drawColor,line width= 0.4pt,line join=round,line cap=round] (149.03, 75.49) -- (153.53, 79.99);

\path[draw=drawColor,line width= 0.4pt,line join=round,line cap=round] (149.03, 79.99) -- (153.53, 75.49);

\path[draw=drawColor,line width= 0.4pt,line join=round,line cap=round] (173.47, 79.40) circle (  2.25);

\path[draw=drawColor,line width= 0.4pt,line join=round,line cap=round] (171.22, 77.15) -- (175.72, 81.65);

\path[draw=drawColor,line width= 0.4pt,line join=round,line cap=round] (171.22, 81.65) -- (175.72, 77.15);

\path[draw=drawColor,line width= 0.4pt,line join=round,line cap=round] (195.66, 81.05) circle (  2.25);

\path[draw=drawColor,line width= 0.4pt,line join=round,line cap=round] (193.41, 78.80) -- (197.91, 83.30);

\path[draw=drawColor,line width= 0.4pt,line join=round,line cap=round] (193.41, 83.30) -- (197.91, 78.80);

\path[draw=drawColor,line width= 0.4pt,line join=round,line cap=round] (217.85, 86.84) circle (  2.25);

\path[draw=drawColor,line width= 0.4pt,line join=round,line cap=round] (215.60, 84.59) -- (220.10, 89.09);

\path[draw=drawColor,line width= 0.4pt,line join=round,line cap=round] (215.60, 89.09) -- (220.10, 84.59);

\path[draw=drawColor,line width= 0.4pt,line join=round,line cap=round] (240.04, 95.12) circle (  2.25);

\path[draw=drawColor,line width= 0.4pt,line join=round,line cap=round] (237.79, 92.87) -- (242.29, 97.37);

\path[draw=drawColor,line width= 0.4pt,line join=round,line cap=round] (237.79, 97.37) -- (242.29, 92.87);

\path[draw=drawColor,line width= 0.4pt,line join=round,line cap=round] (262.23, 98.43) circle (  2.25);

\path[draw=drawColor,line width= 0.4pt,line join=round,line cap=round] (259.98, 96.18) -- (264.48,100.68);

\path[draw=drawColor,line width= 0.4pt,line join=round,line cap=round] (259.98,100.68) -- (264.48, 96.18);

\path[draw=drawColor,line width= 0.4pt,line join=round,line cap=round] (284.42,107.52) circle (  2.25);

\path[draw=drawColor,line width= 0.4pt,line join=round,line cap=round] (282.17,105.27) -- (286.67,109.77);

\path[draw=drawColor,line width= 0.4pt,line join=round,line cap=round] (282.17,109.77) -- (286.67,105.27);

\path[draw=drawColor,line width= 0.4pt,line join=round,line cap=round] (306.61,126.55) circle (  2.25);

\path[draw=drawColor,line width= 0.4pt,line join=round,line cap=round] (304.36,124.30) -- (308.86,128.80);

\path[draw=drawColor,line width= 0.4pt,line join=round,line cap=round] (304.36,128.80) -- (308.86,124.30);

\path[draw=drawColor,line width= 0.4pt,line join=round,line cap=round] (328.80,133.17) circle (  2.25);

\path[draw=drawColor,line width= 0.4pt,line join=round,line cap=round] (326.55,130.92) -- (331.05,135.42);

\path[draw=drawColor,line width= 0.4pt,line join=round,line cap=round] (326.55,135.42) -- (331.05,130.92);

\path[draw=drawColor,line width= 0.4pt,line join=round,line cap=round] (350.99,144.75) circle (  2.25);

\path[draw=drawColor,line width= 0.4pt,line join=round,line cap=round] (348.74,142.50) -- (353.24,147.00);

\path[draw=drawColor,line width= 0.4pt,line join=round,line cap=round] (348.74,147.00) -- (353.24,142.50);

\path[draw=drawColor,line width= 0.4pt,line join=round,line cap=round] (373.18,164.60) circle (  2.25);

\path[draw=drawColor,line width= 0.4pt,line join=round,line cap=round] (370.93,162.35) -- (375.43,166.85);

\path[draw=drawColor,line width= 0.4pt,line join=round,line cap=round] (370.93,166.85) -- (375.43,162.35);

\path[draw=drawColor,line width= 0.4pt,line join=round,line cap=round] (395.38,168.74) circle (  2.25);

\path[draw=drawColor,line width= 0.4pt,line join=round,line cap=round] (393.13,166.49) -- (397.63,170.99);

\path[draw=drawColor,line width= 0.4pt,line join=round,line cap=round] (393.13,170.99) -- (397.63,166.49);
\definecolor{drawColor}{RGB}{205,96,144}

\path[draw=drawColor,line width= 0.4pt,line join=round,line cap=round] ( 68.51, 75.87) -- ( 78.71, 75.49);

\path[draw=drawColor,line width= 0.4pt,line join=round,line cap=round] ( 90.70, 75.04) -- (100.90, 74.66);

\path[draw=drawColor,line width= 0.4pt,line join=round,line cap=round] (112.70, 75.95) -- (123.28, 78.71);

\path[draw=drawColor,line width= 0.4pt,line join=round,line cap=round] (135.08, 80.00) -- (145.28, 79.62);

\path[draw=drawColor,line width= 0.4pt,line join=round,line cap=round] (157.03, 81.11) -- (167.72, 84.30);

\path[draw=drawColor,line width= 0.4pt,line join=round,line cap=round] (179.46, 85.79) -- (189.66, 85.41);

\path[draw=drawColor,line width= 0.4pt,line join=round,line cap=round] (201.21, 87.47) -- (212.30, 92.01);

\path[draw=drawColor,line width= 0.4pt,line join=round,line cap=round] (223.17, 97.06) -- (234.72,103.09);

\path[draw=drawColor,line width= 0.4pt,line join=round,line cap=round] (244.60,109.78) -- (257.68,120.99);

\path[draw=drawColor,line width= 0.4pt,line join=round,line cap=round] (267.04,128.48) -- (279.61,137.85);

\path[draw=drawColor,line width= 0.4pt,line join=round,line cap=round] (289.23,145.03) -- (301.80,154.40);

\path[draw=drawColor,line width= 0.4pt,line join=round,line cap=round] (310.92,162.16) -- (324.49,175.32);

\path[draw=drawColor,line width= 0.4pt,line join=round,line cap=round] (332.66,184.09) -- (347.14,201.37);

\path[draw=drawColor,line width= 0.4pt,line join=round,line cap=round] (355.72,209.66) -- (368.46,219.64);

\path[draw=drawColor,line width= 0.4pt,line join=round,line cap=round] (379.08,224.43) -- (389.48,226.37);

\path[draw=drawColor,line width= 0.4pt,line join=round,line cap=round] ( 60.26, 73.84) rectangle ( 64.76, 78.34);

\path[draw=drawColor,line width= 0.4pt,line join=round,line cap=round] ( 60.26, 73.84) -- ( 64.76, 78.34);

\path[draw=drawColor,line width= 0.4pt,line join=round,line cap=round] ( 60.26, 78.34) -- ( 64.76, 73.84);

\path[draw=drawColor,line width= 0.4pt,line join=round,line cap=round] ( 82.46, 73.01) rectangle ( 86.96, 77.51);

\path[draw=drawColor,line width= 0.4pt,line join=round,line cap=round] ( 82.46, 73.01) -- ( 86.96, 77.51);

\path[draw=drawColor,line width= 0.4pt,line join=round,line cap=round] ( 82.46, 77.51) -- ( 86.96, 73.01);

\path[draw=drawColor,line width= 0.4pt,line join=round,line cap=round] (104.65, 72.19) rectangle (109.15, 76.69);

\path[draw=drawColor,line width= 0.4pt,line join=round,line cap=round] (104.65, 72.19) -- (109.15, 76.69);

\path[draw=drawColor,line width= 0.4pt,line join=round,line cap=round] (104.65, 76.69) -- (109.15, 72.19);

\path[draw=drawColor,line width= 0.4pt,line join=round,line cap=round] (126.84, 77.98) rectangle (131.34, 82.48);

\path[draw=drawColor,line width= 0.4pt,line join=round,line cap=round] (126.84, 77.98) -- (131.34, 82.48);

\path[draw=drawColor,line width= 0.4pt,line join=round,line cap=round] (126.84, 82.48) -- (131.34, 77.98);

\path[draw=drawColor,line width= 0.4pt,line join=round,line cap=round] (149.03, 77.15) rectangle (153.53, 81.65);

\path[draw=drawColor,line width= 0.4pt,line join=round,line cap=round] (149.03, 77.15) -- (153.53, 81.65);

\path[draw=drawColor,line width= 0.4pt,line join=round,line cap=round] (149.03, 81.65) -- (153.53, 77.15);

\path[draw=drawColor,line width= 0.4pt,line join=round,line cap=round] (171.22, 83.77) rectangle (175.72, 88.27);

\path[draw=drawColor,line width= 0.4pt,line join=round,line cap=round] (171.22, 83.77) -- (175.72, 88.27);

\path[draw=drawColor,line width= 0.4pt,line join=round,line cap=round] (171.22, 88.27) -- (175.72, 83.77);

\path[draw=drawColor,line width= 0.4pt,line join=round,line cap=round] (193.41, 82.94) rectangle (197.91, 87.44);

\path[draw=drawColor,line width= 0.4pt,line join=round,line cap=round] (193.41, 82.94) -- (197.91, 87.44);

\path[draw=drawColor,line width= 0.4pt,line join=round,line cap=round] (193.41, 87.44) -- (197.91, 82.94);

\path[draw=drawColor,line width= 0.4pt,line join=round,line cap=round] (215.60, 92.04) rectangle (220.10, 96.54);

\path[draw=drawColor,line width= 0.4pt,line join=round,line cap=round] (215.60, 92.04) -- (220.10, 96.54);

\path[draw=drawColor,line width= 0.4pt,line join=round,line cap=round] (215.60, 96.54) -- (220.10, 92.04);

\path[draw=drawColor,line width= 0.4pt,line join=round,line cap=round] (237.79,103.62) rectangle (242.29,108.12);

\path[draw=drawColor,line width= 0.4pt,line join=round,line cap=round] (237.79,103.62) -- (242.29,108.12);

\path[draw=drawColor,line width= 0.4pt,line join=round,line cap=round] (237.79,108.12) -- (242.29,103.62);

\path[draw=drawColor,line width= 0.4pt,line join=round,line cap=round] (259.98,122.65) rectangle (264.48,127.15);

\path[draw=drawColor,line width= 0.4pt,line join=round,line cap=round] (259.98,122.65) -- (264.48,127.15);

\path[draw=drawColor,line width= 0.4pt,line join=round,line cap=round] (259.98,127.15) -- (264.48,122.65);

\path[draw=drawColor,line width= 0.4pt,line join=round,line cap=round] (282.17,139.19) rectangle (286.67,143.69);

\path[draw=drawColor,line width= 0.4pt,line join=round,line cap=round] (282.17,139.19) -- (286.67,143.69);

\path[draw=drawColor,line width= 0.4pt,line join=round,line cap=round] (282.17,143.69) -- (286.67,139.19);

\path[draw=drawColor,line width= 0.4pt,line join=round,line cap=round] (304.36,155.74) rectangle (308.86,160.24);

\path[draw=drawColor,line width= 0.4pt,line join=round,line cap=round] (304.36,155.74) -- (308.86,160.24);

\path[draw=drawColor,line width= 0.4pt,line join=round,line cap=round] (304.36,160.24) -- (308.86,155.74);

\path[draw=drawColor,line width= 0.4pt,line join=round,line cap=round] (326.55,177.24) rectangle (331.05,181.74);

\path[draw=drawColor,line width= 0.4pt,line join=round,line cap=round] (326.55,177.24) -- (331.05,181.74);

\path[draw=drawColor,line width= 0.4pt,line join=round,line cap=round] (326.55,181.74) -- (331.05,177.24);

\path[draw=drawColor,line width= 0.4pt,line join=round,line cap=round] (348.74,203.71) rectangle (353.24,208.21);

\path[draw=drawColor,line width= 0.4pt,line join=round,line cap=round] (348.74,203.71) -- (353.24,208.21);

\path[draw=drawColor,line width= 0.4pt,line join=round,line cap=round] (348.74,208.21) -- (353.24,203.71);

\path[draw=drawColor,line width= 0.4pt,line join=round,line cap=round] (370.93,221.09) rectangle (375.43,225.59);

\path[draw=drawColor,line width= 0.4pt,line join=round,line cap=round] (370.93,221.09) -- (375.43,225.59);

\path[draw=drawColor,line width= 0.4pt,line join=round,line cap=round] (370.93,225.59) -- (375.43,221.09);

\path[draw=drawColor,line width= 0.4pt,line join=round,line cap=round] (393.13,225.22) rectangle (397.63,229.72);

\path[draw=drawColor,line width= 0.4pt,line join=round,line cap=round] (393.13,225.22) -- (397.63,229.72);

\path[draw=drawColor,line width= 0.4pt,line join=round,line cap=round] (393.13,229.72) -- (397.63,225.22);
\definecolor{drawColor}{RGB}{0,0,0}

\path[draw=drawColor,line width= 0.4pt,dash pattern=on 4pt off 4pt ,line join=round,line cap=round] ( 55.86, 76.09) --
	( 59.35, 76.09) --
	( 62.85, 76.09) --
	( 66.35, 76.09) --
	( 69.84, 76.09) --
	( 73.34, 76.09) --
	( 76.84, 76.09) --
	( 80.33, 76.09) --
	( 83.83, 76.09) --
	( 87.33, 76.09) --
	( 90.82, 76.09) --
	( 94.32, 76.09) --
	( 97.82, 76.09) --
	(101.31, 76.09) --
	(104.81, 76.09) --
	(108.31, 76.09) --
	(111.80, 76.09) --
	(115.30, 76.09) --
	(118.80, 76.09) --
	(122.29, 76.09) --
	(125.79, 76.09) --
	(129.29, 76.09) --
	(132.79, 76.09) --
	(136.28, 76.09) --
	(139.78, 76.09) --
	(143.28, 76.09) --
	(146.77, 76.09) --
	(150.27, 76.09) --
	(153.77, 76.09) --
	(157.26, 76.09) --
	(160.76, 76.09) --
	(164.26, 76.09) --
	(167.75, 76.09) --
	(171.25, 76.09) --
	(174.75, 76.09) --
	(178.24, 76.09) --
	(181.74, 76.09) --
	(185.24, 76.09) --
	(188.73, 76.09) --
	(192.23, 76.09) --
	(195.73, 76.09) --
	(199.22, 76.09) --
	(202.72, 76.09) --
	(206.22, 76.09) --
	(209.71, 76.09) --
	(213.21, 76.09) --
	(216.71, 76.09) --
	(220.20, 76.09) --
	(223.70, 76.09) --
	(227.20, 76.09) --
	(230.69, 76.09) --
	(234.19, 76.09) --
	(237.69, 76.09) --
	(241.18, 76.09) --
	(244.68, 76.09) --
	(248.18, 76.09) --
	(251.67, 76.09) --
	(255.17, 76.09) --
	(258.67, 76.09) --
	(262.16, 76.09) --
	(265.66, 76.09) --
	(269.16, 76.09) --
	(272.65, 76.09) --
	(276.15, 76.09) --
	(279.65, 76.09) --
	(283.14, 76.09) --
	(286.64, 76.09) --
	(290.14, 76.09) --
	(293.63, 76.09) --
	(297.13, 76.09) --
	(300.63, 76.09) --
	(304.12, 76.09) --
	(307.62, 76.09) --
	(311.12, 76.09) --
	(314.61, 76.09) --
	(318.11, 76.09) --
	(321.61, 76.09) --
	(325.10, 76.09) --
	(328.60, 76.09) --
	(332.10, 76.09) --
	(335.60, 76.09) --
	(339.09, 76.09) --
	(342.59, 76.09) --
	(346.09, 76.09) --
	(349.58, 76.09) --
	(353.08, 76.09) --
	(356.58, 76.09) --
	(360.07, 76.09) --
	(363.57, 76.09) --
	(367.07, 76.09) --
	(370.56, 76.09) --
	(374.06, 76.09) --
	(377.56, 76.09) --
	(381.05, 76.09) --
	(384.55, 76.09) --
	(388.05, 76.09) --
	(391.54, 76.09) --
	(395.04, 76.09) --
	(398.54, 76.09) --
	(402.03, 76.09);
\definecolor{drawColor}{RGB}{139,0,0}
\definecolor{fillColor}{RGB}{139,0,0}

\path[draw=drawColor,line width= 0.4pt,line join=round,line cap=round,fill=fillColor] (421.11,227.88) circle (  2.25);
\definecolor{fillColor}{RGB}{255,99,71}

\path[fill=fillColor] (421.11,219.38) --
	(424.15,214.13) --
	(418.08,214.13) --
	cycle;
\definecolor{fillColor}{RGB}{108,166,205}

\path[fill=fillColor] (418.86,201.63) --
	(423.36,201.63) --
	(423.36,206.13) --
	(418.86,206.13) --
	cycle;
\definecolor{drawColor}{RGB}{139,137,137}

\path[draw=drawColor,line width= 0.4pt,line join=round,line cap=round] (417.93,191.88) -- (424.30,191.88);

\path[draw=drawColor,line width= 0.4pt,line join=round,line cap=round] (421.11,188.70) -- (421.11,195.06);
\definecolor{drawColor}{RGB}{84,139,84}

\path[draw=drawColor,line width= 0.4pt,line join=round,line cap=round] (418.86,177.63) -- (423.36,182.13);

\path[draw=drawColor,line width= 0.4pt,line join=round,line cap=round] (418.86,182.13) -- (423.36,177.63);
\definecolor{drawColor}{RGB}{238,180,34}

\path[draw=drawColor,line width= 0.4pt,line join=round,line cap=round] (421.11,167.88) circle (  2.25);

\path[draw=drawColor,line width= 0.4pt,line join=round,line cap=round] (418.86,165.63) -- (423.36,170.13);

\path[draw=drawColor,line width= 0.4pt,line join=round,line cap=round] (418.86,170.13) -- (423.36,165.63);
\definecolor{drawColor}{RGB}{205,96,144}

\path[draw=drawColor,line width= 0.4pt,line join=round,line cap=round] (418.86,153.63) rectangle (423.36,158.13);

\path[draw=drawColor,line width= 0.4pt,line join=round,line cap=round] (418.86,153.63) -- (423.36,158.13);

\path[draw=drawColor,line width= 0.4pt,line join=round,line cap=round] (418.86,158.13) -- (423.36,153.63);
\definecolor{drawColor}{RGB}{0,0,0}

\node[text=drawColor,anchor=base west,inner sep=0pt, outer sep=0pt, scale=  1.00] at (430.11,224.44) {HypoRF};

\node[text=drawColor,anchor=base west,inner sep=0pt, outer sep=0pt, scale=  1.00] at (430.11,212.44) {Binomial};

\node[text=drawColor,anchor=base west,inner sep=0pt, outer sep=0pt, scale=  1.00] at (430.11,200.44) {ME-full};

\node[text=drawColor,anchor=base west,inner sep=0pt, outer sep=0pt, scale=  1.00] at (430.11,188.44) {MMDboot};

\node[text=drawColor,anchor=base west,inner sep=0pt, outer sep=0pt, scale=  1.00] at (430.11,176.44) {MMD-full};

\node[text=drawColor,anchor=base west,inner sep=0pt, outer sep=0pt, scale=  1.00] at (430.11,164.44) {LDA};

\node[text=drawColor,anchor=base west,inner sep=0pt, outer sep=0pt, scale=  1.00] at (430.11,152.44) {CPT-RF};
\end{scope}
\end{tikzpicture}

%% file: 1b_plot_gaussian_meanshift_d2_Normapprox_F_LDA_with_Cai.tex
\begin{tikzpicture}[x=1pt,y=1pt]
\definecolor{fillColor}{RGB}{255,255,255}
\path[use as bounding box,fill=fillColor,fill opacity=0.00] (0,0) rectangle (505.89,289.08);
\begin{scope}
\path[clip] (  0.00,  0.00) rectangle (505.89,289.08);
\definecolor{drawColor}{RGB}{139,0,0}

\path[draw=drawColor,line width= 0.4pt,line join=round,line cap=round] ( 68.50, 75.64) -- ( 78.72, 74.88);

\path[draw=drawColor,line width= 0.4pt,line join=round,line cap=round] ( 90.70, 74.66) -- (100.90, 75.04);

\path[draw=drawColor,line width= 0.4pt,line join=round,line cap=round] (112.83, 76.15) -- (123.15, 77.69);

\path[draw=drawColor,line width= 0.4pt,line join=round,line cap=round] (134.99, 79.67) -- (145.38, 81.61);

\path[draw=drawColor,line width= 0.4pt,line join=round,line cap=round] (157.13, 84.02) -- (167.61, 86.36);

\path[draw=drawColor,line width= 0.4pt,line join=round,line cap=round] (178.36, 91.14) -- (190.76, 99.92);

\path[draw=drawColor,line width= 0.4pt,line join=round,line cap=round] (200.30,107.19) -- (213.21,117.78);

\path[draw=drawColor,line width= 0.4pt,line join=round,line cap=round] (220.92,126.74) -- (236.97,153.66);

\path[draw=drawColor,line width= 0.4pt,line join=round,line cap=round] (243.69,163.57) -- (258.58,183.00);

\path[draw=drawColor,line width= 0.4pt,line join=round,line cap=round] (265.88,192.53) -- (280.77,211.96);

\path[draw=drawColor,line width= 0.4pt,line join=round,line cap=round] (290.17,218.43) -- (300.86,221.62);

\path[draw=drawColor,line width= 0.4pt,line join=round,line cap=round] (312.16,225.61) -- (323.25,230.16);

\path[draw=drawColor,line width= 0.4pt,line join=round,line cap=round] (334.80,232.66) -- (345.00,233.04);

\path[draw=drawColor,line width= 0.4pt,line join=round,line cap=round] (356.99,233.26) -- (367.18,233.26);

\path[draw=drawColor,line width= 0.4pt,line join=round,line cap=round] (379.18,233.26) -- (389.38,233.26);
\definecolor{fillColor}{RGB}{139,0,0}

\path[draw=drawColor,line width= 0.4pt,line join=round,line cap=round,fill=fillColor] ( 62.51, 76.09) circle (  2.25);

\path[draw=drawColor,line width= 0.4pt,line join=round,line cap=round,fill=fillColor] ( 84.71, 74.44) circle (  2.25);

\path[draw=drawColor,line width= 0.4pt,line join=round,line cap=round,fill=fillColor] (106.90, 75.26) circle (  2.25);

\path[draw=drawColor,line width= 0.4pt,line join=round,line cap=round,fill=fillColor] (129.09, 78.57) circle (  2.25);

\path[draw=drawColor,line width= 0.4pt,line join=round,line cap=round,fill=fillColor] (151.28, 82.71) circle (  2.25);

\path[draw=drawColor,line width= 0.4pt,line join=round,line cap=round,fill=fillColor] (173.47, 87.67) circle (  2.25);

\path[draw=drawColor,line width= 0.4pt,line join=round,line cap=round,fill=fillColor] (195.66,103.39) circle (  2.25);

\path[draw=drawColor,line width= 0.4pt,line join=round,line cap=round,fill=fillColor] (217.85,121.59) circle (  2.25);

\path[draw=drawColor,line width= 0.4pt,line join=round,line cap=round,fill=fillColor] (240.04,158.81) circle (  2.25);

\path[draw=drawColor,line width= 0.4pt,line join=round,line cap=round,fill=fillColor] (262.23,187.77) circle (  2.25);

\path[draw=drawColor,line width= 0.4pt,line join=round,line cap=round,fill=fillColor] (284.42,216.72) circle (  2.25);

\path[draw=drawColor,line width= 0.4pt,line join=round,line cap=round,fill=fillColor] (306.61,223.34) circle (  2.25);

\path[draw=drawColor,line width= 0.4pt,line join=round,line cap=round,fill=fillColor] (328.80,232.43) circle (  2.25);

\path[draw=drawColor,line width= 0.4pt,line join=round,line cap=round,fill=fillColor] (350.99,233.26) circle (  2.25);

\path[draw=drawColor,line width= 0.4pt,line join=round,line cap=round,fill=fillColor] (373.18,233.26) circle (  2.25);

\path[draw=drawColor,line width= 0.4pt,line join=round,line cap=round,fill=fillColor] (395.38,233.26) circle (  2.25);
\end{scope}
\begin{scope}
\path[clip] (  0.00,  0.00) rectangle (505.89,289.08);
\definecolor{drawColor}{RGB}{0,0,0}

\path[draw=drawColor,line width= 0.4pt,line join=round,line cap=round] ( 62.51, 61.20) -- (395.38, 61.20);

\path[draw=drawColor,line width= 0.4pt,line join=round,line cap=round] ( 62.51, 61.20) -- ( 62.51, 55.20);

\path[draw=drawColor,line width= 0.4pt,line join=round,line cap=round] (129.09, 61.20) -- (129.09, 55.20);

\path[draw=drawColor,line width= 0.4pt,line join=round,line cap=round] (195.66, 61.20) -- (195.66, 55.20);

\path[draw=drawColor,line width= 0.4pt,line join=round,line cap=round] (262.23, 61.20) -- (262.23, 55.20);

\path[draw=drawColor,line width= 0.4pt,line join=round,line cap=round] (328.80, 61.20) -- (328.80, 55.20);

\path[draw=drawColor,line width= 0.4pt,line join=round,line cap=round] (395.38, 61.20) -- (395.38, 55.20);

\node[text=drawColor,anchor=base,inner sep=0pt, outer sep=0pt, scale=  1.00] at ( 62.51, 39.60) {0.0};

\node[text=drawColor,anchor=base,inner sep=0pt, outer sep=0pt, scale=  1.00] at (129.09, 39.60) {0.2};

\node[text=drawColor,anchor=base,inner sep=0pt, outer sep=0pt, scale=  1.00] at (195.66, 39.60) {0.4};

\node[text=drawColor,anchor=base,inner sep=0pt, outer sep=0pt, scale=  1.00] at (262.23, 39.60) {0.6};

\node[text=drawColor,anchor=base,inner sep=0pt, outer sep=0pt, scale=  1.00] at (328.80, 39.60) {0.8};

\node[text=drawColor,anchor=base,inner sep=0pt, outer sep=0pt, scale=  1.00] at (395.38, 39.60) {1.0};

\path[draw=drawColor,line width= 0.4pt,line join=round,line cap=round] ( 49.20, 67.82) -- ( 49.20,233.26);

\path[draw=drawColor,line width= 0.4pt,line join=round,line cap=round] ( 49.20, 67.82) -- ( 43.20, 67.82);

\path[draw=drawColor,line width= 0.4pt,line join=round,line cap=round] ( 49.20,100.91) -- ( 43.20,100.91);

\path[draw=drawColor,line width= 0.4pt,line join=round,line cap=round] ( 49.20,134.00) -- ( 43.20,134.00);

\path[draw=drawColor,line width= 0.4pt,line join=round,line cap=round] ( 49.20,167.08) -- ( 43.20,167.08);

\path[draw=drawColor,line width= 0.4pt,line join=round,line cap=round] ( 49.20,200.17) -- ( 43.20,200.17);

\path[draw=drawColor,line width= 0.4pt,line join=round,line cap=round] ( 49.20,233.26) -- ( 43.20,233.26);

\node[text=drawColor,rotate= 90.00,anchor=base,inner sep=0pt, outer sep=0pt, scale=  1.00] at ( 34.80, 67.82) {0.0};

\node[text=drawColor,rotate= 90.00,anchor=base,inner sep=0pt, outer sep=0pt, scale=  1.00] at ( 34.80,100.91) {0.2};

\node[text=drawColor,rotate= 90.00,anchor=base,inner sep=0pt, outer sep=0pt, scale=  1.00] at ( 34.80,134.00) {0.4};

\node[text=drawColor,rotate= 90.00,anchor=base,inner sep=0pt, outer sep=0pt, scale=  1.00] at ( 34.80,167.08) {0.6};

\node[text=drawColor,rotate= 90.00,anchor=base,inner sep=0pt, outer sep=0pt, scale=  1.00] at ( 34.80,200.17) {0.8};

\node[text=drawColor,rotate= 90.00,anchor=base,inner sep=0pt, outer sep=0pt, scale=  1.00] at ( 34.80,233.26) {1.0};

\path[draw=drawColor,line width= 0.4pt,line join=round,line cap=round] ( 49.20, 61.20) --
	(408.69, 61.20) --
	(408.69,239.88) --
	( 49.20,239.88) --
	( 49.20, 61.20);
\end{scope}
\begin{scope}
\path[clip] (  0.00,  0.00) rectangle (505.89,289.08);
\definecolor{drawColor}{RGB}{0,0,0}

\node[text=drawColor,anchor=base,inner sep=0pt, outer sep=0pt, scale=  1.00] at (228.94, 15.60) {$\delta$};

\node[text=drawColor,rotate= 90.00,anchor=base,inner sep=0pt, outer sep=0pt, scale=  1.00] at ( 10.80,150.54) {Power};
\definecolor{drawColor}{RGB}{255,99,71}

\path[draw=drawColor,line width= 0.4pt,line join=round,line cap=round] ( 68.45, 77.69) -- ( 78.77, 76.15);

\path[draw=drawColor,line width= 0.4pt,line join=round,line cap=round] ( 90.69, 75.71) -- (100.91, 76.47);

\path[draw=drawColor,line width= 0.4pt,line join=round,line cap=round] (112.88, 77.36) -- (123.10, 78.13);

\path[draw=drawColor,line width= 0.4pt,line join=round,line cap=round] (135.08, 78.80) -- (145.28, 79.18);

\path[draw=drawColor,line width= 0.4pt,line join=round,line cap=round] (156.75, 81.85) -- (167.99, 86.88);

\path[draw=drawColor,line width= 0.4pt,line join=round,line cap=round] (179.43, 89.99) -- (189.70, 91.14);

\path[draw=drawColor,line width= 0.4pt,line join=round,line cap=round] (201.06, 94.42) -- (212.45, 99.94);

\path[draw=drawColor,line width= 0.4pt,line join=round,line cap=round] (221.85,107.03) -- (236.04,122.91);

\path[draw=drawColor,line width= 0.4pt,line join=round,line cap=round] (245.02,130.72) -- (257.25,138.92);

\path[draw=drawColor,line width= 0.4pt,line join=round,line cap=round] (265.35,147.39) -- (281.30,173.54);

\path[draw=drawColor,line width= 0.4pt,line join=round,line cap=round] (288.98,182.57) -- (302.06,193.79);

\path[draw=drawColor,line width= 0.4pt,line join=round,line cap=round] (311.34,201.39) -- (324.08,211.36);

\path[draw=drawColor,line width= 0.4pt,line join=round,line cap=round] (334.12,217.84) -- (345.67,223.87);

\path[draw=drawColor,line width= 0.4pt,line join=round,line cap=round] (356.96,227.31) -- (367.22,228.46);

\path[draw=drawColor,line width= 0.4pt,line join=round,line cap=round] (379.08,230.23) -- (389.48,232.16);
\definecolor{fillColor}{RGB}{255,99,71}

\path[fill=fillColor] ( 62.51, 82.07) --
	( 65.54, 76.82) --
	( 59.48, 76.82) --
	cycle;

\path[fill=fillColor] ( 84.71, 78.76) --
	( 87.74, 73.51) --
	( 81.67, 73.51) --
	cycle;

\path[fill=fillColor] (106.90, 80.42) --
	(109.93, 75.17) --
	(103.87, 75.17) --
	cycle;

\path[fill=fillColor] (129.09, 82.07) --
	(132.12, 76.82) --
	(126.06, 76.82) --
	cycle;

\path[fill=fillColor] (151.28, 82.90) --
	(154.31, 77.65) --
	(148.25, 77.65) --
	cycle;

\path[fill=fillColor] (173.47, 92.82) --
	(176.50, 87.58) --
	(170.44, 87.58) --
	cycle;

\path[fill=fillColor] (195.66, 95.31) --
	(198.69, 90.06) --
	(192.63, 90.06) --
	cycle;

\path[fill=fillColor] (217.85,106.06) --
	(220.88,100.81) --
	(214.82,100.81) --
	cycle;

\path[fill=fillColor] (240.04,130.88) --
	(243.07,125.63) --
	(237.01,125.63) --
	cycle;

\path[fill=fillColor] (262.23,145.77) --
	(265.26,140.52) --
	(259.20,140.52) --
	cycle;

\path[fill=fillColor] (284.42,182.16) --
	(287.45,176.92) --
	(281.39,176.92) --
	cycle;

\path[fill=fillColor] (306.61,201.19) --
	(309.64,195.94) --
	(303.58,195.94) --
	cycle;

\path[fill=fillColor] (328.80,218.56) --
	(331.83,213.31) --
	(325.77,213.31) --
	cycle;

\path[fill=fillColor] (350.99,230.14) --
	(354.02,224.89) --
	(347.96,224.89) --
	cycle;

\path[fill=fillColor] (373.18,232.63) --
	(376.22,227.38) --
	(370.15,227.38) --
	cycle;

\path[fill=fillColor] (395.38,236.76) --
	(398.41,231.51) --
	(392.35,231.51) --
	cycle;
\definecolor{drawColor}{RGB}{108,166,205}

\path[draw=drawColor,line width= 0.4pt,line join=round,line cap=round] ( 68.51, 75.26) -- ( 78.71, 75.26);

\path[draw=drawColor,line width= 0.4pt,line join=round,line cap=round] ( 90.69, 74.82) -- (100.91, 74.05);

\path[draw=drawColor,line width= 0.4pt,line join=round,line cap=round] (112.88, 74.05) -- (123.10, 74.82);

\path[draw=drawColor,line width= 0.4pt,line join=round,line cap=round] (135.09, 75.26) -- (145.28, 75.26);

\path[draw=drawColor,line width= 0.4pt,line join=round,line cap=round] (157.28, 75.26) -- (167.47, 75.26);

\path[draw=drawColor,line width= 0.4pt,line join=round,line cap=round] (179.43, 75.93) -- (189.70, 77.08);

\path[draw=drawColor,line width= 0.4pt,line join=round,line cap=round] (201.51, 79.05) -- (211.99, 81.40);

\path[draw=drawColor,line width= 0.4pt,line join=round,line cap=round] (223.81, 83.37) -- (234.08, 84.52);

\path[draw=drawColor,line width= 0.4pt,line join=round,line cap=round] (245.94, 86.29) -- (256.33, 88.23);

\path[draw=drawColor,line width= 0.4pt,line join=round,line cap=round] (267.38, 92.40) -- (279.27, 99.49);

\path[draw=drawColor,line width= 0.4pt,line join=round,line cap=round] (290.41,103.01) -- (300.63,103.77);

\path[draw=drawColor,line width= 0.4pt,line join=round,line cap=round] (312.01,106.83) -- (323.40,112.35);

\path[draw=drawColor,line width= 0.4pt,line join=round,line cap=round] (334.20,117.59) -- (345.59,123.11);

\path[draw=drawColor,line width= 0.4pt,line join=round,line cap=round] (355.72,129.42) -- (368.46,139.40);

\path[draw=drawColor,line width= 0.4pt,line join=round,line cap=round] (377.74,147.00) -- (390.82,158.22);
\definecolor{fillColor}{RGB}{108,166,205}

\path[fill=fillColor] ( 60.26, 73.01) --
	( 64.76, 73.01) --
	( 64.76, 77.51) --
	( 60.26, 77.51) --
	cycle;

\path[fill=fillColor] ( 82.46, 73.01) --
	( 86.96, 73.01) --
	( 86.96, 77.51) --
	( 82.46, 77.51) --
	cycle;

\path[fill=fillColor] (104.65, 71.36) --
	(109.15, 71.36) --
	(109.15, 75.86) --
	(104.65, 75.86) --
	cycle;

\path[fill=fillColor] (126.84, 73.01) --
	(131.34, 73.01) --
	(131.34, 77.51) --
	(126.84, 77.51) --
	cycle;

\path[fill=fillColor] (149.03, 73.01) --
	(153.53, 73.01) --
	(153.53, 77.51) --
	(149.03, 77.51) --
	cycle;

\path[fill=fillColor] (171.22, 73.01) --
	(175.72, 73.01) --
	(175.72, 77.51) --
	(171.22, 77.51) --
	cycle;

\path[fill=fillColor] (193.41, 75.49) --
	(197.91, 75.49) --
	(197.91, 79.99) --
	(193.41, 79.99) --
	cycle;

\path[fill=fillColor] (215.60, 80.46) --
	(220.10, 80.46) --
	(220.10, 84.96) --
	(215.60, 84.96) --
	cycle;

\path[fill=fillColor] (237.79, 82.94) --
	(242.29, 82.94) --
	(242.29, 87.44) --
	(237.79, 87.44) --
	cycle;

\path[fill=fillColor] (259.98, 87.08) --
	(264.48, 87.08) --
	(264.48, 91.58) --
	(259.98, 91.58) --
	cycle;

\path[fill=fillColor] (282.17,100.31) --
	(286.67,100.31) --
	(286.67,104.81) --
	(282.17,104.81) --
	cycle;

\path[fill=fillColor] (304.36,101.97) --
	(308.86,101.97) --
	(308.86,106.47) --
	(304.36,106.47) --
	cycle;

\path[fill=fillColor] (326.55,112.72) --
	(331.05,112.72) --
	(331.05,117.22) --
	(326.55,117.22) --
	cycle;

\path[fill=fillColor] (348.74,123.47) --
	(353.24,123.47) --
	(353.24,127.97) --
	(348.74,127.97) --
	cycle;

\path[fill=fillColor] (370.93,140.85) --
	(375.43,140.85) --
	(375.43,145.35) --
	(370.93,145.35) --
	cycle;

\path[fill=fillColor] (393.13,159.87) --
	(397.63,159.87) --
	(397.63,164.37) --
	(393.13,164.37) --
	cycle;
\definecolor{drawColor}{RGB}{139,137,137}

\path[draw=drawColor,line width= 0.4pt,line join=round,line cap=round] ( 68.51, 76.92) -- ( 78.71, 76.92);

\path[draw=drawColor,line width= 0.4pt,line join=round,line cap=round] ( 90.71, 76.92) -- (100.90, 76.92);

\path[draw=drawColor,line width= 0.4pt,line join=round,line cap=round] (112.89, 77.14) -- (123.09, 77.52);

\path[draw=drawColor,line width= 0.4pt,line join=round,line cap=round] (135.07, 78.19) -- (145.29, 78.95);

\path[draw=drawColor,line width= 0.4pt,line join=round,line cap=round] (157.26, 79.84) -- (167.48, 80.61);

\path[draw=drawColor,line width= 0.4pt,line join=round,line cap=round] (179.32, 82.36) -- (189.80, 84.71);

\path[draw=drawColor,line width= 0.4pt,line join=round,line cap=round] (201.59, 86.90) -- (211.92, 88.44);

\path[draw=drawColor,line width= 0.4pt,line join=round,line cap=round] (223.54, 91.23) -- (234.35, 94.86);

\path[draw=drawColor,line width= 0.4pt,line join=round,line cap=round] (245.79, 98.49) -- (256.48,101.67);

\path[draw=drawColor,line width= 0.4pt,line join=round,line cap=round] (267.78,105.66) -- (278.87,110.21);

\path[draw=drawColor,line width= 0.4pt,line join=round,line cap=round] (290.28,113.80) -- (300.76,116.14);

\path[draw=drawColor,line width= 0.4pt,line join=round,line cap=round] (312.16,119.73) -- (323.25,124.27);

\path[draw=drawColor,line width= 0.4pt,line join=round,line cap=round] (334.12,129.33) -- (345.67,135.36);

\path[draw=drawColor,line width= 0.4pt,line join=round,line cap=round] (355.55,142.04) -- (368.63,153.25);

\path[draw=drawColor,line width= 0.4pt,line join=round,line cap=round] (378.42,160.09) -- (390.14,166.64);

\path[draw=drawColor,line width= 0.4pt,line join=round,line cap=round] ( 59.33, 76.92) -- ( 65.70, 76.92);

\path[draw=drawColor,line width= 0.4pt,line join=round,line cap=round] ( 62.51, 73.74) -- ( 62.51, 80.10);

\path[draw=drawColor,line width= 0.4pt,line join=round,line cap=round] ( 81.52, 76.92) -- ( 87.89, 76.92);

\path[draw=drawColor,line width= 0.4pt,line join=round,line cap=round] ( 84.71, 73.74) -- ( 84.71, 80.10);

\path[draw=drawColor,line width= 0.4pt,line join=round,line cap=round] (103.71, 76.92) -- (110.08, 76.92);

\path[draw=drawColor,line width= 0.4pt,line join=round,line cap=round] (106.90, 73.74) -- (106.90, 80.10);

\path[draw=drawColor,line width= 0.4pt,line join=round,line cap=round] (125.90, 77.74) -- (132.27, 77.74);

\path[draw=drawColor,line width= 0.4pt,line join=round,line cap=round] (129.09, 74.56) -- (129.09, 80.93);

\path[draw=drawColor,line width= 0.4pt,line join=round,line cap=round] (148.10, 79.40) -- (154.46, 79.40);

\path[draw=drawColor,line width= 0.4pt,line join=round,line cap=round] (151.28, 76.22) -- (151.28, 82.58);

\path[draw=drawColor,line width= 0.4pt,line join=round,line cap=round] (170.29, 81.05) -- (176.65, 81.05);

\path[draw=drawColor,line width= 0.4pt,line join=round,line cap=round] (173.47, 77.87) -- (173.47, 84.24);

\path[draw=drawColor,line width= 0.4pt,line join=round,line cap=round] (192.48, 86.02) -- (198.84, 86.02);

\path[draw=drawColor,line width= 0.4pt,line join=round,line cap=round] (195.66, 82.83) -- (195.66, 89.20);

\path[draw=drawColor,line width= 0.4pt,line join=round,line cap=round] (214.67, 89.33) -- (221.03, 89.33);

\path[draw=drawColor,line width= 0.4pt,line join=round,line cap=round] (217.85, 86.14) -- (217.85, 92.51);

\path[draw=drawColor,line width= 0.4pt,line join=round,line cap=round] (236.86, 96.77) -- (243.22, 96.77);

\path[draw=drawColor,line width= 0.4pt,line join=round,line cap=round] (240.04, 93.59) -- (240.04, 99.95);

\path[draw=drawColor,line width= 0.4pt,line join=round,line cap=round] (259.05,103.39) -- (265.41,103.39);

\path[draw=drawColor,line width= 0.4pt,line join=round,line cap=round] (262.23,100.21) -- (262.23,106.57);

\path[draw=drawColor,line width= 0.4pt,line join=round,line cap=round] (281.24,112.49) -- (287.60,112.49);

\path[draw=drawColor,line width= 0.4pt,line join=round,line cap=round] (284.42,109.31) -- (284.42,115.67);

\path[draw=drawColor,line width= 0.4pt,line join=round,line cap=round] (303.43,117.45) -- (309.79,117.45);

\path[draw=drawColor,line width= 0.4pt,line join=round,line cap=round] (306.61,114.27) -- (306.61,120.63);

\path[draw=drawColor,line width= 0.4pt,line join=round,line cap=round] (325.62,126.55) -- (331.99,126.55);

\path[draw=drawColor,line width= 0.4pt,line join=round,line cap=round] (328.80,123.37) -- (328.80,129.73);

\path[draw=drawColor,line width= 0.4pt,line join=round,line cap=round] (347.81,138.13) -- (354.18,138.13);

\path[draw=drawColor,line width= 0.4pt,line join=round,line cap=round] (350.99,134.95) -- (350.99,141.31);

\path[draw=drawColor,line width= 0.4pt,line join=round,line cap=round] (370.00,157.16) -- (376.37,157.16);

\path[draw=drawColor,line width= 0.4pt,line join=round,line cap=round] (373.18,153.98) -- (373.18,160.34);

\path[draw=drawColor,line width= 0.4pt,line join=round,line cap=round] (392.19,169.57) -- (398.56,169.57);

\path[draw=drawColor,line width= 0.4pt,line join=round,line cap=round] (395.38,166.38) -- (395.38,172.75);
\definecolor{drawColor}{RGB}{84,139,84}

\path[draw=drawColor,line width= 0.4pt,line join=round,line cap=round] ( 68.51, 74.44) -- ( 78.71, 74.44);

\path[draw=drawColor,line width= 0.4pt,line join=round,line cap=round] ( 90.71, 74.44) -- (100.90, 74.44);

\path[draw=drawColor,line width= 0.4pt,line join=round,line cap=round] (112.89, 74.66) -- (123.09, 75.04);

\path[draw=drawColor,line width= 0.4pt,line join=round,line cap=round] (135.05, 75.93) -- (145.31, 77.08);

\path[draw=drawColor,line width= 0.4pt,line join=round,line cap=round] (157.26, 78.19) -- (167.48, 78.95);

\path[draw=drawColor,line width= 0.4pt,line join=round,line cap=round] (179.32, 80.71) -- (189.80, 83.05);

\path[draw=drawColor,line width= 0.4pt,line join=round,line cap=round] (201.41, 86.08) -- (212.10, 89.27);

\path[draw=drawColor,line width= 0.4pt,line join=round,line cap=round] (223.60, 92.69) -- (234.29, 95.88);

\path[draw=drawColor,line width= 0.4pt,line join=round,line cap=round] (245.19,100.67) -- (257.08,107.76);

\path[draw=drawColor,line width= 0.4pt,line join=round,line cap=round] (267.21,114.18) -- (279.44,122.38);

\path[draw=drawColor,line width= 0.4pt,line join=round,line cap=round] (288.42,130.20) -- (302.61,146.07);

\path[draw=drawColor,line width= 0.4pt,line join=round,line cap=round] (310.54,155.08) -- (324.88,171.65);

\path[draw=drawColor,line width= 0.4pt,line join=round,line cap=round] (333.11,180.36) -- (346.69,193.52);

\path[draw=drawColor,line width= 0.4pt,line join=round,line cap=round] (356.06,200.90) -- (368.12,208.54);

\path[draw=drawColor,line width= 0.4pt,line join=round,line cap=round] (378.42,214.68) -- (390.14,221.23);

\path[draw=drawColor,line width= 0.4pt,line join=round,line cap=round] ( 60.26, 72.19) -- ( 64.76, 76.69);

\path[draw=drawColor,line width= 0.4pt,line join=round,line cap=round] ( 60.26, 76.69) -- ( 64.76, 72.19);

\path[draw=drawColor,line width= 0.4pt,line join=round,line cap=round] ( 82.46, 72.19) -- ( 86.96, 76.69);

\path[draw=drawColor,line width= 0.4pt,line join=round,line cap=round] ( 82.46, 76.69) -- ( 86.96, 72.19);

\path[draw=drawColor,line width= 0.4pt,line join=round,line cap=round] (104.65, 72.19) -- (109.15, 76.69);

\path[draw=drawColor,line width= 0.4pt,line join=round,line cap=round] (104.65, 76.69) -- (109.15, 72.19);

\path[draw=drawColor,line width= 0.4pt,line join=round,line cap=round] (126.84, 73.01) -- (131.34, 77.51);

\path[draw=drawColor,line width= 0.4pt,line join=round,line cap=round] (126.84, 77.51) -- (131.34, 73.01);

\path[draw=drawColor,line width= 0.4pt,line join=round,line cap=round] (149.03, 75.49) -- (153.53, 79.99);

\path[draw=drawColor,line width= 0.4pt,line join=round,line cap=round] (149.03, 79.99) -- (153.53, 75.49);

\path[draw=drawColor,line width= 0.4pt,line join=round,line cap=round] (171.22, 77.15) -- (175.72, 81.65);

\path[draw=drawColor,line width= 0.4pt,line join=round,line cap=round] (171.22, 81.65) -- (175.72, 77.15);

\path[draw=drawColor,line width= 0.4pt,line join=round,line cap=round] (193.41, 82.11) -- (197.91, 86.61);

\path[draw=drawColor,line width= 0.4pt,line join=round,line cap=round] (193.41, 86.61) -- (197.91, 82.11);

\path[draw=drawColor,line width= 0.4pt,line join=round,line cap=round] (215.60, 88.73) -- (220.10, 93.23);

\path[draw=drawColor,line width= 0.4pt,line join=round,line cap=round] (215.60, 93.23) -- (220.10, 88.73);

\path[draw=drawColor,line width= 0.4pt,line join=round,line cap=round] (237.79, 95.35) -- (242.29, 99.85);

\path[draw=drawColor,line width= 0.4pt,line join=round,line cap=round] (237.79, 99.85) -- (242.29, 95.35);

\path[draw=drawColor,line width= 0.4pt,line join=round,line cap=round] (259.98,108.58) -- (264.48,113.08);

\path[draw=drawColor,line width= 0.4pt,line join=round,line cap=round] (259.98,113.08) -- (264.48,108.58);

\path[draw=drawColor,line width= 0.4pt,line join=round,line cap=round] (282.17,123.47) -- (286.67,127.97);

\path[draw=drawColor,line width= 0.4pt,line join=round,line cap=round] (282.17,127.97) -- (286.67,123.47);

\path[draw=drawColor,line width= 0.4pt,line join=round,line cap=round] (304.36,148.29) -- (308.86,152.79);

\path[draw=drawColor,line width= 0.4pt,line join=round,line cap=round] (304.36,152.79) -- (308.86,148.29);

\path[draw=drawColor,line width= 0.4pt,line join=round,line cap=round] (326.55,173.93) -- (331.05,178.43);

\path[draw=drawColor,line width= 0.4pt,line join=round,line cap=round] (326.55,178.43) -- (331.05,173.93);

\path[draw=drawColor,line width= 0.4pt,line join=round,line cap=round] (348.74,195.44) -- (353.24,199.94);

\path[draw=drawColor,line width= 0.4pt,line join=round,line cap=round] (348.74,199.94) -- (353.24,195.44);

\path[draw=drawColor,line width= 0.4pt,line join=round,line cap=round] (370.93,209.50) -- (375.43,214.00);

\path[draw=drawColor,line width= 0.4pt,line join=round,line cap=round] (370.93,214.00) -- (375.43,209.50);

\path[draw=drawColor,line width= 0.4pt,line join=round,line cap=round] (393.13,221.91) -- (397.63,226.41);

\path[draw=drawColor,line width= 0.4pt,line join=round,line cap=round] (393.13,226.41) -- (397.63,221.91);
\definecolor{drawColor}{RGB}{238,180,34}

\path[draw=drawColor,line width= 0.4pt,line join=round,line cap=round] ( 68.51, 74.66) -- ( 78.71, 75.04);

\path[draw=drawColor,line width= 0.4pt,line join=round,line cap=round] ( 90.70, 75.04) -- (100.90, 74.66);

\path[draw=drawColor,line width= 0.4pt,line join=round,line cap=round] (112.88, 74.88) -- (123.10, 75.64);

\path[draw=drawColor,line width= 0.4pt,line join=round,line cap=round] (135.07, 76.54) -- (145.29, 77.30);

\path[draw=drawColor,line width= 0.4pt,line join=round,line cap=round] (157.21, 78.63) -- (167.53, 80.17);

\path[draw=drawColor,line width= 0.4pt,line join=round,line cap=round] (179.43, 81.72) -- (189.70, 82.87);

\path[draw=drawColor,line width= 0.4pt,line join=round,line cap=round] (201.59, 84.42) -- (211.92, 85.96);

\path[draw=drawColor,line width= 0.4pt,line join=round,line cap=round] (223.75, 87.94) -- (234.14, 89.88);

\path[draw=drawColor,line width= 0.4pt,line join=round,line cap=round] (245.79, 92.69) -- (256.48, 95.88);

\path[draw=drawColor,line width= 0.4pt,line join=round,line cap=round] (267.63,100.21) -- (279.02,105.74);

\path[draw=drawColor,line width= 0.4pt,line join=round,line cap=round] (290.23,109.87) -- (300.81,112.63);

\path[draw=drawColor,line width= 0.4pt,line join=round,line cap=round] (311.77,117.22) -- (323.65,124.30);

\path[draw=drawColor,line width= 0.4pt,line join=round,line cap=round] (333.79,130.72) -- (346.01,138.92);

\path[draw=drawColor,line width= 0.4pt,line join=round,line cap=round] (356.23,145.20) -- (367.95,151.75);

\path[draw=drawColor,line width= 0.4pt,line join=round,line cap=round] (378.34,157.75) -- (390.22,164.84);

\path[draw=drawColor,line width= 0.4pt,line join=round,line cap=round] ( 62.51, 74.44) circle (  2.25);

\path[draw=drawColor,line width= 0.4pt,line join=round,line cap=round] ( 60.26, 72.19) -- ( 64.76, 76.69);

\path[draw=drawColor,line width= 0.4pt,line join=round,line cap=round] ( 60.26, 76.69) -- ( 64.76, 72.19);

\path[draw=drawColor,line width= 0.4pt,line join=round,line cap=round] ( 84.71, 75.26) circle (  2.25);

\path[draw=drawColor,line width= 0.4pt,line join=round,line cap=round] ( 82.46, 73.01) -- ( 86.96, 77.51);

\path[draw=drawColor,line width= 0.4pt,line join=round,line cap=round] ( 82.46, 77.51) -- ( 86.96, 73.01);

\path[draw=drawColor,line width= 0.4pt,line join=round,line cap=round] (106.90, 74.44) circle (  2.25);

\path[draw=drawColor,line width= 0.4pt,line join=round,line cap=round] (104.65, 72.19) -- (109.15, 76.69);

\path[draw=drawColor,line width= 0.4pt,line join=round,line cap=round] (104.65, 76.69) -- (109.15, 72.19);

\path[draw=drawColor,line width= 0.4pt,line join=round,line cap=round] (129.09, 76.09) circle (  2.25);

\path[draw=drawColor,line width= 0.4pt,line join=round,line cap=round] (126.84, 73.84) -- (131.34, 78.34);

\path[draw=drawColor,line width= 0.4pt,line join=round,line cap=round] (126.84, 78.34) -- (131.34, 73.84);

\path[draw=drawColor,line width= 0.4pt,line join=round,line cap=round] (151.28, 77.74) circle (  2.25);

\path[draw=drawColor,line width= 0.4pt,line join=round,line cap=round] (149.03, 75.49) -- (153.53, 79.99);

\path[draw=drawColor,line width= 0.4pt,line join=round,line cap=round] (149.03, 79.99) -- (153.53, 75.49);

\path[draw=drawColor,line width= 0.4pt,line join=round,line cap=round] (173.47, 81.05) circle (  2.25);

\path[draw=drawColor,line width= 0.4pt,line join=round,line cap=round] (171.22, 78.80) -- (175.72, 83.30);

\path[draw=drawColor,line width= 0.4pt,line join=round,line cap=round] (171.22, 83.30) -- (175.72, 78.80);

\path[draw=drawColor,line width= 0.4pt,line join=round,line cap=round] (195.66, 83.53) circle (  2.25);

\path[draw=drawColor,line width= 0.4pt,line join=round,line cap=round] (193.41, 81.28) -- (197.91, 85.78);

\path[draw=drawColor,line width= 0.4pt,line join=round,line cap=round] (193.41, 85.78) -- (197.91, 81.28);

\path[draw=drawColor,line width= 0.4pt,line join=round,line cap=round] (217.85, 86.84) circle (  2.25);

\path[draw=drawColor,line width= 0.4pt,line join=round,line cap=round] (215.60, 84.59) -- (220.10, 89.09);

\path[draw=drawColor,line width= 0.4pt,line join=round,line cap=round] (215.60, 89.09) -- (220.10, 84.59);

\path[draw=drawColor,line width= 0.4pt,line join=round,line cap=round] (240.04, 90.98) circle (  2.25);

\path[draw=drawColor,line width= 0.4pt,line join=round,line cap=round] (237.79, 88.73) -- (242.29, 93.23);

\path[draw=drawColor,line width= 0.4pt,line join=round,line cap=round] (237.79, 93.23) -- (242.29, 88.73);

\path[draw=drawColor,line width= 0.4pt,line join=round,line cap=round] (262.23, 97.60) circle (  2.25);

\path[draw=drawColor,line width= 0.4pt,line join=round,line cap=round] (259.98, 95.35) -- (264.48, 99.85);

\path[draw=drawColor,line width= 0.4pt,line join=round,line cap=round] (259.98, 99.85) -- (264.48, 95.35);

\path[draw=drawColor,line width= 0.4pt,line join=round,line cap=round] (284.42,108.35) circle (  2.25);

\path[draw=drawColor,line width= 0.4pt,line join=round,line cap=round] (282.17,106.10) -- (286.67,110.60);

\path[draw=drawColor,line width= 0.4pt,line join=round,line cap=round] (282.17,110.60) -- (286.67,106.10);

\path[draw=drawColor,line width= 0.4pt,line join=round,line cap=round] (306.61,114.14) circle (  2.25);

\path[draw=drawColor,line width= 0.4pt,line join=round,line cap=round] (304.36,111.89) -- (308.86,116.39);

\path[draw=drawColor,line width= 0.4pt,line join=round,line cap=round] (304.36,116.39) -- (308.86,111.89);

\path[draw=drawColor,line width= 0.4pt,line join=round,line cap=round] (328.80,127.38) circle (  2.25);

\path[draw=drawColor,line width= 0.4pt,line join=round,line cap=round] (326.55,125.13) -- (331.05,129.63);

\path[draw=drawColor,line width= 0.4pt,line join=round,line cap=round] (326.55,129.63) -- (331.05,125.13);

\path[draw=drawColor,line width= 0.4pt,line join=round,line cap=round] (350.99,142.27) circle (  2.25);

\path[draw=drawColor,line width= 0.4pt,line join=round,line cap=round] (348.74,140.02) -- (353.24,144.52);

\path[draw=drawColor,line width= 0.4pt,line join=round,line cap=round] (348.74,144.52) -- (353.24,140.02);

\path[draw=drawColor,line width= 0.4pt,line join=round,line cap=round] (373.18,154.68) circle (  2.25);

\path[draw=drawColor,line width= 0.4pt,line join=round,line cap=round] (370.93,152.43) -- (375.43,156.93);

\path[draw=drawColor,line width= 0.4pt,line join=round,line cap=round] (370.93,156.93) -- (375.43,152.43);

\path[draw=drawColor,line width= 0.4pt,line join=round,line cap=round] (395.38,167.91) circle (  2.25);

\path[draw=drawColor,line width= 0.4pt,line join=round,line cap=round] (393.13,165.66) -- (397.63,170.16);

\path[draw=drawColor,line width= 0.4pt,line join=round,line cap=round] (393.13,170.16) -- (397.63,165.66);
\definecolor{drawColor}{RGB}{205,96,144}

\path[draw=drawColor,line width= 0.4pt,line join=round,line cap=round] ( 68.51, 76.09) -- ( 78.71, 76.09);

\path[draw=drawColor,line width= 0.4pt,line join=round,line cap=round] ( 90.70, 76.31) -- (100.90, 76.69);

\path[draw=drawColor,line width= 0.4pt,line join=round,line cap=round] (112.89, 77.14) -- (123.09, 77.52);

\path[draw=drawColor,line width= 0.4pt,line join=round,line cap=round] (135.02, 78.63) -- (145.34, 80.17);

\path[draw=drawColor,line width= 0.4pt,line join=round,line cap=round] (157.13, 82.36) -- (167.61, 84.71);

\path[draw=drawColor,line width= 0.4pt,line join=round,line cap=round] (179.02, 88.29) -- (190.11, 92.84);

\path[draw=drawColor,line width= 0.4pt,line join=round,line cap=round] (199.73, 99.52) -- (213.78,114.70);

\path[draw=drawColor,line width= 0.4pt,line join=round,line cap=round] (221.92,123.51) -- (235.97,138.69);

\path[draw=drawColor,line width= 0.4pt,line join=round,line cap=round] (243.44,148.04) -- (258.83,170.41);

\path[draw=drawColor,line width= 0.4pt,line join=round,line cap=round] (266.16,179.89) -- (280.50,196.46);

\path[draw=drawColor,line width= 0.4pt,line join=round,line cap=round] (289.40,204.34) -- (301.63,212.55);

\path[draw=drawColor,line width= 0.4pt,line join=round,line cap=round] (312.23,217.99) -- (323.18,222.07);

\path[draw=drawColor,line width= 0.4pt,line join=round,line cap=round] (334.61,225.68) -- (345.19,228.44);

\path[draw=drawColor,line width= 0.4pt,line join=round,line cap=round] (356.96,230.62) -- (367.22,231.77);

\path[draw=drawColor,line width= 0.4pt,line join=round,line cap=round] (379.18,232.66) -- (389.38,233.04);

\path[draw=drawColor,line width= 0.4pt,line join=round,line cap=round] ( 60.26, 73.84) rectangle ( 64.76, 78.34);

\path[draw=drawColor,line width= 0.4pt,line join=round,line cap=round] ( 60.26, 73.84) -- ( 64.76, 78.34);

\path[draw=drawColor,line width= 0.4pt,line join=round,line cap=round] ( 60.26, 78.34) -- ( 64.76, 73.84);

\path[draw=drawColor,line width= 0.4pt,line join=round,line cap=round] ( 82.46, 73.84) rectangle ( 86.96, 78.34);

\path[draw=drawColor,line width= 0.4pt,line join=round,line cap=round] ( 82.46, 73.84) -- ( 86.96, 78.34);

\path[draw=drawColor,line width= 0.4pt,line join=round,line cap=round] ( 82.46, 78.34) -- ( 86.96, 73.84);

\path[draw=drawColor,line width= 0.4pt,line join=round,line cap=round] (104.65, 74.67) rectangle (109.15, 79.17);

\path[draw=drawColor,line width= 0.4pt,line join=round,line cap=round] (104.65, 74.67) -- (109.15, 79.17);

\path[draw=drawColor,line width= 0.4pt,line join=round,line cap=round] (104.65, 79.17) -- (109.15, 74.67);

\path[draw=drawColor,line width= 0.4pt,line join=round,line cap=round] (126.84, 75.49) rectangle (131.34, 79.99);

\path[draw=drawColor,line width= 0.4pt,line join=round,line cap=round] (126.84, 75.49) -- (131.34, 79.99);

\path[draw=drawColor,line width= 0.4pt,line join=round,line cap=round] (126.84, 79.99) -- (131.34, 75.49);

\path[draw=drawColor,line width= 0.4pt,line join=round,line cap=round] (149.03, 78.80) rectangle (153.53, 83.30);

\path[draw=drawColor,line width= 0.4pt,line join=round,line cap=round] (149.03, 78.80) -- (153.53, 83.30);

\path[draw=drawColor,line width= 0.4pt,line join=round,line cap=round] (149.03, 83.30) -- (153.53, 78.80);

\path[draw=drawColor,line width= 0.4pt,line join=round,line cap=round] (171.22, 83.77) rectangle (175.72, 88.27);

\path[draw=drawColor,line width= 0.4pt,line join=round,line cap=round] (171.22, 83.77) -- (175.72, 88.27);

\path[draw=drawColor,line width= 0.4pt,line join=round,line cap=round] (171.22, 88.27) -- (175.72, 83.77);

\path[draw=drawColor,line width= 0.4pt,line join=round,line cap=round] (193.41, 92.87) rectangle (197.91, 97.37);

\path[draw=drawColor,line width= 0.4pt,line join=round,line cap=round] (193.41, 92.87) -- (197.91, 97.37);

\path[draw=drawColor,line width= 0.4pt,line join=round,line cap=round] (193.41, 97.37) -- (197.91, 92.87);

\path[draw=drawColor,line width= 0.4pt,line join=round,line cap=round] (215.60,116.86) rectangle (220.10,121.36);

\path[draw=drawColor,line width= 0.4pt,line join=round,line cap=round] (215.60,116.86) -- (220.10,121.36);

\path[draw=drawColor,line width= 0.4pt,line join=round,line cap=round] (215.60,121.36) -- (220.10,116.86);

\path[draw=drawColor,line width= 0.4pt,line join=round,line cap=round] (237.79,140.85) rectangle (242.29,145.35);

\path[draw=drawColor,line width= 0.4pt,line join=round,line cap=round] (237.79,140.85) -- (242.29,145.35);

\path[draw=drawColor,line width= 0.4pt,line join=round,line cap=round] (237.79,145.35) -- (242.29,140.85);

\path[draw=drawColor,line width= 0.4pt,line join=round,line cap=round] (259.98,173.11) rectangle (264.48,177.61);

\path[draw=drawColor,line width= 0.4pt,line join=round,line cap=round] (259.98,173.11) -- (264.48,177.61);

\path[draw=drawColor,line width= 0.4pt,line join=round,line cap=round] (259.98,177.61) -- (264.48,173.11);

\path[draw=drawColor,line width= 0.4pt,line join=round,line cap=round] (282.17,198.75) rectangle (286.67,203.25);

\path[draw=drawColor,line width= 0.4pt,line join=round,line cap=round] (282.17,198.75) -- (286.67,203.25);

\path[draw=drawColor,line width= 0.4pt,line join=round,line cap=round] (282.17,203.25) -- (286.67,198.75);

\path[draw=drawColor,line width= 0.4pt,line join=round,line cap=round] (304.36,213.64) rectangle (308.86,218.14);

\path[draw=drawColor,line width= 0.4pt,line join=round,line cap=round] (304.36,213.64) -- (308.86,218.14);

\path[draw=drawColor,line width= 0.4pt,line join=round,line cap=round] (304.36,218.14) -- (308.86,213.64);

\path[draw=drawColor,line width= 0.4pt,line join=round,line cap=round] (326.55,221.91) rectangle (331.05,226.41);

\path[draw=drawColor,line width= 0.4pt,line join=round,line cap=round] (326.55,221.91) -- (331.05,226.41);

\path[draw=drawColor,line width= 0.4pt,line join=round,line cap=round] (326.55,226.41) -- (331.05,221.91);

\path[draw=drawColor,line width= 0.4pt,line join=round,line cap=round] (348.74,227.70) rectangle (353.24,232.20);

\path[draw=drawColor,line width= 0.4pt,line join=round,line cap=round] (348.74,227.70) -- (353.24,232.20);

\path[draw=drawColor,line width= 0.4pt,line join=round,line cap=round] (348.74,232.20) -- (353.24,227.70);

\path[draw=drawColor,line width= 0.4pt,line join=round,line cap=round] (370.93,230.18) rectangle (375.43,234.68);

\path[draw=drawColor,line width= 0.4pt,line join=round,line cap=round] (370.93,230.18) -- (375.43,234.68);

\path[draw=drawColor,line width= 0.4pt,line join=round,line cap=round] (370.93,234.68) -- (375.43,230.18);

\path[draw=drawColor,line width= 0.4pt,line join=round,line cap=round] (393.13,231.01) rectangle (397.63,235.51);

\path[draw=drawColor,line width= 0.4pt,line join=round,line cap=round] (393.13,231.01) -- (397.63,235.51);

\path[draw=drawColor,line width= 0.4pt,line join=round,line cap=round] (393.13,235.51) -- (397.63,231.01);
\definecolor{drawColor}{RGB}{0,0,0}

\path[draw=drawColor,line width= 0.4pt,dash pattern=on 4pt off 4pt ,line join=round,line cap=round] ( 55.86, 76.09) --
	( 59.35, 76.09) --
	( 62.85, 76.09) --
	( 66.35, 76.09) --
	( 69.84, 76.09) --
	( 73.34, 76.09) --
	( 76.84, 76.09) --
	( 80.33, 76.09) --
	( 83.83, 76.09) --
	( 87.33, 76.09) --
	( 90.82, 76.09) --
	( 94.32, 76.09) --
	( 97.82, 76.09) --
	(101.31, 76.09) --
	(104.81, 76.09) --
	(108.31, 76.09) --
	(111.80, 76.09) --
	(115.30, 76.09) --
	(118.80, 76.09) --
	(122.29, 76.09) --
	(125.79, 76.09) --
	(129.29, 76.09) --
	(132.79, 76.09) --
	(136.28, 76.09) --
	(139.78, 76.09) --
	(143.28, 76.09) --
	(146.77, 76.09) --
	(150.27, 76.09) --
	(153.77, 76.09) --
	(157.26, 76.09) --
	(160.76, 76.09) --
	(164.26, 76.09) --
	(167.75, 76.09) --
	(171.25, 76.09) --
	(174.75, 76.09) --
	(178.24, 76.09) --
	(181.74, 76.09) --
	(185.24, 76.09) --
	(188.73, 76.09) --
	(192.23, 76.09) --
	(195.73, 76.09) --
	(199.22, 76.09) --
	(202.72, 76.09) --
	(206.22, 76.09) --
	(209.71, 76.09) --
	(213.21, 76.09) --
	(216.71, 76.09) --
	(220.20, 76.09) --
	(223.70, 76.09) --
	(227.20, 76.09) --
	(230.69, 76.09) --
	(234.19, 76.09) --
	(237.69, 76.09) --
	(241.18, 76.09) --
	(244.68, 76.09) --
	(248.18, 76.09) --
	(251.67, 76.09) --
	(255.17, 76.09) --
	(258.67, 76.09) --
	(262.16, 76.09) --
	(265.66, 76.09) --
	(269.16, 76.09) --
	(272.65, 76.09) --
	(276.15, 76.09) --
	(279.65, 76.09) --
	(283.14, 76.09) --
	(286.64, 76.09) --
	(290.14, 76.09) --
	(293.63, 76.09) --
	(297.13, 76.09) --
	(300.63, 76.09) --
	(304.12, 76.09) --
	(307.62, 76.09) --
	(311.12, 76.09) --
	(314.61, 76.09) --
	(318.11, 76.09) --
	(321.61, 76.09) --
	(325.10, 76.09) --
	(328.60, 76.09) --
	(332.10, 76.09) --
	(335.60, 76.09) --
	(339.09, 76.09) --
	(342.59, 76.09) --
	(346.09, 76.09) --
	(349.58, 76.09) --
	(353.08, 76.09) --
	(356.58, 76.09) --
	(360.07, 76.09) --
	(363.57, 76.09) --
	(367.07, 76.09) --
	(370.56, 76.09) --
	(374.06, 76.09) --
	(377.56, 76.09) --
	(381.05, 76.09) --
	(384.55, 76.09) --
	(388.05, 76.09) --
	(391.54, 76.09) --
	(395.04, 76.09) --
	(398.54, 76.09) --
	(402.03, 76.09);
\definecolor{drawColor}{RGB}{139,0,0}
\definecolor{fillColor}{RGB}{139,0,0}

\path[draw=drawColor,line width= 0.4pt,line join=round,line cap=round,fill=fillColor] (421.11,227.88) circle (  2.25);
\definecolor{fillColor}{RGB}{255,99,71}

\path[fill=fillColor] (421.11,219.38) --
	(424.15,214.13) --
	(418.08,214.13) --
	cycle;
\definecolor{fillColor}{RGB}{108,166,205}

\path[fill=fillColor] (418.86,201.63) --
	(423.36,201.63) --
	(423.36,206.13) --
	(418.86,206.13) --
	cycle;
\definecolor{drawColor}{RGB}{139,137,137}

\path[draw=drawColor,line width= 0.4pt,line join=round,line cap=round] (417.93,191.88) -- (424.30,191.88);

\path[draw=drawColor,line width= 0.4pt,line join=round,line cap=round] (421.11,188.70) -- (421.11,195.06);
\definecolor{drawColor}{RGB}{84,139,84}

\path[draw=drawColor,line width= 0.4pt,line join=round,line cap=round] (418.86,177.63) -- (423.36,182.13);

\path[draw=drawColor,line width= 0.4pt,line join=round,line cap=round] (418.86,182.13) -- (423.36,177.63);
\definecolor{drawColor}{RGB}{238,180,34}

\path[draw=drawColor,line width= 0.4pt,line join=round,line cap=round] (421.11,167.88) circle (  2.25);

\path[draw=drawColor,line width= 0.4pt,line join=round,line cap=round] (418.86,165.63) -- (423.36,170.13);

\path[draw=drawColor,line width= 0.4pt,line join=round,line cap=round] (418.86,170.13) -- (423.36,165.63);
\definecolor{drawColor}{RGB}{205,96,144}

\path[draw=drawColor,line width= 0.4pt,line join=round,line cap=round] (418.86,153.63) rectangle (423.36,158.13);

\path[draw=drawColor,line width= 0.4pt,line join=round,line cap=round] (418.86,153.63) -- (423.36,158.13);

\path[draw=drawColor,line width= 0.4pt,line join=round,line cap=round] (418.86,158.13) -- (423.36,153.63);
\definecolor{drawColor}{RGB}{0,0,0}

\node[text=drawColor,anchor=base west,inner sep=0pt, outer sep=0pt, scale=  1.00] at (430.11,224.44) {HypoRF};

\node[text=drawColor,anchor=base west,inner sep=0pt, outer sep=0pt, scale=  1.00] at (430.11,212.44) {Binomial};

\node[text=drawColor,anchor=base west,inner sep=0pt, outer sep=0pt, scale=  1.00] at (430.11,200.44) {ME-full};

\node[text=drawColor,anchor=base west,inner sep=0pt, outer sep=0pt, scale=  1.00] at (430.11,188.44) {MMDboot};

\node[text=drawColor,anchor=base west,inner sep=0pt, outer sep=0pt, scale=  1.00] at (430.11,176.44) {MMD-full};

\node[text=drawColor,anchor=base west,inner sep=0pt, outer sep=0pt, scale=  1.00] at (430.11,164.44) {LDA};

\node[text=drawColor,anchor=base west,inner sep=0pt, outer sep=0pt, scale=  1.00] at (430.11,152.44) {CPT-RF};
\end{scope}
\end{tikzpicture}

%% file: 3a_plot_gaussian_dep_K100_Normapprox_F_with_Cai.tex
\begin{tikzpicture}[x=1pt,y=1pt]
\definecolor{fillColor}{RGB}{255,255,255}
\path[use as bounding box,fill=fillColor,fill opacity=0.00] (0,0) rectangle (505.89,289.08);
\begin{scope}
\path[clip] (  0.00,  0.00) rectangle (505.89,289.08);
\definecolor{drawColor}{RGB}{139,0,0}

\path[draw=drawColor,line width= 0.4pt,line join=round,line cap=round] ( 68.26, 80.17) -- ( 78.96, 76.98);

\path[draw=drawColor,line width= 0.4pt,line join=round,line cap=round] ( 90.67, 75.93) -- (100.93, 77.08);

\path[draw=drawColor,line width= 0.4pt,line join=round,line cap=round] (112.58, 79.65) -- (123.40, 83.28);

\path[draw=drawColor,line width= 0.4pt,line join=round,line cap=round] (133.81, 88.89) -- (146.55, 98.86);

\path[draw=drawColor,line width= 0.4pt,line join=round,line cap=round] (156.26,105.90) -- (168.49,114.11);

\path[draw=drawColor,line width= 0.4pt,line join=round,line cap=round] (177.86,121.54) -- (191.27,134.04);

\path[draw=drawColor,line width= 0.4pt,line join=round,line cap=round] (200.30,141.94) -- (213.21,152.53);

\path[draw=drawColor,line width= 0.4pt,line join=round,line cap=round] (222.83,159.67) -- (235.06,167.88);

\path[draw=drawColor,line width= 0.4pt,line join=round,line cap=round] (242.97,176.46) -- (259.30,205.69);

\path[draw=drawColor,line width= 0.4pt,line join=round,line cap=round] (267.47,213.86) -- (279.18,220.41);

\path[draw=drawColor,line width= 0.4pt,line join=round,line cap=round] (290.23,224.85) -- (300.81,227.61);

\path[draw=drawColor,line width= 0.4pt,line join=round,line cap=round] (312.58,229.79) -- (322.84,230.94);

\path[draw=drawColor,line width= 0.4pt,line join=round,line cap=round] (334.80,231.83) -- (345.00,232.21);

\path[draw=drawColor,line width= 0.4pt,line join=round,line cap=round] (356.99,232.66) -- (367.19,233.04);

\path[draw=drawColor,line width= 0.4pt,line join=round,line cap=round] (379.18,233.26) -- (389.38,233.26);
\definecolor{fillColor}{RGB}{139,0,0}

\path[draw=drawColor,line width= 0.4pt,line join=round,line cap=round,fill=fillColor] ( 62.51, 81.88) circle (  2.25);

\path[draw=drawColor,line width= 0.4pt,line join=round,line cap=round,fill=fillColor] ( 84.71, 75.26) circle (  2.25);

\path[draw=drawColor,line width= 0.4pt,line join=round,line cap=round,fill=fillColor] (106.90, 77.74) circle (  2.25);

\path[draw=drawColor,line width= 0.4pt,line join=round,line cap=round,fill=fillColor] (129.09, 85.19) circle (  2.25);

\path[draw=drawColor,line width= 0.4pt,line join=round,line cap=round,fill=fillColor] (151.28,102.56) circle (  2.25);

\path[draw=drawColor,line width= 0.4pt,line join=round,line cap=round,fill=fillColor] (173.47,117.45) circle (  2.25);

\path[draw=drawColor,line width= 0.4pt,line join=round,line cap=round,fill=fillColor] (195.66,138.13) circle (  2.25);

\path[draw=drawColor,line width= 0.4pt,line join=round,line cap=round,fill=fillColor] (217.85,156.33) circle (  2.25);

\path[draw=drawColor,line width= 0.4pt,line join=round,line cap=round,fill=fillColor] (240.04,171.22) circle (  2.25);

\path[draw=drawColor,line width= 0.4pt,line join=round,line cap=round,fill=fillColor] (262.23,210.93) circle (  2.25);

\path[draw=drawColor,line width= 0.4pt,line join=round,line cap=round,fill=fillColor] (284.42,223.34) circle (  2.25);

\path[draw=drawColor,line width= 0.4pt,line join=round,line cap=round,fill=fillColor] (306.61,229.13) circle (  2.25);

\path[draw=drawColor,line width= 0.4pt,line join=round,line cap=round,fill=fillColor] (328.80,231.61) circle (  2.25);

\path[draw=drawColor,line width= 0.4pt,line join=round,line cap=round,fill=fillColor] (350.99,232.43) circle (  2.25);

\path[draw=drawColor,line width= 0.4pt,line join=round,line cap=round,fill=fillColor] (373.18,233.26) circle (  2.25);

\path[draw=drawColor,line width= 0.4pt,line join=round,line cap=round,fill=fillColor] (395.38,233.26) circle (  2.25);
\end{scope}
\begin{scope}
\path[clip] (  0.00,  0.00) rectangle (505.89,289.08);
\definecolor{drawColor}{RGB}{0,0,0}

\path[draw=drawColor,line width= 0.4pt,line join=round,line cap=round] ( 62.51, 61.20) -- (395.38, 61.20);

\path[draw=drawColor,line width= 0.4pt,line join=round,line cap=round] ( 62.51, 61.20) -- ( 62.51, 55.20);

\path[draw=drawColor,line width= 0.4pt,line join=round,line cap=round] (173.47, 61.20) -- (173.47, 55.20);

\path[draw=drawColor,line width= 0.4pt,line join=round,line cap=round] (284.42, 61.20) -- (284.42, 55.20);

\path[draw=drawColor,line width= 0.4pt,line join=round,line cap=round] (395.38, 61.20) -- (395.38, 55.20);

\node[text=drawColor,anchor=base,inner sep=0pt, outer sep=0pt, scale=  1.00] at ( 62.51, 39.60) {0.00};

\node[text=drawColor,anchor=base,inner sep=0pt, outer sep=0pt, scale=  1.00] at (173.47, 39.60) {0.05};

\node[text=drawColor,anchor=base,inner sep=0pt, outer sep=0pt, scale=  1.00] at (284.42, 39.60) {0.10};

\node[text=drawColor,anchor=base,inner sep=0pt, outer sep=0pt, scale=  1.00] at (395.38, 39.60) {0.15};

\path[draw=drawColor,line width= 0.4pt,line join=round,line cap=round] ( 49.20, 67.82) -- ( 49.20,233.26);

\path[draw=drawColor,line width= 0.4pt,line join=round,line cap=round] ( 49.20, 67.82) -- ( 43.20, 67.82);

\path[draw=drawColor,line width= 0.4pt,line join=round,line cap=round] ( 49.20,100.91) -- ( 43.20,100.91);

\path[draw=drawColor,line width= 0.4pt,line join=round,line cap=round] ( 49.20,134.00) -- ( 43.20,134.00);

\path[draw=drawColor,line width= 0.4pt,line join=round,line cap=round] ( 49.20,167.08) -- ( 43.20,167.08);

\path[draw=drawColor,line width= 0.4pt,line join=round,line cap=round] ( 49.20,200.17) -- ( 43.20,200.17);

\path[draw=drawColor,line width= 0.4pt,line join=round,line cap=round] ( 49.20,233.26) -- ( 43.20,233.26);

\node[text=drawColor,rotate= 90.00,anchor=base,inner sep=0pt, outer sep=0pt, scale=  1.00] at ( 34.80, 67.82) {0.0};

\node[text=drawColor,rotate= 90.00,anchor=base,inner sep=0pt, outer sep=0pt, scale=  1.00] at ( 34.80,100.91) {0.2};

\node[text=drawColor,rotate= 90.00,anchor=base,inner sep=0pt, outer sep=0pt, scale=  1.00] at ( 34.80,134.00) {0.4};

\node[text=drawColor,rotate= 90.00,anchor=base,inner sep=0pt, outer sep=0pt, scale=  1.00] at ( 34.80,167.08) {0.6};

\node[text=drawColor,rotate= 90.00,anchor=base,inner sep=0pt, outer sep=0pt, scale=  1.00] at ( 34.80,200.17) {0.8};

\node[text=drawColor,rotate= 90.00,anchor=base,inner sep=0pt, outer sep=0pt, scale=  1.00] at ( 34.80,233.26) {1.0};

\path[draw=drawColor,line width= 0.4pt,line join=round,line cap=round] ( 49.20, 61.20) --
	(408.69, 61.20) --
	(408.69,239.88) --
	( 49.20,239.88) --
	( 49.20, 61.20);
\end{scope}
\begin{scope}
\path[clip] (  0.00,  0.00) rectangle (505.89,289.08);
\definecolor{drawColor}{RGB}{0,0,0}

\node[text=drawColor,anchor=base,inner sep=0pt, outer sep=0pt, scale=  1.00] at (228.94, 15.60) {$\rho$};

\node[text=drawColor,rotate= 90.00,anchor=base,inner sep=0pt, outer sep=0pt, scale=  1.00] at ( 10.80,150.54) {Power};
\definecolor{drawColor}{RGB}{255,99,71}

\path[draw=drawColor,line width= 0.4pt,line join=round,line cap=round] ( 68.41, 78.84) -- ( 78.81, 80.78);

\path[draw=drawColor,line width= 0.4pt,line join=round,line cap=round] ( 90.64, 81.00) -- (100.96, 79.46);

\path[draw=drawColor,line width= 0.4pt,line join=round,line cap=round] (112.75, 79.88) -- (123.23, 82.23);

\path[draw=drawColor,line width= 0.4pt,line join=round,line cap=round] (134.99, 84.63) -- (145.38, 86.57);

\path[draw=drawColor,line width= 0.4pt,line join=round,line cap=round] (157.21, 88.56) -- (167.53, 90.10);

\path[draw=drawColor,line width= 0.4pt,line join=round,line cap=round] (178.62, 94.05) -- (190.51,101.14);

\path[draw=drawColor,line width= 0.4pt,line join=round,line cap=round] (200.98,106.99) -- (212.53,113.02);

\path[draw=drawColor,line width= 0.4pt,line join=round,line cap=round] (223.09,118.72) -- (234.80,125.28);

\path[draw=drawColor,line width= 0.4pt,line join=round,line cap=round] (243.97,132.74) -- (258.30,149.31);

\path[draw=drawColor,line width= 0.4pt,line join=round,line cap=round] (266.16,158.39) -- (280.50,174.96);

\path[draw=drawColor,line width= 0.4pt,line join=round,line cap=round] (290.32,180.59) -- (300.71,182.53);

\path[draw=drawColor,line width= 0.4pt,line join=round,line cap=round] (311.42,187.22) -- (323.99,196.59);

\path[draw=drawColor,line width= 0.4pt,line join=round,line cap=round] (334.43,202.27) -- (345.37,206.35);

\path[draw=drawColor,line width= 0.4pt,line join=round,line cap=round] (356.23,211.37) -- (367.95,217.93);

\path[draw=drawColor,line width= 0.4pt,line join=round,line cap=round] (379.12,221.74) -- (389.44,223.28);
\definecolor{fillColor}{RGB}{255,99,71}

\path[fill=fillColor] ( 62.51, 81.24) --
	( 65.54, 75.99) --
	( 59.48, 75.99) --
	cycle;

\path[fill=fillColor] ( 84.71, 85.38) --
	( 87.74, 80.13) --
	( 81.67, 80.13) --
	cycle;

\path[fill=fillColor] (106.90, 82.07) --
	(109.93, 76.82) --
	(103.87, 76.82) --
	cycle;

\path[fill=fillColor] (129.09, 87.03) --
	(132.12, 81.79) --
	(126.06, 81.79) --
	cycle;

\path[fill=fillColor] (151.28, 91.17) --
	(154.31, 85.92) --
	(148.25, 85.92) --
	cycle;

\path[fill=fillColor] (173.47, 94.48) --
	(176.50, 89.23) --
	(170.44, 89.23) --
	cycle;

\path[fill=fillColor] (195.66,107.71) --
	(198.69,102.47) --
	(192.63,102.47) --
	cycle;

\path[fill=fillColor] (217.85,119.30) --
	(220.88,114.05) --
	(214.82,114.05) --
	cycle;

\path[fill=fillColor] (240.04,131.70) --
	(243.07,126.46) --
	(237.01,126.46) --
	cycle;

\path[fill=fillColor] (262.23,157.35) --
	(265.26,152.10) --
	(259.20,152.10) --
	cycle;

\path[fill=fillColor] (284.42,182.99) --
	(287.45,177.74) --
	(281.39,177.74) --
	cycle;

\path[fill=fillColor] (306.61,187.13) --
	(309.64,181.88) --
	(303.58,181.88) --
	cycle;

\path[fill=fillColor] (328.80,203.67) --
	(331.83,198.42) --
	(325.77,198.42) --
	cycle;

\path[fill=fillColor] (350.99,211.94) --
	(354.02,206.70) --
	(347.96,206.70) --
	cycle;

\path[fill=fillColor] (373.18,224.35) --
	(376.22,219.10) --
	(370.15,219.10) --
	cycle;

\path[fill=fillColor] (395.38,227.66) --
	(398.41,222.41) --
	(392.35,222.41) --
	cycle;
\definecolor{drawColor}{RGB}{108,166,205}

\path[draw=drawColor,line width= 0.4pt,line join=round,line cap=round] ( 68.32, 80.37) -- ( 78.90, 77.60);

\path[draw=drawColor,line width= 0.4pt,line join=round,line cap=round] ( 90.39, 78.00) -- (101.21, 81.63);

\path[draw=drawColor,line width= 0.4pt,line join=round,line cap=round] (112.58, 85.44) -- (123.40, 89.07);

\path[draw=drawColor,line width= 0.4pt,line join=round,line cap=round] (133.73, 94.78) -- (146.64,105.37);

\path[draw=drawColor,line width= 0.4pt,line join=round,line cap=round] (156.26,112.52) -- (168.49,120.73);

\path[draw=drawColor,line width= 0.4pt,line join=round,line cap=round] (176.81,129.05) -- (192.32,152.17);

\path[draw=drawColor,line width= 0.4pt,line join=round,line cap=round] (199.89,161.41) -- (213.62,175.24);

\path[draw=drawColor,line width= 0.4pt,line join=round,line cap=round] (222.75,182.96) -- (235.14,191.74);

\path[draw=drawColor,line width= 0.4pt,line join=round,line cap=round] (244.94,198.68) -- (257.33,207.46);

\path[draw=drawColor,line width= 0.4pt,line join=round,line cap=round] (267.85,213.02) -- (278.80,217.10);

\path[draw=drawColor,line width= 0.4pt,line join=round,line cap=round] (290.04,221.30) -- (300.99,225.38);

\path[draw=drawColor,line width= 0.4pt,line join=round,line cap=round] (312.58,228.14) -- (322.84,229.29);

\path[draw=drawColor,line width= 0.4pt,line join=round,line cap=round] (334.79,230.40) -- (345.01,231.16);

\path[draw=drawColor,line width= 0.4pt,line join=round,line cap=round] (356.98,232.05) -- (367.20,232.82);

\path[draw=drawColor,line width= 0.4pt,line join=round,line cap=round] (379.18,233.26) -- (389.38,233.26);
\definecolor{fillColor}{RGB}{108,166,205}

\path[fill=fillColor] ( 60.26, 79.63) --
	( 64.76, 79.63) --
	( 64.76, 84.13) --
	( 60.26, 84.13) --
	cycle;

\path[fill=fillColor] ( 82.46, 73.84) --
	( 86.96, 73.84) --
	( 86.96, 78.34) --
	( 82.46, 78.34) --
	cycle;

\path[fill=fillColor] (104.65, 81.28) --
	(109.15, 81.28) --
	(109.15, 85.78) --
	(104.65, 85.78) --
	cycle;

\path[fill=fillColor] (126.84, 88.73) --
	(131.34, 88.73) --
	(131.34, 93.23) --
	(126.84, 93.23) --
	cycle;

\path[fill=fillColor] (149.03,106.93) --
	(153.53,106.93) --
	(153.53,111.43) --
	(149.03,111.43) --
	cycle;

\path[fill=fillColor] (171.22,121.82) --
	(175.72,121.82) --
	(175.72,126.32) --
	(171.22,126.32) --
	cycle;

\path[fill=fillColor] (193.41,154.91) --
	(197.91,154.91) --
	(197.91,159.41) --
	(193.41,159.41) --
	cycle;

\path[fill=fillColor] (215.60,177.24) --
	(220.10,177.24) --
	(220.10,181.74) --
	(215.60,181.74) --
	cycle;

\path[fill=fillColor] (237.79,192.96) --
	(242.29,192.96) --
	(242.29,197.46) --
	(237.79,197.46) --
	cycle;

\path[fill=fillColor] (259.98,208.68) --
	(264.48,208.68) --
	(264.48,213.18) --
	(259.98,213.18) --
	cycle;

\path[fill=fillColor] (282.17,216.95) --
	(286.67,216.95) --
	(286.67,221.45) --
	(282.17,221.45) --
	cycle;

\path[fill=fillColor] (304.36,225.22) --
	(308.86,225.22) --
	(308.86,229.72) --
	(304.36,229.72) --
	cycle;

\path[fill=fillColor] (326.55,227.70) --
	(331.05,227.70) --
	(331.05,232.20) --
	(326.55,232.20) --
	cycle;

\path[fill=fillColor] (348.74,229.36) --
	(353.24,229.36) --
	(353.24,233.86) --
	(348.74,233.86) --
	cycle;

\path[fill=fillColor] (370.93,231.01) --
	(375.43,231.01) --
	(375.43,235.51) --
	(370.93,235.51) --
	cycle;

\path[fill=fillColor] (393.13,231.01) --
	(397.63,231.01) --
	(397.63,235.51) --
	(393.13,235.51) --
	cycle;
\definecolor{drawColor}{RGB}{139,137,137}

\path[draw=drawColor,line width= 0.4pt,line join=round,line cap=round] ( 68.37, 75.75) -- ( 78.85, 78.09);

\path[draw=drawColor,line width= 0.4pt,line join=round,line cap=round] ( 90.69, 79.84) -- (100.91, 80.61);

\path[draw=drawColor,line width= 0.4pt,line join=round,line cap=round] (111.96, 84.27) -- (124.02, 91.90);

\path[draw=drawColor,line width= 0.4pt,line join=round,line cap=round] (132.37,100.14) -- (147.99,124.01);

\path[draw=drawColor,line width= 0.4pt,line join=round,line cap=round] (154.80,133.89) -- (169.95,154.78);

\path[draw=drawColor,line width= 0.4pt,line join=round,line cap=round] (176.99,164.50) -- (192.14,185.39);

\path[draw=drawColor,line width= 0.4pt,line join=round,line cap=round] (199.73,194.65) -- (213.78,209.83);

\path[draw=drawColor,line width= 0.4pt,line join=round,line cap=round] (222.83,217.58) -- (235.06,225.78);

\path[draw=drawColor,line width= 0.4pt,line join=round,line cap=round] (246.00,229.79) -- (256.27,230.94);

\path[draw=drawColor,line width= 0.4pt,line join=round,line cap=round] (268.21,232.05) -- (278.44,232.82);

\path[draw=drawColor,line width= 0.4pt,line join=round,line cap=round] (290.42,233.26) -- (300.61,233.26);

\path[draw=drawColor,line width= 0.4pt,line join=round,line cap=round] (312.61,233.26) -- (322.80,233.26);

\path[draw=drawColor,line width= 0.4pt,line join=round,line cap=round] (334.80,233.26) -- (344.99,233.26);

\path[draw=drawColor,line width= 0.4pt,line join=round,line cap=round] (356.99,233.26) -- (367.18,233.26);

\path[draw=drawColor,line width= 0.4pt,line join=round,line cap=round] (379.18,233.26) -- (389.38,233.26);

\path[draw=drawColor,line width= 0.4pt,line join=round,line cap=round] ( 59.33, 74.44) -- ( 65.70, 74.44);

\path[draw=drawColor,line width= 0.4pt,line join=round,line cap=round] ( 62.51, 71.25) -- ( 62.51, 77.62);

\path[draw=drawColor,line width= 0.4pt,line join=round,line cap=round] ( 81.52, 79.40) -- ( 87.89, 79.40);

\path[draw=drawColor,line width= 0.4pt,line join=round,line cap=round] ( 84.71, 76.22) -- ( 84.71, 82.58);

\path[draw=drawColor,line width= 0.4pt,line join=round,line cap=round] (103.71, 81.05) -- (110.08, 81.05);

\path[draw=drawColor,line width= 0.4pt,line join=round,line cap=round] (106.90, 77.87) -- (106.90, 84.24);

\path[draw=drawColor,line width= 0.4pt,line join=round,line cap=round] (125.90, 95.12) -- (132.27, 95.12);

\path[draw=drawColor,line width= 0.4pt,line join=round,line cap=round] (129.09, 91.93) -- (129.09, 98.30);

\path[draw=drawColor,line width= 0.4pt,line join=round,line cap=round] (148.10,129.03) -- (154.46,129.03);

\path[draw=drawColor,line width= 0.4pt,line join=round,line cap=round] (151.28,125.85) -- (151.28,132.21);

\path[draw=drawColor,line width= 0.4pt,line join=round,line cap=round] (170.29,159.64) -- (176.65,159.64);

\path[draw=drawColor,line width= 0.4pt,line join=round,line cap=round] (173.47,156.46) -- (173.47,162.82);

\path[draw=drawColor,line width= 0.4pt,line join=round,line cap=round] (192.48,190.25) -- (198.84,190.25);

\path[draw=drawColor,line width= 0.4pt,line join=round,line cap=round] (195.66,187.06) -- (195.66,193.43);

\path[draw=drawColor,line width= 0.4pt,line join=round,line cap=round] (214.67,214.24) -- (221.03,214.24);

\path[draw=drawColor,line width= 0.4pt,line join=round,line cap=round] (217.85,211.05) -- (217.85,217.42);

\path[draw=drawColor,line width= 0.4pt,line join=round,line cap=round] (236.86,229.13) -- (243.22,229.13);

\path[draw=drawColor,line width= 0.4pt,line join=round,line cap=round] (240.04,225.94) -- (240.04,232.31);

\path[draw=drawColor,line width= 0.4pt,line join=round,line cap=round] (259.05,231.61) -- (265.41,231.61);

\path[draw=drawColor,line width= 0.4pt,line join=round,line cap=round] (262.23,228.43) -- (262.23,234.79);

\path[draw=drawColor,line width= 0.4pt,line join=round,line cap=round] (281.24,233.26) -- (287.60,233.26);

\path[draw=drawColor,line width= 0.4pt,line join=round,line cap=round] (284.42,230.08) -- (284.42,236.44);

\path[draw=drawColor,line width= 0.4pt,line join=round,line cap=round] (303.43,233.26) -- (309.79,233.26);

\path[draw=drawColor,line width= 0.4pt,line join=round,line cap=round] (306.61,230.08) -- (306.61,236.44);

\path[draw=drawColor,line width= 0.4pt,line join=round,line cap=round] (325.62,233.26) -- (331.99,233.26);

\path[draw=drawColor,line width= 0.4pt,line join=round,line cap=round] (328.80,230.08) -- (328.80,236.44);

\path[draw=drawColor,line width= 0.4pt,line join=round,line cap=round] (347.81,233.26) -- (354.18,233.26);

\path[draw=drawColor,line width= 0.4pt,line join=round,line cap=round] (350.99,230.08) -- (350.99,236.44);

\path[draw=drawColor,line width= 0.4pt,line join=round,line cap=round] (370.00,233.26) -- (376.37,233.26);

\path[draw=drawColor,line width= 0.4pt,line join=round,line cap=round] (373.18,230.08) -- (373.18,236.44);

\path[draw=drawColor,line width= 0.4pt,line join=round,line cap=round] (392.19,233.26) -- (398.56,233.26);

\path[draw=drawColor,line width= 0.4pt,line join=round,line cap=round] (395.38,230.08) -- (395.38,236.44);
\definecolor{drawColor}{RGB}{84,139,84}

\path[draw=drawColor,line width= 0.4pt,line join=round,line cap=round] ( 68.51, 75.04) -- ( 78.71, 74.66);

\path[draw=drawColor,line width= 0.4pt,line join=round,line cap=round] ( 90.51, 75.95) -- (101.09, 78.71);

\path[draw=drawColor,line width= 0.4pt,line join=round,line cap=round] (112.79, 79.13) -- (123.19, 77.19);

\path[draw=drawColor,line width= 0.4pt,line join=round,line cap=round] (134.56, 78.54) -- (145.80, 83.57);

\path[draw=drawColor,line width= 0.4pt,line join=round,line cap=round] (156.26, 89.36) -- (168.49, 97.56);

\path[draw=drawColor,line width= 0.4pt,line join=round,line cap=round] (179.43,101.57) -- (189.70,102.72);

\path[draw=drawColor,line width= 0.4pt,line join=round,line cap=round] (199.73,107.79) -- (213.78,122.97);

\path[draw=drawColor,line width= 0.4pt,line join=round,line cap=round] (223.47,129.47) -- (234.42,133.55);

\path[draw=drawColor,line width= 0.4pt,line join=round,line cap=round] (244.19,139.98) -- (258.08,154.48);

\path[draw=drawColor,line width= 0.4pt,line join=round,line cap=round] (266.79,162.72) -- (279.87,173.93);

\path[draw=drawColor,line width= 0.4pt,line join=round,line cap=round] (288.50,182.24) -- (302.54,197.42);

\path[draw=drawColor,line width= 0.4pt,line join=round,line cap=round] (311.68,205.04) -- (323.74,212.68);

\path[draw=drawColor,line width= 0.4pt,line join=round,line cap=round] (334.43,217.99) -- (345.37,222.07);

\path[draw=drawColor,line width= 0.4pt,line join=round,line cap=round] (356.80,225.68) -- (367.38,228.44);

\path[draw=drawColor,line width= 0.4pt,line join=round,line cap=round] (379.17,230.40) -- (389.39,231.16);

\path[draw=drawColor,line width= 0.4pt,line join=round,line cap=round] ( 60.26, 73.01) -- ( 64.76, 77.51);

\path[draw=drawColor,line width= 0.4pt,line join=round,line cap=round] ( 60.26, 77.51) -- ( 64.76, 73.01);

\path[draw=drawColor,line width= 0.4pt,line join=round,line cap=round] ( 82.46, 72.19) -- ( 86.96, 76.69);

\path[draw=drawColor,line width= 0.4pt,line join=round,line cap=round] ( 82.46, 76.69) -- ( 86.96, 72.19);

\path[draw=drawColor,line width= 0.4pt,line join=round,line cap=round] (104.65, 77.98) -- (109.15, 82.48);

\path[draw=drawColor,line width= 0.4pt,line join=round,line cap=round] (104.65, 82.48) -- (109.15, 77.98);

\path[draw=drawColor,line width= 0.4pt,line join=round,line cap=round] (126.84, 73.84) -- (131.34, 78.34);

\path[draw=drawColor,line width= 0.4pt,line join=round,line cap=round] (126.84, 78.34) -- (131.34, 73.84);

\path[draw=drawColor,line width= 0.4pt,line join=round,line cap=round] (149.03, 83.77) -- (153.53, 88.27);

\path[draw=drawColor,line width= 0.4pt,line join=round,line cap=round] (149.03, 88.27) -- (153.53, 83.77);

\path[draw=drawColor,line width= 0.4pt,line join=round,line cap=round] (171.22, 98.66) -- (175.72,103.16);

\path[draw=drawColor,line width= 0.4pt,line join=round,line cap=round] (171.22,103.16) -- (175.72, 98.66);

\path[draw=drawColor,line width= 0.4pt,line join=round,line cap=round] (193.41,101.14) -- (197.91,105.64);

\path[draw=drawColor,line width= 0.4pt,line join=round,line cap=round] (193.41,105.64) -- (197.91,101.14);

\path[draw=drawColor,line width= 0.4pt,line join=round,line cap=round] (215.60,125.13) -- (220.10,129.63);

\path[draw=drawColor,line width= 0.4pt,line join=round,line cap=round] (215.60,129.63) -- (220.10,125.13);

\path[draw=drawColor,line width= 0.4pt,line join=round,line cap=round] (237.79,133.40) -- (242.29,137.90);

\path[draw=drawColor,line width= 0.4pt,line join=round,line cap=round] (237.79,137.90) -- (242.29,133.40);

\path[draw=drawColor,line width= 0.4pt,line join=round,line cap=round] (259.98,156.56) -- (264.48,161.06);

\path[draw=drawColor,line width= 0.4pt,line join=round,line cap=round] (259.98,161.06) -- (264.48,156.56);

\path[draw=drawColor,line width= 0.4pt,line join=round,line cap=round] (282.17,175.59) -- (286.67,180.09);

\path[draw=drawColor,line width= 0.4pt,line join=round,line cap=round] (282.17,180.09) -- (286.67,175.59);

\path[draw=drawColor,line width= 0.4pt,line join=round,line cap=round] (304.36,199.58) -- (308.86,204.08);

\path[draw=drawColor,line width= 0.4pt,line join=round,line cap=round] (304.36,204.08) -- (308.86,199.58);

\path[draw=drawColor,line width= 0.4pt,line join=round,line cap=round] (326.55,213.64) -- (331.05,218.14);

\path[draw=drawColor,line width= 0.4pt,line join=round,line cap=round] (326.55,218.14) -- (331.05,213.64);

\path[draw=drawColor,line width= 0.4pt,line join=round,line cap=round] (348.74,221.91) -- (353.24,226.41);

\path[draw=drawColor,line width= 0.4pt,line join=round,line cap=round] (348.74,226.41) -- (353.24,221.91);

\path[draw=drawColor,line width= 0.4pt,line join=round,line cap=round] (370.93,227.70) -- (375.43,232.20);

\path[draw=drawColor,line width= 0.4pt,line join=round,line cap=round] (370.93,232.20) -- (375.43,227.70);

\path[draw=drawColor,line width= 0.4pt,line join=round,line cap=round] (393.13,229.36) -- (397.63,233.86);

\path[draw=drawColor,line width= 0.4pt,line join=round,line cap=round] (393.13,233.86) -- (397.63,229.36);
\definecolor{drawColor}{RGB}{205,96,144}

\path[draw=drawColor,line width= 0.4pt,line join=round,line cap=round] ( 68.51, 78.57) -- ( 78.71, 78.57);

\path[draw=drawColor,line width= 0.4pt,line join=round,line cap=round] ( 90.71, 78.57) -- (100.90, 78.57);

\path[draw=drawColor,line width= 0.4pt,line join=round,line cap=round] (112.86, 79.24) -- (123.12, 80.39);

\path[draw=drawColor,line width= 0.4pt,line join=round,line cap=round] (133.48, 85.14) -- (146.89, 97.64);

\path[draw=drawColor,line width= 0.4pt,line join=round,line cap=round] (155.92,105.54) -- (168.83,116.13);

\path[draw=drawColor,line width= 0.4pt,line join=round,line cap=round] (178.11,123.74) -- (191.02,134.33);

\path[draw=drawColor,line width= 0.4pt,line join=round,line cap=round] (199.12,143.03) -- (214.39,164.66);

\path[draw=drawColor,line width= 0.4pt,line join=round,line cap=round] (221.37,174.42) -- (236.52,195.32);

\path[draw=drawColor,line width= 0.4pt,line join=round,line cap=round] (244.94,203.64) -- (257.33,212.42);

\path[draw=drawColor,line width= 0.4pt,line join=round,line cap=round] (267.13,219.36) -- (279.53,228.14);

\path[draw=drawColor,line width= 0.4pt,line join=round,line cap=round] (290.42,231.83) -- (300.62,232.21);

\path[draw=drawColor,line width= 0.4pt,line join=round,line cap=round] (312.61,232.66) -- (322.81,233.04);

\path[draw=drawColor,line width= 0.4pt,line join=round,line cap=round] (334.80,233.26) -- (344.99,233.26);

\path[draw=drawColor,line width= 0.4pt,line join=round,line cap=round] (356.99,233.26) -- (367.18,233.26);

\path[draw=drawColor,line width= 0.4pt,line join=round,line cap=round] (379.18,233.26) -- (389.38,233.26);

\path[draw=drawColor,line width= 0.4pt,line join=round,line cap=round] ( 60.26, 76.32) rectangle ( 64.76, 80.82);

\path[draw=drawColor,line width= 0.4pt,line join=round,line cap=round] ( 60.26, 76.32) -- ( 64.76, 80.82);

\path[draw=drawColor,line width= 0.4pt,line join=round,line cap=round] ( 60.26, 80.82) -- ( 64.76, 76.32);

\path[draw=drawColor,line width= 0.4pt,line join=round,line cap=round] ( 82.46, 76.32) rectangle ( 86.96, 80.82);

\path[draw=drawColor,line width= 0.4pt,line join=round,line cap=round] ( 82.46, 76.32) -- ( 86.96, 80.82);

\path[draw=drawColor,line width= 0.4pt,line join=round,line cap=round] ( 82.46, 80.82) -- ( 86.96, 76.32);

\path[draw=drawColor,line width= 0.4pt,line join=round,line cap=round] (104.65, 76.32) rectangle (109.15, 80.82);

\path[draw=drawColor,line width= 0.4pt,line join=round,line cap=round] (104.65, 76.32) -- (109.15, 80.82);

\path[draw=drawColor,line width= 0.4pt,line join=round,line cap=round] (104.65, 80.82) -- (109.15, 76.32);

\path[draw=drawColor,line width= 0.4pt,line join=round,line cap=round] (126.84, 78.80) rectangle (131.34, 83.30);

\path[draw=drawColor,line width= 0.4pt,line join=round,line cap=round] (126.84, 78.80) -- (131.34, 83.30);

\path[draw=drawColor,line width= 0.4pt,line join=round,line cap=round] (126.84, 83.30) -- (131.34, 78.80);

\path[draw=drawColor,line width= 0.4pt,line join=round,line cap=round] (149.03, 99.48) rectangle (153.53,103.98);

\path[draw=drawColor,line width= 0.4pt,line join=round,line cap=round] (149.03, 99.48) -- (153.53,103.98);

\path[draw=drawColor,line width= 0.4pt,line join=round,line cap=round] (149.03,103.98) -- (153.53, 99.48);

\path[draw=drawColor,line width= 0.4pt,line join=round,line cap=round] (171.22,117.68) rectangle (175.72,122.18);

\path[draw=drawColor,line width= 0.4pt,line join=round,line cap=round] (171.22,117.68) -- (175.72,122.18);

\path[draw=drawColor,line width= 0.4pt,line join=round,line cap=round] (171.22,122.18) -- (175.72,117.68);

\path[draw=drawColor,line width= 0.4pt,line join=round,line cap=round] (193.41,135.88) rectangle (197.91,140.38);

\path[draw=drawColor,line width= 0.4pt,line join=round,line cap=round] (193.41,135.88) -- (197.91,140.38);

\path[draw=drawColor,line width= 0.4pt,line join=round,line cap=round] (193.41,140.38) -- (197.91,135.88);

\path[draw=drawColor,line width= 0.4pt,line join=round,line cap=round] (215.60,167.32) rectangle (220.10,171.82);

\path[draw=drawColor,line width= 0.4pt,line join=round,line cap=round] (215.60,167.32) -- (220.10,171.82);

\path[draw=drawColor,line width= 0.4pt,line join=round,line cap=round] (215.60,171.82) -- (220.10,167.32);

\path[draw=drawColor,line width= 0.4pt,line join=round,line cap=round] (237.79,197.92) rectangle (242.29,202.42);

\path[draw=drawColor,line width= 0.4pt,line join=round,line cap=round] (237.79,197.92) -- (242.29,202.42);

\path[draw=drawColor,line width= 0.4pt,line join=round,line cap=round] (237.79,202.42) -- (242.29,197.92);

\path[draw=drawColor,line width= 0.4pt,line join=round,line cap=round] (259.98,213.64) rectangle (264.48,218.14);

\path[draw=drawColor,line width= 0.4pt,line join=round,line cap=round] (259.98,213.64) -- (264.48,218.14);

\path[draw=drawColor,line width= 0.4pt,line join=round,line cap=round] (259.98,218.14) -- (264.48,213.64);

\path[draw=drawColor,line width= 0.4pt,line join=round,line cap=round] (282.17,229.36) rectangle (286.67,233.86);

\path[draw=drawColor,line width= 0.4pt,line join=round,line cap=round] (282.17,229.36) -- (286.67,233.86);

\path[draw=drawColor,line width= 0.4pt,line join=round,line cap=round] (282.17,233.86) -- (286.67,229.36);

\path[draw=drawColor,line width= 0.4pt,line join=round,line cap=round] (304.36,230.18) rectangle (308.86,234.68);

\path[draw=drawColor,line width= 0.4pt,line join=round,line cap=round] (304.36,230.18) -- (308.86,234.68);

\path[draw=drawColor,line width= 0.4pt,line join=round,line cap=round] (304.36,234.68) -- (308.86,230.18);

\path[draw=drawColor,line width= 0.4pt,line join=round,line cap=round] (326.55,231.01) rectangle (331.05,235.51);

\path[draw=drawColor,line width= 0.4pt,line join=round,line cap=round] (326.55,231.01) -- (331.05,235.51);

\path[draw=drawColor,line width= 0.4pt,line join=round,line cap=round] (326.55,235.51) -- (331.05,231.01);

\path[draw=drawColor,line width= 0.4pt,line join=round,line cap=round] (348.74,231.01) rectangle (353.24,235.51);

\path[draw=drawColor,line width= 0.4pt,line join=round,line cap=round] (348.74,231.01) -- (353.24,235.51);

\path[draw=drawColor,line width= 0.4pt,line join=round,line cap=round] (348.74,235.51) -- (353.24,231.01);

\path[draw=drawColor,line width= 0.4pt,line join=round,line cap=round] (370.93,231.01) rectangle (375.43,235.51);

\path[draw=drawColor,line width= 0.4pt,line join=round,line cap=round] (370.93,231.01) -- (375.43,235.51);

\path[draw=drawColor,line width= 0.4pt,line join=round,line cap=round] (370.93,235.51) -- (375.43,231.01);

\path[draw=drawColor,line width= 0.4pt,line join=round,line cap=round] (393.13,231.01) rectangle (397.63,235.51);

\path[draw=drawColor,line width= 0.4pt,line join=round,line cap=round] (393.13,231.01) -- (397.63,235.51);

\path[draw=drawColor,line width= 0.4pt,line join=round,line cap=round] (393.13,235.51) -- (397.63,231.01);
\definecolor{drawColor}{RGB}{0,0,0}

\path[draw=drawColor,line width= 0.4pt,dash pattern=on 4pt off 4pt ,line join=round,line cap=round] ( 58.08, 76.09) --
	( 61.53, 76.09) --
	( 64.98, 76.09) --
	( 68.43, 76.09) --
	( 71.88, 76.09) --
	( 75.34, 76.09) --
	( 78.79, 76.09) --
	( 82.24, 76.09) --
	( 85.69, 76.09) --
	( 89.14, 76.09) --
	( 92.60, 76.09) --
	( 96.05, 76.09) --
	( 99.50, 76.09) --
	(102.95, 76.09) --
	(106.40, 76.09) --
	(109.85, 76.09) --
	(113.31, 76.09) --
	(116.76, 76.09) --
	(120.21, 76.09) --
	(123.66, 76.09) --
	(127.11, 76.09) --
	(130.57, 76.09) --
	(134.02, 76.09) --
	(137.47, 76.09) --
	(140.92, 76.09) --
	(144.37, 76.09) --
	(147.83, 76.09) --
	(151.28, 76.09) --
	(154.73, 76.09) --
	(158.18, 76.09) --
	(161.63, 76.09) --
	(165.08, 76.09) --
	(168.54, 76.09) --
	(171.99, 76.09) --
	(175.44, 76.09) --
	(178.89, 76.09) --
	(182.34, 76.09) --
	(185.80, 76.09) --
	(189.25, 76.09) --
	(192.70, 76.09) --
	(196.15, 76.09) --
	(199.60, 76.09) --
	(203.06, 76.09) --
	(206.51, 76.09) --
	(209.96, 76.09) --
	(213.41, 76.09) --
	(216.86, 76.09) --
	(220.32, 76.09) --
	(223.77, 76.09) --
	(227.22, 76.09) --
	(230.67, 76.09) --
	(234.12, 76.09) --
	(237.57, 76.09) --
	(241.03, 76.09) --
	(244.48, 76.09) --
	(247.93, 76.09) --
	(251.38, 76.09) --
	(254.83, 76.09) --
	(258.29, 76.09) --
	(261.74, 76.09) --
	(265.19, 76.09) --
	(268.64, 76.09) --
	(272.09, 76.09) --
	(275.55, 76.09) --
	(279.00, 76.09) --
	(282.45, 76.09) --
	(285.90, 76.09) --
	(289.35, 76.09) --
	(292.81, 76.09) --
	(296.26, 76.09) --
	(299.71, 76.09) --
	(303.16, 76.09) --
	(306.61, 76.09) --
	(310.06, 76.09) --
	(313.52, 76.09) --
	(316.97, 76.09) --
	(320.42, 76.09) --
	(323.87, 76.09) --
	(327.32, 76.09) --
	(330.78, 76.09) --
	(334.23, 76.09) --
	(337.68, 76.09) --
	(341.13, 76.09) --
	(344.58, 76.09) --
	(348.04, 76.09) --
	(351.49, 76.09) --
	(354.94, 76.09) --
	(358.39, 76.09) --
	(361.84, 76.09) --
	(365.29, 76.09) --
	(368.75, 76.09) --
	(372.20, 76.09) --
	(375.65, 76.09) --
	(379.10, 76.09) --
	(382.55, 76.09) --
	(386.01, 76.09) --
	(389.46, 76.09) --
	(392.91, 76.09) --
	(396.36, 76.09) --
	(399.81, 76.09);
\definecolor{drawColor}{RGB}{139,0,0}
\definecolor{fillColor}{RGB}{139,0,0}

\path[draw=drawColor,line width= 0.4pt,line join=round,line cap=round,fill=fillColor] (421.11,227.88) circle (  2.25);
\definecolor{fillColor}{RGB}{255,99,71}

\path[fill=fillColor] (421.11,219.38) --
	(424.15,214.13) --
	(418.08,214.13) --
	cycle;
\definecolor{fillColor}{RGB}{108,166,205}

\path[fill=fillColor] (418.86,201.63) --
	(423.36,201.63) --
	(423.36,206.13) --
	(418.86,206.13) --
	cycle;
\definecolor{drawColor}{RGB}{139,137,137}

\path[draw=drawColor,line width= 0.4pt,line join=round,line cap=round] (417.93,191.88) -- (424.30,191.88);

\path[draw=drawColor,line width= 0.4pt,line join=round,line cap=round] (421.11,188.70) -- (421.11,195.06);
\definecolor{drawColor}{RGB}{84,139,84}

\path[draw=drawColor,line width= 0.4pt,line join=round,line cap=round] (418.86,177.63) -- (423.36,182.13);

\path[draw=drawColor,line width= 0.4pt,line join=round,line cap=round] (418.86,182.13) -- (423.36,177.63);
\definecolor{drawColor}{RGB}{205,96,144}

\path[draw=drawColor,line width= 0.4pt,line join=round,line cap=round] (418.86,165.63) rectangle (423.36,170.13);

\path[draw=drawColor,line width= 0.4pt,line join=round,line cap=round] (418.86,165.63) -- (423.36,170.13);

\path[draw=drawColor,line width= 0.4pt,line join=round,line cap=round] (418.86,170.13) -- (423.36,165.63);
\definecolor{drawColor}{RGB}{0,0,0}

\node[text=drawColor,anchor=base west,inner sep=0pt, outer sep=0pt, scale=  1.00] at (430.11,224.44) {HypoRF};

\node[text=drawColor,anchor=base west,inner sep=0pt, outer sep=0pt, scale=  1.00] at (430.11,212.44) {Binomial};

\node[text=drawColor,anchor=base west,inner sep=0pt, outer sep=0pt, scale=  1.00] at (430.11,200.44) {ME-full};

\node[text=drawColor,anchor=base west,inner sep=0pt, outer sep=0pt, scale=  1.00] at (430.11,188.44) {MMDboot};

\node[text=drawColor,anchor=base west,inner sep=0pt, outer sep=0pt, scale=  1.00] at (430.11,176.44) {MMD-full};

\node[text=drawColor,anchor=base west,inner sep=0pt, outer sep=0pt, scale=  1.00] at (430.11,164.44) {CPT-RF};
\end{scope}
\end{tikzpicture}

%% file: 3b_plot_gaussian_dep_K100_Normapprox_F_with_Cai.tex
\begin{tikzpicture}[x=1pt,y=1pt]
\definecolor{fillColor}{RGB}{255,255,255}
\path[use as bounding box,fill=fillColor,fill opacity=0.00] (0,0) rectangle (505.89,289.08);
\begin{scope}
\path[clip] (  0.00,  0.00) rectangle (505.89,289.08);
\definecolor{drawColor}{RGB}{139,0,0}

\path[draw=drawColor,line width= 0.4pt,line join=round,line cap=round] ( 68.45, 79.34) -- ( 78.77, 77.80);

\path[draw=drawColor,line width= 0.4pt,line join=round,line cap=round] ( 90.70, 76.69) -- (100.90, 76.31);

\path[draw=drawColor,line width= 0.4pt,line join=round,line cap=round] (112.89, 76.31) -- (123.09, 76.69);

\path[draw=drawColor,line width= 0.4pt,line join=round,line cap=round] (134.99, 78.02) -- (145.38, 79.95);

\path[draw=drawColor,line width= 0.4pt,line join=round,line cap=round] (157.28, 81.05) -- (167.47, 81.05);

\path[draw=drawColor,line width= 0.4pt,line join=round,line cap=round] (179.43, 81.72) -- (189.70, 82.87);

\path[draw=drawColor,line width= 0.4pt,line join=round,line cap=round] (200.98, 86.31) -- (212.53, 92.34);

\path[draw=drawColor,line width= 0.4pt,line join=round,line cap=round] (223.81, 95.78) -- (234.08, 96.93);

\path[draw=drawColor,line width= 0.4pt,line join=round,line cap=round] (246.04, 97.37) -- (256.24, 96.99);

\path[draw=drawColor,line width= 0.4pt,line join=round,line cap=round] (267.47, 99.70) -- (279.18,106.25);

\path[draw=drawColor,line width= 0.4pt,line join=round,line cap=round] (289.15,112.88) -- (301.89,122.85);

\path[draw=drawColor,line width= 0.4pt,line join=round,line cap=round] (312.30,128.46) -- (323.11,132.09);

\path[draw=drawColor,line width= 0.4pt,line join=round,line cap=round] (333.96,137.07) -- (345.84,144.16);

\path[draw=drawColor,line width= 0.4pt,line join=round,line cap=round] (354.78,151.89) -- (369.40,169.87);

\path[draw=drawColor,line width= 0.4pt,line join=round,line cap=round] (378.99,176.04) -- (389.57,178.81);
\definecolor{fillColor}{RGB}{139,0,0}

\path[draw=drawColor,line width= 0.4pt,line join=round,line cap=round,fill=fillColor] ( 62.51, 80.23) circle (  2.25);

\path[draw=drawColor,line width= 0.4pt,line join=round,line cap=round,fill=fillColor] ( 84.71, 76.92) circle (  2.25);

\path[draw=drawColor,line width= 0.4pt,line join=round,line cap=round,fill=fillColor] (106.90, 76.09) circle (  2.25);

\path[draw=drawColor,line width= 0.4pt,line join=round,line cap=round,fill=fillColor] (129.09, 76.92) circle (  2.25);

\path[draw=drawColor,line width= 0.4pt,line join=round,line cap=round,fill=fillColor] (151.28, 81.05) circle (  2.25);

\path[draw=drawColor,line width= 0.4pt,line join=round,line cap=round,fill=fillColor] (173.47, 81.05) circle (  2.25);

\path[draw=drawColor,line width= 0.4pt,line join=round,line cap=round,fill=fillColor] (195.66, 83.53) circle (  2.25);

\path[draw=drawColor,line width= 0.4pt,line join=round,line cap=round,fill=fillColor] (217.85, 95.12) circle (  2.25);

\path[draw=drawColor,line width= 0.4pt,line join=round,line cap=round,fill=fillColor] (240.04, 97.60) circle (  2.25);

\path[draw=drawColor,line width= 0.4pt,line join=round,line cap=round,fill=fillColor] (262.23, 96.77) circle (  2.25);

\path[draw=drawColor,line width= 0.4pt,line join=round,line cap=round,fill=fillColor] (284.42,109.18) circle (  2.25);

\path[draw=drawColor,line width= 0.4pt,line join=round,line cap=round,fill=fillColor] (306.61,126.55) circle (  2.25);

\path[draw=drawColor,line width= 0.4pt,line join=round,line cap=round,fill=fillColor] (328.80,134.00) circle (  2.25);

\path[draw=drawColor,line width= 0.4pt,line join=round,line cap=round,fill=fillColor] (350.99,147.23) circle (  2.25);

\path[draw=drawColor,line width= 0.4pt,line join=round,line cap=round,fill=fillColor] (373.18,174.53) circle (  2.25);

\path[draw=drawColor,line width= 0.4pt,line join=round,line cap=round,fill=fillColor] (395.38,180.32) circle (  2.25);
\end{scope}
\begin{scope}
\path[clip] (  0.00,  0.00) rectangle (505.89,289.08);
\definecolor{drawColor}{RGB}{0,0,0}

\path[draw=drawColor,line width= 0.4pt,line join=round,line cap=round] ( 62.51, 61.20) -- (328.80, 61.20);

\path[draw=drawColor,line width= 0.4pt,line join=round,line cap=round] ( 62.51, 61.20) -- ( 62.51, 55.20);

\path[draw=drawColor,line width= 0.4pt,line join=round,line cap=round] (151.28, 61.20) -- (151.28, 55.20);

\path[draw=drawColor,line width= 0.4pt,line join=round,line cap=round] (240.04, 61.20) -- (240.04, 55.20);

\path[draw=drawColor,line width= 0.4pt,line join=round,line cap=round] (328.80, 61.20) -- (328.80, 55.20);

\node[text=drawColor,anchor=base,inner sep=0pt, outer sep=0pt, scale=  1.00] at ( 62.51, 39.60) {0.0};

\node[text=drawColor,anchor=base,inner sep=0pt, outer sep=0pt, scale=  1.00] at (151.28, 39.60) {0.1};

\node[text=drawColor,anchor=base,inner sep=0pt, outer sep=0pt, scale=  1.00] at (240.04, 39.60) {0.2};

\node[text=drawColor,anchor=base,inner sep=0pt, outer sep=0pt, scale=  1.00] at (328.80, 39.60) {0.3};

\path[draw=drawColor,line width= 0.4pt,line join=round,line cap=round] ( 49.20, 67.82) -- ( 49.20,233.26);

\path[draw=drawColor,line width= 0.4pt,line join=round,line cap=round] ( 49.20, 67.82) -- ( 43.20, 67.82);

\path[draw=drawColor,line width= 0.4pt,line join=round,line cap=round] ( 49.20,100.91) -- ( 43.20,100.91);

\path[draw=drawColor,line width= 0.4pt,line join=round,line cap=round] ( 49.20,134.00) -- ( 43.20,134.00);

\path[draw=drawColor,line width= 0.4pt,line join=round,line cap=round] ( 49.20,167.08) -- ( 43.20,167.08);

\path[draw=drawColor,line width= 0.4pt,line join=round,line cap=round] ( 49.20,200.17) -- ( 43.20,200.17);

\path[draw=drawColor,line width= 0.4pt,line join=round,line cap=round] ( 49.20,233.26) -- ( 43.20,233.26);

\node[text=drawColor,rotate= 90.00,anchor=base,inner sep=0pt, outer sep=0pt, scale=  1.00] at ( 34.80, 67.82) {0.0};

\node[text=drawColor,rotate= 90.00,anchor=base,inner sep=0pt, outer sep=0pt, scale=  1.00] at ( 34.80,100.91) {0.2};

\node[text=drawColor,rotate= 90.00,anchor=base,inner sep=0pt, outer sep=0pt, scale=  1.00] at ( 34.80,134.00) {0.4};

\node[text=drawColor,rotate= 90.00,anchor=base,inner sep=0pt, outer sep=0pt, scale=  1.00] at ( 34.80,167.08) {0.6};

\node[text=drawColor,rotate= 90.00,anchor=base,inner sep=0pt, outer sep=0pt, scale=  1.00] at ( 34.80,200.17) {0.8};

\node[text=drawColor,rotate= 90.00,anchor=base,inner sep=0pt, outer sep=0pt, scale=  1.00] at ( 34.80,233.26) {1.0};

\path[draw=drawColor,line width= 0.4pt,line join=round,line cap=round] ( 49.20, 61.20) --
	(408.69, 61.20) --
	(408.69,239.88) --
	( 49.20,239.88) --
	( 49.20, 61.20);
\end{scope}
\begin{scope}
\path[clip] (  0.00,  0.00) rectangle (505.89,289.08);
\definecolor{drawColor}{RGB}{0,0,0}

\node[text=drawColor,anchor=base,inner sep=0pt, outer sep=0pt, scale=  1.00] at (228.94, 15.60) {$\rho$};

\node[text=drawColor,rotate= 90.00,anchor=base,inner sep=0pt, outer sep=0pt, scale=  1.00] at ( 10.80,150.54) {Power};
\definecolor{drawColor}{RGB}{255,99,71}

\path[draw=drawColor,line width= 0.4pt,line join=round,line cap=round] ( 68.41, 77.19) -- ( 78.81, 79.13);

\path[draw=drawColor,line width= 0.4pt,line join=round,line cap=round] ( 90.64, 79.34) -- (100.96, 77.80);

\path[draw=drawColor,line width= 0.4pt,line join=round,line cap=round] (112.75, 78.23) -- (123.23, 80.57);

\path[draw=drawColor,line width= 0.4pt,line join=round,line cap=round] (135.05, 81.21) -- (145.31, 80.07);

\path[draw=drawColor,line width= 0.4pt,line join=round,line cap=round] (157.24, 80.07) -- (167.51, 81.21);

\path[draw=drawColor,line width= 0.4pt,line join=round,line cap=round] (179.40, 81.00) -- (189.72, 79.46);

\path[draw=drawColor,line width= 0.4pt,line join=round,line cap=round] (201.46, 80.09) -- (212.04, 82.85);

\path[draw=drawColor,line width= 0.4pt,line join=round,line cap=round] (223.78, 83.48) -- (234.11, 81.94);

\path[draw=drawColor,line width= 0.4pt,line join=round,line cap=round] (245.94, 82.15) -- (256.33, 84.09);

\path[draw=drawColor,line width= 0.4pt,line join=round,line cap=round] (267.55, 87.97) -- (279.10, 93.99);

\path[draw=drawColor,line width= 0.4pt,line join=round,line cap=round] (290.17, 95.06) -- (300.86, 91.87);

\path[draw=drawColor,line width= 0.4pt,line join=round,line cap=round] (311.59, 93.50) -- (323.82,101.70);

\path[draw=drawColor,line width= 0.4pt,line join=round,line cap=round] (334.74,105.93) -- (345.06,107.47);

\path[draw=drawColor,line width= 0.4pt,line join=round,line cap=round] (355.72,112.05) -- (368.46,122.02);

\path[draw=drawColor,line width= 0.4pt,line join=round,line cap=round] (379.08,126.82) -- (389.48,128.76);
\definecolor{fillColor}{RGB}{255,99,71}

\path[fill=fillColor] ( 62.51, 79.59) --
	( 65.54, 74.34) --
	( 59.48, 74.34) --
	cycle;

\path[fill=fillColor] ( 84.71, 83.73) --
	( 87.74, 78.48) --
	( 81.67, 78.48) --
	cycle;

\path[fill=fillColor] (106.90, 80.42) --
	(109.93, 75.17) --
	(103.87, 75.17) --
	cycle;

\path[fill=fillColor] (129.09, 85.38) --
	(132.12, 80.13) --
	(126.06, 80.13) --
	cycle;

\path[fill=fillColor] (151.28, 82.90) --
	(154.31, 77.65) --
	(148.25, 77.65) --
	cycle;

\path[fill=fillColor] (173.47, 85.38) --
	(176.50, 80.13) --
	(170.44, 80.13) --
	cycle;

\path[fill=fillColor] (195.66, 82.07) --
	(198.69, 76.82) --
	(192.63, 76.82) --
	cycle;

\path[fill=fillColor] (217.85, 87.86) --
	(220.88, 82.61) --
	(214.82, 82.61) --
	cycle;

\path[fill=fillColor] (240.04, 84.55) --
	(243.07, 79.30) --
	(237.01, 79.30) --
	cycle;

\path[fill=fillColor] (262.23, 88.69) --
	(265.26, 83.44) --
	(259.20, 83.44) --
	cycle;

\path[fill=fillColor] (284.42,100.27) --
	(287.45, 95.02) --
	(281.39, 95.02) --
	cycle;

\path[fill=fillColor] (306.61, 93.65) --
	(309.64, 88.40) --
	(303.58, 88.40) --
	cycle;

\path[fill=fillColor] (328.80,108.54) --
	(331.83,103.29) --
	(325.77,103.29) --
	cycle;

\path[fill=fillColor] (350.99,111.85) --
	(354.02,106.60) --
	(347.96,106.60) --
	cycle;

\path[fill=fillColor] (373.18,129.22) --
	(376.22,123.97) --
	(370.15,123.97) --
	cycle;

\path[fill=fillColor] (395.38,133.36) --
	(398.41,128.11) --
	(392.35,128.11) --
	cycle;
\definecolor{drawColor}{RGB}{108,166,205}

\path[draw=drawColor,line width= 0.4pt,line join=round,line cap=round] ( 68.51, 79.62) -- ( 78.71, 80.00);

\path[draw=drawColor,line width= 0.4pt,line join=round,line cap=round] ( 90.69, 79.78) -- (100.91, 79.02);

\path[draw=drawColor,line width= 0.4pt,line join=round,line cap=round] (112.65, 76.86) -- (123.34, 73.67);

\path[draw=drawColor,line width= 0.4pt,line join=round,line cap=round] (134.64, 74.23) -- (145.73, 78.78);

\path[draw=drawColor,line width= 0.4pt,line join=round,line cap=round] (157.24, 80.39) -- (167.51, 79.24);

\path[draw=drawColor,line width= 0.4pt,line join=round,line cap=round] (179.40, 79.46) -- (189.72, 81.00);

\path[draw=drawColor,line width= 0.4pt,line join=round,line cap=round] (201.56, 82.98) -- (211.95, 84.92);

\path[draw=drawColor,line width= 0.4pt,line join=round,line cap=round] (223.83, 85.57) -- (234.06, 84.81);

\path[draw=drawColor,line width= 0.4pt,line join=round,line cap=round] (245.94, 85.46) -- (256.33, 87.40);

\path[draw=drawColor,line width= 0.4pt,line join=round,line cap=round] (268.21, 88.94) -- (278.44, 89.71);

\path[draw=drawColor,line width= 0.4pt,line join=round,line cap=round] (290.17, 91.87) -- (300.86, 95.06);

\path[draw=drawColor,line width= 0.4pt,line join=round,line cap=round] (312.61, 96.99) -- (322.81, 97.37);

\path[draw=drawColor,line width= 0.4pt,line join=round,line cap=round] (333.19,101.69) -- (346.60,114.19);

\path[draw=drawColor,line width= 0.4pt,line join=round,line cap=round] (355.30,122.45) -- (368.88,135.61);

\path[draw=drawColor,line width= 0.4pt,line join=round,line cap=round] (377.41,144.04) -- (391.15,157.86);
\definecolor{fillColor}{RGB}{108,166,205}

\path[fill=fillColor] ( 60.26, 77.15) --
	( 64.76, 77.15) --
	( 64.76, 81.65) --
	( 60.26, 81.65) --
	cycle;

\path[fill=fillColor] ( 82.46, 77.98) --
	( 86.96, 77.98) --
	( 86.96, 82.48) --
	( 82.46, 82.48) --
	cycle;

\path[fill=fillColor] (104.65, 76.32) --
	(109.15, 76.32) --
	(109.15, 80.82) --
	(104.65, 80.82) --
	cycle;

\path[fill=fillColor] (126.84, 69.70) --
	(131.34, 69.70) --
	(131.34, 74.20) --
	(126.84, 74.20) --
	cycle;

\path[fill=fillColor] (149.03, 78.80) --
	(153.53, 78.80) --
	(153.53, 83.30) --
	(149.03, 83.30) --
	cycle;

\path[fill=fillColor] (171.22, 76.32) --
	(175.72, 76.32) --
	(175.72, 80.82) --
	(171.22, 80.82) --
	cycle;

\path[fill=fillColor] (193.41, 79.63) --
	(197.91, 79.63) --
	(197.91, 84.13) --
	(193.41, 84.13) --
	cycle;

\path[fill=fillColor] (215.60, 83.77) --
	(220.10, 83.77) --
	(220.10, 88.27) --
	(215.60, 88.27) --
	cycle;

\path[fill=fillColor] (237.79, 82.11) --
	(242.29, 82.11) --
	(242.29, 86.61) --
	(237.79, 86.61) --
	cycle;

\path[fill=fillColor] (259.98, 86.25) --
	(264.48, 86.25) --
	(264.48, 90.75) --
	(259.98, 90.75) --
	cycle;

\path[fill=fillColor] (282.17, 87.90) --
	(286.67, 87.90) --
	(286.67, 92.40) --
	(282.17, 92.40) --
	cycle;

\path[fill=fillColor] (304.36, 94.52) --
	(308.86, 94.52) --
	(308.86, 99.02) --
	(304.36, 99.02) --
	cycle;

\path[fill=fillColor] (326.55, 95.35) --
	(331.05, 95.35) --
	(331.05, 99.85) --
	(326.55, 99.85) --
	cycle;

\path[fill=fillColor] (348.74,116.03) --
	(353.24,116.03) --
	(353.24,120.53) --
	(348.74,120.53) --
	cycle;

\path[fill=fillColor] (370.93,137.54) --
	(375.43,137.54) --
	(375.43,142.04) --
	(370.93,142.04) --
	cycle;

\path[fill=fillColor] (393.13,159.87) --
	(397.63,159.87) --
	(397.63,164.37) --
	(393.13,164.37) --
	cycle;
\definecolor{drawColor}{RGB}{139,137,137}

\path[draw=drawColor,line width= 0.4pt,line join=round,line cap=round] ( 68.32, 76.78) -- ( 78.90, 79.54);

\path[draw=drawColor,line width= 0.4pt,line join=round,line cap=round] ( 90.69, 80.61) -- (100.91, 79.84);

\path[draw=drawColor,line width= 0.4pt,line join=round,line cap=round] (112.83, 78.51) -- (123.15, 76.97);

\path[draw=drawColor,line width= 0.4pt,line join=round,line cap=round] (134.84, 77.80) -- (145.53, 80.99);

\path[draw=drawColor,line width= 0.4pt,line join=round,line cap=round] (157.27, 82.48) -- (167.47, 82.10);

\path[draw=drawColor,line width= 0.4pt,line join=round,line cap=round] (179.46, 82.10) -- (189.66, 82.48);

\path[draw=drawColor,line width= 0.4pt,line join=round,line cap=round] (201.21, 84.98) -- (212.30, 89.53);

\path[draw=drawColor,line width= 0.4pt,line join=round,line cap=round] (223.75, 92.91) -- (234.14, 94.84);

\path[draw=drawColor,line width= 0.4pt,line join=round,line cap=round] (245.36, 98.72) -- (256.91,104.75);

\path[draw=drawColor,line width= 0.4pt,line join=round,line cap=round] (268.09,106.21) -- (278.57,103.87);

\path[draw=drawColor,line width= 0.4pt,line join=round,line cap=round] (288.73,106.74) -- (302.30,119.89);

\path[draw=drawColor,line width= 0.4pt,line join=round,line cap=round] (312.61,123.85) -- (322.81,123.47);

\path[draw=drawColor,line width= 0.4pt,line join=round,line cap=round] (332.59,127.90) -- (347.21,145.88);

\path[draw=drawColor,line width= 0.4pt,line join=round,line cap=round] (354.85,155.14) -- (369.33,172.41);

\path[draw=drawColor,line width= 0.4pt,line join=round,line cap=round] (378.08,180.48) -- (390.48,189.26);

\path[draw=drawColor,line width= 0.4pt,line join=round,line cap=round] ( 59.33, 75.26) -- ( 65.70, 75.26);

\path[draw=drawColor,line width= 0.4pt,line join=round,line cap=round] ( 62.51, 72.08) -- ( 62.51, 78.44);

\path[draw=drawColor,line width= 0.4pt,line join=round,line cap=round] ( 81.52, 81.05) -- ( 87.89, 81.05);

\path[draw=drawColor,line width= 0.4pt,line join=round,line cap=round] ( 84.71, 77.87) -- ( 84.71, 84.24);

\path[draw=drawColor,line width= 0.4pt,line join=round,line cap=round] (103.71, 79.40) -- (110.08, 79.40);

\path[draw=drawColor,line width= 0.4pt,line join=round,line cap=round] (106.90, 76.22) -- (106.90, 82.58);

\path[draw=drawColor,line width= 0.4pt,line join=round,line cap=round] (125.90, 76.09) -- (132.27, 76.09);

\path[draw=drawColor,line width= 0.4pt,line join=round,line cap=round] (129.09, 72.91) -- (129.09, 79.27);

\path[draw=drawColor,line width= 0.4pt,line join=round,line cap=round] (148.10, 82.71) -- (154.46, 82.71);

\path[draw=drawColor,line width= 0.4pt,line join=round,line cap=round] (151.28, 79.53) -- (151.28, 85.89);

\path[draw=drawColor,line width= 0.4pt,line join=round,line cap=round] (170.29, 81.88) -- (176.65, 81.88);

\path[draw=drawColor,line width= 0.4pt,line join=round,line cap=round] (173.47, 78.70) -- (173.47, 85.06);

\path[draw=drawColor,line width= 0.4pt,line join=round,line cap=round] (192.48, 82.71) -- (198.84, 82.71);

\path[draw=drawColor,line width= 0.4pt,line join=round,line cap=round] (195.66, 79.53) -- (195.66, 85.89);

\path[draw=drawColor,line width= 0.4pt,line join=round,line cap=round] (214.67, 91.81) -- (221.03, 91.81);

\path[draw=drawColor,line width= 0.4pt,line join=round,line cap=round] (217.85, 88.63) -- (217.85, 94.99);

\path[draw=drawColor,line width= 0.4pt,line join=round,line cap=round] (236.86, 95.94) -- (243.22, 95.94);

\path[draw=drawColor,line width= 0.4pt,line join=round,line cap=round] (240.04, 92.76) -- (240.04, 99.13);

\path[draw=drawColor,line width= 0.4pt,line join=round,line cap=round] (259.05,107.52) -- (265.41,107.52);

\path[draw=drawColor,line width= 0.4pt,line join=round,line cap=round] (262.23,104.34) -- (262.23,110.71);

\path[draw=drawColor,line width= 0.4pt,line join=round,line cap=round] (281.24,102.56) -- (287.60,102.56);

\path[draw=drawColor,line width= 0.4pt,line join=round,line cap=round] (284.42, 99.38) -- (284.42,105.74);

\path[draw=drawColor,line width= 0.4pt,line join=round,line cap=round] (303.43,124.07) -- (309.79,124.07);

\path[draw=drawColor,line width= 0.4pt,line join=round,line cap=round] (306.61,120.89) -- (306.61,127.25);

\path[draw=drawColor,line width= 0.4pt,line join=round,line cap=round] (325.62,123.24) -- (331.99,123.24);

\path[draw=drawColor,line width= 0.4pt,line join=round,line cap=round] (328.80,120.06) -- (328.80,126.42);

\path[draw=drawColor,line width= 0.4pt,line join=round,line cap=round] (347.81,150.54) -- (354.18,150.54);

\path[draw=drawColor,line width= 0.4pt,line join=round,line cap=round] (350.99,147.36) -- (350.99,153.72);

\path[draw=drawColor,line width= 0.4pt,line join=round,line cap=round] (370.00,177.01) -- (376.37,177.01);

\path[draw=drawColor,line width= 0.4pt,line join=round,line cap=round] (373.18,173.83) -- (373.18,180.19);

\path[draw=drawColor,line width= 0.4pt,line join=round,line cap=round] (392.19,192.73) -- (398.56,192.73);

\path[draw=drawColor,line width= 0.4pt,line join=round,line cap=round] (395.38,189.55) -- (395.38,195.91);
\definecolor{drawColor}{RGB}{84,139,84}

\path[draw=drawColor,line width= 0.4pt,line join=round,line cap=round] ( 68.51, 73.83) -- ( 78.71, 74.21);

\path[draw=drawColor,line width= 0.4pt,line join=round,line cap=round] ( 90.69, 73.99) -- (100.91, 73.23);

\path[draw=drawColor,line width= 0.4pt,line join=round,line cap=round] (112.88, 73.23) -- (123.10, 73.99);

\path[draw=drawColor,line width= 0.4pt,line join=round,line cap=round] (135.07, 74.88) -- (145.29, 75.64);

\path[draw=drawColor,line width= 0.4pt,line join=round,line cap=round] (157.26, 76.54) -- (167.48, 77.30);

\path[draw=drawColor,line width= 0.4pt,line join=round,line cap=round] (179.46, 77.52) -- (189.66, 77.14);

\path[draw=drawColor,line width= 0.4pt,line join=round,line cap=round] (201.66, 76.92) -- (211.85, 76.92);

\path[draw=drawColor,line width= 0.4pt,line join=round,line cap=round] (223.70, 78.23) -- (234.19, 80.57);

\path[draw=drawColor,line width= 0.4pt,line join=round,line cap=round] (246.04, 81.66) -- (256.24, 81.28);

\path[draw=drawColor,line width= 0.4pt,line join=round,line cap=round] (268.21, 80.61) -- (278.44, 79.84);

\path[draw=drawColor,line width= 0.4pt,line join=round,line cap=round] (290.17, 81.11) -- (300.86, 84.30);

\path[draw=drawColor,line width= 0.4pt,line join=round,line cap=round] (312.61, 86.24) -- (322.81, 86.62);

\path[draw=drawColor,line width= 0.4pt,line join=round,line cap=round] (334.74, 87.73) -- (345.06, 89.27);

\path[draw=drawColor,line width= 0.4pt,line join=round,line cap=round] (355.98, 93.50) -- (368.20,101.70);

\path[draw=drawColor,line width= 0.4pt,line join=round,line cap=round] (379.17,105.49) -- (389.39,106.25);

\path[draw=drawColor,line width= 0.4pt,line join=round,line cap=round] ( 60.26, 71.36) -- ( 64.76, 75.86);

\path[draw=drawColor,line width= 0.4pt,line join=round,line cap=round] ( 60.26, 75.86) -- ( 64.76, 71.36);

\path[draw=drawColor,line width= 0.4pt,line join=round,line cap=round] ( 82.46, 72.19) -- ( 86.96, 76.69);

\path[draw=drawColor,line width= 0.4pt,line join=round,line cap=round] ( 82.46, 76.69) -- ( 86.96, 72.19);

\path[draw=drawColor,line width= 0.4pt,line join=round,line cap=round] (104.65, 70.53) -- (109.15, 75.03);

\path[draw=drawColor,line width= 0.4pt,line join=round,line cap=round] (104.65, 75.03) -- (109.15, 70.53);

\path[draw=drawColor,line width= 0.4pt,line join=round,line cap=round] (126.84, 72.19) -- (131.34, 76.69);

\path[draw=drawColor,line width= 0.4pt,line join=round,line cap=round] (126.84, 76.69) -- (131.34, 72.19);

\path[draw=drawColor,line width= 0.4pt,line join=round,line cap=round] (149.03, 73.84) -- (153.53, 78.34);

\path[draw=drawColor,line width= 0.4pt,line join=round,line cap=round] (149.03, 78.34) -- (153.53, 73.84);

\path[draw=drawColor,line width= 0.4pt,line join=round,line cap=round] (171.22, 75.49) -- (175.72, 79.99);

\path[draw=drawColor,line width= 0.4pt,line join=round,line cap=round] (171.22, 79.99) -- (175.72, 75.49);

\path[draw=drawColor,line width= 0.4pt,line join=round,line cap=round] (193.41, 74.67) -- (197.91, 79.17);

\path[draw=drawColor,line width= 0.4pt,line join=round,line cap=round] (193.41, 79.17) -- (197.91, 74.67);

\path[draw=drawColor,line width= 0.4pt,line join=round,line cap=round] (215.60, 74.67) -- (220.10, 79.17);

\path[draw=drawColor,line width= 0.4pt,line join=round,line cap=round] (215.60, 79.17) -- (220.10, 74.67);

\path[draw=drawColor,line width= 0.4pt,line join=round,line cap=round] (237.79, 79.63) -- (242.29, 84.13);

\path[draw=drawColor,line width= 0.4pt,line join=round,line cap=round] (237.79, 84.13) -- (242.29, 79.63);

\path[draw=drawColor,line width= 0.4pt,line join=round,line cap=round] (259.98, 78.80) -- (264.48, 83.30);

\path[draw=drawColor,line width= 0.4pt,line join=round,line cap=round] (259.98, 83.30) -- (264.48, 78.80);

\path[draw=drawColor,line width= 0.4pt,line join=round,line cap=round] (282.17, 77.15) -- (286.67, 81.65);

\path[draw=drawColor,line width= 0.4pt,line join=round,line cap=round] (282.17, 81.65) -- (286.67, 77.15);

\path[draw=drawColor,line width= 0.4pt,line join=round,line cap=round] (304.36, 83.77) -- (308.86, 88.27);

\path[draw=drawColor,line width= 0.4pt,line join=round,line cap=round] (304.36, 88.27) -- (308.86, 83.77);

\path[draw=drawColor,line width= 0.4pt,line join=round,line cap=round] (326.55, 84.59) -- (331.05, 89.09);

\path[draw=drawColor,line width= 0.4pt,line join=round,line cap=round] (326.55, 89.09) -- (331.05, 84.59);

\path[draw=drawColor,line width= 0.4pt,line join=round,line cap=round] (348.74, 87.90) -- (353.24, 92.40);

\path[draw=drawColor,line width= 0.4pt,line join=round,line cap=round] (348.74, 92.40) -- (353.24, 87.90);

\path[draw=drawColor,line width= 0.4pt,line join=round,line cap=round] (370.93,102.79) -- (375.43,107.29);

\path[draw=drawColor,line width= 0.4pt,line join=round,line cap=round] (370.93,107.29) -- (375.43,102.79);

\path[draw=drawColor,line width= 0.4pt,line join=round,line cap=round] (393.13,104.45) -- (397.63,108.95);

\path[draw=drawColor,line width= 0.4pt,line join=round,line cap=round] (393.13,108.95) -- (397.63,104.45);
\definecolor{drawColor}{RGB}{205,96,144}

\path[draw=drawColor,line width= 0.4pt,line join=round,line cap=round] ( 68.51, 76.69) -- ( 78.71, 76.31);

\path[draw=drawColor,line width= 0.4pt,line join=round,line cap=round] ( 90.69, 76.54) -- (100.91, 77.30);

\path[draw=drawColor,line width= 0.4pt,line join=round,line cap=round] (112.79, 76.65) -- (123.19, 74.71);

\path[draw=drawColor,line width= 0.4pt,line join=round,line cap=round] (134.94, 74.92) -- (145.42, 77.26);

\path[draw=drawColor,line width= 0.4pt,line join=round,line cap=round] (157.21, 77.69) -- (167.53, 76.15);

\path[draw=drawColor,line width= 0.4pt,line join=round,line cap=round] (179.02, 77.54) -- (190.11, 82.09);

\path[draw=drawColor,line width= 0.4pt,line join=round,line cap=round] (201.51, 85.67) -- (211.99, 88.02);

\path[draw=drawColor,line width= 0.4pt,line join=round,line cap=round] (223.66, 90.84) -- (234.23, 93.60);

\path[draw=drawColor,line width= 0.4pt,line join=round,line cap=round] (246.04, 95.34) -- (256.24, 95.72);

\path[draw=drawColor,line width= 0.4pt,line join=round,line cap=round] (267.55, 98.72) -- (279.10,104.75);

\path[draw=drawColor,line width= 0.4pt,line join=round,line cap=round] (290.28,108.83) -- (300.76,111.18);

\path[draw=drawColor,line width= 0.4pt,line join=round,line cap=round] (312.01,115.10) -- (323.40,120.63);

\path[draw=drawColor,line width= 0.4pt,line join=round,line cap=round] (333.44,127.05) -- (346.35,137.64);

\path[draw=drawColor,line width= 0.4pt,line join=round,line cap=round] (354.52,146.30) -- (369.66,167.19);

\path[draw=drawColor,line width= 0.4pt,line join=round,line cap=round] (378.81,174.14) -- (389.75,178.22);

\path[draw=drawColor,line width= 0.4pt,line join=round,line cap=round] ( 60.26, 74.67) rectangle ( 64.76, 79.17);

\path[draw=drawColor,line width= 0.4pt,line join=round,line cap=round] ( 60.26, 74.67) -- ( 64.76, 79.17);

\path[draw=drawColor,line width= 0.4pt,line join=round,line cap=round] ( 60.26, 79.17) -- ( 64.76, 74.67);

\path[draw=drawColor,line width= 0.4pt,line join=round,line cap=round] ( 82.46, 73.84) rectangle ( 86.96, 78.34);

\path[draw=drawColor,line width= 0.4pt,line join=round,line cap=round] ( 82.46, 73.84) -- ( 86.96, 78.34);

\path[draw=drawColor,line width= 0.4pt,line join=round,line cap=round] ( 82.46, 78.34) -- ( 86.96, 73.84);

\path[draw=drawColor,line width= 0.4pt,line join=round,line cap=round] (104.65, 75.49) rectangle (109.15, 79.99);

\path[draw=drawColor,line width= 0.4pt,line join=round,line cap=round] (104.65, 75.49) -- (109.15, 79.99);

\path[draw=drawColor,line width= 0.4pt,line join=round,line cap=round] (104.65, 79.99) -- (109.15, 75.49);

\path[draw=drawColor,line width= 0.4pt,line join=round,line cap=round] (126.84, 71.36) rectangle (131.34, 75.86);

\path[draw=drawColor,line width= 0.4pt,line join=round,line cap=round] (126.84, 71.36) -- (131.34, 75.86);

\path[draw=drawColor,line width= 0.4pt,line join=round,line cap=round] (126.84, 75.86) -- (131.34, 71.36);

\path[draw=drawColor,line width= 0.4pt,line join=round,line cap=round] (149.03, 76.32) rectangle (153.53, 80.82);

\path[draw=drawColor,line width= 0.4pt,line join=round,line cap=round] (149.03, 76.32) -- (153.53, 80.82);

\path[draw=drawColor,line width= 0.4pt,line join=round,line cap=round] (149.03, 80.82) -- (153.53, 76.32);

\path[draw=drawColor,line width= 0.4pt,line join=round,line cap=round] (171.22, 73.01) rectangle (175.72, 77.51);

\path[draw=drawColor,line width= 0.4pt,line join=round,line cap=round] (171.22, 73.01) -- (175.72, 77.51);

\path[draw=drawColor,line width= 0.4pt,line join=round,line cap=round] (171.22, 77.51) -- (175.72, 73.01);

\path[draw=drawColor,line width= 0.4pt,line join=round,line cap=round] (193.41, 82.11) rectangle (197.91, 86.61);

\path[draw=drawColor,line width= 0.4pt,line join=round,line cap=round] (193.41, 82.11) -- (197.91, 86.61);

\path[draw=drawColor,line width= 0.4pt,line join=round,line cap=round] (193.41, 86.61) -- (197.91, 82.11);

\path[draw=drawColor,line width= 0.4pt,line join=round,line cap=round] (215.60, 87.08) rectangle (220.10, 91.58);

\path[draw=drawColor,line width= 0.4pt,line join=round,line cap=round] (215.60, 87.08) -- (220.10, 91.58);

\path[draw=drawColor,line width= 0.4pt,line join=round,line cap=round] (215.60, 91.58) -- (220.10, 87.08);

\path[draw=drawColor,line width= 0.4pt,line join=round,line cap=round] (237.79, 92.87) rectangle (242.29, 97.37);

\path[draw=drawColor,line width= 0.4pt,line join=round,line cap=round] (237.79, 92.87) -- (242.29, 97.37);

\path[draw=drawColor,line width= 0.4pt,line join=round,line cap=round] (237.79, 97.37) -- (242.29, 92.87);

\path[draw=drawColor,line width= 0.4pt,line join=round,line cap=round] (259.98, 93.69) rectangle (264.48, 98.19);

\path[draw=drawColor,line width= 0.4pt,line join=round,line cap=round] (259.98, 93.69) -- (264.48, 98.19);

\path[draw=drawColor,line width= 0.4pt,line join=round,line cap=round] (259.98, 98.19) -- (264.48, 93.69);

\path[draw=drawColor,line width= 0.4pt,line join=round,line cap=round] (282.17,105.27) rectangle (286.67,109.77);

\path[draw=drawColor,line width= 0.4pt,line join=round,line cap=round] (282.17,105.27) -- (286.67,109.77);

\path[draw=drawColor,line width= 0.4pt,line join=round,line cap=round] (282.17,109.77) -- (286.67,105.27);

\path[draw=drawColor,line width= 0.4pt,line join=round,line cap=round] (304.36,110.24) rectangle (308.86,114.74);

\path[draw=drawColor,line width= 0.4pt,line join=round,line cap=round] (304.36,110.24) -- (308.86,114.74);

\path[draw=drawColor,line width= 0.4pt,line join=round,line cap=round] (304.36,114.74) -- (308.86,110.24);

\path[draw=drawColor,line width= 0.4pt,line join=round,line cap=round] (326.55,120.99) rectangle (331.05,125.49);

\path[draw=drawColor,line width= 0.4pt,line join=round,line cap=round] (326.55,120.99) -- (331.05,125.49);

\path[draw=drawColor,line width= 0.4pt,line join=round,line cap=round] (326.55,125.49) -- (331.05,120.99);

\path[draw=drawColor,line width= 0.4pt,line join=round,line cap=round] (348.74,139.19) rectangle (353.24,143.69);

\path[draw=drawColor,line width= 0.4pt,line join=round,line cap=round] (348.74,139.19) -- (353.24,143.69);

\path[draw=drawColor,line width= 0.4pt,line join=round,line cap=round] (348.74,143.69) -- (353.24,139.19);

\path[draw=drawColor,line width= 0.4pt,line join=round,line cap=round] (370.93,169.80) rectangle (375.43,174.30);

\path[draw=drawColor,line width= 0.4pt,line join=round,line cap=round] (370.93,169.80) -- (375.43,174.30);

\path[draw=drawColor,line width= 0.4pt,line join=round,line cap=round] (370.93,174.30) -- (375.43,169.80);

\path[draw=drawColor,line width= 0.4pt,line join=round,line cap=round] (393.13,178.07) rectangle (397.63,182.57);

\path[draw=drawColor,line width= 0.4pt,line join=round,line cap=round] (393.13,178.07) -- (397.63,182.57);

\path[draw=drawColor,line width= 0.4pt,line join=round,line cap=round] (393.13,182.57) -- (397.63,178.07);
\definecolor{drawColor}{RGB}{0,0,0}

\path[draw=drawColor,line width= 0.4pt,dash pattern=on 4pt off 4pt ,line join=round,line cap=round] ( 60.74, 76.09) --
	( 64.14, 76.09) --
	( 67.54, 76.09) --
	( 70.93, 76.09) --
	( 74.33, 76.09) --
	( 77.73, 76.09) --
	( 81.13, 76.09) --
	( 84.53, 76.09) --
	( 87.92, 76.09) --
	( 91.32, 76.09) --
	( 94.72, 76.09) --
	( 98.12, 76.09) --
	(101.52, 76.09) --
	(104.91, 76.09) --
	(108.31, 76.09) --
	(111.71, 76.09) --
	(115.11, 76.09) --
	(118.51, 76.09) --
	(121.90, 76.09) --
	(125.30, 76.09) --
	(128.70, 76.09) --
	(132.10, 76.09) --
	(135.50, 76.09) --
	(138.90, 76.09) --
	(142.29, 76.09) --
	(145.69, 76.09) --
	(149.09, 76.09) --
	(152.49, 76.09) --
	(155.89, 76.09) --
	(159.28, 76.09) --
	(162.68, 76.09) --
	(166.08, 76.09) --
	(169.48, 76.09) --
	(172.88, 76.09) --
	(176.27, 76.09) --
	(179.67, 76.09) --
	(183.07, 76.09) --
	(186.47, 76.09) --
	(189.87, 76.09) --
	(193.26, 76.09) --
	(196.66, 76.09) --
	(200.06, 76.09) --
	(203.46, 76.09) --
	(206.86, 76.09) --
	(210.26, 76.09) --
	(213.65, 76.09) --
	(217.05, 76.09) --
	(220.45, 76.09) --
	(223.85, 76.09) --
	(227.25, 76.09) --
	(230.64, 76.09) --
	(234.04, 76.09) --
	(237.44, 76.09) --
	(240.84, 76.09) --
	(244.24, 76.09) --
	(247.63, 76.09) --
	(251.03, 76.09) --
	(254.43, 76.09) --
	(257.83, 76.09) --
	(261.23, 76.09) --
	(264.63, 76.09) --
	(268.02, 76.09) --
	(271.42, 76.09) --
	(274.82, 76.09) --
	(278.22, 76.09) --
	(281.62, 76.09) --
	(285.01, 76.09) --
	(288.41, 76.09) --
	(291.81, 76.09) --
	(295.21, 76.09) --
	(298.61, 76.09) --
	(302.00, 76.09) --
	(305.40, 76.09) --
	(308.80, 76.09) --
	(312.20, 76.09) --
	(315.60, 76.09) --
	(318.99, 76.09) --
	(322.39, 76.09) --
	(325.79, 76.09) --
	(329.19, 76.09) --
	(332.59, 76.09) --
	(335.99, 76.09) --
	(339.38, 76.09) --
	(342.78, 76.09) --
	(346.18, 76.09) --
	(349.58, 76.09) --
	(352.98, 76.09) --
	(356.37, 76.09) --
	(359.77, 76.09) --
	(363.17, 76.09) --
	(366.57, 76.09) --
	(369.97, 76.09) --
	(373.36, 76.09) --
	(376.76, 76.09) --
	(380.16, 76.09) --
	(383.56, 76.09) --
	(386.96, 76.09) --
	(390.35, 76.09) --
	(393.75, 76.09) --
	(397.15, 76.09);
\definecolor{drawColor}{RGB}{139,0,0}
\definecolor{fillColor}{RGB}{139,0,0}

\path[draw=drawColor,line width= 0.4pt,line join=round,line cap=round,fill=fillColor] (421.11,227.88) circle (  2.25);
\definecolor{fillColor}{RGB}{255,99,71}

\path[fill=fillColor] (421.11,219.38) --
	(424.15,214.13) --
	(418.08,214.13) --
	cycle;
\definecolor{fillColor}{RGB}{108,166,205}

\path[fill=fillColor] (418.86,201.63) --
	(423.36,201.63) --
	(423.36,206.13) --
	(418.86,206.13) --
	cycle;
\definecolor{drawColor}{RGB}{139,137,137}

\path[draw=drawColor,line width= 0.4pt,line join=round,line cap=round] (417.93,191.88) -- (424.30,191.88);

\path[draw=drawColor,line width= 0.4pt,line join=round,line cap=round] (421.11,188.70) -- (421.11,195.06);
\definecolor{drawColor}{RGB}{84,139,84}

\path[draw=drawColor,line width= 0.4pt,line join=round,line cap=round] (418.86,177.63) -- (423.36,182.13);

\path[draw=drawColor,line width= 0.4pt,line join=round,line cap=round] (418.86,182.13) -- (423.36,177.63);
\definecolor{drawColor}{RGB}{205,96,144}

\path[draw=drawColor,line width= 0.4pt,line join=round,line cap=round] (418.86,165.63) rectangle (423.36,170.13);

\path[draw=drawColor,line width= 0.4pt,line join=round,line cap=round] (418.86,165.63) -- (423.36,170.13);

\path[draw=drawColor,line width= 0.4pt,line join=round,line cap=round] (418.86,170.13) -- (423.36,165.63);
\definecolor{drawColor}{RGB}{0,0,0}

\node[text=drawColor,anchor=base west,inner sep=0pt, outer sep=0pt, scale=  1.00] at (430.11,224.44) {HypoRF};

\node[text=drawColor,anchor=base west,inner sep=0pt, outer sep=0pt, scale=  1.00] at (430.11,212.44) {Binomial};

\node[text=drawColor,anchor=base west,inner sep=0pt, outer sep=0pt, scale=  1.00] at (430.11,200.44) {ME-full};

\node[text=drawColor,anchor=base west,inner sep=0pt, outer sep=0pt, scale=  1.00] at (430.11,188.44) {MMDboot};

\node[text=drawColor,anchor=base west,inner sep=0pt, outer sep=0pt, scale=  1.00] at (430.11,176.44) {MMD-full};

\node[text=drawColor,anchor=base west,inner sep=0pt, outer sep=0pt, scale=  1.00] at (430.11,164.44) {CPT-RF};
\end{scope}
\end{tikzpicture}

%% file: 3a_plot_copula_K100_Normapprox_F_with_Cai.tex
\begin{tikzpicture}[x=1pt,y=1pt]
\definecolor{fillColor}{RGB}{255,255,255}
\path[use as bounding box,fill=fillColor,fill opacity=0.00] (0,0) rectangle (505.89,289.08);
\begin{scope}
\path[clip] (  0.00,  0.00) rectangle (505.89,289.08);
\definecolor{drawColor}{RGB}{139,0,0}

\path[draw=drawColor,line width= 0.4pt,line join=round,line cap=round] ( 68.51,233.26) -- ( 80.29,233.26);

\path[draw=drawColor,line width= 0.4pt,line join=round,line cap=round] ( 92.29,233.26) -- (104.07,233.26);

\path[draw=drawColor,line width= 0.4pt,line join=round,line cap=round] (116.07,233.26) -- (127.84,233.26);

\path[draw=drawColor,line width= 0.4pt,line join=round,line cap=round] (139.84,233.26) -- (151.62,233.26);

\path[draw=drawColor,line width= 0.4pt,line join=round,line cap=round] (163.61,233.05) -- (175.40,232.64);

\path[draw=drawColor,line width= 0.4pt,line join=round,line cap=round] (187.39,232.23) -- (199.17,231.82);

\path[draw=drawColor,line width= 0.4pt,line join=round,line cap=round] (210.09,228.18) -- (224.02,218.49);

\path[draw=drawColor,line width= 0.4pt,line join=round,line cap=round] (234.89,214.24) -- (246.78,212.58);

\path[draw=drawColor,line width= 0.4pt,line join=round,line cap=round] (257.81,208.57) -- (271.41,200.05);

\path[draw=drawColor,line width= 0.4pt,line join=round,line cap=round] (280.79,192.68) -- (295.97,177.89);

\path[draw=drawColor,line width= 0.4pt,line join=round,line cap=round] (305.67,171.07) -- (318.65,164.75);

\path[draw=drawColor,line width= 0.4pt,line join=round,line cap=round] (329.21,159.07) -- (342.66,151.11);

\path[draw=drawColor,line width= 0.4pt,line join=round,line cap=round] (352.83,144.75) -- (366.59,135.65);

\path[draw=drawColor,line width= 0.4pt,line join=round,line cap=round] (377.38,130.73) -- (389.60,127.33);
\definecolor{fillColor}{RGB}{139,0,0}

\path[draw=drawColor,line width= 0.4pt,line join=round,line cap=round,fill=fillColor] ( 62.51,233.26) circle (  2.25);

\path[draw=drawColor,line width= 0.4pt,line join=round,line cap=round,fill=fillColor] ( 86.29,233.26) circle (  2.25);

\path[draw=drawColor,line width= 0.4pt,line join=round,line cap=round,fill=fillColor] (110.07,233.26) circle (  2.25);

\path[draw=drawColor,line width= 0.4pt,line join=round,line cap=round,fill=fillColor] (133.84,233.26) circle (  2.25);

\path[draw=drawColor,line width= 0.4pt,line join=round,line cap=round,fill=fillColor] (157.62,233.26) circle (  2.25);

\path[draw=drawColor,line width= 0.4pt,line join=round,line cap=round,fill=fillColor] (181.39,232.43) circle (  2.25);

\path[draw=drawColor,line width= 0.4pt,line join=round,line cap=round,fill=fillColor] (205.17,231.61) circle (  2.25);

\path[draw=drawColor,line width= 0.4pt,line join=round,line cap=round,fill=fillColor] (228.94,215.06) circle (  2.25);

\path[draw=drawColor,line width= 0.4pt,line join=round,line cap=round,fill=fillColor] (252.72,211.75) circle (  2.25);

\path[draw=drawColor,line width= 0.4pt,line join=round,line cap=round,fill=fillColor] (276.50,196.86) circle (  2.25);

\path[draw=drawColor,line width= 0.4pt,line join=round,line cap=round,fill=fillColor] (300.27,173.70) circle (  2.25);

\path[draw=drawColor,line width= 0.4pt,line join=round,line cap=round,fill=fillColor] (324.05,162.12) circle (  2.25);

\path[draw=drawColor,line width= 0.4pt,line join=round,line cap=round,fill=fillColor] (347.82,148.06) circle (  2.25);

\path[draw=drawColor,line width= 0.4pt,line join=round,line cap=round,fill=fillColor] (371.60,132.34) circle (  2.25);

\path[draw=drawColor,line width= 0.4pt,line join=round,line cap=round,fill=fillColor] (395.38,125.72) circle (  2.25);
\end{scope}
\begin{scope}
\path[clip] (  0.00,  0.00) rectangle (505.89,289.08);
\definecolor{drawColor}{RGB}{0,0,0}

\path[draw=drawColor,line width= 0.4pt,line join=round,line cap=round] ( 62.51, 61.20) -- (395.38, 61.20);

\path[draw=drawColor,line width= 0.4pt,line join=round,line cap=round] ( 62.51, 61.20) -- ( 62.51, 55.20);

\path[draw=drawColor,line width= 0.4pt,line join=round,line cap=round] (110.07, 61.20) -- (110.07, 55.20);

\path[draw=drawColor,line width= 0.4pt,line join=round,line cap=round] (157.62, 61.20) -- (157.62, 55.20);

\path[draw=drawColor,line width= 0.4pt,line join=round,line cap=round] (205.17, 61.20) -- (205.17, 55.20);

\path[draw=drawColor,line width= 0.4pt,line join=round,line cap=round] (252.72, 61.20) -- (252.72, 55.20);

\path[draw=drawColor,line width= 0.4pt,line join=round,line cap=round] (300.27, 61.20) -- (300.27, 55.20);

\path[draw=drawColor,line width= 0.4pt,line join=round,line cap=round] (347.82, 61.20) -- (347.82, 55.20);

\path[draw=drawColor,line width= 0.4pt,line join=round,line cap=round] (395.38, 61.20) -- (395.38, 55.20);

\node[text=drawColor,anchor=base,inner sep=0pt, outer sep=0pt, scale=  1.00] at ( 62.51, 39.60) {1};

\node[text=drawColor,anchor=base,inner sep=0pt, outer sep=0pt, scale=  1.00] at (110.07, 39.60) {2};

\node[text=drawColor,anchor=base,inner sep=0pt, outer sep=0pt, scale=  1.00] at (157.62, 39.60) {3};

\node[text=drawColor,anchor=base,inner sep=0pt, outer sep=0pt, scale=  1.00] at (205.17, 39.60) {4};

\node[text=drawColor,anchor=base,inner sep=0pt, outer sep=0pt, scale=  1.00] at (252.72, 39.60) {5};

\node[text=drawColor,anchor=base,inner sep=0pt, outer sep=0pt, scale=  1.00] at (300.27, 39.60) {6};

\node[text=drawColor,anchor=base,inner sep=0pt, outer sep=0pt, scale=  1.00] at (347.82, 39.60) {7};

\node[text=drawColor,anchor=base,inner sep=0pt, outer sep=0pt, scale=  1.00] at (395.38, 39.60) {8};

\path[draw=drawColor,line width= 0.4pt,line join=round,line cap=round] ( 49.20, 67.82) -- ( 49.20,233.26);

\path[draw=drawColor,line width= 0.4pt,line join=round,line cap=round] ( 49.20, 67.82) -- ( 43.20, 67.82);

\path[draw=drawColor,line width= 0.4pt,line join=round,line cap=round] ( 49.20,100.91) -- ( 43.20,100.91);

\path[draw=drawColor,line width= 0.4pt,line join=round,line cap=round] ( 49.20,134.00) -- ( 43.20,134.00);

\path[draw=drawColor,line width= 0.4pt,line join=round,line cap=round] ( 49.20,167.08) -- ( 43.20,167.08);

\path[draw=drawColor,line width= 0.4pt,line join=round,line cap=round] ( 49.20,200.17) -- ( 43.20,200.17);

\path[draw=drawColor,line width= 0.4pt,line join=round,line cap=round] ( 49.20,233.26) -- ( 43.20,233.26);

\node[text=drawColor,rotate= 90.00,anchor=base,inner sep=0pt, outer sep=0pt, scale=  1.00] at ( 34.80, 67.82) {0.0};

\node[text=drawColor,rotate= 90.00,anchor=base,inner sep=0pt, outer sep=0pt, scale=  1.00] at ( 34.80,100.91) {0.2};

\node[text=drawColor,rotate= 90.00,anchor=base,inner sep=0pt, outer sep=0pt, scale=  1.00] at ( 34.80,134.00) {0.4};

\node[text=drawColor,rotate= 90.00,anchor=base,inner sep=0pt, outer sep=0pt, scale=  1.00] at ( 34.80,167.08) {0.6};

\node[text=drawColor,rotate= 90.00,anchor=base,inner sep=0pt, outer sep=0pt, scale=  1.00] at ( 34.80,200.17) {0.8};

\node[text=drawColor,rotate= 90.00,anchor=base,inner sep=0pt, outer sep=0pt, scale=  1.00] at ( 34.80,233.26) {1.0};

\path[draw=drawColor,line width= 0.4pt,line join=round,line cap=round] ( 49.20, 61.20) --
	(408.69, 61.20) --
	(408.69,239.88) --
	( 49.20,239.88) --
	( 49.20, 61.20);
\end{scope}
\begin{scope}
\path[clip] (  0.00,  0.00) rectangle (505.89,289.08);
\definecolor{drawColor}{RGB}{0,0,0}

\node[text=drawColor,anchor=base,inner sep=0pt, outer sep=0pt, scale=  1.00] at (228.94, 15.60) {$v$};

\node[text=drawColor,rotate= 90.00,anchor=base,inner sep=0pt, outer sep=0pt, scale=  1.00] at ( 10.80,150.54) {Power};
\definecolor{drawColor}{RGB}{255,99,71}

\path[draw=drawColor,line width= 0.4pt,line join=round,line cap=round] ( 68.51,233.26) -- ( 80.29,233.26);

\path[draw=drawColor,line width= 0.4pt,line join=round,line cap=round] ( 92.29,233.05) -- (104.07,232.64);

\path[draw=drawColor,line width= 0.4pt,line join=round,line cap=round] (115.46,229.81) -- (128.45,223.48);

\path[draw=drawColor,line width= 0.4pt,line join=round,line cap=round] (138.85,217.55) -- (152.61,208.45);

\path[draw=drawColor,line width= 0.4pt,line join=round,line cap=round] (161.49,200.55) -- (177.52,181.59);

\path[draw=drawColor,line width= 0.4pt,line join=round,line cap=round] (184.84,172.10) -- (201.73,148.01);

\path[draw=drawColor,line width= 0.4pt,line join=round,line cap=round] (210.77,140.95) -- (223.34,136.14);

\path[draw=drawColor,line width= 0.4pt,line join=round,line cap=round] (234.82,132.77) -- (246.85,130.26);

\path[draw=drawColor,line width= 0.4pt,line join=round,line cap=round] (258.32,126.89) -- (270.89,122.08);

\path[draw=drawColor,line width= 0.4pt,line join=round,line cap=round] (281.58,116.75) -- (295.19,108.23);

\path[draw=drawColor,line width= 0.4pt,line join=round,line cap=round] (306.27,105.25) -- (318.05,105.66);

\path[draw=drawColor,line width= 0.4pt,line join=round,line cap=round] (329.96,104.84) -- (341.91,102.76);

\path[draw=drawColor,line width= 0.4pt,line join=round,line cap=round] (353.55, 99.94) -- (365.87, 96.08);

\path[draw=drawColor,line width= 0.4pt,line join=round,line cap=round] (377.57, 93.67) -- (389.41, 92.43);
\definecolor{fillColor}{RGB}{255,99,71}

\path[fill=fillColor] ( 62.51,236.76) --
	( 65.54,231.51) --
	( 59.48,231.51) --
	cycle;

\path[fill=fillColor] ( 86.29,236.76) --
	( 89.32,231.51) --
	( 83.26,231.51) --
	cycle;

\path[fill=fillColor] (110.07,235.93) --
	(113.10,230.69) --
	(107.04,230.69) --
	cycle;

\path[fill=fillColor] (133.84,224.35) --
	(136.87,219.10) --
	(130.81,219.10) --
	cycle;

\path[fill=fillColor] (157.62,208.64) --
	(160.65,203.39) --
	(154.59,203.39) --
	cycle;

\path[fill=fillColor] (181.39,180.51) --
	(184.42,175.26) --
	(178.36,175.26) --
	cycle;

\path[fill=fillColor] (205.17,146.59) --
	(208.20,141.35) --
	(202.14,141.35) --
	cycle;

\path[fill=fillColor] (228.94,137.49) --
	(231.98,132.25) --
	(225.91,132.25) --
	cycle;

\path[fill=fillColor] (252.72,132.53) --
	(255.75,127.28) --
	(249.69,127.28) --
	cycle;

\path[fill=fillColor] (276.50,123.43) --
	(279.53,118.18) --
	(273.47,118.18) --
	cycle;

\path[fill=fillColor] (300.27,108.54) --
	(303.30,103.29) --
	(297.24,103.29) --
	cycle;

\path[fill=fillColor] (324.05,109.37) --
	(327.08,104.12) --
	(321.02,104.12) --
	cycle;

\path[fill=fillColor] (347.82,105.23) --
	(350.85, 99.98) --
	(344.79, 99.98) --
	cycle;

\path[fill=fillColor] (371.60, 97.79) --
	(374.63, 92.54) --
	(368.57, 92.54) --
	cycle;

\path[fill=fillColor] (395.38, 95.31) --
	(398.41, 90.06) --
	(392.35, 90.06) --
	cycle;
\definecolor{drawColor}{RGB}{108,166,205}

\path[draw=drawColor,line width= 0.4pt,line join=round,line cap=round] ( 68.51,233.26) -- ( 80.29,233.26);

\path[draw=drawColor,line width= 0.4pt,line join=round,line cap=round] ( 92.29,233.26) -- (104.07,233.26);

\path[draw=drawColor,line width= 0.4pt,line join=round,line cap=round] (116.05,232.85) -- (127.86,232.02);

\path[draw=drawColor,line width= 0.4pt,line join=round,line cap=round] (139.78,230.78) -- (151.67,229.13);

\path[draw=drawColor,line width= 0.4pt,line join=round,line cap=round] (163.60,227.88) -- (175.41,227.06);

\path[draw=drawColor,line width= 0.4pt,line join=round,line cap=round] (186.86,224.17) -- (199.70,218.36);

\path[draw=drawColor,line width= 0.4pt,line join=round,line cap=round] (211.15,216.31) -- (222.96,217.13);

\path[draw=drawColor,line width= 0.4pt,line join=round,line cap=round] (234.48,215.23) -- (247.18,209.93);

\path[draw=drawColor,line width= 0.4pt,line join=round,line cap=round] (258.72,207.41) -- (270.50,207.00);

\path[draw=drawColor,line width= 0.4pt,line join=round,line cap=round] (282.22,205.00) -- (294.55,201.14);

\path[draw=drawColor,line width= 0.4pt,line join=round,line cap=round] (305.67,196.72) -- (318.65,190.39);

\path[draw=drawColor,line width= 0.4pt,line join=round,line cap=round] (329.44,185.14) -- (342.43,178.81);

\path[draw=drawColor,line width= 0.4pt,line join=round,line cap=round] (353.29,173.71) -- (366.13,167.90);

\path[draw=drawColor,line width= 0.4pt,line join=round,line cap=round] (376.99,162.80) -- (389.98,156.48);
\definecolor{fillColor}{RGB}{108,166,205}

\path[fill=fillColor] ( 60.26,231.01) --
	( 64.76,231.01) --
	( 64.76,235.51) --
	( 60.26,235.51) --
	cycle;

\path[fill=fillColor] ( 84.04,231.01) --
	( 88.54,231.01) --
	( 88.54,235.51) --
	( 84.04,235.51) --
	cycle;

\path[fill=fillColor] (107.82,231.01) --
	(112.32,231.01) --
	(112.32,235.51) --
	(107.82,235.51) --
	cycle;

\path[fill=fillColor] (131.59,229.36) --
	(136.09,229.36) --
	(136.09,233.86) --
	(131.59,233.86) --
	cycle;

\path[fill=fillColor] (155.37,226.05) --
	(159.87,226.05) --
	(159.87,230.55) --
	(155.37,230.55) --
	cycle;

\path[fill=fillColor] (179.14,224.39) --
	(183.64,224.39) --
	(183.64,228.89) --
	(179.14,228.89) --
	cycle;

\path[fill=fillColor] (202.92,213.64) --
	(207.42,213.64) --
	(207.42,218.14) --
	(202.92,218.14) --
	cycle;

\path[fill=fillColor] (226.69,215.30) --
	(231.19,215.30) --
	(231.19,219.80) --
	(226.69,219.80) --
	cycle;

\path[fill=fillColor] (250.47,205.37) --
	(254.97,205.37) --
	(254.97,209.87) --
	(250.47,209.87) --
	cycle;

\path[fill=fillColor] (274.25,204.54) --
	(278.75,204.54) --
	(278.75,209.04) --
	(274.25,209.04) --
	cycle;

\path[fill=fillColor] (298.02,197.10) --
	(302.52,197.10) --
	(302.52,201.60) --
	(298.02,201.60) --
	cycle;

\path[fill=fillColor] (321.80,185.52) --
	(326.30,185.52) --
	(326.30,190.02) --
	(321.80,190.02) --
	cycle;

\path[fill=fillColor] (345.57,173.93) --
	(350.07,173.93) --
	(350.07,178.43) --
	(345.57,178.43) --
	cycle;

\path[fill=fillColor] (369.35,163.18) --
	(373.85,163.18) --
	(373.85,167.68) --
	(369.35,167.68) --
	cycle;

\path[fill=fillColor] (393.13,151.60) --
	(397.63,151.60) --
	(397.63,156.10) --
	(393.13,156.10) --
	cycle;
\definecolor{drawColor}{RGB}{139,137,137}

\path[draw=drawColor,line width= 0.4pt,line join=round,line cap=round] ( 68.51,233.26) -- ( 80.29,233.26);

\path[draw=drawColor,line width= 0.4pt,line join=round,line cap=round] ( 92.29,233.26) -- (104.07,233.26);

\path[draw=drawColor,line width= 0.4pt,line join=round,line cap=round] (116.07,233.26) -- (127.84,233.26);

\path[draw=drawColor,line width= 0.4pt,line join=round,line cap=round] (139.84,233.26) -- (151.62,233.26);

\path[draw=drawColor,line width= 0.4pt,line join=round,line cap=round] (163.62,233.26) -- (175.39,233.26);

\path[draw=drawColor,line width= 0.4pt,line join=round,line cap=round] (187.39,233.26) -- (199.17,233.26);

\path[draw=drawColor,line width= 0.4pt,line join=round,line cap=round] (211.17,233.26) -- (222.94,233.26);

\path[draw=drawColor,line width= 0.4pt,line join=round,line cap=round] (234.95,233.26) -- (246.72,233.26);

\path[draw=drawColor,line width= 0.4pt,line join=round,line cap=round] (258.72,233.26) -- (270.50,233.26);

\path[draw=drawColor,line width= 0.4pt,line join=round,line cap=round] (282.50,233.26) -- (294.27,233.26);

\path[draw=drawColor,line width= 0.4pt,line join=round,line cap=round] (306.27,233.26) -- (318.05,233.26);

\path[draw=drawColor,line width= 0.4pt,line join=round,line cap=round] (330.05,233.26) -- (341.82,233.26);

\path[draw=drawColor,line width= 0.4pt,line join=round,line cap=round] (353.82,233.26) -- (365.60,233.26);

\path[draw=drawColor,line width= 0.4pt,line join=round,line cap=round] (377.60,233.26) -- (389.38,233.26);

\path[draw=drawColor,line width= 0.4pt,line join=round,line cap=round] ( 59.33,233.26) -- ( 65.70,233.26);

\path[draw=drawColor,line width= 0.4pt,line join=round,line cap=round] ( 62.51,230.08) -- ( 62.51,236.44);

\path[draw=drawColor,line width= 0.4pt,line join=round,line cap=round] ( 83.11,233.26) -- ( 89.47,233.26);

\path[draw=drawColor,line width= 0.4pt,line join=round,line cap=round] ( 86.29,230.08) -- ( 86.29,236.44);

\path[draw=drawColor,line width= 0.4pt,line join=round,line cap=round] (106.88,233.26) -- (113.25,233.26);

\path[draw=drawColor,line width= 0.4pt,line join=round,line cap=round] (110.07,230.08) -- (110.07,236.44);

\path[draw=drawColor,line width= 0.4pt,line join=round,line cap=round] (130.66,233.26) -- (137.02,233.26);

\path[draw=drawColor,line width= 0.4pt,line join=round,line cap=round] (133.84,230.08) -- (133.84,236.44);

\path[draw=drawColor,line width= 0.4pt,line join=round,line cap=round] (154.44,233.26) -- (160.80,233.26);

\path[draw=drawColor,line width= 0.4pt,line join=round,line cap=round] (157.62,230.08) -- (157.62,236.44);

\path[draw=drawColor,line width= 0.4pt,line join=round,line cap=round] (178.21,233.26) -- (184.58,233.26);

\path[draw=drawColor,line width= 0.4pt,line join=round,line cap=round] (181.39,230.08) -- (181.39,236.44);

\path[draw=drawColor,line width= 0.4pt,line join=round,line cap=round] (201.99,233.26) -- (208.35,233.26);

\path[draw=drawColor,line width= 0.4pt,line join=round,line cap=round] (205.17,230.08) -- (205.17,236.44);

\path[draw=drawColor,line width= 0.4pt,line join=round,line cap=round] (225.76,233.26) -- (232.13,233.26);

\path[draw=drawColor,line width= 0.4pt,line join=round,line cap=round] (228.94,230.08) -- (228.94,236.44);

\path[draw=drawColor,line width= 0.4pt,line join=round,line cap=round] (249.54,233.26) -- (255.90,233.26);

\path[draw=drawColor,line width= 0.4pt,line join=round,line cap=round] (252.72,230.08) -- (252.72,236.44);

\path[draw=drawColor,line width= 0.4pt,line join=round,line cap=round] (273.31,233.26) -- (279.68,233.26);

\path[draw=drawColor,line width= 0.4pt,line join=round,line cap=round] (276.50,230.08) -- (276.50,236.44);

\path[draw=drawColor,line width= 0.4pt,line join=round,line cap=round] (297.09,233.26) -- (303.45,233.26);

\path[draw=drawColor,line width= 0.4pt,line join=round,line cap=round] (300.27,230.08) -- (300.27,236.44);

\path[draw=drawColor,line width= 0.4pt,line join=round,line cap=round] (320.87,233.26) -- (327.23,233.26);

\path[draw=drawColor,line width= 0.4pt,line join=round,line cap=round] (324.05,230.08) -- (324.05,236.44);

\path[draw=drawColor,line width= 0.4pt,line join=round,line cap=round] (344.64,233.26) -- (351.01,233.26);

\path[draw=drawColor,line width= 0.4pt,line join=round,line cap=round] (347.82,230.08) -- (347.82,236.44);

\path[draw=drawColor,line width= 0.4pt,line join=round,line cap=round] (368.42,233.26) -- (374.78,233.26);

\path[draw=drawColor,line width= 0.4pt,line join=round,line cap=round] (371.60,230.08) -- (371.60,236.44);

\path[draw=drawColor,line width= 0.4pt,line join=round,line cap=round] (392.19,233.26) -- (398.56,233.26);

\path[draw=drawColor,line width= 0.4pt,line join=round,line cap=round] (395.38,230.08) -- (395.38,236.44);
\definecolor{drawColor}{RGB}{84,139,84}

\path[draw=drawColor,line width= 0.4pt,line join=round,line cap=round] ( 68.51,233.26) -- ( 80.29,233.26);

\path[draw=drawColor,line width= 0.4pt,line join=round,line cap=round] ( 92.29,233.26) -- (104.07,233.26);

\path[draw=drawColor,line width= 0.4pt,line join=round,line cap=round] (116.07,233.26) -- (127.84,233.26);

\path[draw=drawColor,line width= 0.4pt,line join=round,line cap=round] (139.84,233.26) -- (151.62,233.26);

\path[draw=drawColor,line width= 0.4pt,line join=round,line cap=round] (163.62,233.26) -- (175.39,233.26);

\path[draw=drawColor,line width= 0.4pt,line join=round,line cap=round] (187.39,233.26) -- (199.17,233.26);

\path[draw=drawColor,line width= 0.4pt,line join=round,line cap=round] (211.17,233.26) -- (222.94,233.26);

\path[draw=drawColor,line width= 0.4pt,line join=round,line cap=round] (234.95,233.26) -- (246.72,233.26);

\path[draw=drawColor,line width= 0.4pt,line join=round,line cap=round] (258.72,233.05) -- (270.50,232.64);

\path[draw=drawColor,line width= 0.4pt,line join=round,line cap=round] (282.50,232.43) -- (294.27,232.43);

\path[draw=drawColor,line width= 0.4pt,line join=round,line cap=round] (306.27,232.23) -- (318.05,231.82);

\path[draw=drawColor,line width= 0.4pt,line join=round,line cap=round] (329.96,230.58) -- (341.91,228.50);

\path[draw=drawColor,line width= 0.4pt,line join=round,line cap=round] (353.82,227.26) -- (365.60,226.85);

\path[draw=drawColor,line width= 0.4pt,line join=round,line cap=round] (377.57,226.02) -- (389.41,224.79);

\path[draw=drawColor,line width= 0.4pt,line join=round,line cap=round] ( 60.26,231.01) -- ( 64.76,235.51);

\path[draw=drawColor,line width= 0.4pt,line join=round,line cap=round] ( 60.26,235.51) -- ( 64.76,231.01);

\path[draw=drawColor,line width= 0.4pt,line join=round,line cap=round] ( 84.04,231.01) -- ( 88.54,235.51);

\path[draw=drawColor,line width= 0.4pt,line join=round,line cap=round] ( 84.04,235.51) -- ( 88.54,231.01);

\path[draw=drawColor,line width= 0.4pt,line join=round,line cap=round] (107.82,231.01) -- (112.32,235.51);

\path[draw=drawColor,line width= 0.4pt,line join=round,line cap=round] (107.82,235.51) -- (112.32,231.01);

\path[draw=drawColor,line width= 0.4pt,line join=round,line cap=round] (131.59,231.01) -- (136.09,235.51);

\path[draw=drawColor,line width= 0.4pt,line join=round,line cap=round] (131.59,235.51) -- (136.09,231.01);

\path[draw=drawColor,line width= 0.4pt,line join=round,line cap=round] (155.37,231.01) -- (159.87,235.51);

\path[draw=drawColor,line width= 0.4pt,line join=round,line cap=round] (155.37,235.51) -- (159.87,231.01);

\path[draw=drawColor,line width= 0.4pt,line join=round,line cap=round] (179.14,231.01) -- (183.64,235.51);

\path[draw=drawColor,line width= 0.4pt,line join=round,line cap=round] (179.14,235.51) -- (183.64,231.01);

\path[draw=drawColor,line width= 0.4pt,line join=round,line cap=round] (202.92,231.01) -- (207.42,235.51);

\path[draw=drawColor,line width= 0.4pt,line join=round,line cap=round] (202.92,235.51) -- (207.42,231.01);

\path[draw=drawColor,line width= 0.4pt,line join=round,line cap=round] (226.69,231.01) -- (231.19,235.51);

\path[draw=drawColor,line width= 0.4pt,line join=round,line cap=round] (226.69,235.51) -- (231.19,231.01);

\path[draw=drawColor,line width= 0.4pt,line join=round,line cap=round] (250.47,231.01) -- (254.97,235.51);

\path[draw=drawColor,line width= 0.4pt,line join=round,line cap=round] (250.47,235.51) -- (254.97,231.01);

\path[draw=drawColor,line width= 0.4pt,line join=round,line cap=round] (274.25,230.18) -- (278.75,234.68);

\path[draw=drawColor,line width= 0.4pt,line join=round,line cap=round] (274.25,234.68) -- (278.75,230.18);

\path[draw=drawColor,line width= 0.4pt,line join=round,line cap=round] (298.02,230.18) -- (302.52,234.68);

\path[draw=drawColor,line width= 0.4pt,line join=round,line cap=round] (298.02,234.68) -- (302.52,230.18);

\path[draw=drawColor,line width= 0.4pt,line join=round,line cap=round] (321.80,229.36) -- (326.30,233.86);

\path[draw=drawColor,line width= 0.4pt,line join=round,line cap=round] (321.80,233.86) -- (326.30,229.36);

\path[draw=drawColor,line width= 0.4pt,line join=round,line cap=round] (345.57,225.22) -- (350.07,229.72);

\path[draw=drawColor,line width= 0.4pt,line join=round,line cap=round] (345.57,229.72) -- (350.07,225.22);

\path[draw=drawColor,line width= 0.4pt,line join=round,line cap=round] (369.35,224.39) -- (373.85,228.89);

\path[draw=drawColor,line width= 0.4pt,line join=round,line cap=round] (369.35,228.89) -- (373.85,224.39);

\path[draw=drawColor,line width= 0.4pt,line join=round,line cap=round] (393.13,221.91) -- (397.63,226.41);

\path[draw=drawColor,line width= 0.4pt,line join=round,line cap=round] (393.13,226.41) -- (397.63,221.91);
\definecolor{drawColor}{RGB}{205,96,144}

\path[draw=drawColor,line width= 0.4pt,line join=round,line cap=round] ( 68.51,233.26) -- ( 80.29,233.26);

\path[draw=drawColor,line width= 0.4pt,line join=round,line cap=round] ( 92.29,233.26) -- (104.07,233.26);

\path[draw=drawColor,line width= 0.4pt,line join=round,line cap=round] (116.07,233.26) -- (127.84,233.26);

\path[draw=drawColor,line width= 0.4pt,line join=round,line cap=round] (139.84,233.26) -- (151.62,233.26);

\path[draw=drawColor,line width= 0.4pt,line join=round,line cap=round] (163.62,233.26) -- (175.39,233.26);

\path[draw=drawColor,line width= 0.4pt,line join=round,line cap=round] (187.39,233.26) -- (199.17,233.26);

\path[draw=drawColor,line width= 0.4pt,line join=round,line cap=round] (211.17,233.26) -- (222.94,233.26);

\path[draw=drawColor,line width= 0.4pt,line join=round,line cap=round] (234.94,233.05) -- (246.72,232.64);

\path[draw=drawColor,line width= 0.4pt,line join=round,line cap=round] (258.66,231.61) -- (270.55,229.95);

\path[draw=drawColor,line width= 0.4pt,line join=round,line cap=round] (282.46,228.50) -- (294.30,227.27);

\path[draw=drawColor,line width= 0.4pt,line join=round,line cap=round] (306.15,225.42) -- (318.17,222.91);

\path[draw=drawColor,line width= 0.4pt,line join=round,line cap=round] (330.03,222.10) -- (341.84,222.92);

\path[draw=drawColor,line width= 0.4pt,line join=round,line cap=round] (353.14,220.56) -- (366.28,213.70);

\path[draw=drawColor,line width= 0.4pt,line join=round,line cap=round] (377.20,208.78) -- (389.77,203.97);

\path[draw=drawColor,line width= 0.4pt,line join=round,line cap=round] ( 60.26,231.01) rectangle ( 64.76,235.51);

\path[draw=drawColor,line width= 0.4pt,line join=round,line cap=round] ( 60.26,231.01) -- ( 64.76,235.51);

\path[draw=drawColor,line width= 0.4pt,line join=round,line cap=round] ( 60.26,235.51) -- ( 64.76,231.01);

\path[draw=drawColor,line width= 0.4pt,line join=round,line cap=round] ( 84.04,231.01) rectangle ( 88.54,235.51);

\path[draw=drawColor,line width= 0.4pt,line join=round,line cap=round] ( 84.04,231.01) -- ( 88.54,235.51);

\path[draw=drawColor,line width= 0.4pt,line join=round,line cap=round] ( 84.04,235.51) -- ( 88.54,231.01);

\path[draw=drawColor,line width= 0.4pt,line join=round,line cap=round] (107.82,231.01) rectangle (112.32,235.51);

\path[draw=drawColor,line width= 0.4pt,line join=round,line cap=round] (107.82,231.01) -- (112.32,235.51);

\path[draw=drawColor,line width= 0.4pt,line join=round,line cap=round] (107.82,235.51) -- (112.32,231.01);

\path[draw=drawColor,line width= 0.4pt,line join=round,line cap=round] (131.59,231.01) rectangle (136.09,235.51);

\path[draw=drawColor,line width= 0.4pt,line join=round,line cap=round] (131.59,231.01) -- (136.09,235.51);

\path[draw=drawColor,line width= 0.4pt,line join=round,line cap=round] (131.59,235.51) -- (136.09,231.01);

\path[draw=drawColor,line width= 0.4pt,line join=round,line cap=round] (155.37,231.01) rectangle (159.87,235.51);

\path[draw=drawColor,line width= 0.4pt,line join=round,line cap=round] (155.37,231.01) -- (159.87,235.51);

\path[draw=drawColor,line width= 0.4pt,line join=round,line cap=round] (155.37,235.51) -- (159.87,231.01);

\path[draw=drawColor,line width= 0.4pt,line join=round,line cap=round] (179.14,231.01) rectangle (183.64,235.51);

\path[draw=drawColor,line width= 0.4pt,line join=round,line cap=round] (179.14,231.01) -- (183.64,235.51);

\path[draw=drawColor,line width= 0.4pt,line join=round,line cap=round] (179.14,235.51) -- (183.64,231.01);

\path[draw=drawColor,line width= 0.4pt,line join=round,line cap=round] (202.92,231.01) rectangle (207.42,235.51);

\path[draw=drawColor,line width= 0.4pt,line join=round,line cap=round] (202.92,231.01) -- (207.42,235.51);

\path[draw=drawColor,line width= 0.4pt,line join=round,line cap=round] (202.92,235.51) -- (207.42,231.01);

\path[draw=drawColor,line width= 0.4pt,line join=round,line cap=round] (226.69,231.01) rectangle (231.19,235.51);

\path[draw=drawColor,line width= 0.4pt,line join=round,line cap=round] (226.69,231.01) -- (231.19,235.51);

\path[draw=drawColor,line width= 0.4pt,line join=round,line cap=round] (226.69,235.51) -- (231.19,231.01);

\path[draw=drawColor,line width= 0.4pt,line join=round,line cap=round] (250.47,230.18) rectangle (254.97,234.68);

\path[draw=drawColor,line width= 0.4pt,line join=round,line cap=round] (250.47,230.18) -- (254.97,234.68);

\path[draw=drawColor,line width= 0.4pt,line join=round,line cap=round] (250.47,234.68) -- (254.97,230.18);

\path[draw=drawColor,line width= 0.4pt,line join=round,line cap=round] (274.25,226.88) rectangle (278.75,231.38);

\path[draw=drawColor,line width= 0.4pt,line join=round,line cap=round] (274.25,226.88) -- (278.75,231.38);

\path[draw=drawColor,line width= 0.4pt,line join=round,line cap=round] (274.25,231.38) -- (278.75,226.88);

\path[draw=drawColor,line width= 0.4pt,line join=round,line cap=round] (298.02,224.39) rectangle (302.52,228.89);

\path[draw=drawColor,line width= 0.4pt,line join=round,line cap=round] (298.02,224.39) -- (302.52,228.89);

\path[draw=drawColor,line width= 0.4pt,line join=round,line cap=round] (298.02,228.89) -- (302.52,224.39);

\path[draw=drawColor,line width= 0.4pt,line join=round,line cap=round] (321.80,219.43) rectangle (326.30,223.93);

\path[draw=drawColor,line width= 0.4pt,line join=round,line cap=round] (321.80,219.43) -- (326.30,223.93);

\path[draw=drawColor,line width= 0.4pt,line join=round,line cap=round] (321.80,223.93) -- (326.30,219.43);

\path[draw=drawColor,line width= 0.4pt,line join=round,line cap=round] (345.57,221.09) rectangle (350.07,225.59);

\path[draw=drawColor,line width= 0.4pt,line join=round,line cap=round] (345.57,221.09) -- (350.07,225.59);

\path[draw=drawColor,line width= 0.4pt,line join=round,line cap=round] (345.57,225.59) -- (350.07,221.09);

\path[draw=drawColor,line width= 0.4pt,line join=round,line cap=round] (369.35,208.68) rectangle (373.85,213.18);

\path[draw=drawColor,line width= 0.4pt,line join=round,line cap=round] (369.35,208.68) -- (373.85,213.18);

\path[draw=drawColor,line width= 0.4pt,line join=round,line cap=round] (369.35,213.18) -- (373.85,208.68);

\path[draw=drawColor,line width= 0.4pt,line join=round,line cap=round] (393.13,199.58) rectangle (397.63,204.08);

\path[draw=drawColor,line width= 0.4pt,line join=round,line cap=round] (393.13,199.58) -- (397.63,204.08);

\path[draw=drawColor,line width= 0.4pt,line join=round,line cap=round] (393.13,204.08) -- (397.63,199.58);
\definecolor{drawColor}{RGB}{0,0,0}

\path[draw=drawColor,line width= 0.4pt,dash pattern=on 4pt off 4pt ,line join=round,line cap=round] ( 61.56, 76.09) --
	( 64.94, 76.09) --
	( 68.33, 76.09) --
	( 71.71, 76.09) --
	( 75.09, 76.09) --
	( 78.47, 76.09) --
	( 81.85, 76.09) --
	( 85.23, 76.09) --
	( 88.61, 76.09) --
	( 92.00, 76.09) --
	( 95.38, 76.09) --
	( 98.76, 76.09) --
	(102.14, 76.09) --
	(105.52, 76.09) --
	(108.90, 76.09) --
	(112.29, 76.09) --
	(115.67, 76.09) --
	(119.05, 76.09) --
	(122.43, 76.09) --
	(125.81, 76.09) --
	(129.19, 76.09) --
	(132.57, 76.09) --
	(135.96, 76.09) --
	(139.34, 76.09) --
	(142.72, 76.09) --
	(146.10, 76.09) --
	(149.48, 76.09) --
	(152.86, 76.09) --
	(156.24, 76.09) --
	(159.63, 76.09) --
	(163.01, 76.09) --
	(166.39, 76.09) --
	(169.77, 76.09) --
	(173.15, 76.09) --
	(176.53, 76.09) --
	(179.91, 76.09) --
	(183.30, 76.09) --
	(186.68, 76.09) --
	(190.06, 76.09) --
	(193.44, 76.09) --
	(196.82, 76.09) --
	(200.20, 76.09) --
	(203.58, 76.09) --
	(206.97, 76.09) --
	(210.35, 76.09) --
	(213.73, 76.09) --
	(217.11, 76.09) --
	(220.49, 76.09) --
	(223.87, 76.09) --
	(227.25, 76.09) --
	(230.64, 76.09) --
	(234.02, 76.09) --
	(237.40, 76.09) --
	(240.78, 76.09) --
	(244.16, 76.09) --
	(247.54, 76.09) --
	(250.92, 76.09) --
	(254.31, 76.09) --
	(257.69, 76.09) --
	(261.07, 76.09) --
	(264.45, 76.09) --
	(267.83, 76.09) --
	(271.21, 76.09) --
	(274.59, 76.09) --
	(277.98, 76.09) --
	(281.36, 76.09) --
	(284.74, 76.09) --
	(288.12, 76.09) --
	(291.50, 76.09) --
	(294.88, 76.09) --
	(298.26, 76.09) --
	(301.65, 76.09) --
	(305.03, 76.09) --
	(308.41, 76.09) --
	(311.79, 76.09) --
	(315.17, 76.09) --
	(318.55, 76.09) --
	(321.93, 76.09) --
	(325.32, 76.09) --
	(328.70, 76.09) --
	(332.08, 76.09) --
	(335.46, 76.09) --
	(338.84, 76.09) --
	(342.22, 76.09) --
	(345.60, 76.09) --
	(348.99, 76.09) --
	(352.37, 76.09) --
	(355.75, 76.09) --
	(359.13, 76.09) --
	(362.51, 76.09) --
	(365.89, 76.09) --
	(369.28, 76.09) --
	(372.66, 76.09) --
	(376.04, 76.09) --
	(379.42, 76.09) --
	(382.80, 76.09) --
	(386.18, 76.09) --
	(389.56, 76.09) --
	(392.95, 76.09) --
	(396.33, 76.09);
\definecolor{drawColor}{RGB}{139,0,0}
\definecolor{fillColor}{RGB}{139,0,0}

\path[draw=drawColor,line width= 0.4pt,line join=round,line cap=round,fill=fillColor] (421.11,227.88) circle (  2.25);
\definecolor{fillColor}{RGB}{255,99,71}

\path[fill=fillColor] (421.11,219.38) --
	(424.15,214.13) --
	(418.08,214.13) --
	cycle;
\definecolor{fillColor}{RGB}{108,166,205}

\path[fill=fillColor] (418.86,201.63) --
	(423.36,201.63) --
	(423.36,206.13) --
	(418.86,206.13) --
	cycle;
\definecolor{drawColor}{RGB}{139,137,137}

\path[draw=drawColor,line width= 0.4pt,line join=round,line cap=round] (417.93,191.88) -- (424.30,191.88);

\path[draw=drawColor,line width= 0.4pt,line join=round,line cap=round] (421.11,188.70) -- (421.11,195.06);
\definecolor{drawColor}{RGB}{84,139,84}

\path[draw=drawColor,line width= 0.4pt,line join=round,line cap=round] (418.86,177.63) -- (423.36,182.13);

\path[draw=drawColor,line width= 0.4pt,line join=round,line cap=round] (418.86,182.13) -- (423.36,177.63);
\definecolor{drawColor}{RGB}{205,96,144}

\path[draw=drawColor,line width= 0.4pt,line join=round,line cap=round] (418.86,165.63) rectangle (423.36,170.13);

\path[draw=drawColor,line width= 0.4pt,line join=round,line cap=round] (418.86,165.63) -- (423.36,170.13);

\path[draw=drawColor,line width= 0.4pt,line join=round,line cap=round] (418.86,170.13) -- (423.36,165.63);
\definecolor{drawColor}{RGB}{0,0,0}

\node[text=drawColor,anchor=base west,inner sep=0pt, outer sep=0pt, scale=  1.00] at (430.11,224.44) {HypoRF};

\node[text=drawColor,anchor=base west,inner sep=0pt, outer sep=0pt, scale=  1.00] at (430.11,212.44) {Binomial};

\node[text=drawColor,anchor=base west,inner sep=0pt, outer sep=0pt, scale=  1.00] at (430.11,200.44) {ME-full};

\node[text=drawColor,anchor=base west,inner sep=0pt, outer sep=0pt, scale=  1.00] at (430.11,188.44) {MMDboot};

\node[text=drawColor,anchor=base west,inner sep=0pt, outer sep=0pt, scale=  1.00] at (430.11,176.44) {MMD-full};

\node[text=drawColor,anchor=base west,inner sep=0pt, outer sep=0pt, scale=  1.00] at (430.11,164.44) {CPT-RF};
\end{scope}
\end{tikzpicture}

%% file: 3b_plot_copula_K100_Normapprox_F_with_Cai.tex
\begin{tikzpicture}[x=1pt,y=1pt]
\definecolor{fillColor}{RGB}{255,255,255}
\path[use as bounding box,fill=fillColor,fill opacity=0.00] (0,0) rectangle (505.89,289.08);
\begin{scope}
\path[clip] (  0.00,  0.00) rectangle (505.89,289.08);
\definecolor{drawColor}{RGB}{139,0,0}

\path[draw=drawColor,line width= 0.4pt,line join=round,line cap=round] ( 67.68, 92.06) -- ( 81.13, 84.11);

\path[draw=drawColor,line width= 0.4pt,line join=round,line cap=round] ( 92.26, 80.43) -- (104.10, 79.19);

\path[draw=drawColor,line width= 0.4pt,line join=round,line cap=round] (116.06, 78.78) -- (127.85, 79.19);

\path[draw=drawColor,line width= 0.4pt,line join=round,line cap=round] (139.84, 79.40) -- (151.62, 79.40);

\path[draw=drawColor,line width= 0.4pt,line join=round,line cap=round] (163.60, 78.98) -- (175.41, 78.16);

\path[draw=drawColor,line width= 0.4pt,line join=round,line cap=round] (187.34, 76.92) -- (199.23, 75.26);

\path[draw=drawColor,line width= 0.4pt,line join=round,line cap=round] (211.17, 74.64) -- (222.95, 75.05);

\path[draw=drawColor,line width= 0.4pt,line join=round,line cap=round] (234.93, 74.85) -- (246.74, 74.02);

\path[draw=drawColor,line width= 0.4pt,line join=round,line cap=round] (258.55, 75.03) -- (270.67, 77.98);

\path[draw=drawColor,line width= 0.4pt,line join=round,line cap=round] (282.50, 79.40) -- (294.27, 79.40);

\path[draw=drawColor,line width= 0.4pt,line join=round,line cap=round] (306.26, 79.82) -- (318.06, 80.64);

\path[draw=drawColor,line width= 0.4pt,line join=round,line cap=round] (330.02, 80.43) -- (341.86, 79.19);

\path[draw=drawColor,line width= 0.4pt,line join=round,line cap=round] (353.65, 77.15) -- (365.77, 74.20);

\path[draw=drawColor,line width= 0.4pt,line join=round,line cap=round] (377.51, 73.81) -- (389.46, 75.89);
\definecolor{fillColor}{RGB}{139,0,0}

\path[draw=drawColor,line width= 0.4pt,line join=round,line cap=round,fill=fillColor] ( 62.51, 95.12) circle (  2.25);

\path[draw=drawColor,line width= 0.4pt,line join=round,line cap=round,fill=fillColor] ( 86.29, 81.05) circle (  2.25);

\path[draw=drawColor,line width= 0.4pt,line join=round,line cap=round,fill=fillColor] (110.07, 78.57) circle (  2.25);

\path[draw=drawColor,line width= 0.4pt,line join=round,line cap=round,fill=fillColor] (133.84, 79.40) circle (  2.25);

\path[draw=drawColor,line width= 0.4pt,line join=round,line cap=round,fill=fillColor] (157.62, 79.40) circle (  2.25);

\path[draw=drawColor,line width= 0.4pt,line join=round,line cap=round,fill=fillColor] (181.39, 77.74) circle (  2.25);

\path[draw=drawColor,line width= 0.4pt,line join=round,line cap=round,fill=fillColor] (205.17, 74.44) circle (  2.25);

\path[draw=drawColor,line width= 0.4pt,line join=round,line cap=round,fill=fillColor] (228.94, 75.26) circle (  2.25);

\path[draw=drawColor,line width= 0.4pt,line join=round,line cap=round,fill=fillColor] (252.72, 73.61) circle (  2.25);

\path[draw=drawColor,line width= 0.4pt,line join=round,line cap=round,fill=fillColor] (276.50, 79.40) circle (  2.25);

\path[draw=drawColor,line width= 0.4pt,line join=round,line cap=round,fill=fillColor] (300.27, 79.40) circle (  2.25);

\path[draw=drawColor,line width= 0.4pt,line join=round,line cap=round,fill=fillColor] (324.05, 81.05) circle (  2.25);

\path[draw=drawColor,line width= 0.4pt,line join=round,line cap=round,fill=fillColor] (347.82, 78.57) circle (  2.25);

\path[draw=drawColor,line width= 0.4pt,line join=round,line cap=round,fill=fillColor] (371.60, 72.78) circle (  2.25);

\path[draw=drawColor,line width= 0.4pt,line join=round,line cap=round,fill=fillColor] (395.38, 76.92) circle (  2.25);
\end{scope}
\begin{scope}
\path[clip] (  0.00,  0.00) rectangle (505.89,289.08);
\definecolor{drawColor}{RGB}{0,0,0}

\path[draw=drawColor,line width= 0.4pt,line join=round,line cap=round] ( 62.51, 61.20) -- (395.38, 61.20);

\path[draw=drawColor,line width= 0.4pt,line join=round,line cap=round] ( 62.51, 61.20) -- ( 62.51, 55.20);

\path[draw=drawColor,line width= 0.4pt,line join=round,line cap=round] (110.07, 61.20) -- (110.07, 55.20);

\path[draw=drawColor,line width= 0.4pt,line join=round,line cap=round] (157.62, 61.20) -- (157.62, 55.20);

\path[draw=drawColor,line width= 0.4pt,line join=round,line cap=round] (205.17, 61.20) -- (205.17, 55.20);

\path[draw=drawColor,line width= 0.4pt,line join=round,line cap=round] (252.72, 61.20) -- (252.72, 55.20);

\path[draw=drawColor,line width= 0.4pt,line join=round,line cap=round] (300.27, 61.20) -- (300.27, 55.20);

\path[draw=drawColor,line width= 0.4pt,line join=round,line cap=round] (347.82, 61.20) -- (347.82, 55.20);

\path[draw=drawColor,line width= 0.4pt,line join=round,line cap=round] (395.38, 61.20) -- (395.38, 55.20);

\node[text=drawColor,anchor=base,inner sep=0pt, outer sep=0pt, scale=  1.00] at ( 62.51, 39.60) {1};

\node[text=drawColor,anchor=base,inner sep=0pt, outer sep=0pt, scale=  1.00] at (110.07, 39.60) {2};

\node[text=drawColor,anchor=base,inner sep=0pt, outer sep=0pt, scale=  1.00] at (157.62, 39.60) {3};

\node[text=drawColor,anchor=base,inner sep=0pt, outer sep=0pt, scale=  1.00] at (205.17, 39.60) {4};

\node[text=drawColor,anchor=base,inner sep=0pt, outer sep=0pt, scale=  1.00] at (252.72, 39.60) {5};

\node[text=drawColor,anchor=base,inner sep=0pt, outer sep=0pt, scale=  1.00] at (300.27, 39.60) {6};

\node[text=drawColor,anchor=base,inner sep=0pt, outer sep=0pt, scale=  1.00] at (347.82, 39.60) {7};

\node[text=drawColor,anchor=base,inner sep=0pt, outer sep=0pt, scale=  1.00] at (395.38, 39.60) {8};

\path[draw=drawColor,line width= 0.4pt,line join=round,line cap=round] ( 49.20, 67.82) -- ( 49.20,233.26);

\path[draw=drawColor,line width= 0.4pt,line join=round,line cap=round] ( 49.20, 67.82) -- ( 43.20, 67.82);

\path[draw=drawColor,line width= 0.4pt,line join=round,line cap=round] ( 49.20,100.91) -- ( 43.20,100.91);

\path[draw=drawColor,line width= 0.4pt,line join=round,line cap=round] ( 49.20,134.00) -- ( 43.20,134.00);

\path[draw=drawColor,line width= 0.4pt,line join=round,line cap=round] ( 49.20,167.08) -- ( 43.20,167.08);

\path[draw=drawColor,line width= 0.4pt,line join=round,line cap=round] ( 49.20,200.17) -- ( 43.20,200.17);

\path[draw=drawColor,line width= 0.4pt,line join=round,line cap=round] ( 49.20,233.26) -- ( 43.20,233.26);

\node[text=drawColor,rotate= 90.00,anchor=base,inner sep=0pt, outer sep=0pt, scale=  1.00] at ( 34.80, 67.82) {0.0};

\node[text=drawColor,rotate= 90.00,anchor=base,inner sep=0pt, outer sep=0pt, scale=  1.00] at ( 34.80,100.91) {0.2};

\node[text=drawColor,rotate= 90.00,anchor=base,inner sep=0pt, outer sep=0pt, scale=  1.00] at ( 34.80,134.00) {0.4};

\node[text=drawColor,rotate= 90.00,anchor=base,inner sep=0pt, outer sep=0pt, scale=  1.00] at ( 34.80,167.08) {0.6};

\node[text=drawColor,rotate= 90.00,anchor=base,inner sep=0pt, outer sep=0pt, scale=  1.00] at ( 34.80,200.17) {0.8};

\node[text=drawColor,rotate= 90.00,anchor=base,inner sep=0pt, outer sep=0pt, scale=  1.00] at ( 34.80,233.26) {1.0};

\path[draw=drawColor,line width= 0.4pt,line join=round,line cap=round] ( 49.20, 61.20) --
	(408.69, 61.20) --
	(408.69,239.88) --
	( 49.20,239.88) --
	( 49.20, 61.20);
\end{scope}
\begin{scope}
\path[clip] (  0.00,  0.00) rectangle (505.89,289.08);
\definecolor{drawColor}{RGB}{0,0,0}

\node[text=drawColor,anchor=base,inner sep=0pt, outer sep=0pt, scale=  1.00] at (228.94, 15.60) {$v$};

\node[text=drawColor,rotate= 90.00,anchor=base,inner sep=0pt, outer sep=0pt, scale=  1.00] at ( 10.80,150.54) {Power};
\definecolor{drawColor}{RGB}{255,99,71}

\path[draw=drawColor,line width= 0.4pt,line join=round,line cap=round] ( 68.29, 73.56) -- ( 80.51, 76.96);

\path[draw=drawColor,line width= 0.4pt,line join=round,line cap=round] ( 92.28, 78.99) -- (104.08, 79.81);

\path[draw=drawColor,line width= 0.4pt,line join=round,line cap=round] (116.01, 79.40) -- (127.90, 77.74);

\path[draw=drawColor,line width= 0.4pt,line join=round,line cap=round] (139.84, 76.71) -- (151.62, 76.30);

\path[draw=drawColor,line width= 0.4pt,line join=round,line cap=round] (163.45, 77.51) -- (175.56, 80.46);

\path[draw=drawColor,line width= 0.4pt,line join=round,line cap=round] (187.34, 81.05) -- (199.23, 79.40);

\path[draw=drawColor,line width= 0.4pt,line join=round,line cap=round] (211.14, 77.95) -- (222.98, 76.71);

\path[draw=drawColor,line width= 0.4pt,line join=round,line cap=round] (234.94, 76.30) -- (246.72, 76.71);

\path[draw=drawColor,line width= 0.4pt,line join=round,line cap=round] (258.66, 77.74) -- (270.55, 79.40);

\path[draw=drawColor,line width= 0.4pt,line join=round,line cap=round] (282.46, 79.60) -- (294.30, 78.37);

\path[draw=drawColor,line width= 0.4pt,line join=round,line cap=round] (306.27, 77.54) -- (318.05, 77.13);

\path[draw=drawColor,line width= 0.4pt,line join=round,line cap=round] (330.04, 77.13) -- (341.83, 77.54);

\path[draw=drawColor,line width= 0.4pt,line join=round,line cap=round] (353.77, 76.92) -- (365.66, 75.26);

\path[draw=drawColor,line width= 0.4pt,line join=round,line cap=round] (377.43, 75.86) -- (389.55, 78.81);
\definecolor{fillColor}{RGB}{255,99,71}

\path[fill=fillColor] ( 62.51, 75.45) --
	( 65.54, 70.20) --
	( 59.48, 70.20) --
	cycle;

\path[fill=fillColor] ( 86.29, 82.07) --
	( 89.32, 76.82) --
	( 83.26, 76.82) --
	cycle;

\path[fill=fillColor] (110.07, 83.73) --
	(113.10, 78.48) --
	(107.04, 78.48) --
	cycle;

\path[fill=fillColor] (133.84, 80.42) --
	(136.87, 75.17) --
	(130.81, 75.17) --
	cycle;

\path[fill=fillColor] (157.62, 79.59) --
	(160.65, 74.34) --
	(154.59, 74.34) --
	cycle;

\path[fill=fillColor] (181.39, 85.38) --
	(184.42, 80.13) --
	(178.36, 80.13) --
	cycle;

\path[fill=fillColor] (205.17, 82.07) --
	(208.20, 76.82) --
	(202.14, 76.82) --
	cycle;

\path[fill=fillColor] (228.94, 79.59) --
	(231.98, 74.34) --
	(225.91, 74.34) --
	cycle;

\path[fill=fillColor] (252.72, 80.42) --
	(255.75, 75.17) --
	(249.69, 75.17) --
	cycle;

\path[fill=fillColor] (276.50, 83.73) --
	(279.53, 78.48) --
	(273.47, 78.48) --
	cycle;

\path[fill=fillColor] (300.27, 81.24) --
	(303.30, 75.99) --
	(297.24, 75.99) --
	cycle;

\path[fill=fillColor] (324.05, 80.42) --
	(327.08, 75.17) --
	(321.02, 75.17) --
	cycle;

\path[fill=fillColor] (347.82, 81.24) --
	(350.85, 75.99) --
	(344.79, 75.99) --
	cycle;

\path[fill=fillColor] (371.60, 77.93) --
	(374.63, 72.69) --
	(368.57, 72.69) --
	cycle;

\path[fill=fillColor] (395.38, 83.73) --
	(398.41, 78.48) --
	(392.35, 78.48) --
	cycle;
\definecolor{drawColor}{RGB}{108,166,205}

\path[draw=drawColor,line width= 0.4pt,line join=round,line cap=round] ( 68.39, 77.32) -- ( 80.42, 79.83);

\path[draw=drawColor,line width= 0.4pt,line join=round,line cap=round] ( 92.12, 79.63) -- (104.24, 76.68);

\path[draw=drawColor,line width= 0.4pt,line join=round,line cap=round] (116.05, 74.85) -- (127.86, 74.02);

\path[draw=drawColor,line width= 0.4pt,line join=round,line cap=round] (139.75, 74.64) -- (151.71, 76.72);

\path[draw=drawColor,line width= 0.4pt,line join=round,line cap=round] (163.56, 78.57) -- (175.45, 80.23);

\path[draw=drawColor,line width= 0.4pt,line join=round,line cap=round] (187.34, 81.88) -- (199.23, 83.54);

\path[draw=drawColor,line width= 0.4pt,line join=round,line cap=round] (210.90, 82.57) -- (223.22, 78.71);

\path[draw=drawColor,line width= 0.4pt,line join=round,line cap=round] (234.94, 76.71) -- (246.72, 76.30);

\path[draw=drawColor,line width= 0.4pt,line join=round,line cap=round] (258.69, 76.71) -- (270.53, 77.95);

\path[draw=drawColor,line width= 0.4pt,line join=round,line cap=round] (282.44, 77.74) -- (294.33, 76.09);

\path[draw=drawColor,line width= 0.4pt,line join=round,line cap=round] (306.26, 74.85) -- (318.06, 74.02);

\path[draw=drawColor,line width= 0.4pt,line join=round,line cap=round] (330.02, 74.23) -- (341.86, 75.47);

\path[draw=drawColor,line width= 0.4pt,line join=round,line cap=round] (353.74, 77.12) -- (365.69, 79.20);

\path[draw=drawColor,line width= 0.4pt,line join=round,line cap=round] (377.57, 79.60) -- (389.41, 78.37);
\definecolor{fillColor}{RGB}{108,166,205}

\path[fill=fillColor] ( 60.26, 73.84) --
	( 64.76, 73.84) --
	( 64.76, 78.34) --
	( 60.26, 78.34) --
	cycle;

\path[fill=fillColor] ( 84.04, 78.80) --
	( 88.54, 78.80) --
	( 88.54, 83.30) --
	( 84.04, 83.30) --
	cycle;

\path[fill=fillColor] (107.82, 73.01) --
	(112.32, 73.01) --
	(112.32, 77.51) --
	(107.82, 77.51) --
	cycle;

\path[fill=fillColor] (131.59, 71.36) --
	(136.09, 71.36) --
	(136.09, 75.86) --
	(131.59, 75.86) --
	cycle;

\path[fill=fillColor] (155.37, 75.49) --
	(159.87, 75.49) --
	(159.87, 79.99) --
	(155.37, 79.99) --
	cycle;

\path[fill=fillColor] (179.14, 78.80) --
	(183.64, 78.80) --
	(183.64, 83.30) --
	(179.14, 83.30) --
	cycle;

\path[fill=fillColor] (202.92, 82.11) --
	(207.42, 82.11) --
	(207.42, 86.61) --
	(202.92, 86.61) --
	cycle;

\path[fill=fillColor] (226.69, 74.67) --
	(231.19, 74.67) --
	(231.19, 79.17) --
	(226.69, 79.17) --
	cycle;

\path[fill=fillColor] (250.47, 73.84) --
	(254.97, 73.84) --
	(254.97, 78.34) --
	(250.47, 78.34) --
	cycle;

\path[fill=fillColor] (274.25, 76.32) --
	(278.75, 76.32) --
	(278.75, 80.82) --
	(274.25, 80.82) --
	cycle;

\path[fill=fillColor] (298.02, 73.01) --
	(302.52, 73.01) --
	(302.52, 77.51) --
	(298.02, 77.51) --
	cycle;

\path[fill=fillColor] (321.80, 71.36) --
	(326.30, 71.36) --
	(326.30, 75.86) --
	(321.80, 75.86) --
	cycle;

\path[fill=fillColor] (345.57, 73.84) --
	(350.07, 73.84) --
	(350.07, 78.34) --
	(345.57, 78.34) --
	cycle;

\path[fill=fillColor] (369.35, 77.98) --
	(373.85, 77.98) --
	(373.85, 82.48) --
	(369.35, 82.48) --
	cycle;

\path[fill=fillColor] (393.13, 75.49) --
	(397.63, 75.49) --
	(397.63, 79.99) --
	(393.13, 79.99) --
	cycle;
\definecolor{drawColor}{RGB}{139,137,137}

\path[draw=drawColor,line width= 0.4pt,line join=round,line cap=round] ( 67.12,164.89) -- ( 81.68,152.73);

\path[draw=drawColor,line width= 0.4pt,line join=round,line cap=round] ( 89.79,144.01) -- (106.56,120.67);

\path[draw=drawColor,line width= 0.4pt,line join=round,line cap=round] (116.01,114.97) -- (127.90,113.31);

\path[draw=drawColor,line width= 0.4pt,line join=round,line cap=round] (139.38,110.18) -- (152.08,104.87);

\path[draw=drawColor,line width= 0.4pt,line join=round,line cap=round] (163.53,101.53) -- (175.48, 99.45);

\path[draw=drawColor,line width= 0.4pt,line join=round,line cap=round] (187.34, 97.60) -- (199.23, 95.94);

\path[draw=drawColor,line width= 0.4pt,line join=round,line cap=round] (211.04, 93.89) -- (223.07, 91.38);

\path[draw=drawColor,line width= 0.4pt,line join=round,line cap=round] (234.94, 90.36) -- (246.72, 90.77);

\path[draw=drawColor,line width= 0.4pt,line join=round,line cap=round] (258.45, 89.19) -- (270.77, 85.33);

\path[draw=drawColor,line width= 0.4pt,line join=round,line cap=round] (282.46, 82.91) -- (294.30, 81.68);

\path[draw=drawColor,line width= 0.4pt,line join=round,line cap=round] (306.15, 79.83) -- (318.17, 77.32);

\path[draw=drawColor,line width= 0.4pt,line join=round,line cap=round] (329.77, 77.88) -- (342.10, 81.74);

\path[draw=drawColor,line width= 0.4pt,line join=round,line cap=round] (353.74, 82.51) -- (365.69, 80.43);

\path[draw=drawColor,line width= 0.4pt,line join=round,line cap=round] (377.43, 80.82) -- (389.55, 83.77);

\path[draw=drawColor,line width= 0.4pt,line join=round,line cap=round] ( 59.33,168.74) -- ( 65.70,168.74);

\path[draw=drawColor,line width= 0.4pt,line join=round,line cap=round] ( 62.51,165.56) -- ( 62.51,171.92);

\path[draw=drawColor,line width= 0.4pt,line join=round,line cap=round] ( 83.11,148.89) -- ( 89.47,148.89);

\path[draw=drawColor,line width= 0.4pt,line join=round,line cap=round] ( 86.29,145.70) -- ( 86.29,152.07);

\path[draw=drawColor,line width= 0.4pt,line join=round,line cap=round] (106.88,115.80) -- (113.25,115.80);

\path[draw=drawColor,line width= 0.4pt,line join=round,line cap=round] (110.07,112.61) -- (110.07,118.98);

\path[draw=drawColor,line width= 0.4pt,line join=round,line cap=round] (130.66,112.49) -- (137.02,112.49);

\path[draw=drawColor,line width= 0.4pt,line join=round,line cap=round] (133.84,109.31) -- (133.84,115.67);

\path[draw=drawColor,line width= 0.4pt,line join=round,line cap=round] (154.44,102.56) -- (160.80,102.56);

\path[draw=drawColor,line width= 0.4pt,line join=round,line cap=round] (157.62, 99.38) -- (157.62,105.74);

\path[draw=drawColor,line width= 0.4pt,line join=round,line cap=round] (178.21, 98.43) -- (184.58, 98.43);

\path[draw=drawColor,line width= 0.4pt,line join=round,line cap=round] (181.39, 95.24) -- (181.39,101.61);

\path[draw=drawColor,line width= 0.4pt,line join=round,line cap=round] (201.99, 95.12) -- (208.35, 95.12);

\path[draw=drawColor,line width= 0.4pt,line join=round,line cap=round] (205.17, 91.93) -- (205.17, 98.30);

\path[draw=drawColor,line width= 0.4pt,line join=round,line cap=round] (225.76, 90.15) -- (232.13, 90.15);

\path[draw=drawColor,line width= 0.4pt,line join=round,line cap=round] (228.94, 86.97) -- (228.94, 93.33);

\path[draw=drawColor,line width= 0.4pt,line join=round,line cap=round] (249.54, 90.98) -- (255.90, 90.98);

\path[draw=drawColor,line width= 0.4pt,line join=round,line cap=round] (252.72, 87.80) -- (252.72, 94.16);

\path[draw=drawColor,line width= 0.4pt,line join=round,line cap=round] (273.31, 83.53) -- (279.68, 83.53);

\path[draw=drawColor,line width= 0.4pt,line join=round,line cap=round] (276.50, 80.35) -- (276.50, 86.72);

\path[draw=drawColor,line width= 0.4pt,line join=round,line cap=round] (297.09, 81.05) -- (303.45, 81.05);

\path[draw=drawColor,line width= 0.4pt,line join=round,line cap=round] (300.27, 77.87) -- (300.27, 84.24);

\path[draw=drawColor,line width= 0.4pt,line join=round,line cap=round] (320.87, 76.09) -- (327.23, 76.09);

\path[draw=drawColor,line width= 0.4pt,line join=round,line cap=round] (324.05, 72.91) -- (324.05, 79.27);

\path[draw=drawColor,line width= 0.4pt,line join=round,line cap=round] (344.64, 83.53) -- (351.01, 83.53);

\path[draw=drawColor,line width= 0.4pt,line join=round,line cap=round] (347.82, 80.35) -- (347.82, 86.72);

\path[draw=drawColor,line width= 0.4pt,line join=round,line cap=round] (368.42, 79.40) -- (374.78, 79.40);

\path[draw=drawColor,line width= 0.4pt,line join=round,line cap=round] (371.60, 76.22) -- (371.60, 82.58);

\path[draw=drawColor,line width= 0.4pt,line join=round,line cap=round] (392.19, 85.19) -- (398.56, 85.19);

\path[draw=drawColor,line width= 0.4pt,line join=round,line cap=round] (395.38, 82.01) -- (395.38, 88.37);
\definecolor{drawColor}{RGB}{84,139,84}

\path[draw=drawColor,line width= 0.4pt,line join=round,line cap=round] ( 68.51, 86.84) -- ( 80.29, 86.84);

\path[draw=drawColor,line width= 0.4pt,line join=round,line cap=round] ( 92.20, 85.82) -- (104.15, 83.74);

\path[draw=drawColor,line width= 0.4pt,line join=round,line cap=round] (116.03, 82.08) -- (127.87, 80.85);

\path[draw=drawColor,line width= 0.4pt,line join=round,line cap=round] (139.84, 80.23) -- (151.62, 80.23);

\path[draw=drawColor,line width= 0.4pt,line join=round,line cap=round] (163.56, 79.40) -- (175.45, 77.74);

\path[draw=drawColor,line width= 0.4pt,line join=round,line cap=round] (187.39, 76.71) -- (199.17, 76.30);

\path[draw=drawColor,line width= 0.4pt,line join=round,line cap=round] (211.17, 76.30) -- (222.95, 76.71);

\path[draw=drawColor,line width= 0.4pt,line join=round,line cap=round] (234.94, 76.71) -- (246.72, 76.30);

\path[draw=drawColor,line width= 0.4pt,line join=round,line cap=round] (258.45, 77.88) -- (270.77, 81.74);

\path[draw=drawColor,line width= 0.4pt,line join=round,line cap=round] (282.16, 81.56) -- (294.61, 77.23);

\path[draw=drawColor,line width= 0.4pt,line join=round,line cap=round] (306.27, 75.26) -- (318.05, 75.26);

\path[draw=drawColor,line width= 0.4pt,line join=round,line cap=round] (330.04, 75.47) -- (341.83, 75.88);

\path[draw=drawColor,line width= 0.4pt,line join=round,line cap=round] (353.79, 76.71) -- (365.63, 77.95);

\path[draw=drawColor,line width= 0.4pt,line join=round,line cap=round] (377.57, 77.95) -- (389.41, 76.71);

\path[draw=drawColor,line width= 0.4pt,line join=round,line cap=round] ( 60.26, 84.59) -- ( 64.76, 89.09);

\path[draw=drawColor,line width= 0.4pt,line join=round,line cap=round] ( 60.26, 89.09) -- ( 64.76, 84.59);

\path[draw=drawColor,line width= 0.4pt,line join=round,line cap=round] ( 84.04, 84.59) -- ( 88.54, 89.09);

\path[draw=drawColor,line width= 0.4pt,line join=round,line cap=round] ( 84.04, 89.09) -- ( 88.54, 84.59);

\path[draw=drawColor,line width= 0.4pt,line join=round,line cap=round] (107.82, 80.46) -- (112.32, 84.96);

\path[draw=drawColor,line width= 0.4pt,line join=round,line cap=round] (107.82, 84.96) -- (112.32, 80.46);

\path[draw=drawColor,line width= 0.4pt,line join=round,line cap=round] (131.59, 77.98) -- (136.09, 82.48);

\path[draw=drawColor,line width= 0.4pt,line join=round,line cap=round] (131.59, 82.48) -- (136.09, 77.98);

\path[draw=drawColor,line width= 0.4pt,line join=round,line cap=round] (155.37, 77.98) -- (159.87, 82.48);

\path[draw=drawColor,line width= 0.4pt,line join=round,line cap=round] (155.37, 82.48) -- (159.87, 77.98);

\path[draw=drawColor,line width= 0.4pt,line join=round,line cap=round] (179.14, 74.67) -- (183.64, 79.17);

\path[draw=drawColor,line width= 0.4pt,line join=round,line cap=round] (179.14, 79.17) -- (183.64, 74.67);

\path[draw=drawColor,line width= 0.4pt,line join=round,line cap=round] (202.92, 73.84) -- (207.42, 78.34);

\path[draw=drawColor,line width= 0.4pt,line join=round,line cap=round] (202.92, 78.34) -- (207.42, 73.84);

\path[draw=drawColor,line width= 0.4pt,line join=round,line cap=round] (226.69, 74.67) -- (231.19, 79.17);

\path[draw=drawColor,line width= 0.4pt,line join=round,line cap=round] (226.69, 79.17) -- (231.19, 74.67);

\path[draw=drawColor,line width= 0.4pt,line join=round,line cap=round] (250.47, 73.84) -- (254.97, 78.34);

\path[draw=drawColor,line width= 0.4pt,line join=round,line cap=round] (250.47, 78.34) -- (254.97, 73.84);

\path[draw=drawColor,line width= 0.4pt,line join=round,line cap=round] (274.25, 81.28) -- (278.75, 85.78);

\path[draw=drawColor,line width= 0.4pt,line join=round,line cap=round] (274.25, 85.78) -- (278.75, 81.28);

\path[draw=drawColor,line width= 0.4pt,line join=round,line cap=round] (298.02, 73.01) -- (302.52, 77.51);

\path[draw=drawColor,line width= 0.4pt,line join=round,line cap=round] (298.02, 77.51) -- (302.52, 73.01);

\path[draw=drawColor,line width= 0.4pt,line join=round,line cap=round] (321.80, 73.01) -- (326.30, 77.51);

\path[draw=drawColor,line width= 0.4pt,line join=round,line cap=round] (321.80, 77.51) -- (326.30, 73.01);

\path[draw=drawColor,line width= 0.4pt,line join=round,line cap=round] (345.57, 73.84) -- (350.07, 78.34);

\path[draw=drawColor,line width= 0.4pt,line join=round,line cap=round] (345.57, 78.34) -- (350.07, 73.84);

\path[draw=drawColor,line width= 0.4pt,line join=round,line cap=round] (369.35, 76.32) -- (373.85, 80.82);

\path[draw=drawColor,line width= 0.4pt,line join=round,line cap=round] (369.35, 80.82) -- (373.85, 76.32);

\path[draw=drawColor,line width= 0.4pt,line join=round,line cap=round] (393.13, 73.84) -- (397.63, 78.34);

\path[draw=drawColor,line width= 0.4pt,line join=round,line cap=round] (393.13, 78.34) -- (397.63, 73.84);
\definecolor{drawColor}{RGB}{205,96,144}

\path[draw=drawColor,line width= 0.4pt,line join=round,line cap=round] ( 67.76, 97.99) -- ( 81.05, 90.59);

\path[draw=drawColor,line width= 0.4pt,line join=round,line cap=round] ( 92.26, 87.05) -- (104.10, 85.81);

\path[draw=drawColor,line width= 0.4pt,line join=round,line cap=round] (115.90, 83.77) -- (128.01, 80.82);

\path[draw=drawColor,line width= 0.4pt,line join=round,line cap=round] (139.84, 79.61) -- (151.62, 80.02);

\path[draw=drawColor,line width= 0.4pt,line join=round,line cap=round] (163.61, 80.02) -- (175.40, 79.61);

\path[draw=drawColor,line width= 0.4pt,line join=round,line cap=round] (187.34, 78.57) -- (199.23, 76.92);

\path[draw=drawColor,line width= 0.4pt,line join=round,line cap=round] (211.17, 76.09) -- (222.94, 76.09);

\path[draw=drawColor,line width= 0.4pt,line join=round,line cap=round] (234.55, 78.23) -- (247.12, 83.04);

\path[draw=drawColor,line width= 0.4pt,line join=round,line cap=round] (258.19, 82.72) -- (271.03, 76.91);

\path[draw=drawColor,line width= 0.4pt,line join=round,line cap=round] (282.46, 75.06) -- (294.30, 76.29);

\path[draw=drawColor,line width= 0.4pt,line join=round,line cap=round] (306.26, 76.50) -- (318.06, 75.68);

\path[draw=drawColor,line width= 0.4pt,line join=round,line cap=round] (330.04, 75.05) -- (341.83, 74.64);

\path[draw=drawColor,line width= 0.4pt,line join=round,line cap=round] (353.82, 74.44) -- (365.60, 74.44);

\path[draw=drawColor,line width= 0.4pt,line join=round,line cap=round] (377.60, 74.64) -- (389.38, 75.05);

\path[draw=drawColor,line width= 0.4pt,line join=round,line cap=round] ( 60.26, 98.66) rectangle ( 64.76,103.16);

\path[draw=drawColor,line width= 0.4pt,line join=round,line cap=round] ( 60.26, 98.66) -- ( 64.76,103.16);

\path[draw=drawColor,line width= 0.4pt,line join=round,line cap=round] ( 60.26,103.16) -- ( 64.76, 98.66);

\path[draw=drawColor,line width= 0.4pt,line join=round,line cap=round] ( 84.04, 85.42) rectangle ( 88.54, 89.92);

\path[draw=drawColor,line width= 0.4pt,line join=round,line cap=round] ( 84.04, 85.42) -- ( 88.54, 89.92);

\path[draw=drawColor,line width= 0.4pt,line join=round,line cap=round] ( 84.04, 89.92) -- ( 88.54, 85.42);

\path[draw=drawColor,line width= 0.4pt,line join=round,line cap=round] (107.82, 82.94) rectangle (112.32, 87.44);

\path[draw=drawColor,line width= 0.4pt,line join=round,line cap=round] (107.82, 82.94) -- (112.32, 87.44);

\path[draw=drawColor,line width= 0.4pt,line join=round,line cap=round] (107.82, 87.44) -- (112.32, 82.94);

\path[draw=drawColor,line width= 0.4pt,line join=round,line cap=round] (131.59, 77.15) rectangle (136.09, 81.65);

\path[draw=drawColor,line width= 0.4pt,line join=round,line cap=round] (131.59, 77.15) -- (136.09, 81.65);

\path[draw=drawColor,line width= 0.4pt,line join=round,line cap=round] (131.59, 81.65) -- (136.09, 77.15);

\path[draw=drawColor,line width= 0.4pt,line join=round,line cap=round] (155.37, 77.98) rectangle (159.87, 82.48);

\path[draw=drawColor,line width= 0.4pt,line join=round,line cap=round] (155.37, 77.98) -- (159.87, 82.48);

\path[draw=drawColor,line width= 0.4pt,line join=round,line cap=round] (155.37, 82.48) -- (159.87, 77.98);

\path[draw=drawColor,line width= 0.4pt,line join=round,line cap=round] (179.14, 77.15) rectangle (183.64, 81.65);

\path[draw=drawColor,line width= 0.4pt,line join=round,line cap=round] (179.14, 77.15) -- (183.64, 81.65);

\path[draw=drawColor,line width= 0.4pt,line join=round,line cap=round] (179.14, 81.65) -- (183.64, 77.15);

\path[draw=drawColor,line width= 0.4pt,line join=round,line cap=round] (202.92, 73.84) rectangle (207.42, 78.34);

\path[draw=drawColor,line width= 0.4pt,line join=round,line cap=round] (202.92, 73.84) -- (207.42, 78.34);

\path[draw=drawColor,line width= 0.4pt,line join=round,line cap=round] (202.92, 78.34) -- (207.42, 73.84);

\path[draw=drawColor,line width= 0.4pt,line join=round,line cap=round] (226.69, 73.84) rectangle (231.19, 78.34);

\path[draw=drawColor,line width= 0.4pt,line join=round,line cap=round] (226.69, 73.84) -- (231.19, 78.34);

\path[draw=drawColor,line width= 0.4pt,line join=round,line cap=round] (226.69, 78.34) -- (231.19, 73.84);

\path[draw=drawColor,line width= 0.4pt,line join=round,line cap=round] (250.47, 82.94) rectangle (254.97, 87.44);

\path[draw=drawColor,line width= 0.4pt,line join=round,line cap=round] (250.47, 82.94) -- (254.97, 87.44);

\path[draw=drawColor,line width= 0.4pt,line join=round,line cap=round] (250.47, 87.44) -- (254.97, 82.94);

\path[draw=drawColor,line width= 0.4pt,line join=round,line cap=round] (274.25, 72.19) rectangle (278.75, 76.69);

\path[draw=drawColor,line width= 0.4pt,line join=round,line cap=round] (274.25, 72.19) -- (278.75, 76.69);

\path[draw=drawColor,line width= 0.4pt,line join=round,line cap=round] (274.25, 76.69) -- (278.75, 72.19);

\path[draw=drawColor,line width= 0.4pt,line join=round,line cap=round] (298.02, 74.67) rectangle (302.52, 79.17);

\path[draw=drawColor,line width= 0.4pt,line join=round,line cap=round] (298.02, 74.67) -- (302.52, 79.17);

\path[draw=drawColor,line width= 0.4pt,line join=round,line cap=round] (298.02, 79.17) -- (302.52, 74.67);

\path[draw=drawColor,line width= 0.4pt,line join=round,line cap=round] (321.80, 73.01) rectangle (326.30, 77.51);

\path[draw=drawColor,line width= 0.4pt,line join=round,line cap=round] (321.80, 73.01) -- (326.30, 77.51);

\path[draw=drawColor,line width= 0.4pt,line join=round,line cap=round] (321.80, 77.51) -- (326.30, 73.01);

\path[draw=drawColor,line width= 0.4pt,line join=round,line cap=round] (345.57, 72.19) rectangle (350.07, 76.69);

\path[draw=drawColor,line width= 0.4pt,line join=round,line cap=round] (345.57, 72.19) -- (350.07, 76.69);

\path[draw=drawColor,line width= 0.4pt,line join=round,line cap=round] (345.57, 76.69) -- (350.07, 72.19);

\path[draw=drawColor,line width= 0.4pt,line join=round,line cap=round] (369.35, 72.19) rectangle (373.85, 76.69);

\path[draw=drawColor,line width= 0.4pt,line join=round,line cap=round] (369.35, 72.19) -- (373.85, 76.69);

\path[draw=drawColor,line width= 0.4pt,line join=round,line cap=round] (369.35, 76.69) -- (373.85, 72.19);

\path[draw=drawColor,line width= 0.4pt,line join=round,line cap=round] (393.13, 73.01) rectangle (397.63, 77.51);

\path[draw=drawColor,line width= 0.4pt,line join=round,line cap=round] (393.13, 73.01) -- (397.63, 77.51);

\path[draw=drawColor,line width= 0.4pt,line join=round,line cap=round] (393.13, 77.51) -- (397.63, 73.01);
\definecolor{drawColor}{RGB}{0,0,0}

\path[draw=drawColor,line width= 0.4pt,dash pattern=on 4pt off 4pt ,line join=round,line cap=round] ( 61.56, 76.09) --
	( 64.94, 76.09) --
	( 68.33, 76.09) --
	( 71.71, 76.09) --
	( 75.09, 76.09) --
	( 78.47, 76.09) --
	( 81.85, 76.09) --
	( 85.23, 76.09) --
	( 88.61, 76.09) --
	( 92.00, 76.09) --
	( 95.38, 76.09) --
	( 98.76, 76.09) --
	(102.14, 76.09) --
	(105.52, 76.09) --
	(108.90, 76.09) --
	(112.29, 76.09) --
	(115.67, 76.09) --
	(119.05, 76.09) --
	(122.43, 76.09) --
	(125.81, 76.09) --
	(129.19, 76.09) --
	(132.57, 76.09) --
	(135.96, 76.09) --
	(139.34, 76.09) --
	(142.72, 76.09) --
	(146.10, 76.09) --
	(149.48, 76.09) --
	(152.86, 76.09) --
	(156.24, 76.09) --
	(159.63, 76.09) --
	(163.01, 76.09) --
	(166.39, 76.09) --
	(169.77, 76.09) --
	(173.15, 76.09) --
	(176.53, 76.09) --
	(179.91, 76.09) --
	(183.30, 76.09) --
	(186.68, 76.09) --
	(190.06, 76.09) --
	(193.44, 76.09) --
	(196.82, 76.09) --
	(200.20, 76.09) --
	(203.58, 76.09) --
	(206.97, 76.09) --
	(210.35, 76.09) --
	(213.73, 76.09) --
	(217.11, 76.09) --
	(220.49, 76.09) --
	(223.87, 76.09) --
	(227.25, 76.09) --
	(230.64, 76.09) --
	(234.02, 76.09) --
	(237.40, 76.09) --
	(240.78, 76.09) --
	(244.16, 76.09) --
	(247.54, 76.09) --
	(250.92, 76.09) --
	(254.31, 76.09) --
	(257.69, 76.09) --
	(261.07, 76.09) --
	(264.45, 76.09) --
	(267.83, 76.09) --
	(271.21, 76.09) --
	(274.59, 76.09) --
	(277.98, 76.09) --
	(281.36, 76.09) --
	(284.74, 76.09) --
	(288.12, 76.09) --
	(291.50, 76.09) --
	(294.88, 76.09) --
	(298.26, 76.09) --
	(301.65, 76.09) --
	(305.03, 76.09) --
	(308.41, 76.09) --
	(311.79, 76.09) --
	(315.17, 76.09) --
	(318.55, 76.09) --
	(321.93, 76.09) --
	(325.32, 76.09) --
	(328.70, 76.09) --
	(332.08, 76.09) --
	(335.46, 76.09) --
	(338.84, 76.09) --
	(342.22, 76.09) --
	(345.60, 76.09) --
	(348.99, 76.09) --
	(352.37, 76.09) --
	(355.75, 76.09) --
	(359.13, 76.09) --
	(362.51, 76.09) --
	(365.89, 76.09) --
	(369.28, 76.09) --
	(372.66, 76.09) --
	(376.04, 76.09) --
	(379.42, 76.09) --
	(382.80, 76.09) --
	(386.18, 76.09) --
	(389.56, 76.09) --
	(392.95, 76.09) --
	(396.33, 76.09);
\definecolor{drawColor}{RGB}{139,0,0}
\definecolor{fillColor}{RGB}{139,0,0}

\path[draw=drawColor,line width= 0.4pt,line join=round,line cap=round,fill=fillColor] (421.11,227.88) circle (  2.25);
\definecolor{fillColor}{RGB}{255,99,71}

\path[fill=fillColor] (421.11,219.38) --
	(424.15,214.13) --
	(418.08,214.13) --
	cycle;
\definecolor{fillColor}{RGB}{108,166,205}

\path[fill=fillColor] (418.86,201.63) --
	(423.36,201.63) --
	(423.36,206.13) --
	(418.86,206.13) --
	cycle;
\definecolor{drawColor}{RGB}{139,137,137}

\path[draw=drawColor,line width= 0.4pt,line join=round,line cap=round] (417.93,191.88) -- (424.30,191.88);

\path[draw=drawColor,line width= 0.4pt,line join=round,line cap=round] (421.11,188.70) -- (421.11,195.06);
\definecolor{drawColor}{RGB}{84,139,84}

\path[draw=drawColor,line width= 0.4pt,line join=round,line cap=round] (418.86,177.63) -- (423.36,182.13);

\path[draw=drawColor,line width= 0.4pt,line join=round,line cap=round] (418.86,182.13) -- (423.36,177.63);
\definecolor{drawColor}{RGB}{205,96,144}

\path[draw=drawColor,line width= 0.4pt,line join=round,line cap=round] (418.86,165.63) rectangle (423.36,170.13);

\path[draw=drawColor,line width= 0.4pt,line join=round,line cap=round] (418.86,165.63) -- (423.36,170.13);

\path[draw=drawColor,line width= 0.4pt,line join=round,line cap=round] (418.86,170.13) -- (423.36,165.63);
\definecolor{drawColor}{RGB}{0,0,0}

\node[text=drawColor,anchor=base west,inner sep=0pt, outer sep=0pt, scale=  1.00] at (430.11,224.44) {HypoRF};

\node[text=drawColor,anchor=base west,inner sep=0pt, outer sep=0pt, scale=  1.00] at (430.11,212.44) {Binomial};

\node[text=drawColor,anchor=base west,inner sep=0pt, outer sep=0pt, scale=  1.00] at (430.11,200.44) {ME-full};

\node[text=drawColor,anchor=base west,inner sep=0pt, outer sep=0pt, scale=  1.00] at (430.11,188.44) {MMDboot};

\node[text=drawColor,anchor=base west,inner sep=0pt, outer sep=0pt, scale=  1.00] at (430.11,176.44) {MMD-full};

\node[text=drawColor,anchor=base west,inner sep=0pt, outer sep=0pt, scale=  1.00] at (430.11,164.44) {CPT-RF};
\end{scope}
\end{tikzpicture}

%% file: 6_Bloboriginal_histX_p1.tex
\begin{tikzpicture}[x=1pt,y=1pt]
\definecolor{fillColor}{RGB}{255,255,255}
\path[use as bounding box,fill=fillColor,fill opacity=0.00] (0,0) rectangle (433.62,289.08);
\begin{scope}
\path[clip] (  0.00,  0.00) rectangle (433.62,289.08);
\definecolor{drawColor}{RGB}{0,0,0}

\node[text=drawColor,rotate= 90.00,anchor=base,inner sep=0pt, outer sep=0pt, scale=  1.00] at ( 10.80,150.54) {Frequency};
\end{scope}
\begin{scope}
\path[clip] (  0.00,  0.00) rectangle (433.62,289.08);
\definecolor{drawColor}{RGB}{0,0,0}

\path[draw=drawColor,line width= 0.4pt,line join=round,line cap=round] ( 62.50, 61.20) -- (395.12, 61.20);

\path[draw=drawColor,line width= 0.4pt,line join=round,line cap=round] ( 62.50, 61.20) -- ( 62.50, 55.20);

\path[draw=drawColor,line width= 0.4pt,line join=round,line cap=round] (145.66, 61.20) -- (145.66, 55.20);

\path[draw=drawColor,line width= 0.4pt,line join=round,line cap=round] (228.81, 61.20) -- (228.81, 55.20);

\path[draw=drawColor,line width= 0.4pt,line join=round,line cap=round] (311.96, 61.20) -- (311.96, 55.20);

\path[draw=drawColor,line width= 0.4pt,line join=round,line cap=round] (395.12, 61.20) -- (395.12, 55.20);

\node[text=drawColor,anchor=base,inner sep=0pt, outer sep=0pt, scale=  1.00] at ( 62.50, 39.60) {0};

\node[text=drawColor,anchor=base,inner sep=0pt, outer sep=0pt, scale=  1.00] at (145.66, 39.60) {1};

\node[text=drawColor,anchor=base,inner sep=0pt, outer sep=0pt, scale=  1.00] at (228.81, 39.60) {2};

\node[text=drawColor,anchor=base,inner sep=0pt, outer sep=0pt, scale=  1.00] at (311.96, 39.60) {3};

\node[text=drawColor,anchor=base,inner sep=0pt, outer sep=0pt, scale=  1.00] at (395.12, 39.60) {4};

\path[draw=drawColor,line width= 0.4pt,line join=round,line cap=round] ( 49.20, 67.82) -- ( 49.20,198.43);

\path[draw=drawColor,line width= 0.4pt,line join=round,line cap=round] ( 49.20, 67.82) -- ( 43.20, 67.82);

\path[draw=drawColor,line width= 0.4pt,line join=round,line cap=round] ( 49.20,111.36) -- ( 43.20,111.36);

\path[draw=drawColor,line width= 0.4pt,line join=round,line cap=round] ( 49.20,154.89) -- ( 43.20,154.89);

\path[draw=drawColor,line width= 0.4pt,line join=round,line cap=round] ( 49.20,198.43) -- ( 43.20,198.43);

\node[text=drawColor,rotate= 90.00,anchor=base,inner sep=0pt, outer sep=0pt, scale=  1.00] at ( 34.80, 67.82) {0};

\node[text=drawColor,rotate= 90.00,anchor=base,inner sep=0pt, outer sep=0pt, scale=  1.00] at ( 34.80,111.36) {50};

\node[text=drawColor,rotate= 90.00,anchor=base,inner sep=0pt, outer sep=0pt, scale=  1.00] at ( 34.80,154.89) {100};

\node[text=drawColor,rotate= 90.00,anchor=base,inner sep=0pt, outer sep=0pt, scale=  1.00] at ( 34.80,198.43) {150};
\end{scope}
\begin{scope}
\path[clip] ( 49.20, 61.20) rectangle (408.42,239.88);
\definecolor{drawColor}{RGB}{0,0,0}

\path[draw=drawColor,line width= 0.4pt,line join=round,line cap=round] ( 62.50, 67.82) rectangle (104.08, 70.43);

\path[draw=drawColor,line width= 0.4pt,line join=round,line cap=round] (104.08, 67.82) rectangle (145.66,198.43);

\path[draw=drawColor,line width= 0.4pt,line join=round,line cap=round] (145.66, 67.82) rectangle (187.23,213.23);

\path[draw=drawColor,line width= 0.4pt,line join=round,line cap=round] (187.23, 67.82) rectangle (228.81,207.14);

\path[draw=drawColor,line width= 0.4pt,line join=round,line cap=round] (228.81, 67.82) rectangle (270.39,233.26);

\path[draw=drawColor,line width= 0.4pt,line join=round,line cap=round] (270.39, 67.82) rectangle (311.96,210.62);

\path[draw=drawColor,line width= 0.4pt,line join=round,line cap=round] (311.96, 67.82) rectangle (353.54,209.75);

\path[draw=drawColor,line width= 0.4pt,line join=round,line cap=round] (353.54, 67.82) rectangle (395.12, 70.43);
\end{scope}
\end{tikzpicture}

%% file: 6_Bloboriginal_histY_p1.tex
\begin{tikzpicture}[x=1pt,y=1pt]
\definecolor{fillColor}{RGB}{255,255,255}
\path[use as bounding box,fill=fillColor,fill opacity=0.00] (0,0) rectangle (433.62,289.08);
\begin{scope}
\path[clip] (  0.00,  0.00) rectangle (433.62,289.08);
\definecolor{drawColor}{RGB}{0,0,0}

\node[text=drawColor,rotate= 90.00,anchor=base,inner sep=0pt, outer sep=0pt, scale=  1.00] at ( 10.80,150.54) {Frequency};
\end{scope}
\begin{scope}
\path[clip] (  0.00,  0.00) rectangle (433.62,289.08);
\definecolor{drawColor}{RGB}{0,0,0}

\path[draw=drawColor,line width= 0.4pt,line join=round,line cap=round] ( 62.50, 61.20) -- (395.12, 61.20);

\path[draw=drawColor,line width= 0.4pt,line join=round,line cap=round] ( 62.50, 61.20) -- ( 62.50, 55.20);

\path[draw=drawColor,line width= 0.4pt,line join=round,line cap=round] (145.66, 61.20) -- (145.66, 55.20);

\path[draw=drawColor,line width= 0.4pt,line join=round,line cap=round] (228.81, 61.20) -- (228.81, 55.20);

\path[draw=drawColor,line width= 0.4pt,line join=round,line cap=round] (311.96, 61.20) -- (311.96, 55.20);

\path[draw=drawColor,line width= 0.4pt,line join=round,line cap=round] (395.12, 61.20) -- (395.12, 55.20);

\node[text=drawColor,anchor=base,inner sep=0pt, outer sep=0pt, scale=  1.00] at ( 62.50, 39.60) {0};

\node[text=drawColor,anchor=base,inner sep=0pt, outer sep=0pt, scale=  1.00] at (145.66, 39.60) {1};

\node[text=drawColor,anchor=base,inner sep=0pt, outer sep=0pt, scale=  1.00] at (228.81, 39.60) {2};

\node[text=drawColor,anchor=base,inner sep=0pt, outer sep=0pt, scale=  1.00] at (311.96, 39.60) {3};

\node[text=drawColor,anchor=base,inner sep=0pt, outer sep=0pt, scale=  1.00] at (395.12, 39.60) {4};

\path[draw=drawColor,line width= 0.4pt,line join=round,line cap=round] ( 49.20, 67.82) -- ( 49.20,207.24);

\path[draw=drawColor,line width= 0.4pt,line join=round,line cap=round] ( 49.20, 67.82) -- ( 43.20, 67.82);

\path[draw=drawColor,line width= 0.4pt,line join=round,line cap=round] ( 49.20,114.29) -- ( 43.20,114.29);

\path[draw=drawColor,line width= 0.4pt,line join=round,line cap=round] ( 49.20,160.76) -- ( 43.20,160.76);

\path[draw=drawColor,line width= 0.4pt,line join=round,line cap=round] ( 49.20,207.24) -- ( 43.20,207.24);

\node[text=drawColor,rotate= 90.00,anchor=base,inner sep=0pt, outer sep=0pt, scale=  1.00] at ( 34.80, 67.82) {0};

\node[text=drawColor,rotate= 90.00,anchor=base,inner sep=0pt, outer sep=0pt, scale=  1.00] at ( 34.80,114.29) {50};

\node[text=drawColor,rotate= 90.00,anchor=base,inner sep=0pt, outer sep=0pt, scale=  1.00] at ( 34.80,160.76) {100};

\node[text=drawColor,rotate= 90.00,anchor=base,inner sep=0pt, outer sep=0pt, scale=  1.00] at ( 34.80,207.24) {150};
\end{scope}
\begin{scope}
\path[clip] ( 49.20, 61.20) rectangle (408.42,239.88);
\definecolor{drawColor}{RGB}{0,0,0}

\path[draw=drawColor,line width= 0.4pt,line join=round,line cap=round] ( 62.50, 67.82) rectangle (104.08, 69.68);

\path[draw=drawColor,line width= 0.4pt,line join=round,line cap=round] (104.08, 67.82) rectangle (145.66,205.38);

\path[draw=drawColor,line width= 0.4pt,line join=round,line cap=round] (145.66, 67.82) rectangle (187.23,220.25);

\path[draw=drawColor,line width= 0.4pt,line join=round,line cap=round] (187.23, 67.82) rectangle (228.81,217.46);

\path[draw=drawColor,line width= 0.4pt,line join=round,line cap=round] (228.81, 67.82) rectangle (270.39,232.33);

\path[draw=drawColor,line width= 0.4pt,line join=round,line cap=round] (270.39, 67.82) rectangle (311.96,224.90);

\path[draw=drawColor,line width= 0.4pt,line join=round,line cap=round] (311.96, 67.82) rectangle (353.54,233.26);

\path[draw=drawColor,line width= 0.4pt,line join=round,line cap=round] (353.54, 67.82) rectangle (395.12, 68.75);
\end{scope}
\end{tikzpicture}

%% file: 5_Blob_histX_p1.tex
\begin{tikzpicture}[x=1pt,y=1pt]
\definecolor{fillColor}{RGB}{255,255,255}
\path[use as bounding box,fill=fillColor,fill opacity=0.00] (0,0) rectangle (433.62,289.08);
\begin{scope}
\path[clip] (  0.00,  0.00) rectangle (433.62,289.08);
\definecolor{drawColor}{RGB}{0,0,0}

\node[text=drawColor,rotate= 90.00,anchor=base,inner sep=0pt, outer sep=0pt, scale=  1.00] at ( 10.80,150.54) {Frequency};
\end{scope}
\begin{scope}
\path[clip] (  0.00,  0.00) rectangle (433.62,289.08);
\definecolor{drawColor}{RGB}{0,0,0}

\path[draw=drawColor,line width= 0.4pt,line join=round,line cap=round] ( 62.50, 61.20) -- (339.68, 61.20);

\path[draw=drawColor,line width= 0.4pt,line join=round,line cap=round] ( 62.50, 61.20) -- ( 62.50, 55.20);

\path[draw=drawColor,line width= 0.4pt,line join=round,line cap=round] (154.90, 61.20) -- (154.90, 55.20);

\path[draw=drawColor,line width= 0.4pt,line join=round,line cap=round] (247.29, 61.20) -- (247.29, 55.20);

\path[draw=drawColor,line width= 0.4pt,line join=round,line cap=round] (339.68, 61.20) -- (339.68, 55.20);

\node[text=drawColor,anchor=base,inner sep=0pt, outer sep=0pt, scale=  1.00] at ( 62.50, 39.60) {-10};

\node[text=drawColor,anchor=base,inner sep=0pt, outer sep=0pt, scale=  1.00] at (154.90, 39.60) {-5};

\node[text=drawColor,anchor=base,inner sep=0pt, outer sep=0pt, scale=  1.00] at (247.29, 39.60) {0};

\node[text=drawColor,anchor=base,inner sep=0pt, outer sep=0pt, scale=  1.00] at (339.68, 39.60) {5};

\path[draw=drawColor,line width= 0.4pt,line join=round,line cap=round] ( 49.20, 67.82) -- ( 49.20,211.06);

\path[draw=drawColor,line width= 0.4pt,line join=round,line cap=round] ( 49.20, 67.82) -- ( 43.20, 67.82);

\path[draw=drawColor,line width= 0.4pt,line join=round,line cap=round] ( 49.20,103.63) -- ( 43.20,103.63);

\path[draw=drawColor,line width= 0.4pt,line join=round,line cap=round] ( 49.20,139.44) -- ( 43.20,139.44);

\path[draw=drawColor,line width= 0.4pt,line join=round,line cap=round] ( 49.20,175.25) -- ( 43.20,175.25);

\path[draw=drawColor,line width= 0.4pt,line join=round,line cap=round] ( 49.20,211.06) -- ( 43.20,211.06);

\node[text=drawColor,rotate= 90.00,anchor=base,inner sep=0pt, outer sep=0pt, scale=  1.00] at ( 34.80, 67.82) {0};

\node[text=drawColor,rotate= 90.00,anchor=base,inner sep=0pt, outer sep=0pt, scale=  1.00] at ( 34.80,103.63) {50};

\node[text=drawColor,rotate= 90.00,anchor=base,inner sep=0pt, outer sep=0pt, scale=  1.00] at ( 34.80,139.44) {100};

\node[text=drawColor,rotate= 90.00,anchor=base,inner sep=0pt, outer sep=0pt, scale=  1.00] at ( 34.80,175.25) {150};

\node[text=drawColor,rotate= 90.00,anchor=base,inner sep=0pt, outer sep=0pt, scale=  1.00] at ( 34.80,211.06) {200};
\end{scope}
\begin{scope}
\path[clip] ( 49.20, 61.20) rectangle (408.42,239.88);
\definecolor{drawColor}{RGB}{0,0,0}

\path[draw=drawColor,line width= 0.4pt,line join=round,line cap=round] ( 62.50, 67.82) rectangle ( 99.46, 68.53);

\path[draw=drawColor,line width= 0.4pt,line join=round,line cap=round] ( 99.46, 67.82) rectangle (136.42,113.66);

\path[draw=drawColor,line width= 0.4pt,line join=round,line cap=round] (136.42, 67.82) rectangle (173.37,218.22);

\path[draw=drawColor,line width= 0.4pt,line join=round,line cap=round] (173.37, 67.82) rectangle (210.33,109.36);

\path[draw=drawColor,line width= 0.4pt,line join=round,line cap=round] (210.33, 67.82) rectangle (247.29,190.29);

\path[draw=drawColor,line width= 0.4pt,line join=round,line cap=round] (247.29, 67.82) rectangle (284.25,189.57);

\path[draw=drawColor,line width= 0.4pt,line join=round,line cap=round] (284.25, 67.82) rectangle (321.20,102.20);

\path[draw=drawColor,line width= 0.4pt,line join=round,line cap=round] (321.20, 67.82) rectangle (358.16,233.26);

\path[draw=drawColor,line width= 0.4pt,line join=round,line cap=round] (358.16, 67.82) rectangle (395.12,101.48);
\end{scope}
\end{tikzpicture}

%% file: 5_Blob_histY_p1.tex
\begin{tikzpicture}[x=1pt,y=1pt]
\definecolor{fillColor}{RGB}{255,255,255}
\path[use as bounding box,fill=fillColor,fill opacity=0.00] (0,0) rectangle (433.62,289.08);
\begin{scope}
\path[clip] (  0.00,  0.00) rectangle (433.62,289.08);
\definecolor{drawColor}{RGB}{0,0,0}

\node[text=drawColor,rotate= 90.00,anchor=base,inner sep=0pt, outer sep=0pt, scale=  1.00] at ( 10.80,150.54) {Frequency};
\end{scope}
\begin{scope}
\path[clip] (  0.00,  0.00) rectangle (433.62,289.08);
\definecolor{drawColor}{RGB}{0,0,0}

\path[draw=drawColor,line width= 0.4pt,line join=round,line cap=round] (124.87, 61.20) -- (332.75, 61.20);

\path[draw=drawColor,line width= 0.4pt,line join=round,line cap=round] (124.87, 61.20) -- (124.87, 55.20);

\path[draw=drawColor,line width= 0.4pt,line join=round,line cap=round] (228.81, 61.20) -- (228.81, 55.20);

\path[draw=drawColor,line width= 0.4pt,line join=round,line cap=round] (332.75, 61.20) -- (332.75, 55.20);

\node[text=drawColor,anchor=base,inner sep=0pt, outer sep=0pt, scale=  1.00] at (124.87, 39.60) {-5};

\node[text=drawColor,anchor=base,inner sep=0pt, outer sep=0pt, scale=  1.00] at (228.81, 39.60) {0};

\node[text=drawColor,anchor=base,inner sep=0pt, outer sep=0pt, scale=  1.00] at (332.75, 39.60) {5};

\path[draw=drawColor,line width= 0.4pt,line join=round,line cap=round] ( 49.20, 67.82) -- ( 49.20,233.26);

\path[draw=drawColor,line width= 0.4pt,line join=round,line cap=round] ( 49.20, 67.82) -- ( 43.20, 67.82);

\path[draw=drawColor,line width= 0.4pt,line join=round,line cap=round] ( 49.20,100.91) -- ( 43.20,100.91);

\path[draw=drawColor,line width= 0.4pt,line join=round,line cap=round] ( 49.20,134.00) -- ( 43.20,134.00);

\path[draw=drawColor,line width= 0.4pt,line join=round,line cap=round] ( 49.20,167.08) -- ( 43.20,167.08);

\path[draw=drawColor,line width= 0.4pt,line join=round,line cap=round] ( 49.20,200.17) -- ( 43.20,200.17);

\path[draw=drawColor,line width= 0.4pt,line join=round,line cap=round] ( 49.20,233.26) -- ( 43.20,233.26);

\node[text=drawColor,rotate= 90.00,anchor=base,inner sep=0pt, outer sep=0pt, scale=  1.00] at ( 34.80, 67.82) {0};

\node[text=drawColor,rotate= 90.00,anchor=base,inner sep=0pt, outer sep=0pt, scale=  1.00] at ( 34.80,100.91) {50};

\node[text=drawColor,rotate= 90.00,anchor=base,inner sep=0pt, outer sep=0pt, scale=  1.00] at ( 34.80,134.00) {100};

\node[text=drawColor,rotate= 90.00,anchor=base,inner sep=0pt, outer sep=0pt, scale=  1.00] at ( 34.80,167.08) {150};

\node[text=drawColor,rotate= 90.00,anchor=base,inner sep=0pt, outer sep=0pt, scale=  1.00] at ( 34.80,200.17) {200};

\node[text=drawColor,rotate= 90.00,anchor=base,inner sep=0pt, outer sep=0pt, scale=  1.00] at ( 34.80,233.26) {250};
\end{scope}
\begin{scope}
\path[clip] ( 49.20, 61.20) rectangle (408.42,239.88);
\definecolor{drawColor}{RGB}{0,0,0}

\path[draw=drawColor,line width= 0.4pt,line join=round,line cap=round] ( 62.50, 67.82) rectangle (104.08, 99.58);

\path[draw=drawColor,line width= 0.4pt,line join=round,line cap=round] (104.08, 67.82) rectangle (145.66,217.38);

\path[draw=drawColor,line width= 0.4pt,line join=round,line cap=round] (145.66, 67.82) rectangle (187.23,134.66);

\path[draw=drawColor,line width= 0.4pt,line join=round,line cap=round] (187.23, 67.82) rectangle (228.81,149.22);

\path[draw=drawColor,line width= 0.4pt,line join=round,line cap=round] (228.81, 67.82) rectangle (270.39,145.25);

\path[draw=drawColor,line width= 0.4pt,line join=round,line cap=round] (270.39, 67.82) rectangle (311.96,121.42);

\path[draw=drawColor,line width= 0.4pt,line join=round,line cap=round] (311.96, 67.82) rectangle (353.54,233.26);

\path[draw=drawColor,line width= 0.4pt,line join=round,line cap=round] (353.54, 67.82) rectangle (395.12,103.55);
\end{scope}
\end{tikzpicture}

%% file: 5_plot_Blob_K100_small_Normapprox_F_with_Cai.tex
\begin{tikzpicture}[x=1pt,y=1pt]
\definecolor{fillColor}{RGB}{255,255,255}
\path[use as bounding box,fill=fillColor,fill opacity=0.00] (0,0) rectangle (505.89,289.08);
\begin{scope}
\path[clip] (  0.00,  0.00) rectangle (505.89,289.08);
\definecolor{drawColor}{RGB}{139,0,0}

\path[draw=drawColor,line width= 0.4pt,line join=round,line cap=round] ( 64.96,152.71) -- ( 82.26,191.39);

\path[draw=drawColor,line width= 0.4pt,line join=round,line cap=round] ( 90.33,198.96) -- (101.27,203.04);

\path[draw=drawColor,line width= 0.4pt,line join=round,line cap=round] (112.70,206.65) -- (123.28,209.41);

\path[draw=drawColor,line width= 0.4pt,line join=round,line cap=round] (135.05,211.59) -- (145.31,212.74);

\path[draw=drawColor,line width= 0.4pt,line join=round,line cap=round] (156.97,215.32) -- (167.78,218.95);

\path[draw=drawColor,line width= 0.4pt,line join=round,line cap=round] (179.47,220.85) -- (189.66,220.85);

\path[draw=drawColor,line width= 0.4pt,line join=round,line cap=round] (201.51,222.16) -- (211.99,224.51);

\path[draw=drawColor,line width= 0.4pt,line join=round,line cap=round] (223.78,226.70) -- (234.11,228.24);

\path[draw=drawColor,line width= 0.4pt,line join=round,line cap=round] (246.02,228.68) -- (256.25,227.92);

\path[draw=drawColor,line width= 0.4pt,line join=round,line cap=round] (268.23,227.47) -- (278.42,227.47);

\path[draw=drawColor,line width= 0.4pt,line join=round,line cap=round] (290.41,227.92) -- (300.63,228.68);

\path[draw=drawColor,line width= 0.4pt,line join=round,line cap=round] (312.61,228.90) -- (322.81,228.52);

\path[draw=drawColor,line width= 0.4pt,line join=round,line cap=round] (334.80,228.08) -- (345.00,227.70);

\path[draw=drawColor,line width= 0.4pt,line join=round,line cap=round] (356.93,228.36) -- (367.25,229.90);

\path[draw=drawColor,line width= 0.4pt,line join=round,line cap=round] (379.18,230.56) -- (389.38,230.18);
\definecolor{fillColor}{RGB}{139,0,0}

\path[draw=drawColor,line width= 0.4pt,line join=round,line cap=round,fill=fillColor] ( 62.51,147.23) circle (  2.25);

\path[draw=drawColor,line width= 0.4pt,line join=round,line cap=round,fill=fillColor] ( 84.71,196.86) circle (  2.25);

\path[draw=drawColor,line width= 0.4pt,line join=round,line cap=round,fill=fillColor] (106.90,205.14) circle (  2.25);

\path[draw=drawColor,line width= 0.4pt,line join=round,line cap=round,fill=fillColor] (129.09,210.93) circle (  2.25);

\path[draw=drawColor,line width= 0.4pt,line join=round,line cap=round,fill=fillColor] (151.28,213.41) circle (  2.25);

\path[draw=drawColor,line width= 0.4pt,line join=round,line cap=round,fill=fillColor] (173.47,220.85) circle (  2.25);

\path[draw=drawColor,line width= 0.4pt,line join=round,line cap=round,fill=fillColor] (195.66,220.85) circle (  2.25);

\path[draw=drawColor,line width= 0.4pt,line join=round,line cap=round,fill=fillColor] (217.85,225.82) circle (  2.25);

\path[draw=drawColor,line width= 0.4pt,line join=round,line cap=round,fill=fillColor] (240.04,229.13) circle (  2.25);

\path[draw=drawColor,line width= 0.4pt,line join=round,line cap=round,fill=fillColor] (262.23,227.47) circle (  2.25);

\path[draw=drawColor,line width= 0.4pt,line join=round,line cap=round,fill=fillColor] (284.42,227.47) circle (  2.25);

\path[draw=drawColor,line width= 0.4pt,line join=round,line cap=round,fill=fillColor] (306.61,229.13) circle (  2.25);

\path[draw=drawColor,line width= 0.4pt,line join=round,line cap=round,fill=fillColor] (328.80,228.30) circle (  2.25);

\path[draw=drawColor,line width= 0.4pt,line join=round,line cap=round,fill=fillColor] (350.99,227.47) circle (  2.25);

\path[draw=drawColor,line width= 0.4pt,line join=round,line cap=round,fill=fillColor] (373.18,230.78) circle (  2.25);

\path[draw=drawColor,line width= 0.4pt,line join=round,line cap=round,fill=fillColor] (395.38,229.95) circle (  2.25);
\end{scope}
\begin{scope}
\path[clip] (  0.00,  0.00) rectangle (505.89,289.08);
\definecolor{drawColor}{RGB}{0,0,0}

\path[draw=drawColor,line width= 0.4pt,line join=round,line cap=round] ( 95.80, 61.20) -- (373.18, 61.20);

\path[draw=drawColor,line width= 0.4pt,line join=round,line cap=round] ( 95.80, 61.20) -- ( 95.80, 55.20);

\path[draw=drawColor,line width= 0.4pt,line join=round,line cap=round] (151.28, 61.20) -- (151.28, 55.20);

\path[draw=drawColor,line width= 0.4pt,line join=round,line cap=round] (206.75, 61.20) -- (206.75, 55.20);

\path[draw=drawColor,line width= 0.4pt,line join=round,line cap=round] (262.23, 61.20) -- (262.23, 55.20);

\path[draw=drawColor,line width= 0.4pt,line join=round,line cap=round] (317.71, 61.20) -- (317.71, 55.20);

\path[draw=drawColor,line width= 0.4pt,line join=round,line cap=round] (373.18, 61.20) -- (373.18, 55.20);

\node[text=drawColor,anchor=base,inner sep=0pt, outer sep=0pt, scale=  1.00] at ( 95.80, 39.60) {5};

\node[text=drawColor,anchor=base,inner sep=0pt, outer sep=0pt, scale=  1.00] at (151.28, 39.60) {10};

\node[text=drawColor,anchor=base,inner sep=0pt, outer sep=0pt, scale=  1.00] at (206.75, 39.60) {15};

\node[text=drawColor,anchor=base,inner sep=0pt, outer sep=0pt, scale=  1.00] at (262.23, 39.60) {20};

\node[text=drawColor,anchor=base,inner sep=0pt, outer sep=0pt, scale=  1.00] at (317.71, 39.60) {25};

\node[text=drawColor,anchor=base,inner sep=0pt, outer sep=0pt, scale=  1.00] at (373.18, 39.60) {30};

\path[draw=drawColor,line width= 0.4pt,line join=round,line cap=round] ( 49.20, 67.82) -- ( 49.20,233.26);

\path[draw=drawColor,line width= 0.4pt,line join=round,line cap=round] ( 49.20, 67.82) -- ( 43.20, 67.82);

\path[draw=drawColor,line width= 0.4pt,line join=round,line cap=round] ( 49.20,100.91) -- ( 43.20,100.91);

\path[draw=drawColor,line width= 0.4pt,line join=round,line cap=round] ( 49.20,134.00) -- ( 43.20,134.00);

\path[draw=drawColor,line width= 0.4pt,line join=round,line cap=round] ( 49.20,167.08) -- ( 43.20,167.08);

\path[draw=drawColor,line width= 0.4pt,line join=round,line cap=round] ( 49.20,200.17) -- ( 43.20,200.17);

\path[draw=drawColor,line width= 0.4pt,line join=round,line cap=round] ( 49.20,233.26) -- ( 43.20,233.26);

\node[text=drawColor,rotate= 90.00,anchor=base,inner sep=0pt, outer sep=0pt, scale=  1.00] at ( 34.80, 67.82) {0.0};

\node[text=drawColor,rotate= 90.00,anchor=base,inner sep=0pt, outer sep=0pt, scale=  1.00] at ( 34.80,100.91) {0.2};

\node[text=drawColor,rotate= 90.00,anchor=base,inner sep=0pt, outer sep=0pt, scale=  1.00] at ( 34.80,134.00) {0.4};

\node[text=drawColor,rotate= 90.00,anchor=base,inner sep=0pt, outer sep=0pt, scale=  1.00] at ( 34.80,167.08) {0.6};

\node[text=drawColor,rotate= 90.00,anchor=base,inner sep=0pt, outer sep=0pt, scale=  1.00] at ( 34.80,200.17) {0.8};

\node[text=drawColor,rotate= 90.00,anchor=base,inner sep=0pt, outer sep=0pt, scale=  1.00] at ( 34.80,233.26) {1.0};

\path[draw=drawColor,line width= 0.4pt,line join=round,line cap=round] ( 49.20, 61.20) --
	(408.69, 61.20) --
	(408.69,239.88) --
	( 49.20,239.88) --
	( 49.20, 61.20);
\end{scope}
\begin{scope}
\path[clip] (  0.00,  0.00) rectangle (505.89,289.08);
\definecolor{drawColor}{RGB}{0,0,0}

\node[text=drawColor,anchor=base,inner sep=0pt, outer sep=0pt, scale=  1.00] at (228.95, 15.60) {$p$};

\node[text=drawColor,rotate= 90.00,anchor=base,inner sep=0pt, outer sep=0pt, scale=  1.00] at ( 10.80,150.54) {Power};
\definecolor{drawColor}{RGB}{255,99,71}

\path[draw=drawColor,line width= 0.4pt,line join=round,line cap=round] ( 66.74,139.08) -- ( 80.48,152.90);

\path[draw=drawColor,line width= 0.4pt,line join=round,line cap=round] ( 89.52,160.74) -- (102.09,170.12);

\path[draw=drawColor,line width= 0.4pt,line join=round,line cap=round] (112.65,171.99) -- (123.34,168.80);

\path[draw=drawColor,line width= 0.4pt,line join=round,line cap=round] (135.07,167.53) -- (145.29,168.29);

\path[draw=drawColor,line width= 0.4pt,line join=round,line cap=round] (157.24,169.41) -- (167.51,170.55);

\path[draw=drawColor,line width= 0.4pt,line join=round,line cap=round] (179.47,171.22) -- (189.66,171.22);

\path[draw=drawColor,line width= 0.4pt,line join=round,line cap=round] (199.97,175.40) -- (213.54,188.55);

\path[draw=drawColor,line width= 0.4pt,line join=round,line cap=round] (223.33,190.28) -- (234.56,185.25);

\path[draw=drawColor,line width= 0.4pt,line join=round,line cap=round] (246.02,183.25) -- (256.25,184.01);

\path[draw=drawColor,line width= 0.4pt,line join=round,line cap=round] (267.92,186.36) -- (278.73,189.99);

\path[draw=drawColor,line width= 0.4pt,line join=round,line cap=round] (290.36,191.02) -- (300.68,189.48);

\path[draw=drawColor,line width= 0.4pt,line join=round,line cap=round] (312.61,188.82) -- (322.81,189.20);

\path[draw=drawColor,line width= 0.4pt,line join=round,line cap=round] (334.49,191.33) -- (345.31,194.96);

\path[draw=drawColor,line width= 0.4pt,line join=round,line cap=round] (356.98,196.42) -- (367.20,195.66);

\path[draw=drawColor,line width= 0.4pt,line join=round,line cap=round] (379.15,195.88) -- (389.41,197.02);
\definecolor{fillColor}{RGB}{255,99,71}

\path[fill=fillColor] ( 62.51,138.32) --
	( 65.54,133.07) --
	( 59.48,133.07) --
	cycle;

\path[fill=fillColor] ( 84.71,160.66) --
	( 87.74,155.41) --
	( 81.67,155.41) --
	cycle;

\path[fill=fillColor] (106.90,177.20) --
	(109.93,171.95) --
	(103.87,171.95) --
	cycle;

\path[fill=fillColor] (129.09,170.58) --
	(132.12,165.33) --
	(126.06,165.33) --
	cycle;

\path[fill=fillColor] (151.28,172.24) --
	(154.31,166.99) --
	(148.25,166.99) --
	cycle;

\path[fill=fillColor] (173.47,174.72) --
	(176.50,169.47) --
	(170.44,169.47) --
	cycle;

\path[fill=fillColor] (195.66,174.72) --
	(198.69,169.47) --
	(192.63,169.47) --
	cycle;

\path[fill=fillColor] (217.85,196.23) --
	(220.88,190.98) --
	(214.82,190.98) --
	cycle;

\path[fill=fillColor] (240.04,186.30) --
	(243.07,181.05) --
	(237.01,181.05) --
	cycle;

\path[fill=fillColor] (262.23,187.96) --
	(265.26,182.71) --
	(259.20,182.71) --
	cycle;

\path[fill=fillColor] (284.42,195.40) --
	(287.45,190.15) --
	(281.39,190.15) --
	cycle;

\path[fill=fillColor] (306.61,192.09) --
	(309.64,186.84) --
	(303.58,186.84) --
	cycle;

\path[fill=fillColor] (328.80,192.92) --
	(331.83,187.67) --
	(325.77,187.67) --
	cycle;

\path[fill=fillColor] (350.99,200.36) --
	(354.02,195.11) --
	(347.96,195.11) --
	cycle;

\path[fill=fillColor] (373.18,198.71) --
	(376.22,193.46) --
	(370.15,193.46) --
	cycle;

\path[fill=fillColor] (395.38,201.19) --
	(398.41,195.94) --
	(392.35,195.94) --
	cycle;
\definecolor{drawColor}{RGB}{108,166,205}

\path[draw=drawColor,line width= 0.4pt,line join=round,line cap=round] ( 66.59,105.60) -- ( 80.63, 90.42);

\path[draw=drawColor,line width= 0.4pt,line join=round,line cap=round] ( 90.70, 86.24) -- (100.90, 86.62);

\path[draw=drawColor,line width= 0.4pt,line join=round,line cap=round] (112.88, 86.40) -- (123.10, 85.64);

\path[draw=drawColor,line width= 0.4pt,line join=round,line cap=round] (134.64, 87.47) -- (145.73, 92.01);

\path[draw=drawColor,line width= 0.4pt,line join=round,line cap=round] (156.90, 92.19) -- (167.85, 88.11);

\path[draw=drawColor,line width= 0.4pt,line join=round,line cap=round] (178.19, 89.72) -- (190.93, 99.69);

\path[draw=drawColor,line width= 0.4pt,line join=round,line cap=round] (201.62,102.72) -- (211.89,101.57);

\path[draw=drawColor,line width= 0.4pt,line join=round,line cap=round] (223.78,101.79) -- (234.11,103.33);

\path[draw=drawColor,line width= 0.4pt,line join=round,line cap=round] (245.73,102.31) -- (256.54, 98.68);

\path[draw=drawColor,line width= 0.4pt,line join=round,line cap=round] (266.62,100.86) -- (280.03,113.36);

\path[draw=drawColor,line width= 0.4pt,line join=round,line cap=round] (290.42,117.45) -- (300.61,117.45);

\path[draw=drawColor,line width= 0.4pt,line join=round,line cap=round] (312.51,116.35) -- (322.90,114.41);

\path[draw=drawColor,line width= 0.4pt,line join=round,line cap=round] (333.03,117.57) -- (346.77,131.39);

\path[draw=drawColor,line width= 0.4pt,line join=round,line cap=round] (356.55,133.37) -- (367.63,128.83);

\path[draw=drawColor,line width= 0.4pt,line join=round,line cap=round] (378.74,128.83) -- (389.82,133.37);
\definecolor{fillColor}{RGB}{108,166,205}

\path[fill=fillColor] ( 60.26,107.76) --
	( 64.76,107.76) --
	( 64.76,112.26) --
	( 60.26,112.26) --
	cycle;

\path[fill=fillColor] ( 82.46, 83.77) --
	( 86.96, 83.77) --
	( 86.96, 88.27) --
	( 82.46, 88.27) --
	cycle;

\path[fill=fillColor] (104.65, 84.59) --
	(109.15, 84.59) --
	(109.15, 89.09) --
	(104.65, 89.09) --
	cycle;

\path[fill=fillColor] (126.84, 82.94) --
	(131.34, 82.94) --
	(131.34, 87.44) --
	(126.84, 87.44) --
	cycle;

\path[fill=fillColor] (149.03, 92.04) --
	(153.53, 92.04) --
	(153.53, 96.54) --
	(149.03, 96.54) --
	cycle;

\path[fill=fillColor] (171.22, 83.77) --
	(175.72, 83.77) --
	(175.72, 88.27) --
	(171.22, 88.27) --
	cycle;

\path[fill=fillColor] (193.41,101.14) --
	(197.91,101.14) --
	(197.91,105.64) --
	(193.41,105.64) --
	cycle;

\path[fill=fillColor] (215.60, 98.66) --
	(220.10, 98.66) --
	(220.10,103.16) --
	(215.60,103.16) --
	cycle;

\path[fill=fillColor] (237.79,101.97) --
	(242.29,101.97) --
	(242.29,106.47) --
	(237.79,106.47) --
	cycle;

\path[fill=fillColor] (259.98, 94.52) --
	(264.48, 94.52) --
	(264.48, 99.02) --
	(259.98, 99.02) --
	cycle;

\path[fill=fillColor] (282.17,115.20) --
	(286.67,115.20) --
	(286.67,119.70) --
	(282.17,119.70) --
	cycle;

\path[fill=fillColor] (304.36,115.20) --
	(308.86,115.20) --
	(308.86,119.70) --
	(304.36,119.70) --
	cycle;

\path[fill=fillColor] (326.55,111.07) --
	(331.05,111.07) --
	(331.05,115.57) --
	(326.55,115.57) --
	cycle;

\path[fill=fillColor] (348.74,133.40) --
	(353.24,133.40) --
	(353.24,137.90) --
	(348.74,137.90) --
	cycle;

\path[fill=fillColor] (370.93,124.30) --
	(375.43,124.30) --
	(375.43,128.80) --
	(370.93,128.80) --
	cycle;

\path[fill=fillColor] (393.13,133.40) --
	(397.63,133.40) --
	(397.63,137.90) --
	(393.13,137.90) --
	cycle;
\definecolor{drawColor}{RGB}{139,137,137}

\path[draw=drawColor,line width= 0.4pt,line join=round,line cap=round] ( 65.80,152.14) -- ( 81.42,128.26);

\path[draw=drawColor,line width= 0.4pt,line join=round,line cap=round] ( 89.60,119.77) -- (102.00,110.99);

\path[draw=drawColor,line width= 0.4pt,line join=round,line cap=round] (111.45,103.62) -- (124.53, 92.40);

\path[draw=drawColor,line width= 0.4pt,line join=round,line cap=round] (135.02, 89.38) -- (145.34, 90.92);

\path[draw=drawColor,line width= 0.4pt,line join=round,line cap=round] (157.03, 90.09) -- (167.72, 86.90);

\path[draw=drawColor,line width= 0.4pt,line join=round,line cap=round] (179.47, 85.19) -- (189.66, 85.19);

\path[draw=drawColor,line width= 0.4pt,line join=round,line cap=round] (201.35, 83.28) -- (212.16, 79.65);

\path[draw=drawColor,line width= 0.4pt,line join=round,line cap=round] (223.78, 78.63) -- (234.11, 80.17);

\path[draw=drawColor,line width= 0.4pt,line join=round,line cap=round] (246.04, 81.28) -- (256.24, 81.66);

\path[draw=drawColor,line width= 0.4pt,line join=round,line cap=round] (268.19, 81.21) -- (278.46, 80.07);

\path[draw=drawColor,line width= 0.4pt,line join=round,line cap=round] (290.41, 79.84) -- (300.63, 80.61);

\path[draw=drawColor,line width= 0.4pt,line join=round,line cap=round] (312.60, 81.50) -- (322.82, 82.26);

\path[draw=drawColor,line width= 0.4pt,line join=round,line cap=round] (334.79, 83.15) -- (345.01, 83.92);

\path[draw=drawColor,line width= 0.4pt,line join=round,line cap=round] (356.96, 85.03) -- (367.22, 86.18);

\path[draw=drawColor,line width= 0.4pt,line join=round,line cap=round] (378.66, 84.39) -- (389.90, 79.37);

\path[draw=drawColor,line width= 0.4pt,line join=round,line cap=round] ( 59.33,157.16) -- ( 65.70,157.16);

\path[draw=drawColor,line width= 0.4pt,line join=round,line cap=round] ( 62.51,153.98) -- ( 62.51,160.34);

\path[draw=drawColor,line width= 0.4pt,line join=round,line cap=round] ( 81.52,123.24) -- ( 87.89,123.24);

\path[draw=drawColor,line width= 0.4pt,line join=round,line cap=round] ( 84.71,120.06) -- ( 84.71,126.42);

\path[draw=drawColor,line width= 0.4pt,line join=round,line cap=round] (103.71,107.52) -- (110.08,107.52);

\path[draw=drawColor,line width= 0.4pt,line join=round,line cap=round] (106.90,104.34) -- (106.90,110.71);

\path[draw=drawColor,line width= 0.4pt,line join=round,line cap=round] (125.90, 88.50) -- (132.27, 88.50);

\path[draw=drawColor,line width= 0.4pt,line join=round,line cap=round] (129.09, 85.32) -- (129.09, 91.68);

\path[draw=drawColor,line width= 0.4pt,line join=round,line cap=round] (148.10, 91.81) -- (154.46, 91.81);

\path[draw=drawColor,line width= 0.4pt,line join=round,line cap=round] (151.28, 88.63) -- (151.28, 94.99);

\path[draw=drawColor,line width= 0.4pt,line join=round,line cap=round] (170.29, 85.19) -- (176.65, 85.19);

\path[draw=drawColor,line width= 0.4pt,line join=round,line cap=round] (173.47, 82.01) -- (173.47, 88.37);

\path[draw=drawColor,line width= 0.4pt,line join=round,line cap=round] (192.48, 85.19) -- (198.84, 85.19);

\path[draw=drawColor,line width= 0.4pt,line join=round,line cap=round] (195.66, 82.01) -- (195.66, 88.37);

\path[draw=drawColor,line width= 0.4pt,line join=round,line cap=round] (214.67, 77.74) -- (221.03, 77.74);

\path[draw=drawColor,line width= 0.4pt,line join=round,line cap=round] (217.85, 74.56) -- (217.85, 80.93);

\path[draw=drawColor,line width= 0.4pt,line join=round,line cap=round] (236.86, 81.05) -- (243.22, 81.05);

\path[draw=drawColor,line width= 0.4pt,line join=round,line cap=round] (240.04, 77.87) -- (240.04, 84.24);

\path[draw=drawColor,line width= 0.4pt,line join=round,line cap=round] (259.05, 81.88) -- (265.41, 81.88);

\path[draw=drawColor,line width= 0.4pt,line join=round,line cap=round] (262.23, 78.70) -- (262.23, 85.06);

\path[draw=drawColor,line width= 0.4pt,line join=round,line cap=round] (281.24, 79.40) -- (287.60, 79.40);

\path[draw=drawColor,line width= 0.4pt,line join=round,line cap=round] (284.42, 76.22) -- (284.42, 82.58);

\path[draw=drawColor,line width= 0.4pt,line join=round,line cap=round] (303.43, 81.05) -- (309.79, 81.05);

\path[draw=drawColor,line width= 0.4pt,line join=round,line cap=round] (306.61, 77.87) -- (306.61, 84.24);

\path[draw=drawColor,line width= 0.4pt,line join=round,line cap=round] (325.62, 82.71) -- (331.99, 82.71);

\path[draw=drawColor,line width= 0.4pt,line join=round,line cap=round] (328.80, 79.53) -- (328.80, 85.89);

\path[draw=drawColor,line width= 0.4pt,line join=round,line cap=round] (347.81, 84.36) -- (354.18, 84.36);

\path[draw=drawColor,line width= 0.4pt,line join=round,line cap=round] (350.99, 81.18) -- (350.99, 87.54);

\path[draw=drawColor,line width= 0.4pt,line join=round,line cap=round] (370.00, 86.84) -- (376.37, 86.84);

\path[draw=drawColor,line width= 0.4pt,line join=round,line cap=round] (373.18, 83.66) -- (373.18, 90.03);

\path[draw=drawColor,line width= 0.4pt,line join=round,line cap=round] (392.19, 76.92) -- (398.56, 76.92);

\path[draw=drawColor,line width= 0.4pt,line join=round,line cap=round] (395.38, 73.74) -- (395.38, 80.10);
\definecolor{drawColor}{RGB}{205,96,144}

\path[draw=drawColor,line width= 0.4pt,line join=round,line cap=round] ( 66.59,187.21) -- ( 80.63,202.39);

\path[draw=drawColor,line width= 0.4pt,line join=round,line cap=round] ( 89.86,209.86) -- (101.74,216.95);

\path[draw=drawColor,line width= 0.4pt,line join=round,line cap=round] (112.86,219.36) -- (123.12,218.21);

\path[draw=drawColor,line width= 0.4pt,line join=round,line cap=round] (134.94,218.85) -- (145.42,221.20);

\path[draw=drawColor,line width= 0.4pt,line join=round,line cap=round] (157.21,221.62) -- (167.53,220.08);

\path[draw=drawColor,line width= 0.4pt,line join=round,line cap=round] (179.40,220.08) -- (189.72,221.62);

\path[draw=drawColor,line width= 0.4pt,line join=round,line cap=round] (201.51,221.20) -- (211.99,218.85);

\path[draw=drawColor,line width= 0.4pt,line join=round,line cap=round] (223.54,219.45) -- (234.35,223.08);

\path[draw=drawColor,line width= 0.4pt,line join=round,line cap=round] (245.97,224.11) -- (256.30,222.57);

\path[draw=drawColor,line width= 0.4pt,line join=round,line cap=round] (268.19,222.35) -- (278.46,223.50);

\path[draw=drawColor,line width= 0.4pt,line join=round,line cap=round] (290.28,222.85) -- (300.76,220.51);

\path[draw=drawColor,line width= 0.4pt,line join=round,line cap=round] (312.47,220.51) -- (322.95,222.85);

\path[draw=drawColor,line width= 0.4pt,line join=round,line cap=round] (334.79,223.72) -- (345.01,222.95);

\path[draw=drawColor,line width= 0.4pt,line join=round,line cap=round] (356.89,223.61) -- (367.29,225.55);

\path[draw=drawColor,line width= 0.4pt,line join=round,line cap=round] (379.08,225.55) -- (389.48,223.61);

\path[draw=drawColor,line width= 0.4pt,line join=round,line cap=round] ( 60.26,180.55) rectangle ( 64.76,185.05);

\path[draw=drawColor,line width= 0.4pt,line join=round,line cap=round] ( 60.26,180.55) -- ( 64.76,185.05);

\path[draw=drawColor,line width= 0.4pt,line join=round,line cap=round] ( 60.26,185.05) -- ( 64.76,180.55);

\path[draw=drawColor,line width= 0.4pt,line join=round,line cap=round] ( 82.46,204.54) rectangle ( 86.96,209.04);

\path[draw=drawColor,line width= 0.4pt,line join=round,line cap=round] ( 82.46,204.54) -- ( 86.96,209.04);

\path[draw=drawColor,line width= 0.4pt,line join=round,line cap=round] ( 82.46,209.04) -- ( 86.96,204.54);

\path[draw=drawColor,line width= 0.4pt,line join=round,line cap=round] (104.65,217.78) rectangle (109.15,222.28);

\path[draw=drawColor,line width= 0.4pt,line join=round,line cap=round] (104.65,217.78) -- (109.15,222.28);

\path[draw=drawColor,line width= 0.4pt,line join=round,line cap=round] (104.65,222.28) -- (109.15,217.78);

\path[draw=drawColor,line width= 0.4pt,line join=round,line cap=round] (126.84,215.30) rectangle (131.34,219.80);

\path[draw=drawColor,line width= 0.4pt,line join=round,line cap=round] (126.84,215.30) -- (131.34,219.80);

\path[draw=drawColor,line width= 0.4pt,line join=round,line cap=round] (126.84,219.80) -- (131.34,215.30);

\path[draw=drawColor,line width= 0.4pt,line join=round,line cap=round] (149.03,220.26) rectangle (153.53,224.76);

\path[draw=drawColor,line width= 0.4pt,line join=round,line cap=round] (149.03,220.26) -- (153.53,224.76);

\path[draw=drawColor,line width= 0.4pt,line join=round,line cap=round] (149.03,224.76) -- (153.53,220.26);

\path[draw=drawColor,line width= 0.4pt,line join=round,line cap=round] (171.22,216.95) rectangle (175.72,221.45);

\path[draw=drawColor,line width= 0.4pt,line join=round,line cap=round] (171.22,216.95) -- (175.72,221.45);

\path[draw=drawColor,line width= 0.4pt,line join=round,line cap=round] (171.22,221.45) -- (175.72,216.95);

\path[draw=drawColor,line width= 0.4pt,line join=round,line cap=round] (193.41,220.26) rectangle (197.91,224.76);

\path[draw=drawColor,line width= 0.4pt,line join=round,line cap=round] (193.41,220.26) -- (197.91,224.76);

\path[draw=drawColor,line width= 0.4pt,line join=round,line cap=round] (193.41,224.76) -- (197.91,220.26);

\path[draw=drawColor,line width= 0.4pt,line join=round,line cap=round] (215.60,215.30) rectangle (220.10,219.80);

\path[draw=drawColor,line width= 0.4pt,line join=round,line cap=round] (215.60,215.30) -- (220.10,219.80);

\path[draw=drawColor,line width= 0.4pt,line join=round,line cap=round] (215.60,219.80) -- (220.10,215.30);

\path[draw=drawColor,line width= 0.4pt,line join=round,line cap=round] (237.79,222.74) rectangle (242.29,227.24);

\path[draw=drawColor,line width= 0.4pt,line join=round,line cap=round] (237.79,222.74) -- (242.29,227.24);

\path[draw=drawColor,line width= 0.4pt,line join=round,line cap=round] (237.79,227.24) -- (242.29,222.74);

\path[draw=drawColor,line width= 0.4pt,line join=round,line cap=round] (259.98,219.43) rectangle (264.48,223.93);

\path[draw=drawColor,line width= 0.4pt,line join=round,line cap=round] (259.98,219.43) -- (264.48,223.93);

\path[draw=drawColor,line width= 0.4pt,line join=round,line cap=round] (259.98,223.93) -- (264.48,219.43);

\path[draw=drawColor,line width= 0.4pt,line join=round,line cap=round] (282.17,221.91) rectangle (286.67,226.41);

\path[draw=drawColor,line width= 0.4pt,line join=round,line cap=round] (282.17,221.91) -- (286.67,226.41);

\path[draw=drawColor,line width= 0.4pt,line join=round,line cap=round] (282.17,226.41) -- (286.67,221.91);

\path[draw=drawColor,line width= 0.4pt,line join=round,line cap=round] (304.36,216.95) rectangle (308.86,221.45);

\path[draw=drawColor,line width= 0.4pt,line join=round,line cap=round] (304.36,216.95) -- (308.86,221.45);

\path[draw=drawColor,line width= 0.4pt,line join=round,line cap=round] (304.36,221.45) -- (308.86,216.95);

\path[draw=drawColor,line width= 0.4pt,line join=round,line cap=round] (326.55,221.91) rectangle (331.05,226.41);

\path[draw=drawColor,line width= 0.4pt,line join=round,line cap=round] (326.55,221.91) -- (331.05,226.41);

\path[draw=drawColor,line width= 0.4pt,line join=round,line cap=round] (326.55,226.41) -- (331.05,221.91);

\path[draw=drawColor,line width= 0.4pt,line join=round,line cap=round] (348.74,220.26) rectangle (353.24,224.76);

\path[draw=drawColor,line width= 0.4pt,line join=round,line cap=round] (348.74,220.26) -- (353.24,224.76);

\path[draw=drawColor,line width= 0.4pt,line join=round,line cap=round] (348.74,224.76) -- (353.24,220.26);

\path[draw=drawColor,line width= 0.4pt,line join=round,line cap=round] (370.93,224.39) rectangle (375.43,228.89);

\path[draw=drawColor,line width= 0.4pt,line join=round,line cap=round] (370.93,224.39) -- (375.43,228.89);

\path[draw=drawColor,line width= 0.4pt,line join=round,line cap=round] (370.93,228.89) -- (375.43,224.39);

\path[draw=drawColor,line width= 0.4pt,line join=round,line cap=round] (393.13,220.26) rectangle (397.63,224.76);

\path[draw=drawColor,line width= 0.4pt,line join=round,line cap=round] (393.13,220.26) -- (397.63,224.76);

\path[draw=drawColor,line width= 0.4pt,line join=round,line cap=round] (393.13,224.76) -- (397.63,220.26);
\definecolor{drawColor}{RGB}{0,0,0}

\path[draw=drawColor,line width= 0.4pt,dash pattern=on 4pt off 4pt ,line join=round,line cap=round] ( 62.29, 76.09) --
	( 65.66, 76.09) --
	( 69.03, 76.09) --
	( 72.39, 76.09) --
	( 75.76, 76.09) --
	( 79.13, 76.09) --
	( 82.49, 76.09) --
	( 85.86, 76.09) --
	( 89.23, 76.09) --
	( 92.59, 76.09) --
	( 95.96, 76.09) --
	( 99.33, 76.09) --
	(102.69, 76.09) --
	(106.06, 76.09) --
	(109.43, 76.09) --
	(112.79, 76.09) --
	(116.16, 76.09) --
	(119.53, 76.09) --
	(122.89, 76.09) --
	(126.26, 76.09) --
	(129.63, 76.09) --
	(132.99, 76.09) --
	(136.36, 76.09) --
	(139.73, 76.09) --
	(143.09, 76.09) --
	(146.46, 76.09) --
	(149.83, 76.09) --
	(153.19, 76.09) --
	(156.56, 76.09) --
	(159.93, 76.09) --
	(163.29, 76.09) --
	(166.66, 76.09) --
	(170.03, 76.09) --
	(173.39, 76.09) --
	(176.76, 76.09) --
	(180.13, 76.09) --
	(183.49, 76.09) --
	(186.86, 76.09) --
	(190.23, 76.09) --
	(193.59, 76.09) --
	(196.96, 76.09) --
	(200.33, 76.09) --
	(203.69, 76.09) --
	(207.06, 76.09) --
	(210.43, 76.09) --
	(213.79, 76.09) --
	(217.16, 76.09) --
	(220.53, 76.09) --
	(223.89, 76.09) --
	(227.26, 76.09) --
	(230.63, 76.09) --
	(234.00, 76.09) --
	(237.36, 76.09) --
	(240.73, 76.09) --
	(244.10, 76.09) --
	(247.46, 76.09) --
	(250.83, 76.09) --
	(254.20, 76.09) --
	(257.56, 76.09) --
	(260.93, 76.09) --
	(264.30, 76.09) --
	(267.66, 76.09) --
	(271.03, 76.09) --
	(274.40, 76.09) --
	(277.76, 76.09) --
	(281.13, 76.09) --
	(284.50, 76.09) --
	(287.86, 76.09) --
	(291.23, 76.09) --
	(294.60, 76.09) --
	(297.96, 76.09) --
	(301.33, 76.09) --
	(304.70, 76.09) --
	(308.06, 76.09) --
	(311.43, 76.09) --
	(314.80, 76.09) --
	(318.16, 76.09) --
	(321.53, 76.09) --
	(324.90, 76.09) --
	(328.26, 76.09) --
	(331.63, 76.09) --
	(335.00, 76.09) --
	(338.36, 76.09) --
	(341.73, 76.09) --
	(345.10, 76.09) --
	(348.46, 76.09) --
	(351.83, 76.09) --
	(355.20, 76.09) --
	(358.56, 76.09) --
	(361.93, 76.09) --
	(365.30, 76.09) --
	(368.66, 76.09) --
	(372.03, 76.09) --
	(375.40, 76.09) --
	(378.76, 76.09) --
	(382.13, 76.09) --
	(385.50, 76.09) --
	(388.86, 76.09) --
	(392.23, 76.09) --
	(395.60, 76.09);
\definecolor{drawColor}{RGB}{139,0,0}
\definecolor{fillColor}{RGB}{139,0,0}

\path[draw=drawColor,line width= 0.4pt,line join=round,line cap=round,fill=fillColor] (421.11,227.88) circle (  2.25);
\definecolor{fillColor}{RGB}{255,99,71}

\path[fill=fillColor] (421.11,219.38) --
	(424.15,214.13) --
	(418.08,214.13) --
	cycle;
\definecolor{fillColor}{RGB}{108,166,205}

\path[fill=fillColor] (418.86,201.63) --
	(423.36,201.63) --
	(423.36,206.13) --
	(418.86,206.13) --
	cycle;
\definecolor{drawColor}{RGB}{139,137,137}

\path[draw=drawColor,line width= 0.4pt,line join=round,line cap=round] (417.93,191.88) -- (424.30,191.88);

\path[draw=drawColor,line width= 0.4pt,line join=round,line cap=round] (421.11,188.70) -- (421.11,195.06);
\definecolor{drawColor}{RGB}{84,139,84}

\path[draw=drawColor,line width= 0.4pt,line join=round,line cap=round] (418.86,177.63) -- (423.36,182.13);

\path[draw=drawColor,line width= 0.4pt,line join=round,line cap=round] (418.86,182.13) -- (423.36,177.63);
\definecolor{drawColor}{RGB}{205,96,144}

\path[draw=drawColor,line width= 0.4pt,line join=round,line cap=round] (418.86,165.63) rectangle (423.36,170.13);

\path[draw=drawColor,line width= 0.4pt,line join=round,line cap=round] (418.86,165.63) -- (423.36,170.13);

\path[draw=drawColor,line width= 0.4pt,line join=round,line cap=round] (418.86,170.13) -- (423.36,165.63);
\definecolor{drawColor}{RGB}{0,0,0}

\node[text=drawColor,anchor=base west,inner sep=0pt, outer sep=0pt, scale=  1.00] at (430.11,224.44) {HypoRF};

\node[text=drawColor,anchor=base west,inner sep=0pt, outer sep=0pt, scale=  1.00] at (430.11,212.44) {Binomial};

\node[text=drawColor,anchor=base west,inner sep=0pt, outer sep=0pt, scale=  1.00] at (430.11,200.44) {ME-full};

\node[text=drawColor,anchor=base west,inner sep=0pt, outer sep=0pt, scale=  1.00] at (430.11,188.44) {MMDboot};

\node[text=drawColor,anchor=base west,inner sep=0pt, outer sep=0pt, scale=  1.00] at (430.11,176.44) {MMD-full};

\node[text=drawColor,anchor=base west,inner sep=0pt, outer sep=0pt, scale=  1.00] at (430.11,164.44) {CPT-RF};
\end{scope}
\end{tikzpicture}